\documentclass[12pt,a4paper]{JHEP}
\usepackage{epsfig}
\usepackage{array,cite,amsmath,amssymb,axodraw}
\newcommand{\newc}{\newcommand}
\newc{\inmath}[1] {\ifmmode#1\else$#1$\fi}
\newc{\definmath}[2] {\def#1{\ifmmode#2\else$#2$\fi}}
\newc{\gev}{\,GeV}
\newc{\mev}{\,MeV}
\newc{\ra}{\rightarrow}
\definmath{\rpv}{\mathrm{\not\!R_p}}
\definmath{\rp}{\mathrm{R_p}}
\newc{\real}{\mathcal{R}e}
\newc{\alsm}{{\displaystyle \sum_{\alpha=1,2}}}
\newc{\besm}{{\displaystyle \sum_{\beta=1,2}}}
\newc{\al}{\alpha}
\definmath{\lampp}{\lambda^{\prime \prime}}
\newc{\sgn}{\mr{sgn}\,}
\newc{\be}{\beta}
\newc{\ga}{\gamma}
\newc{\de}{\delta}
\newc{\sla}{\!\!\!\!\!\not\:\:\!}
\newc{\slab}{\!\!\!\!\!\not\,\,\,}
\newc{\pmn}{\phantom{-}}
\newc{\slac}{\!\!\!\!\!\!\!\not\,\,\,\,}
\newc{\met}{$\not\!\!E_T$}
\newc{\cw}{\cos\theta_W}
\newc{\sw}{\sin\theta_W}
\newc{\ssw}{\sin^2\theta_W}
\newc{\ccw}{\cos^2\theta_W}
\newc{\cbe}{\cos\beta}
\newc{\sbe}{\sin\beta}
\newc{\ort}{\frac1{\sqrt{2}}}
\newc{\sh}{\hat{s}}
\newc{\uh}{\hat{u}}
\newc{\tha}{\hat{t}}
\newc{\sa}{\sin\al}
\newc{\ca}{\cos\al}
\newc{\mz}{M_{\mr{Z}}}
\newc{\mw}{M_{\mr{W}}}
\definmath{\bv}{\mathrm{\not\!B}}
\definmath{\lv}{\mathrm{\not\!L}}
\newc{\beq}{\begin{equation}}
\newc{\eeq}{\end{equation}}
\newc{\ie}{{\it i.e.\/}\ }
\definmath{\lam}{\lambda}
\definmath{\cht}{\tilde{\chi}}
\definmath{\chgone}{\cht^+_1}
\definmath{\ntlone}{\cht^0_1}
\definmath{\ntltwo}{\cht^0_2}
\definmath{\sslr} {\tilde{l}_{R}}
\definmath{\glt}{\tilde{g}}
\definmath{\upt}{\tilde{u}}
\definmath{\qkt}{\tilde{q}}
\definmath{\elt}{\tilde{\ell}}
\definmath{\hgt}{\tilde{H}}
\definmath{\nut}{\tilde{\nu}}
\definmath{\dnt}{\tilde{d}}
\definmath{\ftl}{\mr{\tilde{f}}}
\definmath{\psb}{\bar{\psi}}
\definmath{\rtt}{\sqrt{2}}
\definmath{\mut}{\tilde{\mu}}
\newc{\mr}{\mathrm}
\newc{\bath}{\bar{\theta}}
\newc{\tht}{\theta}
\newc{\JC}{{\bf J}}
\newc{\lra}{\longrightarrow}
\newc{\eg}{{\it e.g.\  }}
\newc{\barr}{\begin{eqnarray}}
\newc{\earr}{\end{eqnarray}}
\newc{\me}{\mathcal{M}}
\definmath{\dbm}{\partial_\mu}
\definmath{\dbmu}{\stackrel{\leftrightarrow\  }{\partial^\mu}}
\definmath{\sgm}{\sigma_\mu}
\newc{\captionB}[2]{\caption[{#1}]{{\small {#2}}}}
\title{Spin Correlations in Monte Carlo Simulations}

\author{Peter Richardson\\
Cavendish Laboratory, University of Cambridge, Madingley Road,
        Cambridge, CB3\nolinebreak\  \nolinebreak{0HE,} UK, and\\
DAMTP, University of Cambridge, Centre for Mathematical Sciences, Wilberforce Road, Cambridge,
         \mbox{CB3~0WA,~UK}.
}
\abstract{
  We show that the algorithm originally proposed by Collins and Knowles for spin
  correlations in the QCD parton shower can be used in order to include spin
  correlations between the production and decay of heavy particles in  Monte Carlo event
  generators. This allows correlations to be included while maintaining the step-by-step
  approach of the Monte Carlo event generation process.

  We present examples of this approach for both the Standard and Minimal Supersymmetric
  Standard Models. A merger of this algorithm and that used in the parton shower is
  discussed in order to include all correlations in the perturbative phase of 
  event generation. Finally we present all the results needed to implement this algorithm
  for the Standard and Minimal Supersymmetric Standard Models.
}
\keywords{ Spin and Polarization Effects, Standard Model, Beyond Standard Model}
\preprint{Cavendish HEP-2001-13\\
	  DAMTP-2001-83}
\begin{document}
\section{Introduction}

  In modern particle physics experiments it is important to have a Monte Carlo
  simulation which accurately predicts the results of the experiment for
  both those processes which have already been observed and any new physics
  which may be discovered.
  As higher energies are probed this often means the production of 
  heavy particles which decay before hadronizing giving decay products
  which are detected, \eg the top quark. 

  As these particles are often fermions the distributions of the decay products
  are affected by correlations between the production and decay of the fermion.
  In most Monte Carlo event generators these correlations are neglected.
  In the Standard Model~(SM) this only applies to the top quark and tau lepton.
  However,  
  in most models of new physics, for example supersymmetry, there
  are heavy fermions which decay and correlations
  between the production and decay of these fermions may be important
  \cite{Moortgat-Pick:1997ny,Moortgat-Pick:1998sk,Moortgat-Pick:1999di,
	Moortgat-Pick:2000uz}.

  There have been a number of calculations of the spin correlation effects
  for specific processes, but these calculations
  all require the hard collision process and the decay of the heavy fermions
  to be performed in the same step which is problematic in Monte Carlo 
  event generators.

  There are however methods
  \cite{Collins:1988cp,Knowles:1988cu,Knowles:1988hu,Knowles:1988vs}
  which have previously been used in order to
  correlate all the partons produced in the QCD parton shower.
  In this paper we shall show how a very similar algorithm can be used in order to
  correlate the spins of all the heavy particle decays in the event while
  still maintaining both the step-by-step approach of the Monte Carlo event
  generator and an algorithm whose complexity grows only linearly with the
  number of particles.

  In the next section we will review the ideas of Monte Carlo simulations
  and discuss the current implementation of heavy particle decays in event
  generators. This is followed by a discussion of the algorithm we will
  use in Section~\ref{sect:algor}. We provide several examples of the
  results of this algorithm for both Standard Model 
  and supersymmetric~(SUSY) processes in Section~\ref{sect:egs}.
  We then discuss how to incorporate spin correlations in the decay of heavy
  particles as well as the spin correlations inside and between jets
  which are already included in some Monte Carlo event generators
  \cite{Corcella:2000bw,Corcella:2001pi} using
  the algorithm of 
  \cite{Collins:1988cp,Knowles:1988cu,Knowles:1988hu,Knowles:1988vs}.
   Finally, we present the necessary results in order to implement this algorithm
  for both Standard and Minimal Supersymmetric Standard Model processes.

%
%  Section on Monte Carlo
%
\section{Monte Carlo Event Generation} 
\label{sect:monte}

\begin{figure}
\begin{center} \begin{picture}(360,190)(0,40)
\SetOffset(85,-15)
%Hard Process
\ArrowLine(0,180)(30,150)
\ArrowLine(30,150)(0,120)
\Photon(30,150)(70,150){5}{5}
\ArrowLine(70,150)(100,180)
\ArrowLine(100,120)(70,150)
\Gluon(100,180)(130,180){-3}{3}
\Gluon(100,120)(160,120){ 3}{6}
\Gluon(130,180)(160,195){ 3}{3}
\Gluon(130,180)(160,165){-3}{3}
\ArrowLine(100,180)(130,210)
\ArrowLine(130,90)(100,120)
\Photon(130,210)(160,240){3}{3}
\ArrowLine(130,210)(160,210)
\Photon(130,90)(160,60){-3}{3}
\ArrowLine(160,90)(130,90)
\ArrowLine(160,60)(175,75)
\ArrowLine(175,45)(160,60)
\ArrowLine(160,210)(175,210)
\ArrowLine(175,205)(160,195)
\ArrowLine(160,195)(175,185)
\ArrowLine(175,175)(160,165)
\ArrowLine(160,165)(175,155)
\ArrowLine(175,135)(160,120)
\ArrowLine(160,120)(175,105)
\ArrowLine(175,90)(160,90)
\ArrowLine(160,240)(175,255)
\ArrowLine(175,225)(160,240)
\LongArrow(177,207.5)(200,215)
\LongArrow(177,207.5)(200,200)
\LongArrow(177,180)(200,187.7)
\LongArrow(177,180)(200,172.5)
\LongArrow(177,145)(200,152.5)
\LongArrow(177,145)(200,137.5)
\LongArrow(177,97.5)(200,105)
\LongArrow(177,97.5)(200,90)
\GOval(177,207.5)(5,3)(0){0.7}
\GOval(177,180)(7,3)(0){0.7}
\GOval(177,145)(15,3)(0){0.7}
\GOval(177,97.5)(10,3)(0){0.7}
%Labels
\Text(20,175)[]{$\mr{e^-}$}
\Text(20,130)[]{$\mr{e^+}$}
\Text(50,165)[]{$\mr{Z_0/\gamma}$}
\Text(85,175)[]{$\mr{t}$}
\Text(85,125)[]{$\mr{\bar{t}}$}
\Text(115,205)[]{$\mr{t}$}
\Text(115,95)[]{$\mr{\bar{t}}$}
\Text(155,213)[b]{$\mr{b}$}
\Text(155,93)[b]{$\mr{\bar{b}}$}
\Text(140,235)[]{$\mr{W}^+$}
\Text(140,65)[]{$\mr{W}^-$}
\Text(178,255)[l]{$\nu$}
\Text(178,225)[l]{$\ell^+$}
\Text(178,45)[l]{$\bar{\nu}$}
\Text(178,75)[l]{$\ell^-$}
\Text(202,153)[l]{$\left.\rule{0mm}{25mm}\right\}$ hadrons}
\end{picture}
\end{center}
\caption{Example of a Monte Carlo event. This example shows the production of
	$\mr{t\bar{t}}$ in $\mr{e}^+\mr{e}^-$ collisions followed by the
	semi-leptonic decay of the top quarks.
        The cluster hadronization model \cite{Webber:1984if} is shown
	where the gluons left after the parton-shower phase
	are non-perturbatively split into quark-antiquark pairs.
	The quarks and antiquarks are then paired into colour-singlet clusters
	using the colour flow information in the event. These clusters decay to give
	the observed hadrons.}
\label{fig:monteeg}
\end{figure}
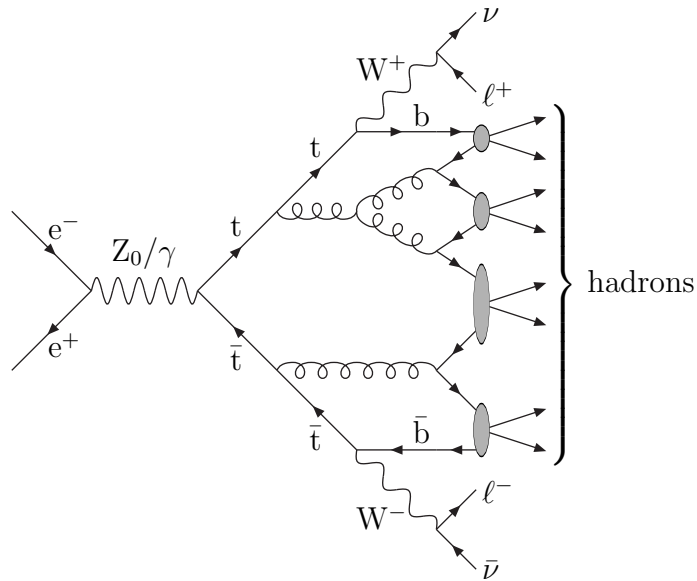

  There are a number of general purpose Monte Carlo event generators available
  \cite{Sjostrand:1994yb,Sjostrand:2000wi,Baer:1999sp,Corcella:2000bw,Corcella:2001pi}.
  The structure of the event generation procedure  is basically the same in
  all these programs. The differences are in the algorithms 
  used in the different steps of generating the event.
  In general the Monte Carlo event generation process can be divided
  into five phases:
\begin{enumerate}

  \item The hard process where the particles in the hard collision and
 	their momenta are generated, usually according to the leading-order
 	matrix element. This can be of either the incoming fundamental
   	particles in lepton collisions or of a parton extracted from a hadron
 	in hadron-initiated processes. In the example event shown in
	Fig.\,\ref{fig:monteeg} the hard process is $\mr{e^+e^-\ra t\bar{t}}$.

  \item The parton-shower phase where the coloured particles in the event are
  	perturbatively evolved from the hard scale of the collision to the
  	infrared cut-off. This is done for both the particles produced in
  	the collision, the final-state shower, and the initial partons
        involved in the collision for processes with incoming hadrons,
	the initial-state shower. In the example shown in 
        Fig.\,\ref{fig:monteeg} the top quarks radiate gluons and the gluons
        branch.
 
  \item Those particles which decay before hadronizing,
  	\eg the top quark and SUSY particles,
        are then decayed.
        Any coloured particles produced in
        these secondary decays are evolved by the parton-shower algorithm. 
        These decays 
        are usually performed according to a calculated branching ratio 
        and often use a matrix element to give the momenta of the decay products.  
        The example in Fig.\,\ref{fig:monteeg} shows the semi-leptonic
        decay of both top quarks.

  \item A hadronization phase in which the partons left after the perturbative
 	evolution are formed into the observed hadrons. For
 	processes with hadrons in the initial state after the removal of the
 	partons in the hard process, we are left with a hadron remnant.
        This remnant is also formed into hadrons by the hadronization model.
        In the example shown in Fig.\,\ref{fig:monteeg} the cluster model 
        \cite{Webber:1984if}, which is used in HERWIG 
	\cite{Corcella:2000bw,Corcella:2001pi}, is shown.

\item Those unstable hadrons which are produced in the hadronization phase
      must also
      be decayed. These decays are usually performed using the experimentally
      measured branching ratios and a phase-space
      distribution for the momenta of the decay products. Any coloured
      particles produced in these decays are evolved according to the 
      parton-shower algorithm and hadronized. This procedure
      is repeated until all the unstable particles have been decayed.
\end{enumerate}
%
% Monte Carlo Event generator picture
% 
%  This needs to be in black and white and probably be a lot better
%
%\FIGURE{

%
%  End of the figure
%
  In the Standard Model the only quark which decays before hadronizing is the
  top quark. It is therefore possible to include $\mr{t\bar{t}}$ production, including spin effects,
  as a $2\to6$ process, \ie including the decay of the top quarks, in a Monte Carlo
  event generator. However this leads to problems when including QCD radiation 
  from the top quark before it decays. 

  In all Monte Carlo event generators the particles are
  produced and then
  QCD radiation is generated via the parton-shower algorithm. Any unstable
  particles are then decayed. This is the main problem for the inclusion
  of spin correlations, \ie we want to be able generate the production and
  decay of the particles as separate processes rather than in one step
  so that QCD radiation can be generated for the particles before they decay.

  The problem is somewhat different for the inclusion of tau decays due to 
  the large number of tau decay modes. This makes it impossible
  to include all the possible decay channels as $2\to n$ body processes.

  The situation in SUSY models is similar to that for tau decays.
  In order to facilitate the experimental search for supersymmetry a number of
  Monte Carlo event generators have either been extended to include supersymmetric
  processes 
  \cite{Mrenna:1997hu,Sjostrand:1994yb,Sjostrand:2000wi,Baer:1999sp,SUSYimplement,Corcella:2000bw,Corcella:2001pi},
  or written specifically for the study of supersymmetry \cite{Katsanevas:1997fb,Ghodbane:1999va}.

  These event generators have become increasingly sophisticated in their treatment
  of supersymmetric processes. 
  In general, the decays of the supersymmetric particles produced in the hard
  collision process are assumed to take place independently. The
  event generators differ in how these decays are performed.
  While HERWIG \cite{SUSYimplement,Corcella:2000bw,Corcella:2001pi} continues
  to use  a phase-space distribution for
  the three-body decays of the supersymmetric particles ISAJET~\cite{Baer:1999sp}, 
  PYTHIA~\cite{Mrenna:1997hu,Sjostrand:1994yb,Sjostrand:2000wi}
   and
  SUSYGEN~\cite{Katsanevas:1997fb,Ghodbane:1999va}
  use the full three-body matrix element.

  A number of studies have also been performed in which spin
  correlation effects are included, and these have been shown to be important
  for a future linear collider
  \cite{Moortgat-Pick:1997ny,Moortgat-Pick:1998sk,Moortgat-Pick:1999di,
	Moortgat-Pick:2000uz}. Some spin correlation effects have been  
  included in SUSYGEN\cite{Katsanevas:1997fb,Ghodbane:1999va}. However in order to 
  accomplished this both the production process and decays must be generated
  at the same time which limits both the number of processes and final-state
  particles that can
  be simulated in this way.

  In general SUSY models, particularly in hadron collisions,  very complicated decay
  chains can occur and it is impossible to generate them all as $2\to n$  particle
  processes. We therefore need an algorithm which both preserves the step-by-step
  approach of the traditional Monte Carlo event generators and a complexity
  which does not grow exponentially with the number of final-state particles.
  
  In this paper we will show that a completely general algorithm can be implemented
  which takes all these effects into account while still maintaining the step-by-step
  approach of the Monte Carlo event generation process.
  The results from 
  the implementation of this algorithm in the HERWIG event generator are then
  compared to matrix element calculations from a number of observables. 

%
%  Section on Ian's algorithm
%
\section{Spin Correlation Algorithm}
\label{sect:algor}

  The spin correlation algorithm we will use is essentially identical to that presented
  in \cite{Collins:1988cp,Knowles:1988cu,Knowles:1988hu,Knowles:1988vs} for the
  case of spin correlations in the QCD parton-shower phase of the Monte Carlo
  event generation process. However, the algorithm presented in 
  \cite{Collins:1988cp,Knowles:1988cu,Knowles:1988hu,Knowles:1988vs} only 
  considers the case of $1\to2$ branchings, as these are all that occur
  in the QCD parton shower.

  The algorithm can be formulated entirely in terms of the matrix elements for 
  the hard process and the various decays, spin density and decay matrices.
  The algorithm is defined as follows:
\begin{enumerate}
\item The momenta of the particles in the hard process are generated according
      to the matrix element\footnote{Throughout this paper we will use the
	Einstein summation convention where repeated indices are summed over.}
\begin{equation}
	\rho^1_{\kappa_1\kappa'_1}\rho^2_{\kappa_2\kappa'_2}
        \mathcal{M}_{\kappa_1\kappa_2;\lam_1\ldots\lam_n}
	\mathcal{M}^*_{\kappa'_1\kappa'_2;\lam'_1\ldots\lam'_n}
        \prod_{i=1,n}D^i_{\lam_i\lam'_i},
\label{eqn:spin1}
\end{equation}
  where $\mathcal{M}_{\kappa_1,\kappa_2;\lam_1\ldots\lam_n}$
  is the matrix element for
  the $2\to n$ body process, $\kappa_i$ is the helicity of the $i$th incoming particle,
  $\lam_i$ is the helicity of the $i$th outgoing particle,
  $\rho^i_{\kappa_i\kappa'_i}$ is the spin density matrix
  for the $i$th incoming particle and $D^i_{\lam_i\lam'_i}$ is the decay matrix for
  the $i$th outgoing particle.

  In the initial stage of the algorithm the momenta of the particles involved in the 
  hard collision are generated according to Eqn.\,\ref{eqn:spin1} with 
  $D^i_{\lam_i\lam'_i}=\delta_{\lam_i\lam'_i}$ and  
  $\rho^i_{\kappa_i\kappa'_i}=\frac12\delta_{\kappa_i\kappa'_i}$
  for unpolarized incoming particles.
  For polarized incoming particles
\begin{equation}
	\rho_{\kappa\kappa'} = \left(\begin{array}{cc}
	\frac12(1+\mathcal{P}_3)& 0 \\
        0 & \frac12(1-\mathcal{P}_3)\end{array}\right),
\end{equation}
   where $\mathcal{P}_3$ is the component of the polarization vector parallel
   to the beam direction. For incoming antiparticles the sign of $\mathcal{P}_3$
   must be changed.
\item One of the outgoing particles is chosen at random and a spin density
      matrix
\begin{equation}
\rho_{\lam_j\lam'_j} = \frac1{N_\rho}
 \rho^1_{\kappa_1\kappa'_1}\rho^2_{\kappa_2\kappa'_2}
	\mathcal{M}_{\kappa_1\kappa_2;\lam_1\ldots\lam_j\ldots\lam_n}
	\mathcal{M}^*_{\kappa'_1\kappa'_2;\lam'_1\ldots\lam'_j\ldots\lam'_n}
	\prod_{i\neq j}D^i_{\lam_i\lam'_i},
\label{eqn:spin2}
\end{equation}
      calculated for the decay of this particle.
      The normalization
\begin{equation}
N_\rho= \rho^1_{\kappa_1\kappa'_1}\rho^2_{\kappa_2\kappa'_2}
	\mathcal{M}_{\kappa_1\kappa_2;\lam_1\ldots\lam_j\ldots\lam_n}
	\mathcal{M}^*_{\kappa'_1\kappa'_2;\lam'_1\ldots\lam_j\ldots\lam'_n}
	\prod_{i\neq j}D^i_{\lam_i\lam'_i},
\label{eqn:spin2b}
\end{equation}
 is chosen so that the trace of the spin density matrix is one.

\item The decay mode of this particle is  selected according to the branching ratios
      and the momenta of the particles produced in the $n$-body 
      decay generated according to
\begin{equation}
\rho_{\lam_0\lam'_0}\mathcal{M}_{\lam_0;\lam_1\ldots\lam_n}
	\mathcal{M}^*_{\lam'_0;\lam'_1\ldots\lam'_n}\prod_{i=1,n} D^i_{\lam_i\lam'_i},
\label{eqn:spin3}
\end{equation}
   where $\lam_0$ is the helicity of the decaying particle and $\lam_i$ is the
   helicity of the $i$th decay product.
   As before the decay matrices are taken to be 
   \mbox{$D^i_{\lam_i\lam'_i}=\delta_{\lam_i\lam'_i}$} for the initial step
   of the algorithm. 

\item One of the particles produced in this decay is selected and a 
      spin density matrix
\begin{equation}
\rho_{\lam_j\lam'_j} = \frac1{N_{D\rho}}\rho_{\lam_0\lam'_0}
		\mathcal{M}_{\lam_0;\lam_1\ldots\lam_j\ldots\lam_n}
		\mathcal{M}^*_{\lam'_0;\lam'_1\ldots\lam'_j\ldots\lam'_n}
	        \prod_{i\neq j} D^i_{\lam_i\lam'_i}
\label{eqn:spin3b}	
\end{equation}
      calculated. Again the normalization
\begin{equation}
N_{D\rho} = \rho_{\lam_0\lam'_0}\mathcal{M}_{\lam_0;\lam_1\ldots\lam_j\ldots\lam_n}
		\mathcal{M}^*_{\lam'_0;\lam'_1\ldots\lam_j\ldots\lam'_n}
	        \prod_{i\neq j} D^i_{\lam_i\lam'_i},
\label{eqn:spin3c}
\end{equation}
      is chosen such that the spin density matrix has unit trace.
      This spin density matrix is used as the input to the third
      step of the algorithm.

\item The third and fourth steps are repeated until the particle
      selected in step four is stable.
      When this occurs the decay matrix 
      for this stable particle is set to $D^i_{\lam_i\lam'_i}=\delta_{\lam_i\lam'_i}$
      and another particle from that decay selected to be decayed next
      using a spin density matrix for this particle calculated
      using Eqn.\,\ref{eqn:spin3b}.
      When all the particles produced in a given decay have been developed a
      decay matrix
\begin{equation}
    D_{\lam_0\lam'_0} = \frac1{N_D}\mathcal{M}_{\lam_0;\lam_1\ldots\lam_n}
	\mathcal{M}^*_{\lam'_0;\lam'_1\ldots\lam'_n}\prod_{i=1,n} D^i_{\lam_i\lam'_i},
\label{eqn:spin4}
\end{equation}
      is calculated for the decay. Again the normalization 
\begin{equation}
  N_D = \mathcal{M}_{\lam_0;\lam_1\ldots\lam_n}
	\mathcal{M}^*_{\lam_0;\lam'_1\ldots\lam'_n}\prod_{i=1,n} D^i_{\lam_i\lam'_i},
\end{equation}
      is chosen such that the trace of the decay matrix is one.     

\item A new particle is selected from the decay which produced the 
      decaying particle.
      This new particle has a spin density matrix given by Eqn.\,\ref{eqn:spin3b}
      using the decay matrix calculated from Eqn.\,\ref{eqn:spin4} 
      for the particles which have already been
      decayed rather than the identity. In this way the decay products of this
      particle will have the correct correlations with the decay products of the
      other particles produced in the same decay. This step is repeated
      until all the particles produced in a decay are developed.
      Eqn.\,\ref{eqn:spin4} is then used to compute the decay matrix for this decay
      and the previous decay in the chain is developed.
      
\item Eventually this will give a decay matrix for the particle produced in the hard
      process. At this point a new particle from the hard process
      is selected to be decayed
      with a spin density matrix given by Eqn.\,\ref{eqn:spin2} with
      the identity replaced by the calculated decay matrix for those
      particles which have already been decayed. In this way the
      correlations between the decays of the particles produced in the hard process
      are generated. This is repeated until all the particles produced in the	
      hard process have been decayed. 
\end{enumerate}

  At each point in the algorithm the normalization is chosen such that the trace of
  the spin density and decay matrices is one. This is necessary in order to maintain the
  probabilistic interpretation of the spin density matrices and proves to be
  usefully for the decay matrices.
   
%
%  Section contain some examples
%
\section{Examples}
\label{sect:egs}

  There are many quantities for which it is interesting to calculate the
  spin correlations. In this section we will merely show a few examples of the method
  we propose for including the spin correlations and comparisons with the
  full analytic result for these quantities. Due to the complexities in calculating these
  full results this will limit us to relatively simple observables,
  although in principle much more complicated quantities can be calculated.
  The quantities have been chosen to both illustrate the technique and
  be phenomenogically important. 

%
% squark decay chain
%
\subsection[${\rm\tilde{q}}_L\ra {\rm q} \tilde{\chi}^0_2 \ra {\rm q} \ell^\pm\tilde{\ell}_R^\mp$]
	   {\boldmath{${\mr\qkt}_L\ra {\rm q}\cht^0_2\ra {\rm q}\ell^\pm\elt_R^\mp$}}
\label{sect:eg1}
  In hadron-hadron collisions the coloured SUSY particles, \ie the squarks and gluinos,
  are preferentially produced and therefore the main source of electroweak
  gaugino and slepton production can be in the cascade decays of the coloured sparticles.

  In many models the decay chain 
  ${\mr\qkt}_L\ra {\rm q}\cht^0_2\ra {\rm q}\ell^\pm\elt_R^\mp\ra{\rm q}
	\ell^+\ell^-\cht^0_1$, Fig.\,\ref{fig:squarkdecay}, 
  is important because by studying edges in the
  $\ell^+\ell^-$, ${\rm q}\ell^\pm$ and ${\rm q}\ell^+\ell^-$ mass
  distributions the masses of the lightest two neutralinos, squark and
  slepton can be reconstructed 
  \cite{Hinchliffe:1997iu,atlastdr,Branson:2001pj,Allanach:2000kt,Hinchliffe:2001bm}.

\begin{figure}
\begin{center}
\begin{picture}(300,100)
\DashArrowLine(0,70)(30,70){5}
\ArrowLine(30,70)(60,100)
\ArrowLine(60,40)(30,70)
\ArrowLine(90,70)(60,40)
\DashArrowLine(60,40)(90,10){5}
\Text(-2,70)[r]{${\rm\qkt}_L$}
\Text(62,100)[l]{$\rm{q}$}
\Text(43,53)[rt]{$\cht^0_2$}
\Text(92,70)[l]{$\ell^+$}
\Text(92,10)[l]{$\elt^-$}
\SetOffset(200,0)
\DashArrowLine(0,70)(30,70){5}
\ArrowLine(30,70)(60,100)
\ArrowLine(60,40)(30,70)
\ArrowLine(60,40)(90,70)
\DashArrowLine(90,10)(60,40){5}
\Text(-2,70)[r]{${\rm\qkt}_L$}
\Text(62,100)[l]{$\rm{q}$}
\Text(43,53)[rt]{$\cht^0_2$}
\Text(92,70)[l]{$\ell^-$}
\Text(92,10)[l]{$\elt^+$}
\end{picture}
\end{center}
\caption{Feynman Diagrams for the decay
         ${\rm \qkt}_L\ra{\rm q} \cht^0_2\ra{\rm q}\ell^\pm\elt^\mp_R$.}
\label{fig:squarkdecay}
\end{figure}

  However, if we consider the spins of the particles involved the quark produced in the
  squark decay will be left-handed in the massless limit. 
  The lepton produced in the neutralino decay will be right-handed.
  This will lead to very different decay distributions for the charge conjugate
  decay modes of the neutralino. 

  This difference is most noticeable in distribution of the mass of the quark and
  the lepton produced in 
  the neutralino decay, Fig.\,\ref{fig:squarkegh}.
  The results in Fig.\,\ref{fig:squarkegh} were generated at
  SUGRA point 5 \cite{atlastdr,Abdullin:1998pm},
  \ie $M_0=200\,\rm{\gev}$, $M_{1/2}=300\,\rm{\gev}$, $A_0=300\,\rm{\gev}$,
  $\tan\beta=2.1$ and $\sgn\mu+$,
  where $M_0$ is the universal SUSY breaking scalar mass at the Grand Unified 
  Theory~(GUT) scale,
  $M_{1/2}$ is the universal SUSY breaking gaugino mass at the GUT scale,
  $A_0$ is the universal tri-linear soft SUSY breaking parameter at the GUT
  scale, $\sgn\mu$ is the sign of the $\mu$ parameter and $\tan\beta$ is the
  ratio of the two Higgs vacuum expectation values.
  At this SUGRA point the squark mass 
  $M_{{\rm\qkt}_L}=833.8\,\rm{\gev}$, the lightest neutralino mass 
  $M_{\cht^0_1}=115.5\,\rm{\gev}$ and the next-to-lightest neutralino mass
  $M_{\cht^0_2}=213.2\,\rm{\gev}$. The SUSY spectrum was generated using
  ISAJET7.51 \cite{Baer:1999sp}.

  The differences in the shapes of the two distributions in Fig.\,\ref{fig:squarkegh}
  can be understood
  by considering the helicities of the particles, the quark is left-handed
  while the produced lepton/antilepton is right-handed.
  In the case of an antilepton this
  means that if the antilepton and the quark are back-to-back they have no net spin
  and therefore this configuration which gives the edge in the mass distribution
  shown in Fig.\,\ref{fig:squarkegh}a
  is favoured. However, if a lepton is produced this configuration has net
  spin one and therefore cannot be produced in the decay of a scalar so
  the distribution in Fig.\,\ref{fig:squarkegh}b vanishes at the end point.

  Unfortunately in any experiment there is no way to distinguish between a quark and
  an antiquark. Hence the average of the distributions in
  Fig.\,\ref{fig:squarkegh} will be observed. This average is in good agreement
  with the phase-space distribution.

%
%    Squark mass distributions
%
\FIGURE[t]
{
\includegraphics[width=0.62\textwidth,angle=90]{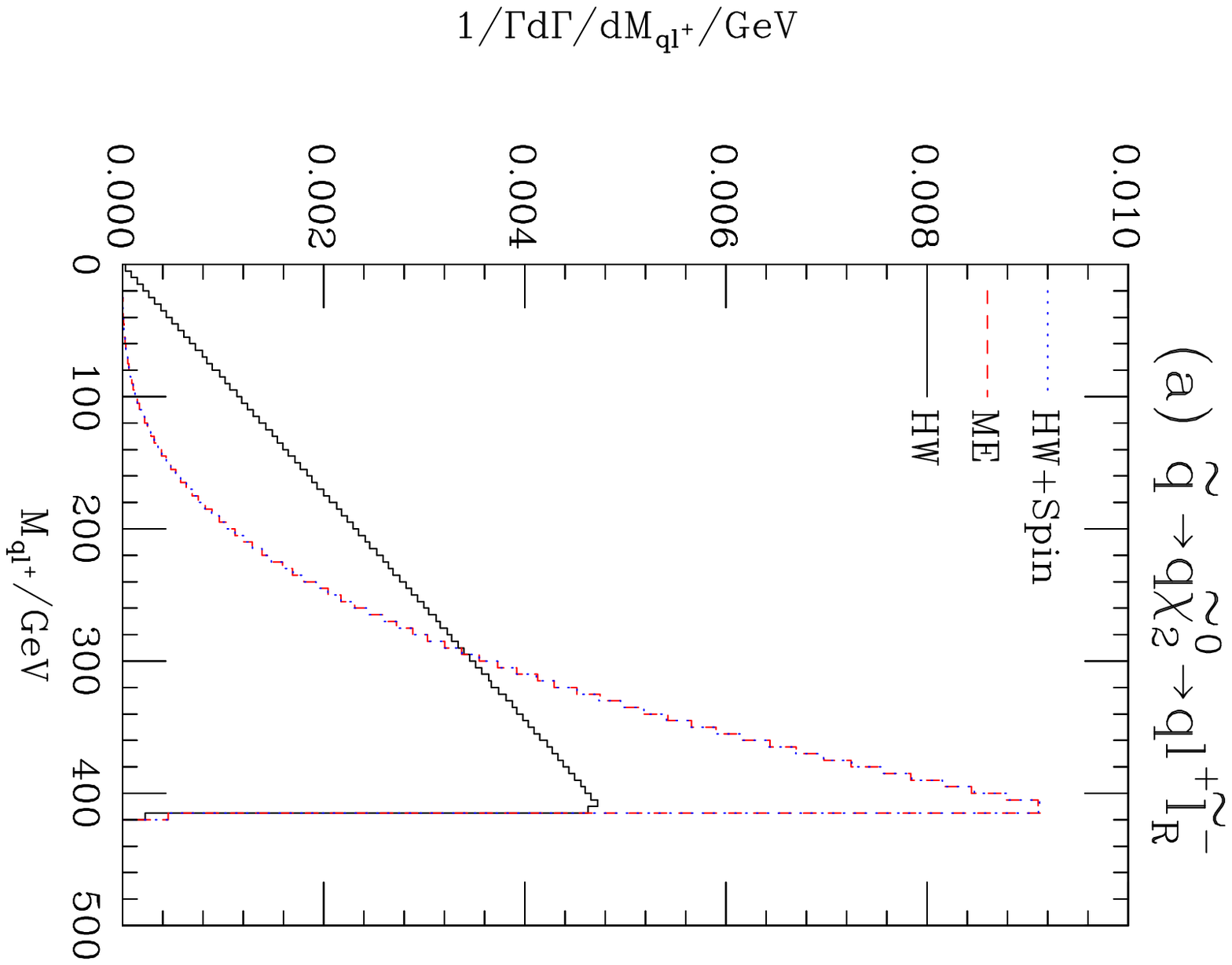}\hfill
\includegraphics[width=0.62\textwidth,angle=90]{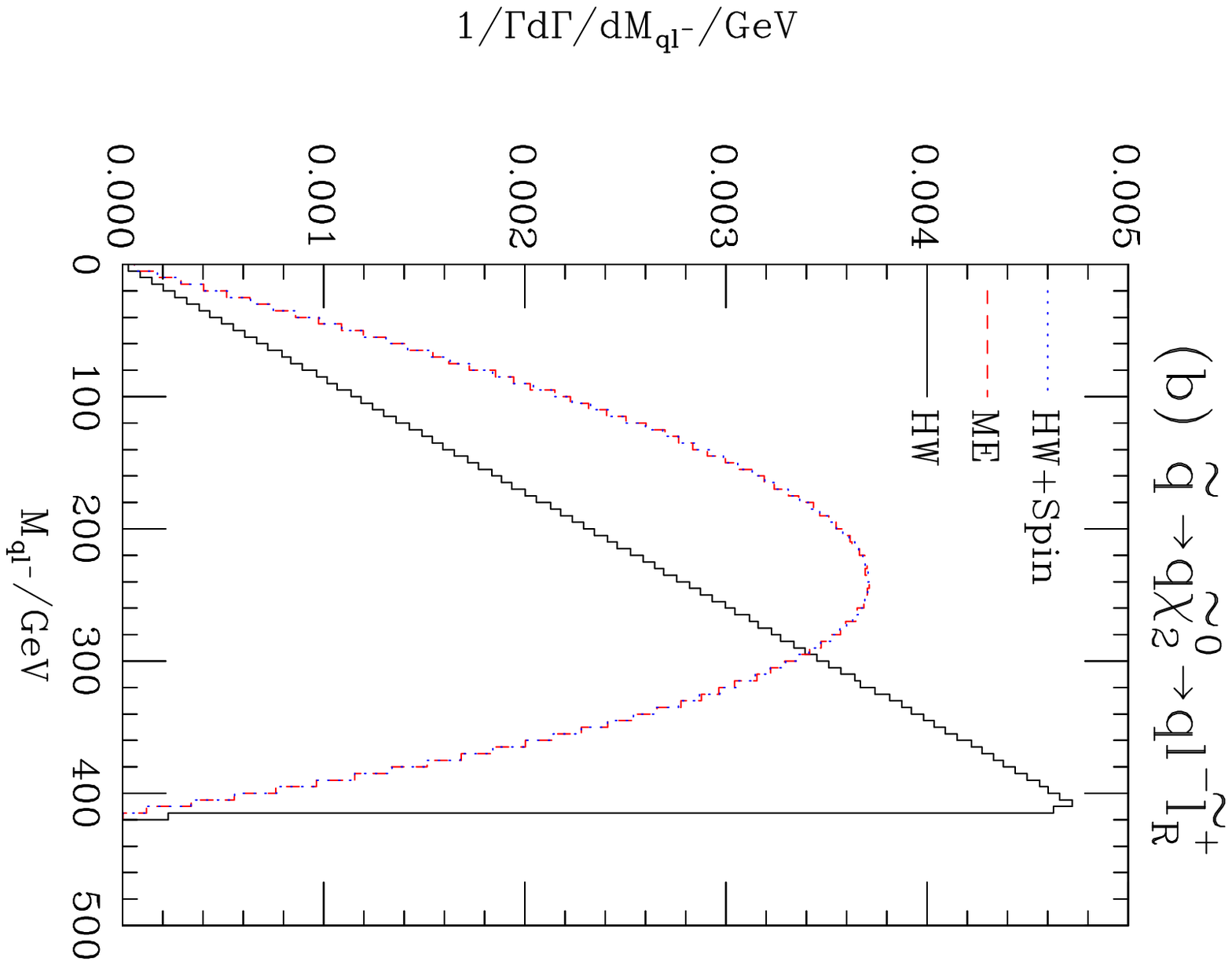}
\caption{Distribution of the mass of quark and lepton produced in the
decay \mbox{${\rm \qkt}_L\ra{\rm q}\cht^0_2\ra{\rm q}\ell^\pm\elt^\mp_R$}.
The solid line gives the result of phase space, the dashed line gives the
full result and the dotted line the result of the spin correlation algorithm.}
\label{fig:squarkegh}
}
% end of figure

  This is the simplest, non-trivial, application of the spin correlation algorithm. 
  Here all that is necessary is that the spin density matrix for the production of
  the neutralino in the scalar decay is used to perform the decay of the neutralino
  to a scalar and a fermion. The matrix elements are therefore relatively simple.
  Indeed the matrix elements are simple enough to allow us to compare the results of the
  algorithm and the full result analytically.
  The matrix element for the 3-body process 
  ${\rm\qkt}_L\ra{\rm q}\cht^0_2\ra {\rm q}\ell^+\elt_R^-$ is given by
\begin{eqnarray}
\mathcal{M}_{{\rm\qkt}_L\ra{\rm q}\cht^0_2\ra {\rm q}\ell^+\elt_R^-} &=&
	 \bar{u}(p_{{\rm q}})a^\lam P_{\lam}\left(p\sla_\chi-M_\chi\right)
		b^{\lam'}P_{\lam'}v(p_\ell),
\end{eqnarray}
  where $p_{\cht}$ is the four-momentum of the neutralino, $p_{{\rm q}}$ is the 
  four-momentum of the quark, $p_\ell$ is the four-momentum of the produced lepton,
  $P_\lam=\frac12(1+\lam\gamma_5)$ ,
  $a^\lam$ and $b^\lam$ are the couplings for the production and decay of the
  neutralino, respectively. In the notation used in Appendix~\ref{app:couplings},
  $a^\lam=a^{*-\lam}_{\cht^0_2{\rm\qkt}_{i1}{\rm q}_i}$ and 
  $b^\lam=a^{\lam}_{\cht^0_2\elt_{i2}\ell_i}$ where in the limit that we neglect
  left/right sfermion mixing $Q^i_{\al\be}=\delta_{\al,\be=1}$ for left squarks
   and $L^i_{\al\be}=\delta_{\al,\be=2}$ for right sleptons.  
  As the width of the neutralino is small compared to its mass we have
  assumed that it is on-mass-shell.

  We can now consider how the spin correlation algorithm attempts 
  to reproduce this result.
  The matrix element for the first step of the process is
\begin{subequations}
\begin{eqnarray}
\mathcal{M}_{{\rm\qkt}_L\ra{\rm q}\cht^0_2} &=& \bar{u}(p_{{\rm q}})a^\lam P_{\lam}v(p_{\cht}),\\
            &=&-\frac1{2\sqrt{p_{\cht} \cdot l_{\cht} p_{{\rm q}} \cdot l_{{\rm q}}}}
		\bar{u}_{-\lam_{{\rm q}}}(l_{{\rm q}})\left(p\sla_{{\rm q}}a^{\lam_{{\rm q}}}+m_{{\rm q}}a^{-\lam_{{\rm q}}}\right)
		(p_{\cht}-m_{\cht})u_{\lam_{\cht}}(l_{\cht}),\ \ \ 
\end{eqnarray}
\end{subequations}
  where $\lam_{\cht}$ is the helicity of the neutralino,
        $\lam_{{\rm q}}$ is the helicity of the quark, 
  $l_{{\rm q}}$ and $l_{\cht}$ are the reference vectors for the quark and neutralino,
  respectively.
  The use of these reference vectors in defining the spinors
  for massive fermions is described in Appendix~\ref{app:spin}.
  The sign of the helicity of the neutralino has be chosen in order
  for the outgoing neutralino to be considered as a particle rather than an antiparticle.

  In the first stage of the algorithm the momenta of the neutralino and quark are
  generated according to the spin averaged matrix element
\begin{equation}
\sum_{\rm{spins}} |\mathcal{M}_{{\rm\qkt}_L\ra{\rm q}\cht^0_2}|^2 = 2\left[
	p_{{\rm q}}\cdot p_{\cht}\left({a^+}^2+{a^-}^2\right)
       -2a^+a^-m_{{\rm q}} m_{\cht}\right].
\label{eqn:squarkaver}
\end{equation}
  As the quark is stable we can average over its spin and produce
  the spin density matrix needed to perform the decay of the neutralino,
\begin{eqnarray}
\rho_{\lam_{\cht}\lam'_{\cht}} &=& \frac1N\frac1{2p_{\cht}\cdot l_{\cht}}
   \bar{u}_{\lam'_{\cht}}(l_{\cht})(p\sla_{\cht}-m_{\cht})\\
&&
	\left[\left(p\sla_{{\rm q}}{a^-}^2+m_{{\rm q}}a^+a^-\right)P_+
             +\left(p\sla_{{\rm q}}{a^+}^2+m_{{\rm q}}a^+a^-\right)P_-\right]
(p\sla_{\cht}-m_{\cht})u_{\lam_{\cht}}(l_{\cht}),\nonumber
\end{eqnarray}
  where the normalization $N$ is chosen to give $\sum_{\lam}\rho_{\lam\lam}=1$
  and is therefore equal to the spin averaged matrix element,
  Eqn.\,\ref{eqn:squarkaver}.

  The matrix element for the decay of the neutralino is given by
\begin{subequations}
\begin{eqnarray}
\mathcal{M}^{\lam_{\cht}}_{\cht^0_2\ra\ell^+\elt_R^-}
 &=& \bar{u}(p_{\cht})b^\lam P_\lam v(p_\ell),\\
       &=& -\frac1{2\sqrt{p_{\cht}\cdot l_{\cht} p_\ell\cdot l_\ell}}
		\bar{u}_{\lam_{\cht}}(l_{\cht})
	\left(p\sla_{\cht}-m_{\cht}\right)
	\left[p\sla_\ell b^{\lam_\ell}-m_{\ell}b^{-\lam_\ell}\right]
	u_{\lam_{\ell}}(l_{\ell}),\ \ \ \ 
\end{eqnarray}
\end{subequations}
  where $p_\ell$ is the four-momentum of the lepton, $l_\ell$ is the reference
  vector used to define the direction of the lepton's spin and $\lam_\ell$
  is the helicity of the lepton.

  As with the quark because the produced lepton is stable we can average over
  its helicities giving
\begin{eqnarray}
\mathcal{M}_{\cht^0_2\ra\ell^+\elt_R^-}^{ \lam_{\cht}}
\mathcal{M}_{\cht^0_2\ra\ell^+\elt_R^-}^{\lam'_{\cht}*}&=&
	\frac1{2p_{\cht}\cdot l_{\cht}} \bar{u}_{\lam_{\cht}}(l_{\cht})
	\left(p\sla_{\cht}-m_{\cht}\right)\\&&
	\left[\left(p\sla_\ell{b^+}^2-m_{\ell}b^+b^-\right)P_-
	+\left(p\sla_\ell{b^-}^2-m_{\ell}b^+b^-\right)P_+\right]\nonumber\\&&
        \left(p\sla_{\cht}-m_{\cht}\right)u_{\lam'_{\cht}}(l_{\cht}).\nonumber	
\end{eqnarray}
  This can now be contracted with the spin density matrix and the
  sum over
  the helicities of the neutralino performed giving
\begin{eqnarray}
\rho_{\lam_{\cht}\lam'_{\cht}}
\lefteqn{\mathcal{M}_{\cht^0_2\ra\ell^+\elt_R^-}^{\lam_{\cht}}
\mathcal{M}_{\cht^0_2\ra\ell^+\elt_R^-}^{\lam'_{\cht}*}=}&\\
&&\frac1N{\rm Tr}\,\left\{\left(p\sla_{\cht}-m_{\cht}\right)
 \left[\left(p\sla_{{\rm q}}{a^-}^2+m_{{\rm q}}a^+a^-\right)P_+
      +\left(p\sla_{{\rm q}}{a^+}^2+m_{{\rm q}}a^+a^-\right)P_-\right]\right.
\nonumber\\&&
\ \ \ \ \ \ \, \ \left.\left(p\sla_{\cht}-m_{\cht}\right)\left[\left(p\sla_\ell{b^+}^2-m_{\ell}b^+b^-\right)P_-
     +\left(p\sla_\ell{b^-}^2-m_{\ell}b^+b^-\right)P_+\right]\right\}.\nonumber
\end{eqnarray}
  The second decay is generated according to this formula. The
  normalization of the spin density matrix is cancelled by the matrix
  element which is used to generate the first decay.
  If we compare the above result with the matrix element for the full
  3-body process we can see that it agrees with the amplitude
  squared. Hence, the decay products will have the same distribution as the full
  three-body matrix element. This can be seen in Fig.\,\ref{fig:squarkegh} where
  the result of the full 3-body matrix element and the spin correlation algorithm are in 
  very good agreement.

%
% Neutralino 2 neutralino 1 production
%
\subsection[$\rm{e^+e^-}\ra\tilde{\chi}^0_2\tilde{\chi}^0_1$]
	{\boldmath{$\rm{e^+e^-}\ra\tilde{\chi}^0_2\tilde{\chi}^0_1$}}
%
%  Neutralino production diagrams in e+e-
%
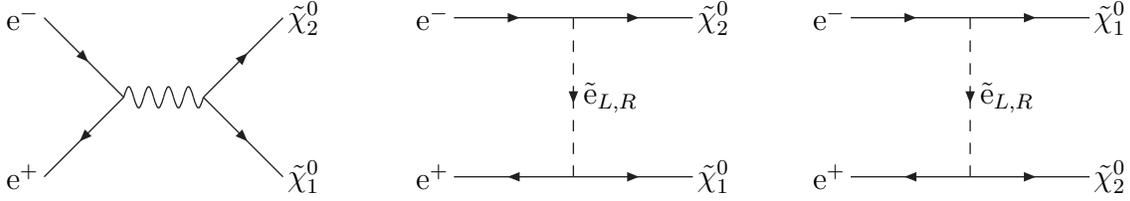
\begin{figure}
\begin{center}
\begin{picture}(300,60)
%
%   First Feynman diagram
%
\SetOffset(-45,0)
\ArrowLine(0,60)(30,30)
\ArrowLine(30,30)(0,0)
\Photon(30,30)(60,30){4}{4}
\ArrowLine(60,30)(90,60)
\ArrowLine(60,30)(90,0)
\Text(-2,60)[r]{$\rm{e}^-$}
\Text(-2, 0)[r]{$\rm{e}^+$}
\Text(92,60)[l]{$\cht^0_2$}
\Text(92, 0)[l]{$\cht^0_1$}
\SetOffset(110,0)
\ArrowLine(0,60)(45,60)
\ArrowLine(45,60)(90,60)
\DashArrowLine(45,60)(45,0){5}
\ArrowLine(45,0)(0,0)
\ArrowLine(45,0)(90,0)
\Text(-2,60)[r]{$\rm{e}^-$}
\Text(-2, 0)[r]{$\rm{e}^+$}
\Text(49,30)[l]{$\tilde{\rm{e}}_{L,R}$}
\Text(92,60)[l]{$\cht^0_2$}
\Text(92, 0)[l]{$\cht^0_1$}
\SetOffset(260,0)
\ArrowLine(0,60)(45,60)
\ArrowLine(45,60)(90,60)
\DashArrowLine(45,60)(45,0){5}
\ArrowLine(45,0)(0,0)
\ArrowLine(45,0)(90,0)
\Text(-2,60)[r]{$\rm{e}^-$}
\Text(-2, 0)[r]{$\rm{e}^+$}
\Text(49,30)[l]{$\tilde{\rm{e}}_{L,R}$}
\Text(92,60)[l]{$\cht^0_1$}
\Text(92, 0)[l]{$\cht^0_2$}
\end{picture}
\end{center}
\caption{Feynman diagrams for the production of $\cht^0_2\cht^0_1$ in
         $\rm{e}^+\rm{e}^-$ collisions.}
\label{fig:cht21prod}
\end{figure}
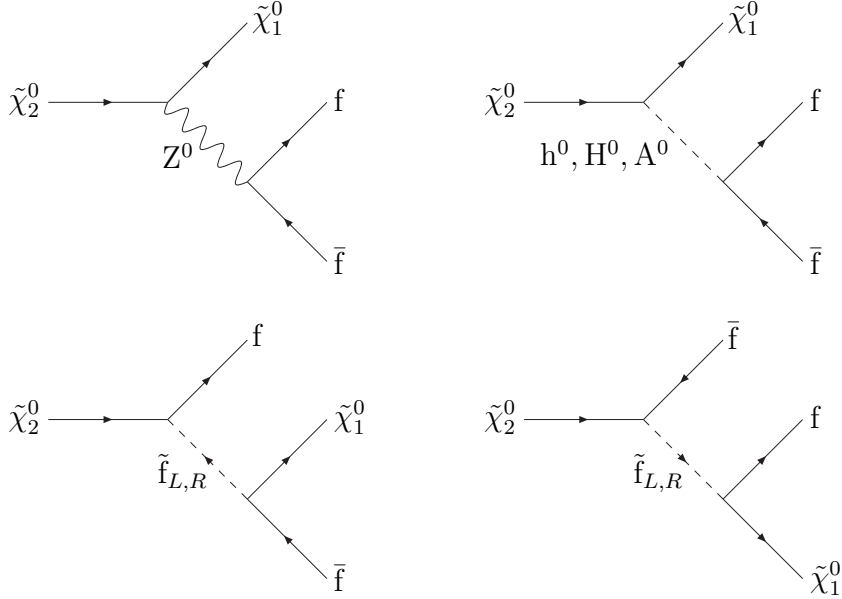
\begin{figure}
\begin{center}
\begin{picture}(300,200)
% first the Z exchage diagram
\SetOffset(20,120)
\SetScale{0.75}
\ArrowLine(-20,60)(40,60)
\ArrowLine(40,60)(80,100)
\Photon(40,60)(80,20){-5}{4.5}
\ArrowLine(80,20)(120,60)
\ArrowLine(120,-20)(80,20)
\Text(-18,46)[r]{$\cht^0_2$}
\Text(62,77)[l]{$\cht^0_1$}
\Text(93,46)[l]{$\rm{f}$}
\Text(93,-15)[l]{$\bar{\rm{f}}$}
\Text(40,25)[r]{$\rm{Z}^0$}
% sfermion exchange diagram
\SetOffset(20,0)
\SetScale{0.75}
\ArrowLine(-20,60)(40,60)
\ArrowLine(40,60)(80,100)
\DashArrowLine(80,20)(40,60){4.5}
\ArrowLine(80,20)(120,60)
\ArrowLine(120,-20)(80,20)
\Text(-18,46)[r]{$\cht^0_2$}
\Text(62,77)[l]{$\rm{f}$}
\Text(93,46)[l]{$\cht^0_1$}
\Text(93,-15)[l]{$\bar{\rm{f}}$}
\Text(45,25)[r]{$\ftl_{L,R}$}
% Higgs exchange diagram
\SetOffset(200,120)
\SetScale{0.75}
\ArrowLine(-20,60)(40,60)
\ArrowLine(40,60)(80,100)
\DashLine(40,60)(80,20){4.5}
\ArrowLine(80,20)(120,60)
\ArrowLine(120,-20)(80,20)
\Text(-18,46)[r]{$\cht^0_2$}
\Text(62,77)[l]{$\cht^0_1$}
\Text(93,46)[l]{$\rm{f}$}
\Text(93,-15)[l]{$\bar{\rm{f}}$}
\Text(40,25)[r]{$\rm{h}^0,\rm{H}^0,\rm{A}^0$}
% sfermion exchange diagram
\SetOffset(200,0)
\SetScale{0.75}
\ArrowLine(-20,60)(40,60)
\ArrowLine(80,100)(40,60)
\DashArrowLine(40,60)(80,20){4.5}
\ArrowLine(80,20)(120,60)
\ArrowLine(80,20)(120,-20)
\Text(-18,46)[r]{$\cht^0_2$}
\Text(62,77)[l]{$\bar{\rm{f}}$}
\Text(93,46)[l]{$\rm{f}$}
\Text(93,-15)[l]{$\cht^0_1$}
\Text(45,25)[r]{$\ftl_{L,R}$}
\end{picture}
\end{center}
\caption{Feynman diagrams for the decay $\cht^0_2\ra\rm{f}\bar{\rm{f}}\cht^0_1$.
	 The exchange of the MSSM Higgs bosons is only important 
	 for the third generation of fermions.}
\label{fig:cht21decay}
\end{figure}

%
%  end of figure
%
  In the MSSM the lightest supersymmetric particle~(LSP), usually taken to be
  the lightest neutralino, is stable and weakly interacting.
  It therefore escapes from the detector without interacting giving missing transverse energy.
  This means that in a future linear collider the production of $\cht^0_2\cht^0_1$,
  Fig.\,\ref{fig:cht21prod}, may
  well be the detectable supersymmetric final state which requires the smallest centre-of-mass energy.
  The production of the second-to-lightest-neutralino will be followed by
  its decay to the lightest neutralino. This decay will be 
  \mbox{$\cht^0_2\ra\cht^0_1\rm{f}\rm{\bar{f}}$} via either real or virtual
  sfermions, Higgs or Z bosons, Fig.\,\ref{fig:cht21decay}.

  There have been a number of studies of spin correlations for this process
  both with \cite{Moortgat-Pick:1999di,Moortgat-Pick:2000uz} and without
  \cite{Moortgat-Pick:1997ny} beam polarization.
  The study of both the decay correlations and polarization effects 
  is important because it would allow the nature of the neutralinos to be determined.

  This is a more complicated example of the use of the spin correlation algorithm than
  the squark decay we studied in the previous section.
  Instead of being produced in a scalar decay the neutralino is produced in a $2\to2$
  process where there are a number of Feynman diagrams and 
  the possibility of polarizing the incoming particles.
  However, as before the spin correlation algorithm only has to compute the
  spin density matrix for the production and use this to generate the decay of the  
  neutralino which may be via either a two or three body process, 
  Fig.\,\ref{fig:cht21decay}.

  The easiest case to consider first is the decay of the neutralino to a lepton
  and a right-handed slepton, as in Section~\ref{sect:eg1}, again
  using SUGRA Point~5. As can be seen in Fig.\,\ref{fig:chitwobeam} there is
  a correlation between the direction of the produced lepton and the beam direction
  and this correlation is affected by the polarization of the beam.
  It should be noted that fully polarized beams
  are not achievable in practice and are only shown in order to illustrate the
  results of the algorithm. As with the results in Section~\ref{sect:eg1} the result
  of the spin correlation algorithm is in good agreement with the result from
  a 4-body matrix element including the decay of both the neutralino and slepton.

%
%   Neutralino Decay diagrams  (there's four of these so in two blocks of two)
%
% end of figure

  It is also possible that neither the gauge boson or sleptons in the diagrams
  in Fig.\,\ref{fig:cht21decay} can be real. 
  As an example we considered a SUGRA point with non-universal gaugino masses
  at the GUT scale in order to decrease the mass difference between the lightest
  two neutralinos. We used the point
  $M_{0}=210\,\rm{\gev}$, $A_0=0\,\rm{\gev}$, $\tan\beta=10$, $M_1=450\,\rm{\gev}$,
  $M_2=350\,\rm{\gev}$, $M_3=350\,\rm{\gev}$, where $M_1$ is the soft SUSY
  breaking mass for the bino at the GUT scale,
  $M_2$ is the soft SUSY breaking mass for the wino at the GUT scale and
  $M_3$ is the soft SUSY breaking mass for the gluino at the GUT scale.
  At this point the lightest neutralino is dominantly bino-like and the
  next-to-lightest neutralino is dominantly wino-like,
  the lightest neutralino mass is $M_{\cht^0_1}=182.4\,\rm{\gev}$,
  the next-to-lightest neutralino mass is $M_{\cht^0_2}=264.2\,\rm{\gev}$,
  the right-slepton mass is $M_{\elt_R}=271.3\,\rm{\gev}$ and
  the left-slepton mass is $M_{\elt_L}=322.5\,\rm{\gev}$.
  This point was chosen so that both the right-handed slepton and the 
  Z boson could be almost on mass-shell in the three body decay
  $\cht^0_2\ra\cht^0_1\ell^+\ell^-$.

\FIGURE[t]
{  
\includegraphics[width=0.45\textwidth,angle=90]{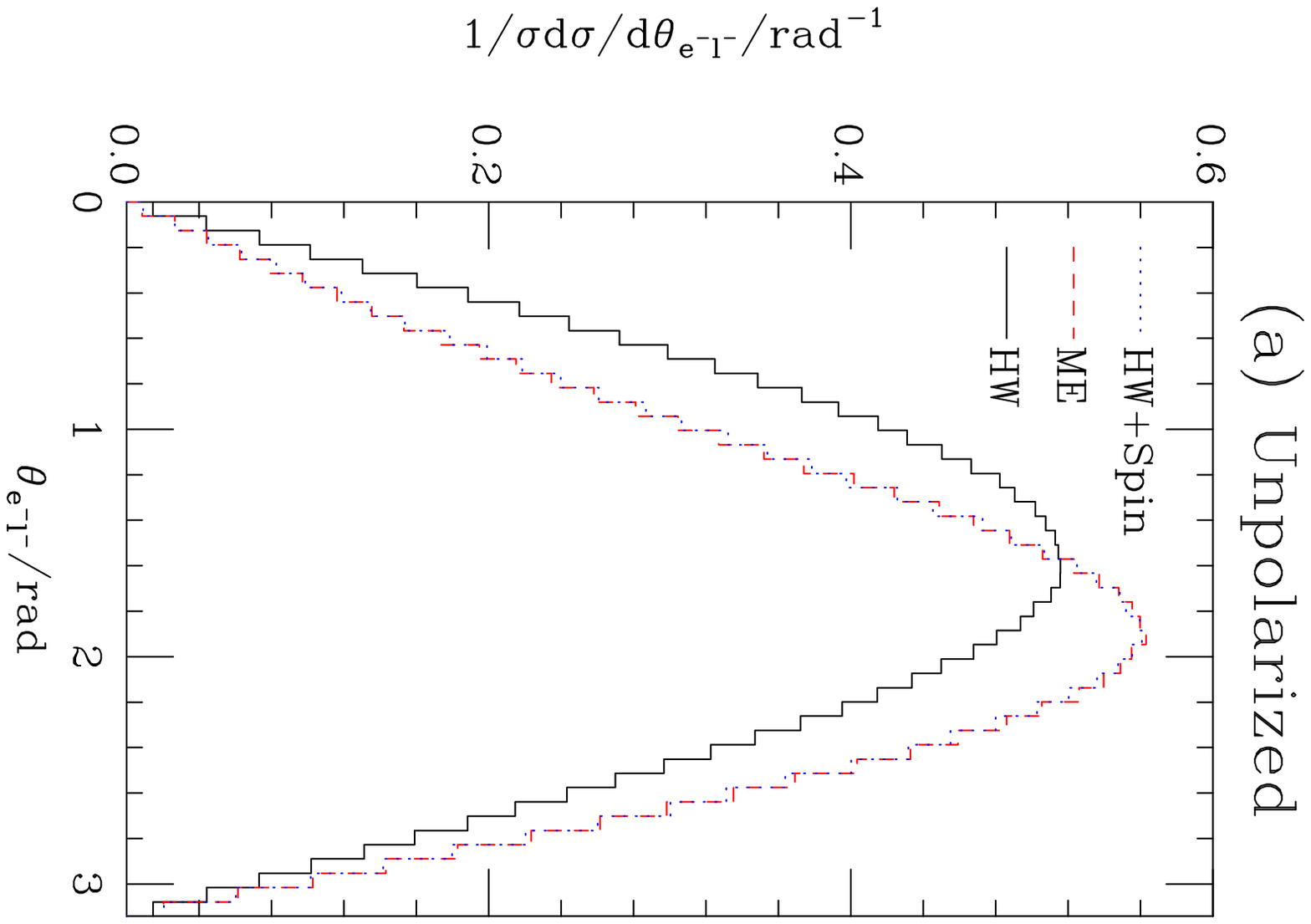}\hfill
\includegraphics[width=0.45\textwidth,angle=90]{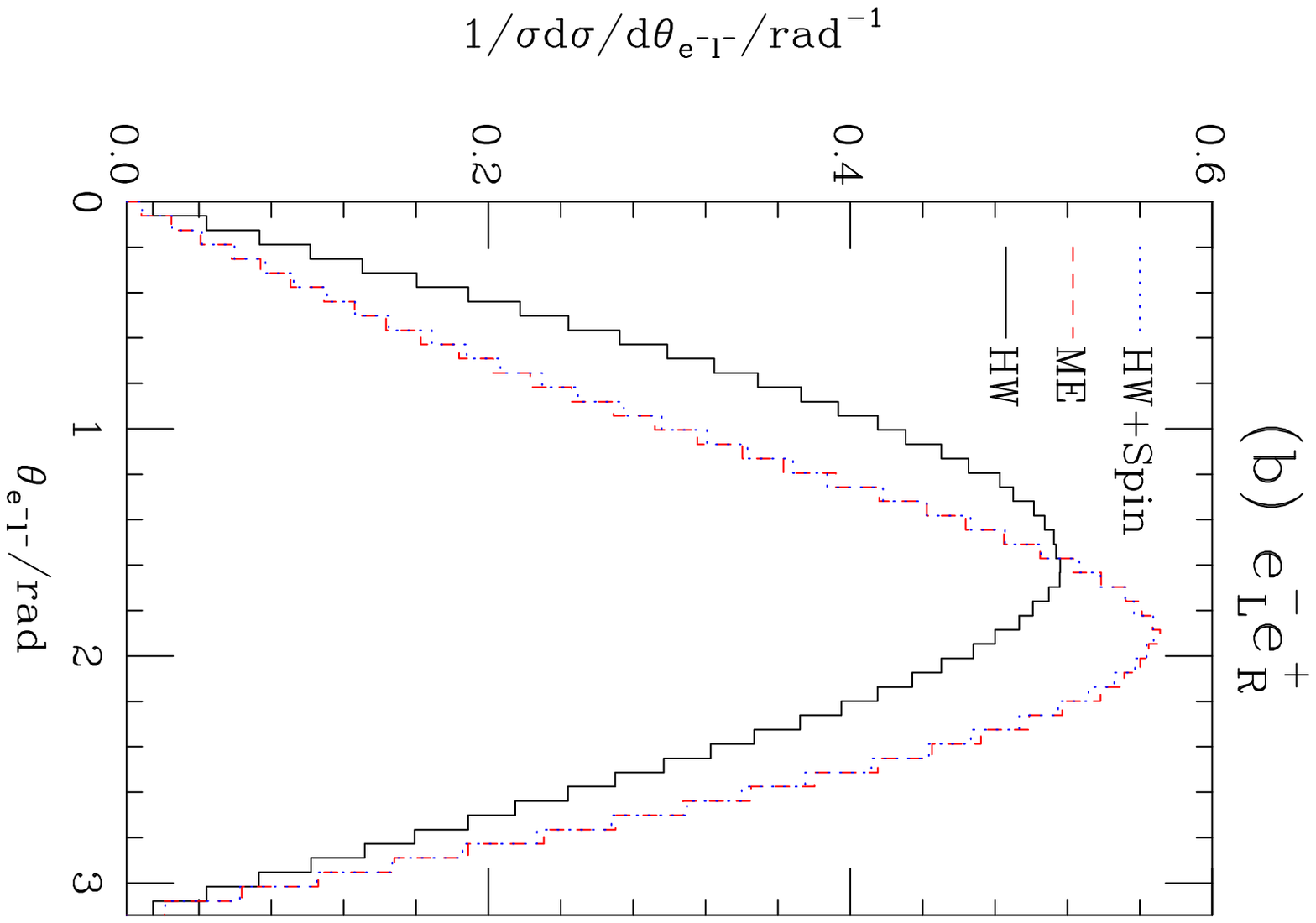}\hfill
\includegraphics[width=0.45\textwidth,angle=90]{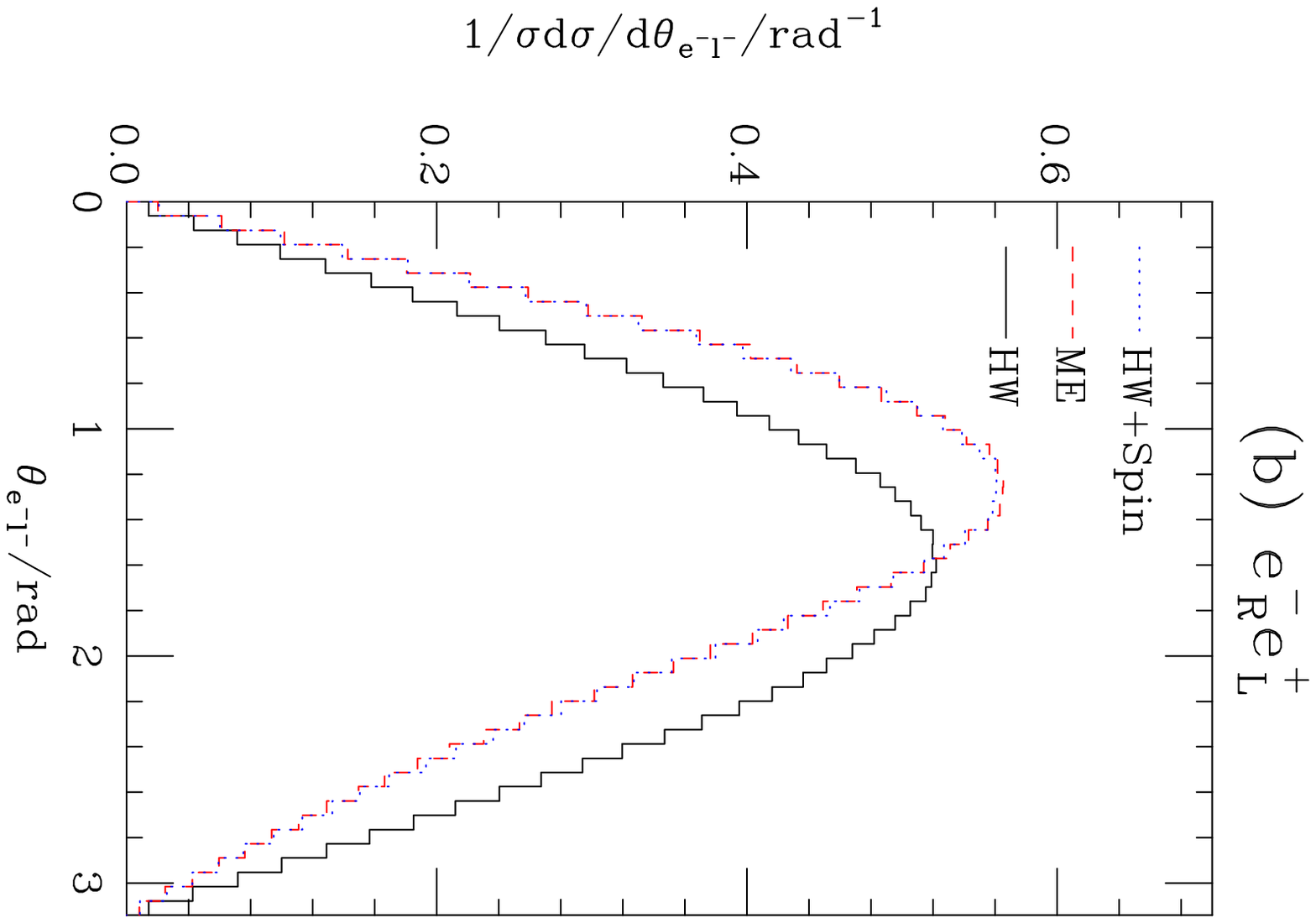}
\caption{Angle between the lepton produced in 
	\mbox{$\rm{e}^+\rm{e}^-\ra\cht^0_2\cht^0_1\ra\elt^+_R\ell^-\cht^0_1$}
	 and the incoming electron beam in the
  	laboratory frame for a centre-of-mass energy of 500\gev\ with 
	{\bf(a)} no polarization, {\bf(b)} negatively polarized electrons and positively polarized positrons 
	and {\bf(c)} positively polarized electrons and negatively polarized positrons.
        The solid line shows the default result from HERWIG which 
	treats the production and decay as independent,
	the dashed line gives the full result from the 4-body
	matrix element and the dotted line the result of the spin correlation algorithm.}
\label{fig:chitwobeam}
}

\begin{figure}[h!!]
\includegraphics[width=0.45\textwidth,angle=90]{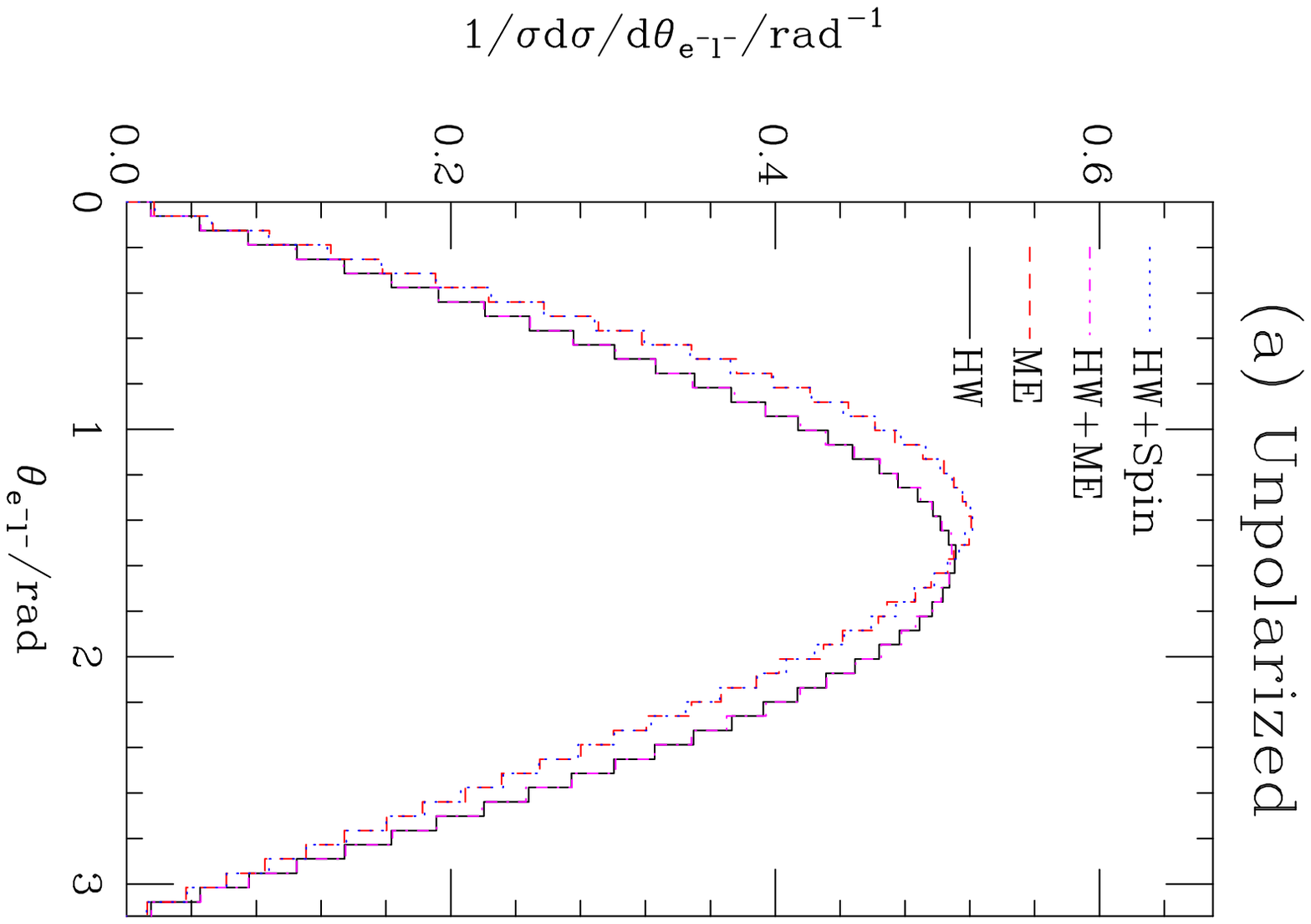}\hfill
\includegraphics[width=0.45\textwidth,angle=90]{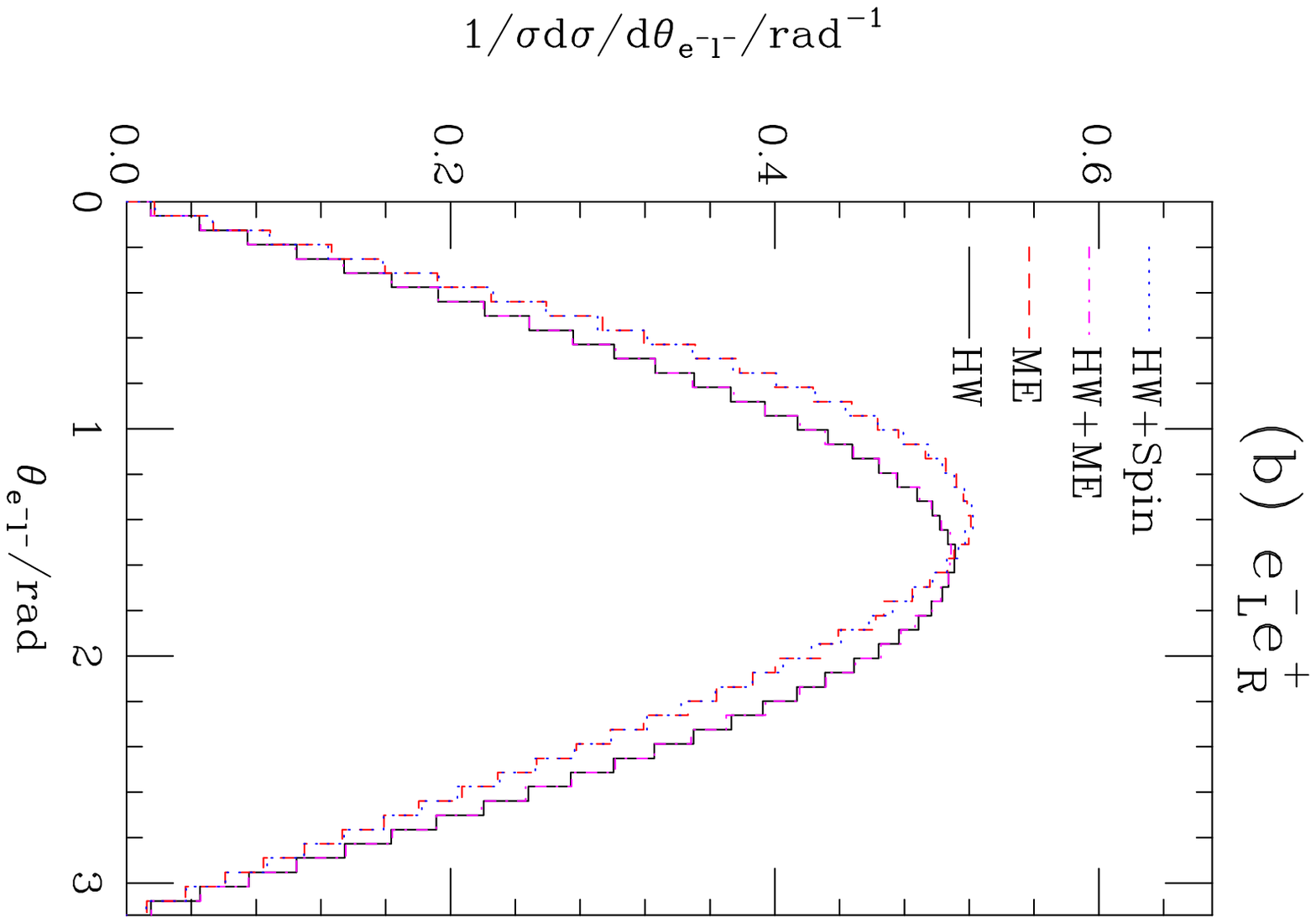}\hfill
\includegraphics[width=0.45\textwidth,angle=90]{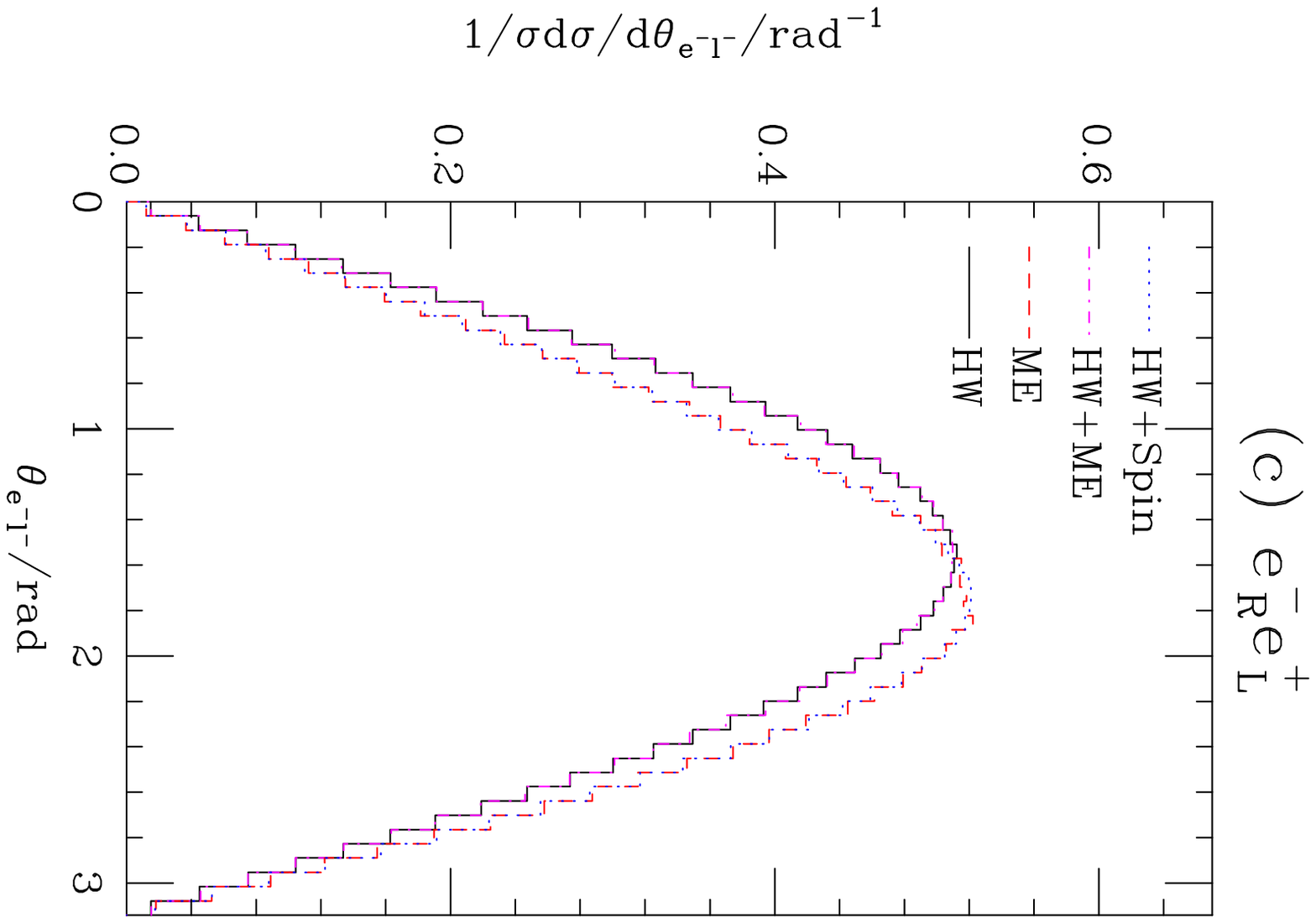}
\caption{Angle between the lepton produced in 
	\mbox{$\rm{e}^+\rm{e}^-\ra\cht^0_2\cht^0_1\ra\ell^+\ell^-\cht^0_1\cht^0_1$}
	 and the incoming electron beam in the
  	laboratory frame for a centre-of-mass energy of 500\gev\ with 
	{\bf(a)} no polarization, {\bf(b)} negatively polarized electrons and positively polarized positrons 
	and {\bf(c)} positively polarized electrons and negatively polarized positrons.
        The solid line shows the default result from HERWIG which 
	treats the production and decay as independent and uses a
	 phase-space distribution for the decay products of the neutralino,
        the dot-dashed line also includes a matrix element for the neutralino decay,
	the dashed line gives the full result from the 4-body
	matrix element and the dotted line the result of the spin correlation algorithm.}
\label{fig:chithreebeam}
%\end{figure}
%\begin{figure}
\vskip 1cm
\includegraphics[width=0.45\textwidth,angle=90]{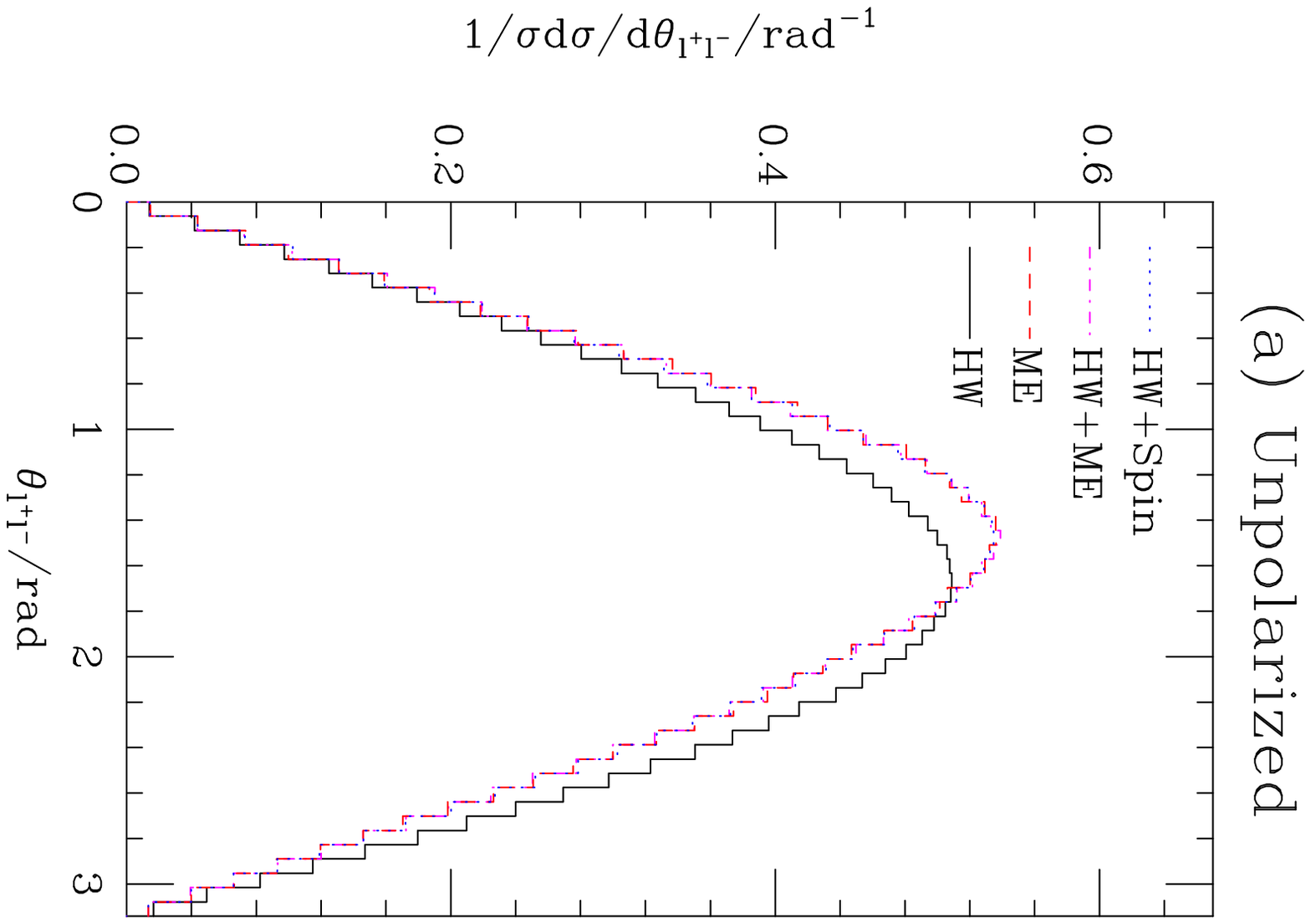}\hfill
\includegraphics[width=0.45\textwidth,angle=90]{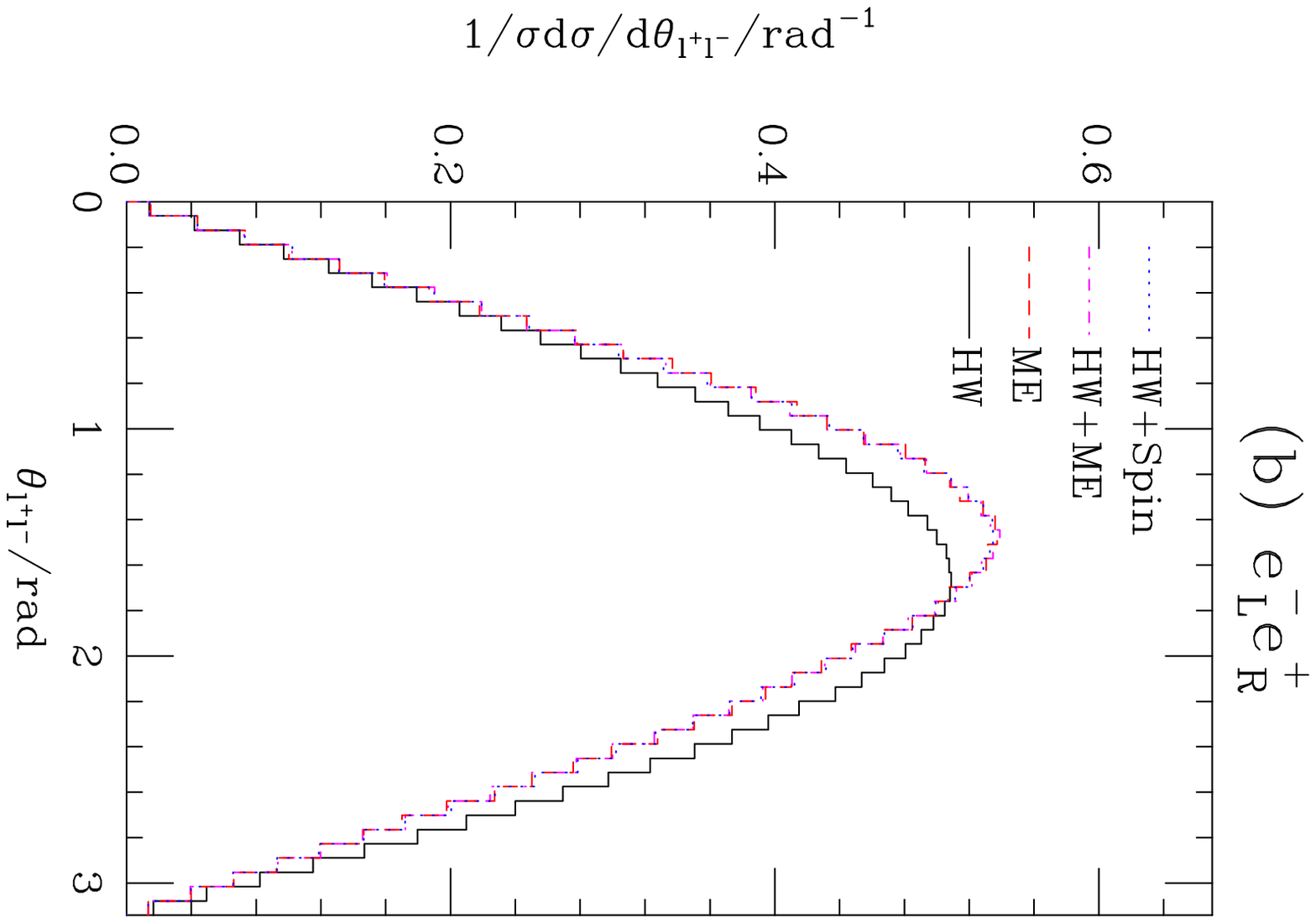}\hfill
\includegraphics[width=0.45\textwidth,angle=90]{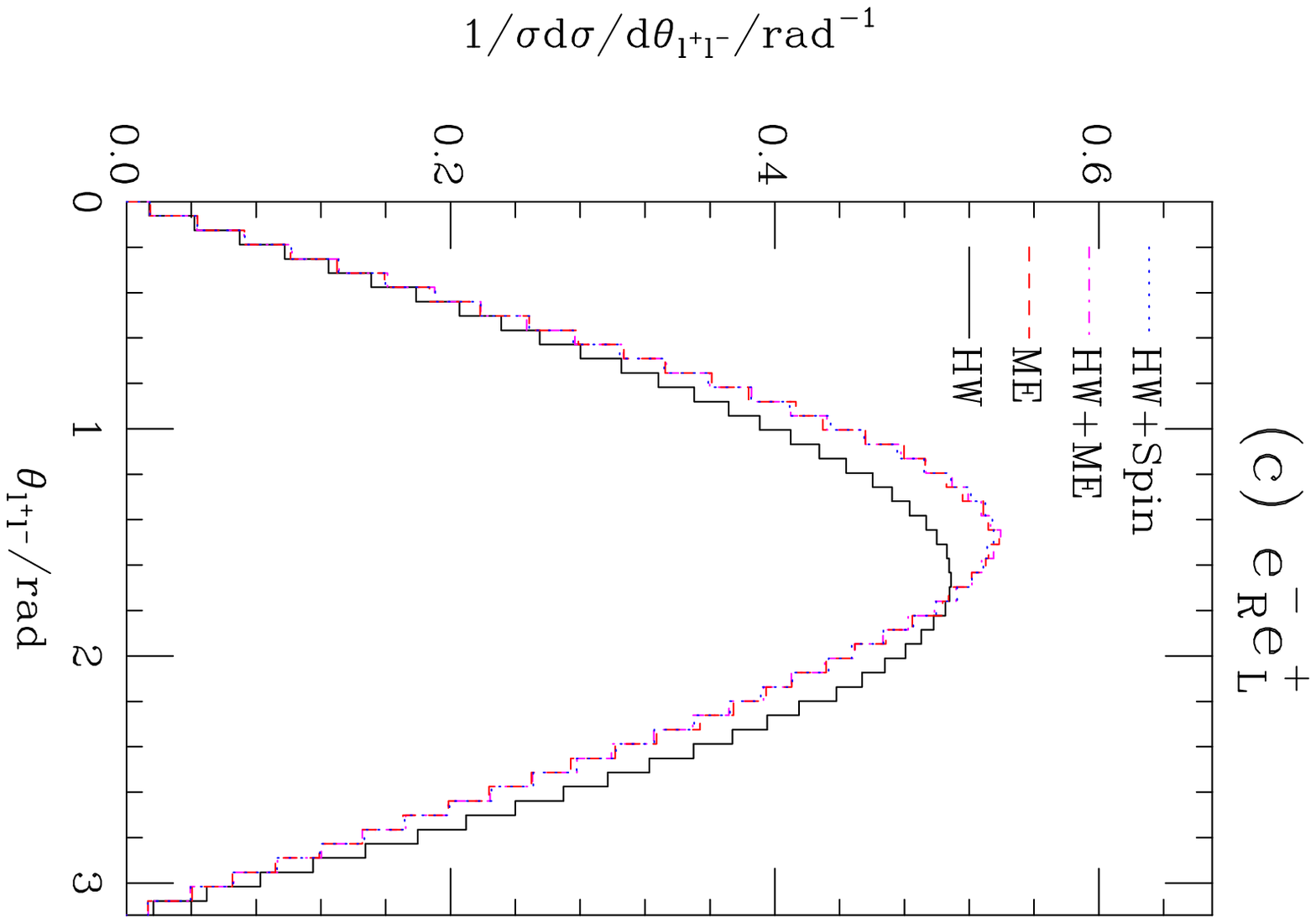}
\caption{Angle between the lepton and antilepton produced in 
	\mbox{$\rm{e}^+\rm{e}^-\ra\cht^0_2\cht^0_1\ra\ell^+\ell^-\cht^0_1\cht^0_1$}
	in the laboratory frame for a centre-of-mass energy of 500\gev\ with 
	{\bf(a)} no polarization, {\bf(b)} negatively polarized electrons and positively polarized positrons 
	and {\bf(c)} positively polarized electrons and negatively polarized positrons.
%        The solid line shows the default result from HERWIG which 
%	treats the production and decay as independent and uses a
%	 phase-space distribution for the decay products of the neutralino,
%        the dot-dashed line also includes a matrix element for the neutralino decay,
%	the dotted line gives the full result from the 4-body
%	matrix element and the dashed line the result of the spin correlation algorithm.
	The lines are described in the caption of Fig.\,\ref{fig:chithreebeam}.
}
\vspace{-1cm}
\label{fig:chithreee+e-}
\end{figure}

  The angle between the lepton produced in the $\cht^0_2$ decay and the beam
  is shown in Fig.\,\ref{fig:chithreebeam} for a three choices of polarization for the
  incoming beams. The angle between the produced leptons is shown in 
  Fig.\,\ref{fig:chithreee+e-} for the same choices of beam polarization.
  The default treatment of this process in the HERWIG event generator is
  that the production and decay of the next-to-lightest neutralino
  take place independently and the neutralino is decayed using a phase-space
  distribution for the decay products. We have therefore shown the result
  of using a matrix element to perform this decay, while still treating the
  production and decay as independent, as well as the result of the spin correlation
  algorithm and a full calculation of the process in 
  Figs.\,\ref{fig:chithreebeam}~and~\ref{fig:chithreee+e-}.

  As can be seen in Fig.\,\ref{fig:chithreebeam} the inclusion of the matrix
  element for the decay on its own has no effect on the angle between the 
  lepton and the beam, whereas the inclusion of this matrix element
  significantly improves the agreement between the
  result of HERWIG and the full result for the 
  distribution of the angles between the
  produced lepton and antilepton, Fig.\,\ref{fig:chithreee+e-}. Indeed
  the inclusion of the matrix element seems to be all that is required
  in order for the Monte Carlo simulation to reproduce the full result for this
  distribution. The full spin correlation algorithm is necessary to reproduce the
  correlation between the beam direction and the direction of the produced
  lepton.
  Again, there is good agreement between the full 4-body matrix element result
  and the spin correlation algorithm for both distributions.

%
%  ttbar in e+e-
%
\subsection[$\rm{e^+e^-\ra t\bar{t}}$]
	{\boldmath{$\rm{e^+e^-\ra t\bar{t}}$}}

  A future linear collider will have both the energy to produce
  top quark pairs and the possibility of polarized incoming beams. An accurate
  measurement of the top quark mass by either scanning the threshold
  or reconstructing the decay products will be a goal of such an experiment.
  There have been a number of studies of spin correlations in top quark
  production at such a machine \cite{Aguilar-Saavedra:2001rg}.

\FIGURE[t]{
\includegraphics[width=0.45\textwidth,angle=90]{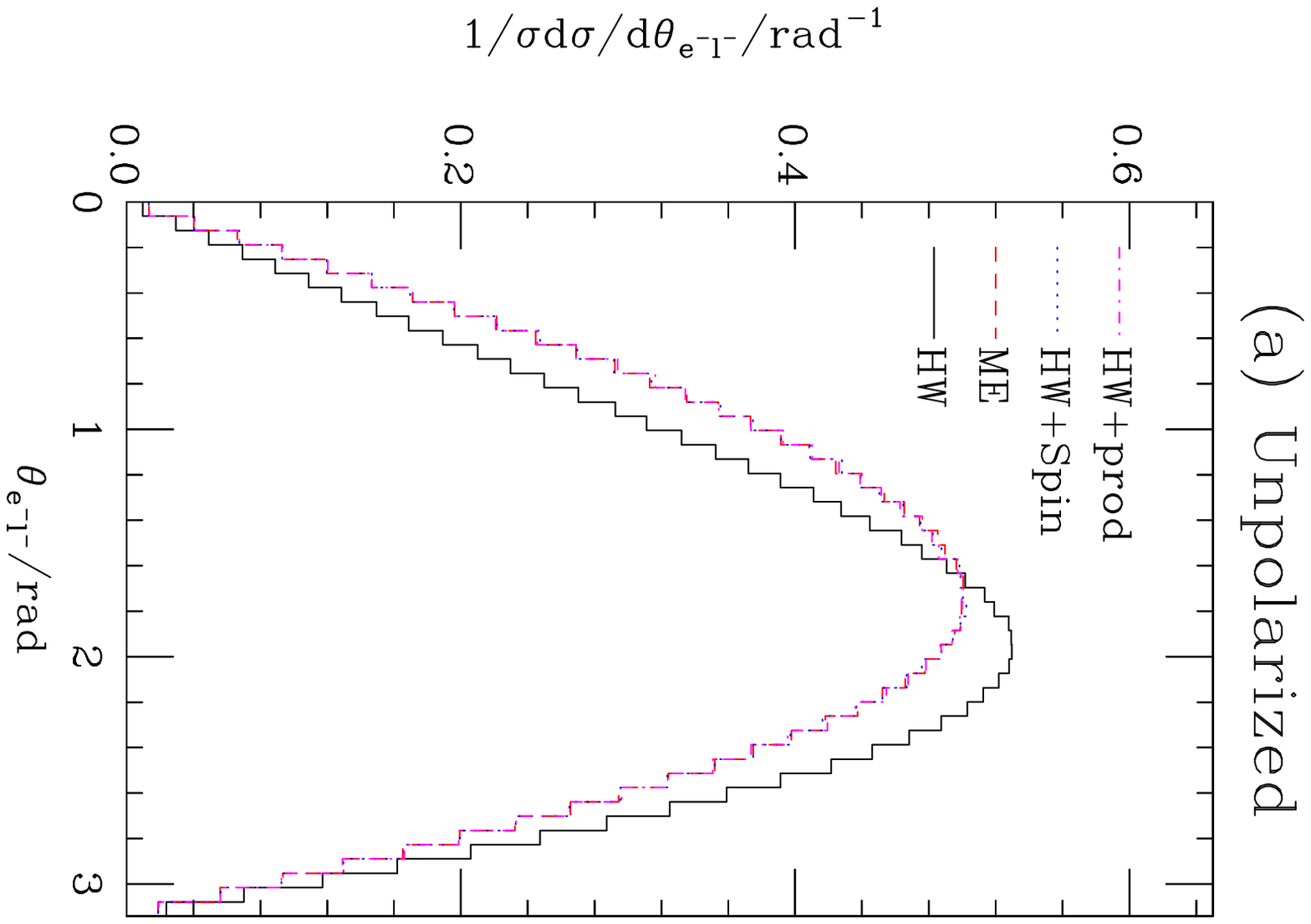}\hfill
\includegraphics[width=0.45\textwidth,angle=90]{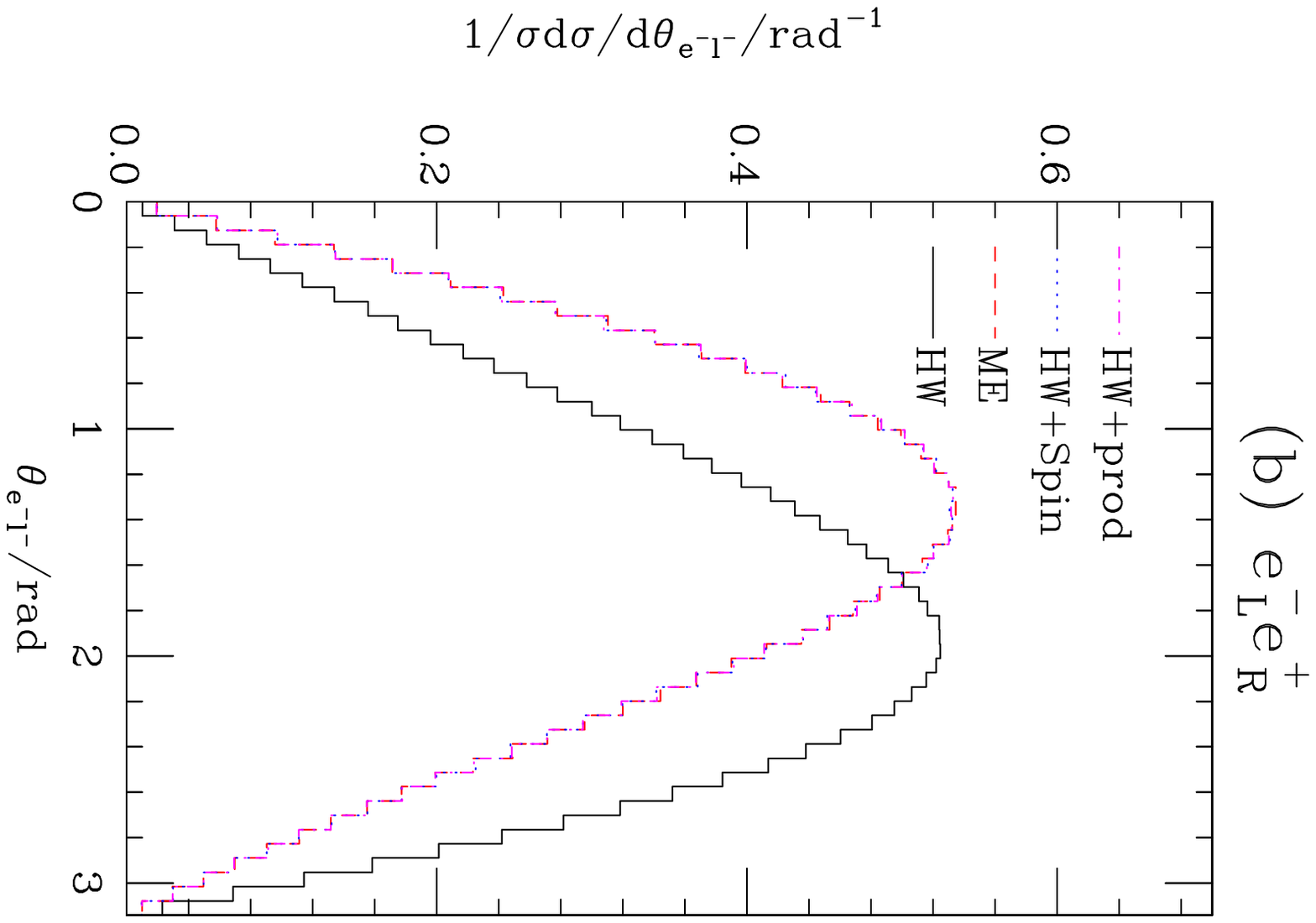}\hfill
\includegraphics[width=0.45\textwidth,angle=90]{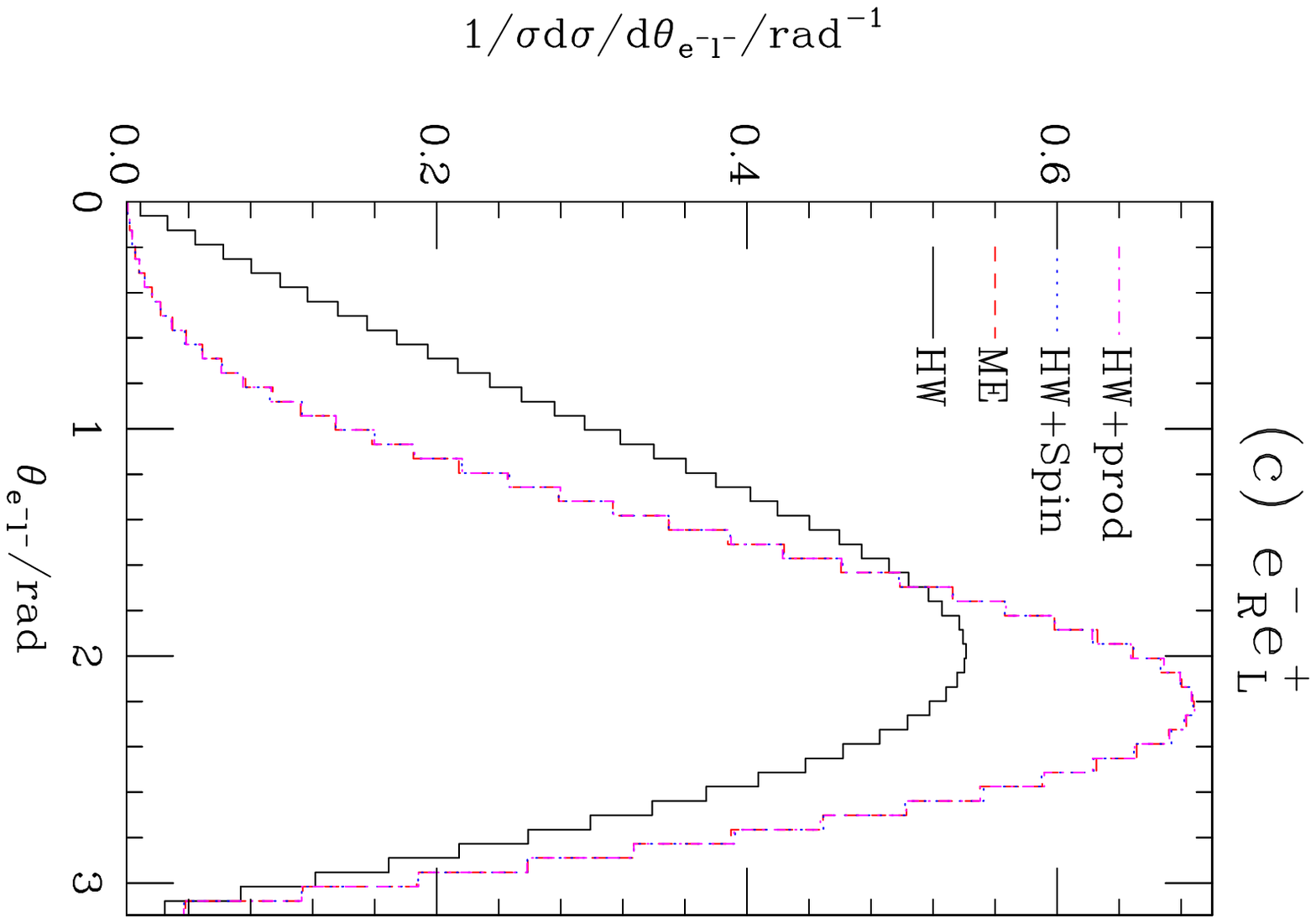}
\caption{Angle between the lepton produced in 
	\mbox{$\rm{e}^+\rm{e}^-\ra\rm{t}\rm{\bar{t}}\ra \rm{b}\rm{\bar{b}}\ell^+
	\nu_\ell\ell^-\bar{\nu}_\ell$} and the incoming electron beam in the
  	laboratory frame for a centre-of-mass energy of 500\gev\ with 
	{\bf(a)} no polarization, {\bf(b)} negatively polarized electrons and
        positively polarized positrons 
	and {\bf(c)} positively polarized electrons and negatively polarized positrons.
        The solid line shows the default result from HERWIG which 
	treats the production and decay as independent but includes a matrix element for
	the weak decay of the top, the dashed line gives the full result from the 6-body
	matrix element, the dotted line the result of the spin correlation algorithm
        and the dot-dashed line the result of the spin correlation algorithm when the
	decay matrix for the first quark is neglected.}
\label{fig:topebeamangle}
}

  Here we are mainly concerned with this process as an example of the application
  of the spin correlation algorithm.
  In this process in addition to the complications of the $2\to2$ process, which
  can have polarization of the incoming particles, there is the problem of
  correlating the decay of the top quarks with each other. In addition
  to using the spin density matrix for the top production to perform the decays the
  decay matrix from the first decay must be used in calculating the 
  spin density matrix used for the decay of the
  second quark. In the default treatment of this process HERWIG performs
  the production and decay of the quarks independently but
  uses the full 3-body matrix element for the quark decays.

\begin{figure}[h!!]
\begin{center}
\includegraphics[width=0.45\textwidth,angle=90]{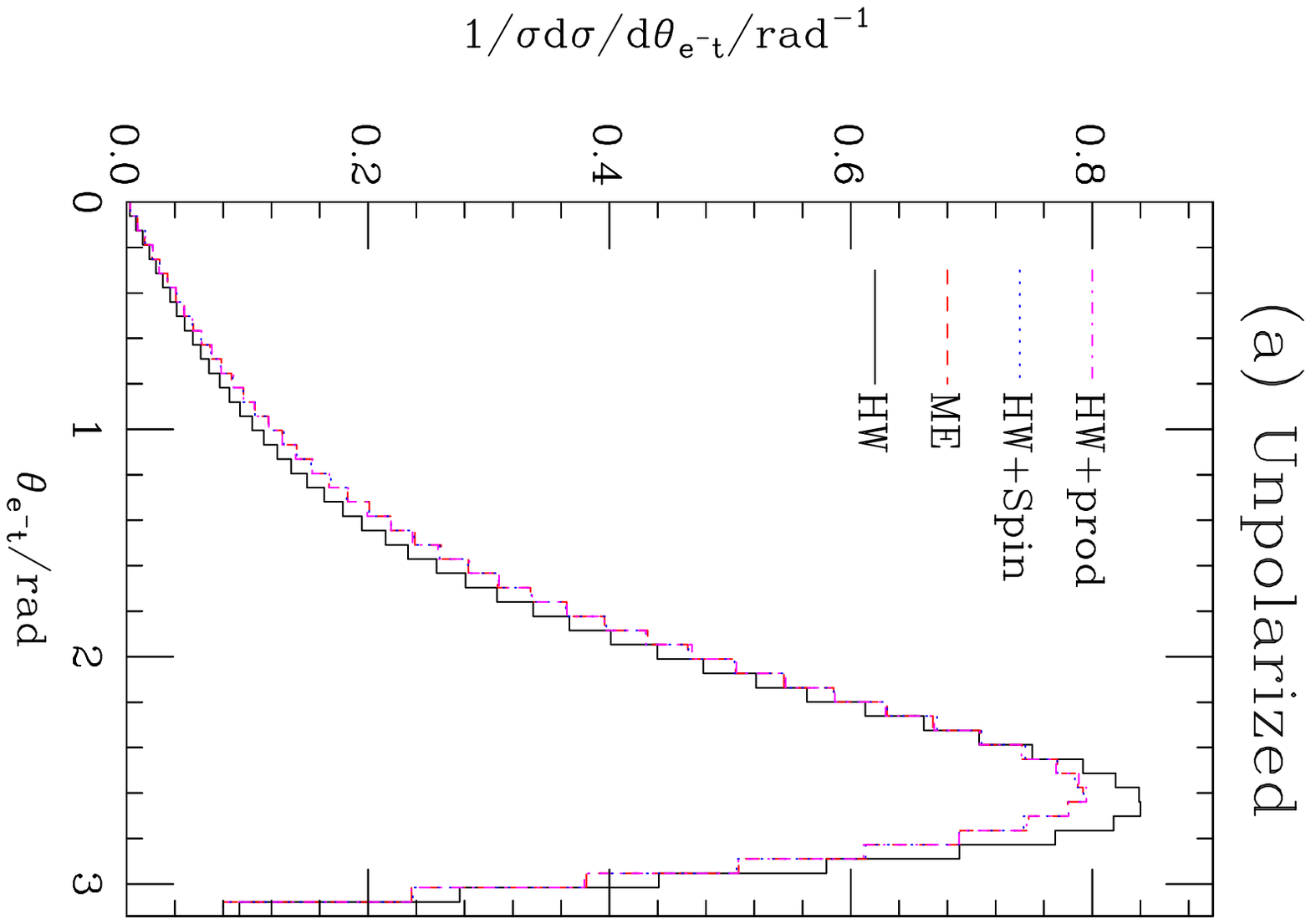}\hfill
\includegraphics[width=0.45\textwidth,angle=90]{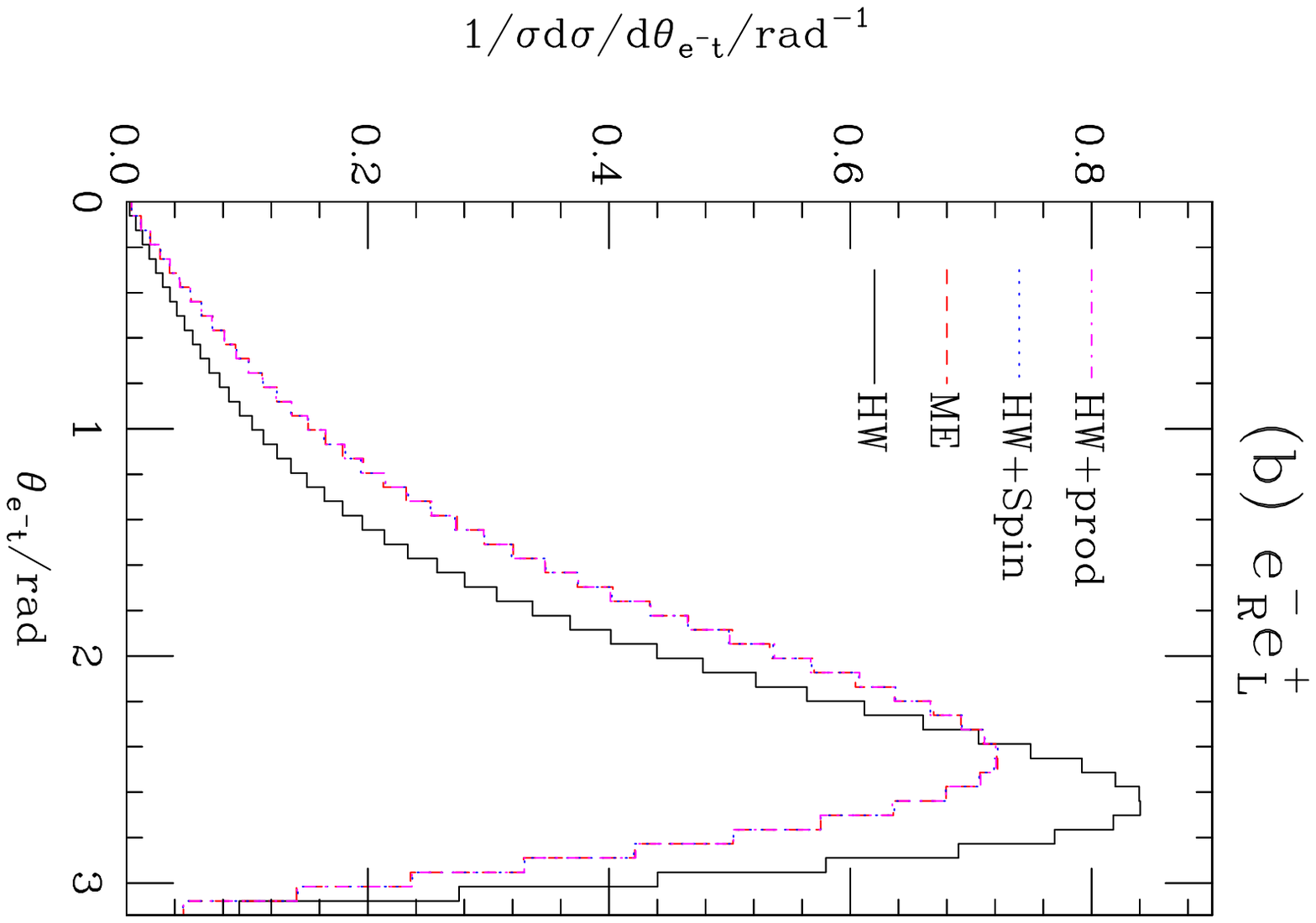}\hfill
\includegraphics[width=0.45\textwidth,angle=90]{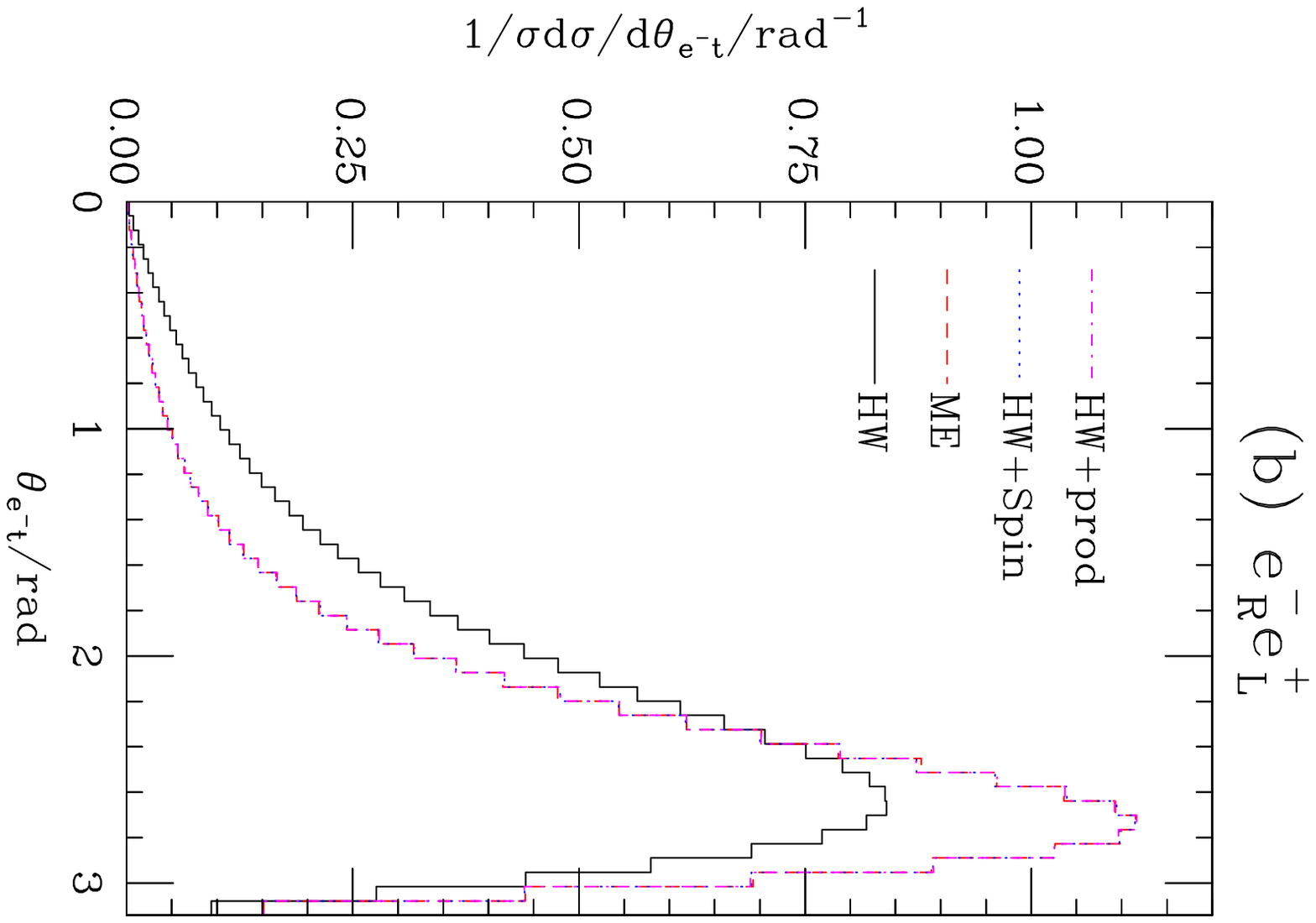}
\caption{Angle between the lepton and top quark produced in 
	\mbox{$\rm{e}^+\rm{e}^-\ra\rm{t}\rm{\bar{t}}\ra \rm{b}\rm{\bar{b}}\ell^+
	\nu_\ell\ell^-\bar{\nu}_\ell$} in the
  	laboratory frame for a centre-of-mass energy of 500\gev\ with 
	{\bf(a)} no polarization, {\bf(b)} negatively polarized electrons and positively
          polarized positrons 
	and {\bf(c)} positively polarized electrons and negatively polarized positrons.
%        The solid line shows the default result from HERWIG which 
%	treats the production and decay as independent but includes a matrix element for
%	the weak decay of the top, the dashed line gives the full result from the 6-body
%	matrix element, the dotted line the result of the spin correlation algorithm
%        and the dotdashed line the result of the spin correlation algorithm when the
%	decay matrix for the first quark is neglected.
	The lines are described in the caption of Fig.\,\ref{fig:topebeamangle}.}
\label{fig:toptopeangle}
\end{center}
%\vspace{2.0cm}
\end{figure}
\begin{figure}
\begin{center}
\includegraphics[width=0.45\textwidth,angle=90]{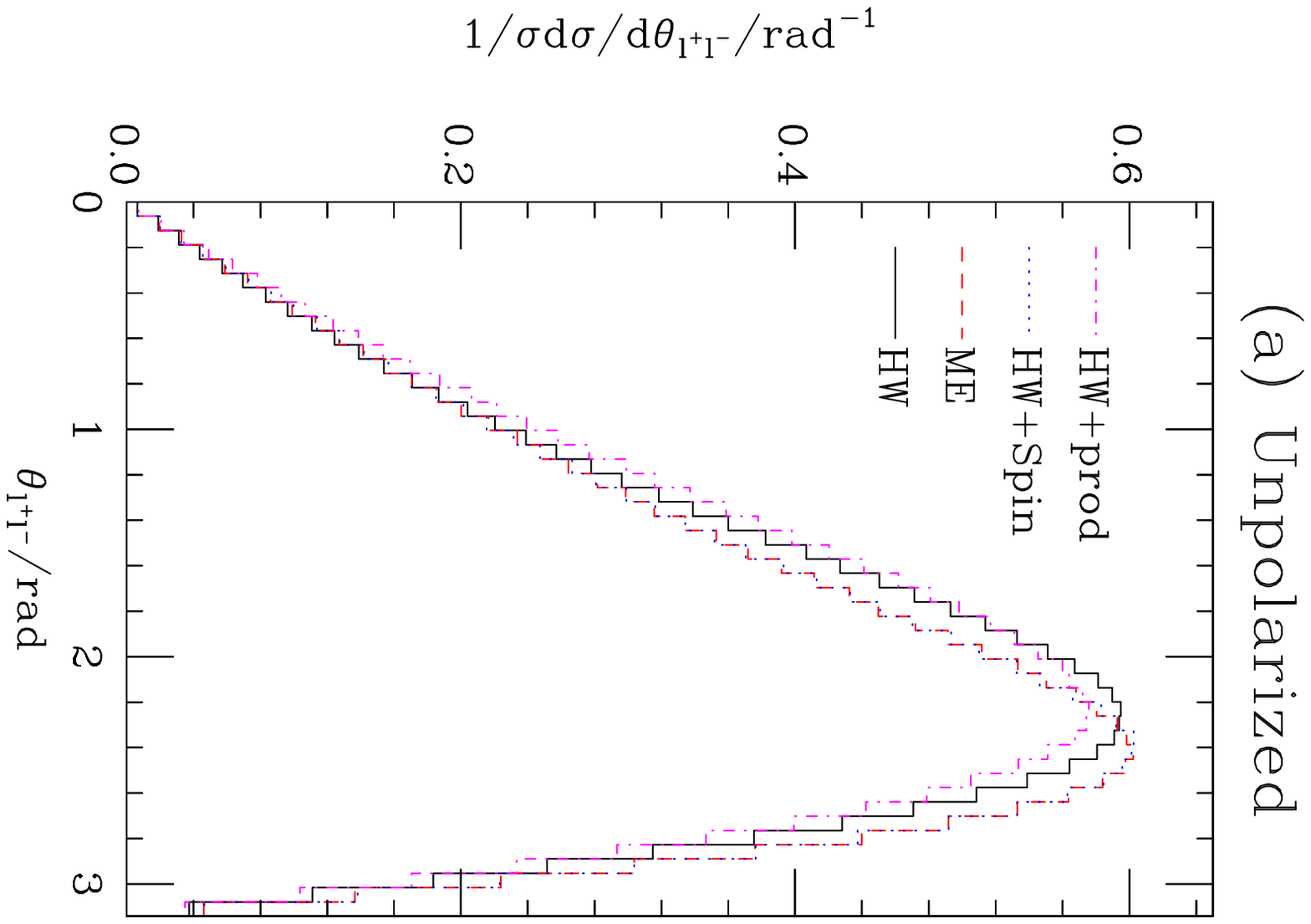}\hfill
\includegraphics[width=0.45\textwidth,angle=90]{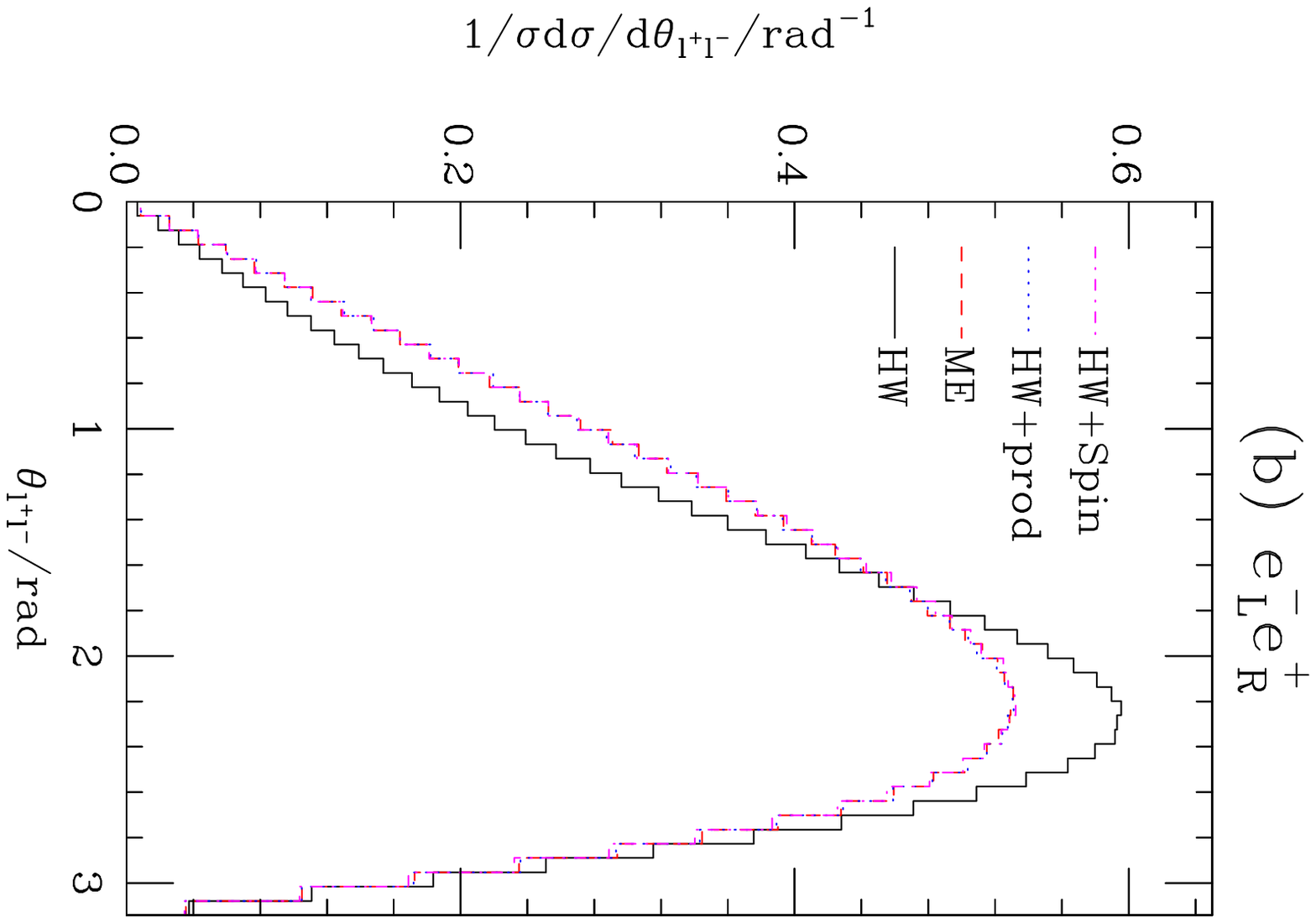}\hfill
\includegraphics[width=0.45\textwidth,angle=90]{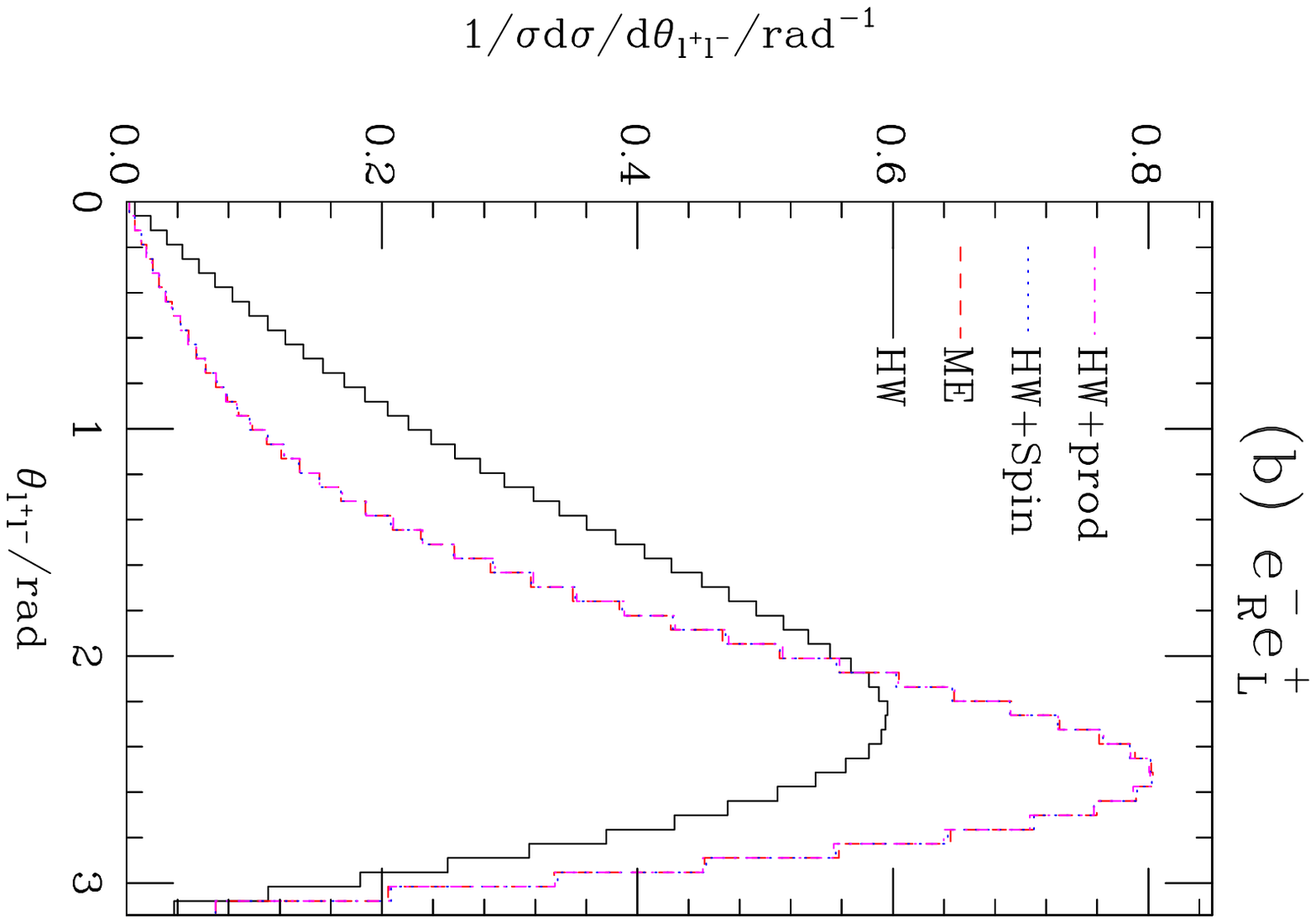}
\caption{Angle between the leptons produced in
 	\mbox{$\rm{e}^+\rm{e}^-\ra\rm{t}\rm{\bar{t}}\ra \rm{b}\rm{\bar{b}}\ell^+
	\nu_\ell\ell^-\bar{\nu}_\ell$} in the laboratory frame for a centre-of-mass
         energy of
	500\gev\  with {\bf(a)} no polarization, {\bf(b)} negatively polarized electrons 
        and positively polarized positrons 
	and {\bf(c)} positively polarized electrons and negatively polarized positrons.
%        The solid line shows the default result from HERWIG which 
%	treats the production and decay as independent but includes a matrix element for
%	the weak decay of the top, the dashed line gives the full result from the 6-body
%	matrix element, the dotted line the result of the spin correlation algorithm
%        and the dotdashed line the result of the spin correlation algorithm when the
%	decay matrix for the first quark is neglected.
	The lines are described in the caption of Fig.\,\ref{fig:topebeamangle}.}
\label{fig:topeeangle}
\end{center}
%\vspace{-1.5cm}
\end{figure}

  Fig.\,\ref{fig:topebeamangle} shows the angle of the lepton produced in the top
  antiquark decay with respect to the incoming electron beam,
  Fig.\,\ref{fig:toptopeangle}
  shows the angle between the lepton and the top quark and
  Fig.\,\ref{fig:topeeangle} shows the angle between the lepton and the antilepton.
  The result of the spin correlation algorithm agrees well with the full
  6-body matrix element for all these distributions. 
  Again the major discrepancies
  between the default treatment in HERWIG and the full result are for
  the correlation of the direction of the outgoing leptons with the beam
  and the correlation
  between the direction of the produced lepton and the second quark in the event.

  In addition to the full result of the spin correlation algorithm we have also
  included the results of the algorithm where the
  identity is used rather than the decay matrix calculated for
  the decay of the first quark when calculating the spin density
  matrix for the decay of the second quark. Neglecting this step of the algorithm
  still gives good agreement with the matrix-element calculation for the correlation
  between the direction of the produced lepton and the beam direction,
  Fig.\,\ref{fig:topebeamangle}, and the correlation between the direction
  of the
  top quark and the lepton, Fig.\,\ref{fig:toptopeangle}.
  This is because these correlations mainly depend on the details
  of the hard production
  process and not how the first quark decayed.

  However, in order to reproduce the result of the matrix element for the
  angle between the produced lepton and antilepton, 
  Fig.\,\ref{fig:topeeangle}, it is essential to include the decay matrix from
  the first quark as this is a correlation between the two decays.
  This effect is most noticeable when there is no beam polarization and both
  possible initial-state spin configurations contribute to the result.

%
%  ttbar in hadron-hadron 
%
\subsection[$\rm{t\bar{t}}$ production in Hadron-Hadron Collisions]
	{\boldmath{$\rm{t\bar{t}}$} production in Hadron-Hadron Collisions}
\label{sect:egtophadron}

  Since the discovery of the top quark at the Tevatron there have been a number of
  studies of spin correlations in top quark pair production for both Run II
  of the Tevatron and the LHC.
  There are two processes which contribute to $\rm{t\bar{t}}$ production in
  hadron collisions
\begin{subequations}
\begin{eqnarray}
\rm{q}\rm{\bar{q}} &\longrightarrow& \rm{t\bar{t}}, \\
\rm{g}\rm{g}       &\longrightarrow& \rm{t\bar{t}}.
\end{eqnarray}
\end{subequations}
  The $\rm{q}\rm{\bar{q}}$ annihilation process presents no additional complications
  and is very similar to the $\rm{e}^+\rm{e}^-$ annihilation we discussed in the previous
  section.

  However the colour flows in the three diagrams for $\rm{t\bar{t}}$ production via 
  gluon scattering, Fig.\,\ref{fig:diaggtt},
  are very different. It is easiest to extract the colour
  matrices from the helicity amplitudes and perform the colour sum/averages separately.
  The colour matrices for the three diagrams, shown in Fig.\,\ref{fig:diaggtt}, are
\begin{subequations}
\begin{eqnarray}
{C_t}^{ab}_{c_1c_2}=&t^a_{c_1c'}t^b_{c'c_2}  & \ \ \ \ \ \ \ \ \ t\rm{-channel},\\
{C_u}^{ab}_{c_1c_2}=&t^b_{c_1c'}t^a_{c'c_2}  & \ \ \ \ \ \ \ \ \ u\rm{-channel},\\
{C_s}^{ab}_{c_1c_2}=&if^{abc}t^c_{c_1c_2}    & \ \ \ \ \ \ \ \ \ s\rm{-channel},
\end{eqnarray}
\end{subequations}
  where $f^{abc}$ is the $SU(3)$ structure constant, $t^a$ is the $SU(3)$ colour
  matrix in the fundamental representation, ${C_{s,t,u}}^{ab}_{c_1c_2}$
  are the colour matrices for the $s$-, $t$- and $u$-channel colour flows
  respectively,
  $a$ is the colour of the first incoming gluon, $b$ is the colour of the
  second incoming gluon, $c_1$ is the colour of the outgoing quark, $c'$ is the
  colour of the intermediate quark and
  $c_2$ is the colour of the outgoing antiquark.

  In general the colour flow for diagrams involving the triple gluon vertex, \ie the
  $s$-channel diagram, is not unique and 
  can be rewritten in terms of the $t$- and
  $u$-channel colour flows using the definition of the structure constants
\begin{equation}
if^{abc}t^c_{c_1c_2} = t^a_{c_1c'}t^b_{c'c_2}-t^b_{c_1c'}t^a_{c'c_2},
\end{equation} 
   giving
\begin{equation}
{C_s}^{ab}_{c_1c_2}={C_t}^{ab}_{c_1c_2}-{C_u}^{ab}_{c_1c_2},
\end{equation}
  so that we only have two different colour flows to deal with.

\begin{figure}
\begin{center}
\begin{picture}(400,100)
\SetScale{0.9}
% first diagrams
\Gluon(0,70)(60,70){4}{5}
\Gluon(60,10)(0,10){4}{5}
\ArrowLine(60,10)(60,70)
\ArrowLine(60,70)(120,70)
\ArrowLine(120,10)(60,10)
\Text(-2,64)[r]{$\rm{g}$}
\Text(-2,9)[r]{$\rm{g}$}
\Text(110,9)[l]{$\rm{\bar{t}}$}
\Text(110,64)[l]{$\rm{t}$}
\SetOffset(160,0)
\Gluon(0,70)(60,10){4}{6}
\Gluon(0,10)(60,70){-4}{6}
\ArrowLine(60,10)(60,70)
\ArrowLine(60,70)(120,70)
\ArrowLine(120,10)(60,10)
\Text(-2,64)[r]{$\rm{g}$}
\Text(-2,9)[r]{$\rm{g}$}
\Text(110,9)[l]{$\rm{\bar{t}}$}
\Text(110,64)[l]{$\rm{t}$}
\SetOffset(320,8)
\SetScale{0.70}
\Gluon(40,40)(0,0){4}{5}
\Gluon(0,80)(40,40){4}{5}
\Gluon(40,40)(80,40){5}{5}
\ArrowLine(80,40)(120,80)
\ArrowLine(120,0)(80,40)
\Text(-2,55)[r]{$\rm{g}$}
\Text(-2, 0)[r]{$\rm{g}$}
\Text(87,55)[l]{$\rm{t}$}
\Text(87, 0)[l]{$\rm{\bar{t}}$}
\end{picture}
\end{center}
\caption{Feynman diagrams for $\rm{g}\rm{g}\ra\rm{t}\rm{\bar{t}}$.}
\label{fig:diaggtt}
\end{figure}

  In principle the presence of different colour flows could be a significant problem to 
  the algorithm we are using.
  We should keep track of the colour flow in the
  production processes and decays and calculate spin density matrices for each of the
  flows and sum over them. If the decays have one than one colour
  flow, in addition to the different colour flows in the hard process,
  this could lead to an increase in the
  complexity of the algorithm with the number of final state particles, \ie
  the number of colour flows would grow exponentially.
  However, in both the Standard and Minimal Supersymmetric Standard Models, provided
  we only consider three-body decays as we are doing, there can only
  be different colour flows in the hard production process and it is therefore
  sufficient to replace Eqn.\,\ref{eqn:spin1} with
\begin{equation}
\sum_{a,b=1,C_N}C_{ab}	\rho^1_{\kappa_1\kappa'_1}\rho^2_{\kappa_2\kappa'_2}
        \mathcal{M}^{a}_{\kappa_1\kappa_2;\lam_1\ldots\lam_n}
	\mathcal{M}^{b*}_{\kappa'_1\kappa'_2;\lam'_1\ldots\lam'_n}
        \prod_{i=1,n}D^i_{\lam_i\lam'_i},
\label{eqn:spin1*}
\end{equation}
  where $\mathcal{M}^a$ is the matrix element for
  colour flow $a$, $C_N$ is the number of
  colour flows and $C_{ab}$ is the colour factor for the matrix elements.
  Eqns.\,\ref{eqn:spin2}~and~\ref{eqn:spin2b} must also be modified in the same way.

  For example in ${\rm t}{\rm\bar{t}}$ collisions
\begin{subequations}
\begin{eqnarray}
C_{tt} =& \frac1{(N^2_c-1)^2}t^a_{c_1c'}t^b_{c'c_2}t^a_{c''c_1}t^b_{c_2c''} 
        &= \pmn\frac1{4N_c},\\
C_{uu} =& \frac1{(N^2_c-1)^2}t^b_{c_1c'}t^a_{c'c_2}t^b_{c''c_1}t^a_{c_2c''}
        &= \pmn\frac1{4N_c},\\
C_{tu} =& \frac1{(N^2_c-1)^2}t^a_{c_1c'}t^b_{c'c_2}t^b_{c''c_1}t^a_{c_2c''}
        &=-\frac1{4N_c(N^2_c-1)},
\end{eqnarray}
  the colour factor 
  $C_{ut}=C_{ut}$.\footnote{The sign of $C_{tu}$ is different to that given in 
				Table\,\ref{tab:SMcross} because it proved easier to
				 absorb this sign as a change of sign of the $u$-channel
				 matrix element.}
\end{subequations}

  Another option which would be necessary if the decays have more than
  one possible colour flow, in for example R-parity violating SUSY models 
  \cite{Dreiner:1999qz,Richardson:2000nt}, is to select one of the
  colour flows in the same way as is already done for the production of QCD
  radiation \cite{Odagiri:1998ep}.
  In this procedure we would first select a colour flow, $a$, using \cite{Odagiri:1998ep}
\begin{equation}
%|\mathcal{M}_{\rm{full},c_a}|^2=
 C_{aa}
	\rho^1_{\kappa_1\kappa'_1}\rho^2_{\kappa_2\kappa'_2}
        \mathcal{M}^{a }_{ \kappa_1 \kappa_2;\lam_1\ldots\lam_n}
	\mathcal{M}^{a*}_{\kappa'_1\kappa'_2;\lam_1\ldots\lam_n}
        A_{a},
\label{eqn:spin1**}
\end{equation}
  where
\begin{equation}
 A_{a}=
\frac{\sum_{b,c=1,C_N}C_{bc}
	\rho^1_{\kappa_1\kappa'_1}\rho^2_{\kappa_2\kappa'_2}
	\mathcal{M}^{b }_{ \kappa_1 \kappa_2;\lam_1\ldots\lam_n}
	\mathcal{M}^{c*}_{\kappa'_1\kappa'_2;\lam_1\ldots\lam_n}}
     {\sum_{b=1,C_N}C_{bb}
	\mathcal{M}^{b }_{ \kappa_1 \kappa_2;\lam_1\ldots\lam_n}
	\mathcal{M}^{b*}_{\kappa'_1\kappa'_2;\lam_1\ldots\lam_n}}.
\end{equation}
  This procedure ensures that the sum over the redefined colour flows reproduces
  the full result, Eqn.\,\ref{eqn:spin1*}. 
  This is used to select the colour flow $a$ for the process and
  instead of Eqn.\,\ref{eqn:spin1} to give the distributions
  of the particles produced in the hard process.
  The matrix element for the selected colour flow is then used with Eqn.\,\ref{eqn:spin2}
  to calculate the spin density matrices for the decays of the particles produced in
  the hard process.

  In models where there is more than one colour flow in a decay we would 
  select the colour flow for the decay and generate the momenta of the
  decay products according to   
\begin{equation}
\rho_{\lam_0\lam'_0}C_{aa}\mathcal{M}^{a}_{\lam_0;\lam_1\ldots\lam_n}
	\mathcal{M}^{a*}_{\lam'_0;\lam_1\ldots\lam_n}E_{a},
\label{eqn:spin3**}
\end{equation}
  where
\begin{equation}
E_{a}=\frac{\sum_{b,c=1,C_N}C_{bc}\mathcal{M}^{b}_{\lam_0;\lam_1\ldots\lam_n}
	                          \mathcal{M}^{c*}_{\lam'_0;\lam_1\ldots\lam_n}}
{\sum_{b=1,C_N}C_{bb}\mathcal{M}^{b}_{\lam_0;\lam_1\ldots\lam_n}
	\mathcal{M}^{b*}_{\lam'_0;\lam_1\ldots\lam_n}}.
\end{equation}
  The matrix element for the selected colour flow is then used to calculate the
  spin density matrices for the particles produced in the decay and
  the decay matrix for this decay once all the particles produced in the decay
  have been developed.

  This approach neglects the interference between the different colour flows in
  generating the spin correlations, just as they are neglected when
  generating the QCD radiation. As this procedure is not necessary for the
  processes we are considering we will use the first procedure in processes where the
  hard process has more than one colour flow.

  There have been many studies of spin correlations in hadron-hadron collisions. In order
  to study the results of the spin correlation algorithm we have chosen to 
  follow the approach of \cite{Brandenburg:1996df,Brandenburg:1996wu} in which the
  following observables were defined for top quark pair production followed by
  $\rm{t}\ra{\rm b}{\rm W}^+\ra {\rm b}\rm{q}\rm{\bar{q}}$ and
  $\rm{\bar{t}}\ra{\rm \bar{b}}{\rm W}^-\ra{\rm \bar{b}}\ell^-\bar{\nu}$:
\begin{subequations}
\begin{eqnarray}
\mathcal{O}_1&=& \hat{{\bf q}}^*_b \cdot \hat{{\bf q}}_{\ell^-};\\
\mathcal{O}_2&=& \left(\hat{{\bf q}}^*_b \cdot \hat{{\bf p}}_p\right)
		 \left(\hat{{\bf q}}_{\ell^-}\cdot\hat{{\bf p}}_p\right);\\
\mathcal{O}_3&=& \left(\hat{{\bf q}}^*_b \cdot \hat{{\bf k}}_t\right)
     		 \left(\hat{{\bf q}}_{\ell^-}\cdot \hat{{\bf k}}_t\right);\\
\mathcal{O}_4&=& \frac12\left[ \left(\hat{{\bf q}}^*_b      \cdot\hat{{\bf p}}_p\right)
                               \left(\hat{{\bf q}}_{\ell^-} \cdot\hat{{\bf k}}_t\right)
                              +\left(\hat{{\bf q}}^*_b      \cdot\hat{{\bf k}}_t\right)
			       \left(\hat{{\bf q}}_{\ell^-} \cdot\hat{{\bf p}}_p\right)
			      \right];\\
\mathcal{O}_5&=& \sgn y_t \, \, \mathcal{O}_4;\\
\mathcal{O}_6&=& \left(\hat{{\bf p}}_p\times\hat{{\bf q}}^*_b\right) \cdot
		 \left(\hat{{\bf p}}_p\times\hat{{\bf q}}_{\ell^-}\right)
		=\mathcal{O}_1-\mathcal{O}_2.
\end{eqnarray}
\end{subequations}
   All the hatted vectors are unit vectors and the quantities without an
   asterisk are measured in the laboratory frame,
         $\hat{{\bf p}}_p$ is the beam direction,
	 $\hat{{\bf q}}_{\ell^-}$ is the direction of the charged lepton,
         $\hat{{\bf k}}_t$ is the direction of the top quark
         and $y_t$ is the rapidity of the top quark.
   The asterisked quantities  are measured in the top quark rest frame and
   $\hat{{\bf q}}^*_b$ is the direction of the bottom quark.
   In a similar way the quantities $\bar{\mathcal{O}}_{1-6}$ are defined for
   the charge conjugate decay modes.
   The observable $\mathcal{O}_4$ is zero when integrated over a symmetric
   rapidity interval which is why the observable $\mathcal{O}_5$ is defined.

\begin{table}
\begin{center}
{\footnotesize
\begin{tabular}{|c|c|c|c|c|c|c|}
\hline
&\multicolumn{3}{c|}{$\mathcal{O}$/\%}&\multicolumn{3}{c|}{$\bar{\mathcal{O}}$/\%}\\
\cline{2-7}
 &Full & Spin 1 & Spin 2 & Full & Spin 1 & Spin 2 \\
\hline
$\mathcal{O}_1$ & $\pmn2.46\pm0.04$ &
                  $\pmn2.45\pm0.02$ & $\pmn2.41\pm0.02$
                & $\pmn2.42\pm0.04$ &
                  $\pmn2.46\pm0.02$ & $\pmn2.44\pm0.02$\\
\hline									       
$\mathcal{O}_2$ & $\pmn2.94\pm0.02$ &
	          $\pmn2.92\pm0.01$ & $\pmn2.90\pm0.01$
                & $\pmn2.91\pm0.02$ &
	          $\pmn2.91\pm0.01$ & $\pmn2.89\pm0.01$\\
\hline									       
$\mathcal{O}_3$ & $\pmn1.68\pm0.02$ &
	          $\pmn1.67\pm0.01$ & $\pmn1.64\pm0.01$
                & $\pmn1.67\pm0.02$ &
	          $\pmn1.68\pm0.01$ & $\pmn1.65\pm0.01$\\
\hline									       
$\mathcal{O}_4$ & $\pmn0.03\pm0.02$ &
	          $\pmn0.01\pm0.01$ & $\pmn0.01\pm0.01$
                & $   -0.01\pm0.02$ &
	          $\pmn0.00\pm0.01$ & $   -0.01\pm0.01$\\
\hline									       
$\mathcal{O}_5$ & $\pmn1.95\pm0.02$ &
	          $\pmn1.94\pm0.01$ & $\pmn1.92\pm0.01$
                &$\pmn1.92\pm0.02$ &
	          $\pmn1.93\pm0.01$ & $\pmn1.92\pm0.01$\\
\hline									       
$\mathcal{O}_6$ & $   -0.48\pm0.03$ &
	          $   -0.47\pm0.02$ & $   -0.46\pm0.02$
                & $   -0.49\pm0.03$ &
	          $   -0.45\pm0.02$ & $   -0.48\pm0.02$\\
\hline
\end{tabular}}
\end{center}
\caption{Average values of the observables $\mathcal{O}_{1-6}$ and 
	$\bar{\mathcal{O}}_{1-6}$ at Run II of the Tevatron. The top (antitop) quark
         was required to have rapidity $|y|<2.0$ and transverse momentum 
	$p_T>15$\,GeV for the observables
	$\mathcal{O}_{1-6}$ and $\bar{\mathcal{O}}_{1-6}$, respectively.
        Spin 1 gives the results of the spin correlation algorithm using 
        Eqn.\,\ref{eqn:spin1*} and spin 2 gives the results of the spin correlation
        algorithm using Eqn.\,\ref{eqn:spin1**}.}
\label{tab:tthadtevatron}
\end{table}

  The results for these observables at Run II of the Tevatron with a centre-of-mass
  energy of 2\,TeV are shown in Table\,\ref{tab:tthadtevatron}
  where for events with \mbox{$\rm{t}\ra{\rm b}{\rm W}^+\ra {\rm b}\rm{q}\rm{\bar{q}}$}
  and \mbox{$\rm{\bar{t}}\ra{\rm \bar{b}}{\rm W}^-\ra{\rm \bar{b}}\ell^-\bar{\nu}$}
  we have required the top quark to have 
  rapidity \mbox{$|y_t|<2.0$} and transverse momentum
  \mbox{$p_T>15$\,GeV}. For the charge conjugate decay modes the
  antitop quark was required to have rapidity \mbox{$|y_{\bar{t}}|<2.0$} and
  transverse momentum \mbox{$p_T>15$\,GeV}.

  The results for the same observables at the LHC with a centre-of-mass energy of
  14\,TeV are shown in Table\,\ref{tab:tthadLHC}.
  For events with \mbox{$\rm{t}\ra{\rm b}{\rm W}^+\ra {\rm b}\rm{q}\rm{\bar{q}}$}
  and \mbox{$\rm{\bar{t}}\ra{\rm \bar{b}}{\rm W}^-\ra{\rm \bar{b}}\ell^-\bar{\nu}$}
  we have required the top quark to have rapidity 
  \mbox{$|y_t|<3.0$} and transverse momentum 
  \mbox{$p_T>20$\,GeV}. For the charge conjugate decay modes the
  antitop quark was required to have rapidity \mbox{$|y_{\bar{t}}|<3.0$}
  and transverse momentum \mbox{$p_T>20$\,GeV}.
\begin{table} 
\begin{center}
{\footnotesize
\begin{tabular}{|c|c|c|c|c|c|c|}
\hline 
&\multicolumn{3}{c|}{$\mathcal{O}$/\%}&\multicolumn{3}{c|}{$\bar{\mathcal{O}}$/\%}\\
\cline{2-7}
& Full & Spin 1& Spin 2& Full & Spin 1& Spin 2\\
\hline
$\mathcal{O}_1$ & $   -2.01\pm0.02$ &
	          $   -2.04\pm0.02$ & $   -2.04\pm0.02$ 
                & $   -2.03\pm0.02$ &
	          $   -2.04\pm0.02$ & $   -2.02\pm0.02$ \\
\hline		   	
$\mathcal{O}_2$ & $   -0.16\pm0.02$ &
	          $   -0.16\pm0.02$ & $   -0.18\pm0.02$ 
                & $   -0.18\pm0.02$ &
	          $   -0.19\pm0.02$ & $   -0.16\pm0.02$  \\
\hline		   	
$\mathcal{O}_3$ & $   -0.55\pm0.02$ &
	          $   -0.56\pm0.02$ & $   -0.57\pm0.01$ 
                & $   -0.57\pm0.02$ &
	          $   -0.59\pm0.02$ & $   -0.57\pm0.01$ \\
\hline		   	
$\mathcal{O}_4$ & $\pmn0.00\pm0.02$ &
	          $   -0.01\pm0.01$ & $   -0.02\pm0.01$ 
                & $   -0.02\pm0.02$ &
	          $\pmn0.00\pm0.01$ & $\pmn0.01\pm0.01$ \\
\hline		   	
$\mathcal{O}_5$ & $   -0.23\pm0.02$ &
	          $   -0.25\pm0.01$ & $   -0.25\pm0.01$ 
                & $   -0.26\pm0.02$ &
	          $   -0.28\pm0.01$ & $   -0.25\pm0.01$ \\
\hline		   	
$\mathcal{O}_6$ & $   -1.85\pm0.02$ &
	          $   -1.87\pm0.02$ & $   -1.86\pm0.01$ 
                & $   -1.85\pm0.02$ &
	          $   -1.85\pm0.02$ & $   -1.85\pm0.01$ \\
\hline
\end{tabular}}
\end{center}
\caption{Average values of the observables $\mathcal{O}_{1-6}$ and 
	$\bar{\mathcal{O}}_{1-6}$ at the LHC. The top (antitop) quark
         was required to have rapidity $|y|<3.0$ and 
        transverse momentum $p_T>20$\,GeV for the observables
	$\mathcal{O}_{1-6}$ and $\bar{\mathcal{O}}_{1-6}$, respectively.
        Spin 1 gives the results of the spin correlation algorithm using 
        Eqn.\,\ref{eqn:spin1*} and spin 2 gives the results of the spin correlation
        algorithm using Eqn.\,\ref{eqn:spin1**}.}
\label{tab:tthadLHC}
\end{table}

  All the results were generated using a top quark mass of \mbox{175\,GeV} 
  and the default parton distribution functions in HERWIG6.3 \cite{Corcella:2001pi}
  which are the average of the central and higher gluon leading-order fits
  of \cite{Martin:1998np}.

  As can be seen in both Tables\,\ref{tab:tthadtevatron}~and~\ref{tab:tthadLHC}
  the results of the spin correlation algorithm, with either method
  of handling the different colour flows, are in good agreement with 
  the result of the 6-body matrix element. 
  The default result from HERWIG
  for all these observables was consistent with zero. This is because these observables
  are designed to be sensitive to the spin correlations and zero in the absence of such 
  correlations.

 At the Tevatron the most significant of these observables is
  $\mathcal{O}_2$ whereas the observable $\mathcal{O}_6$ is the most important
  at the LHC. At the LHC 
  this is because while the average value of $\mathcal{O}_1$ is greater 
  the fluctuations are also greater which gives a greater statistical sensitivity
  for $\mathcal{O}_6$ \cite{Brandenburg:1996df,Brandenburg:1996wu}.

%
%  chargino pair in e+e-
%
\subsection[$\rm{e^+e^-\ra \tilde{\chi}^+\tilde{\chi}^-}$]
	{\boldmath{$\rm{e^+e^-\ra \tilde{\chi}^+\tilde{\chi}^-}$}}

\begin{figure}[h!!]
\begin{center}
\includegraphics[width=0.45\textwidth,angle=90]{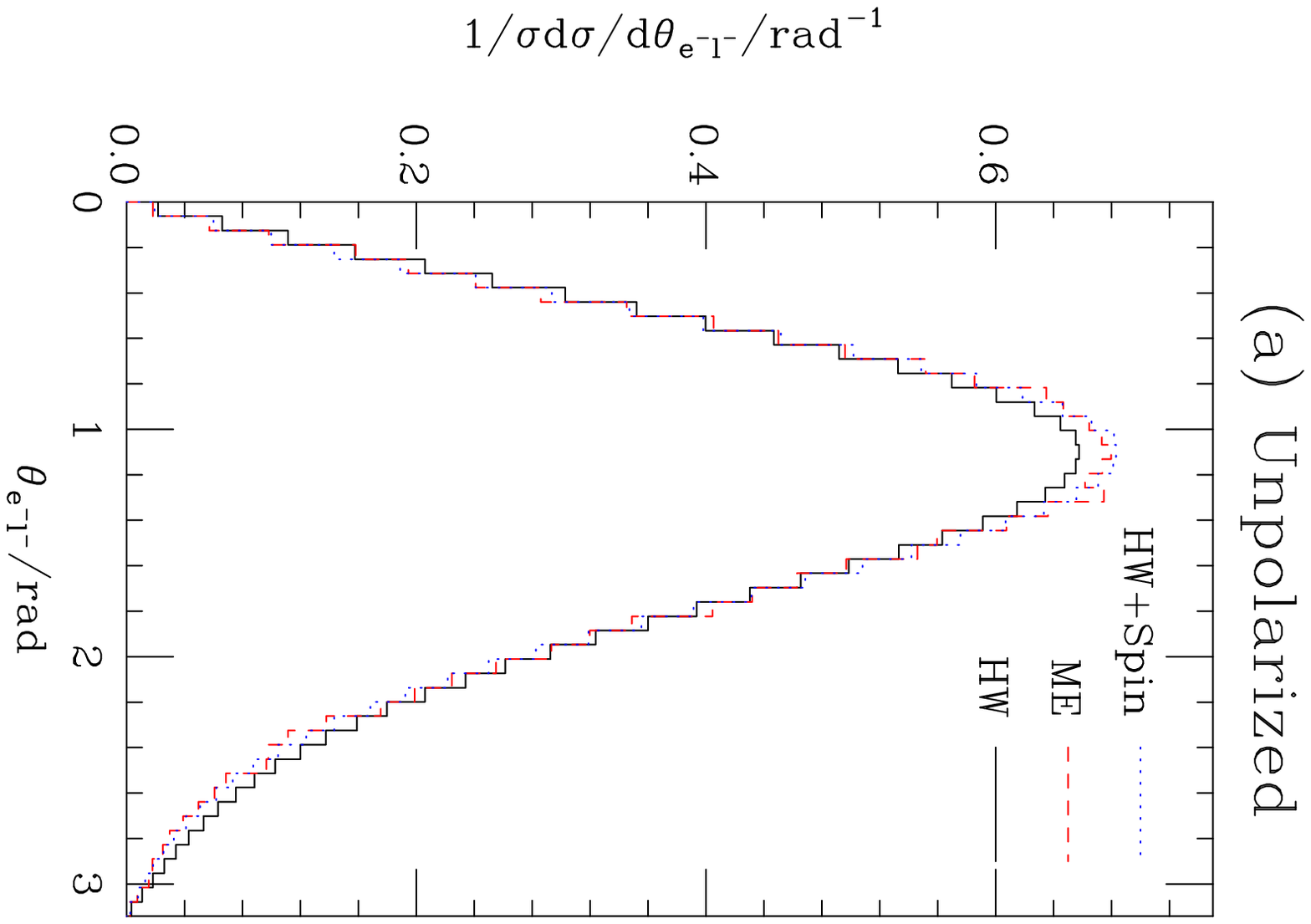}\hfill
\includegraphics[width=0.45\textwidth,angle=90]{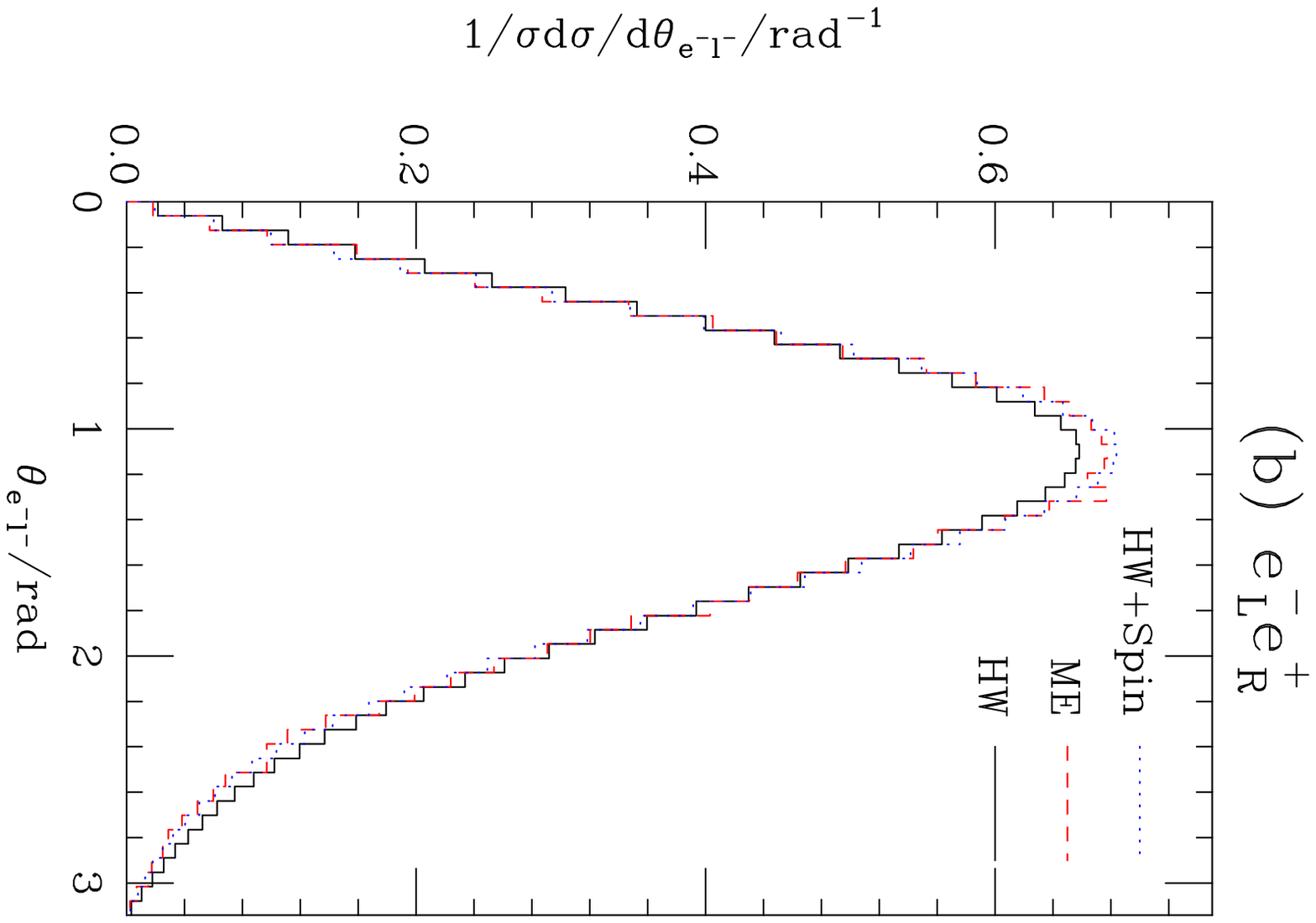}\hfill
\includegraphics[width=0.45\textwidth,angle=90]{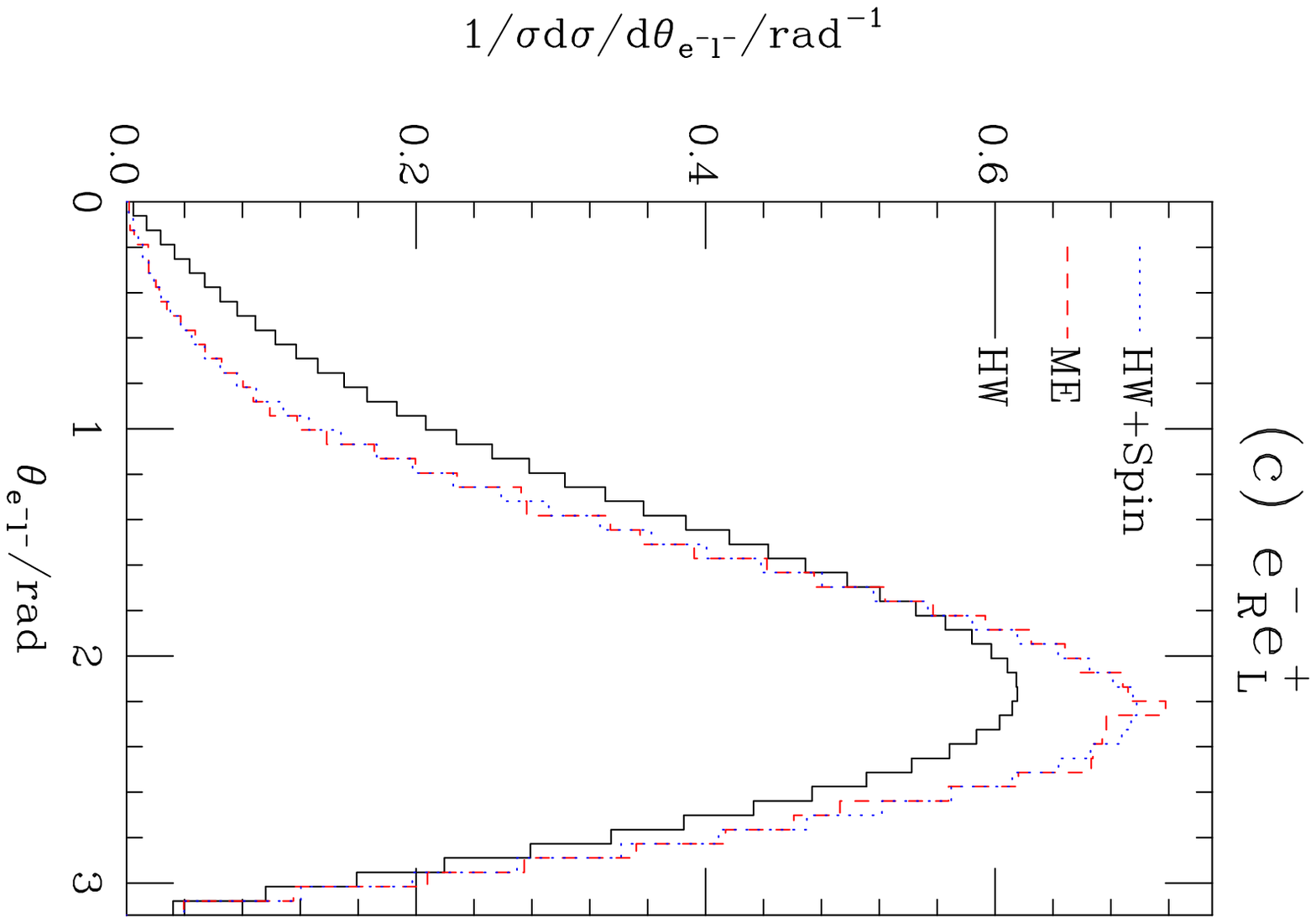}
\caption{Angle between the lepton produced in
 	\mbox{$\rm{e}^+\rm{e}^-\ra\cht^+_1\cht^-_1\ra\ell^+\nu_\ell\cht^0_1
	    \ell^-\bar{\nu}_\ell\cht^0_1$} and the incoming electron beam 
	in the laboratory frame for a centre-of-mass energy of
	500\gev\  with {\bf(a)} no polarization, {\bf(b)} negatively polarized electrons 
        and positively polarized positrons 
	and {\bf(c)} positively polarized electrons and negatively polarized positrons.
        The solid line shows the default result from HERWIG,
	the dashed line gives the full result from the 6-body
	matrix element and the dotted line the result of the spin correlation algorithm.}
\label{fig:chi+11eangle}
\end{center}
\vspace{1.cm}
%\end{figure}
%\begin{figure}
\begin{center}
\includegraphics[width=0.45\textwidth,angle=90]{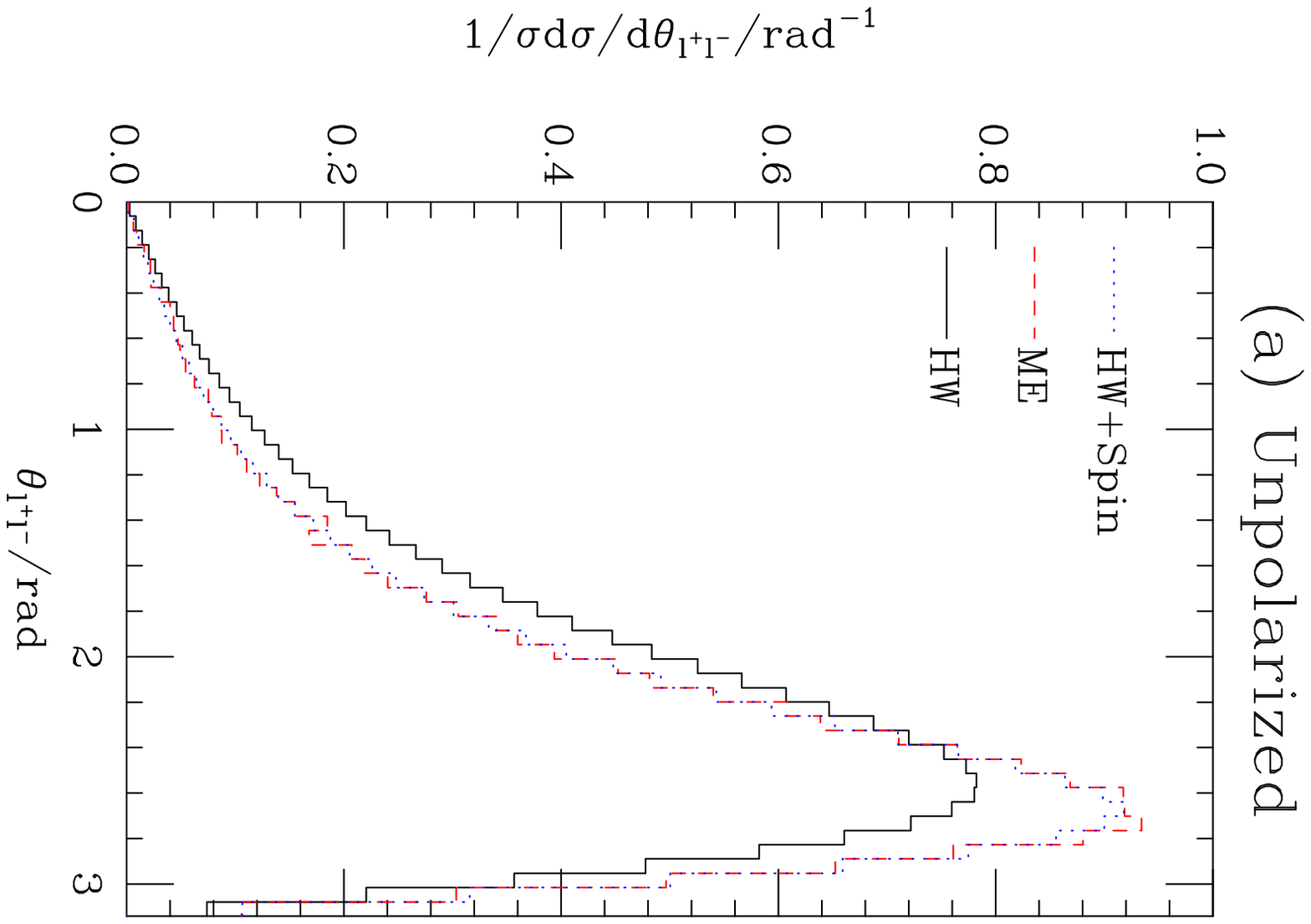}\hfill
\includegraphics[width=0.45\textwidth,angle=90]{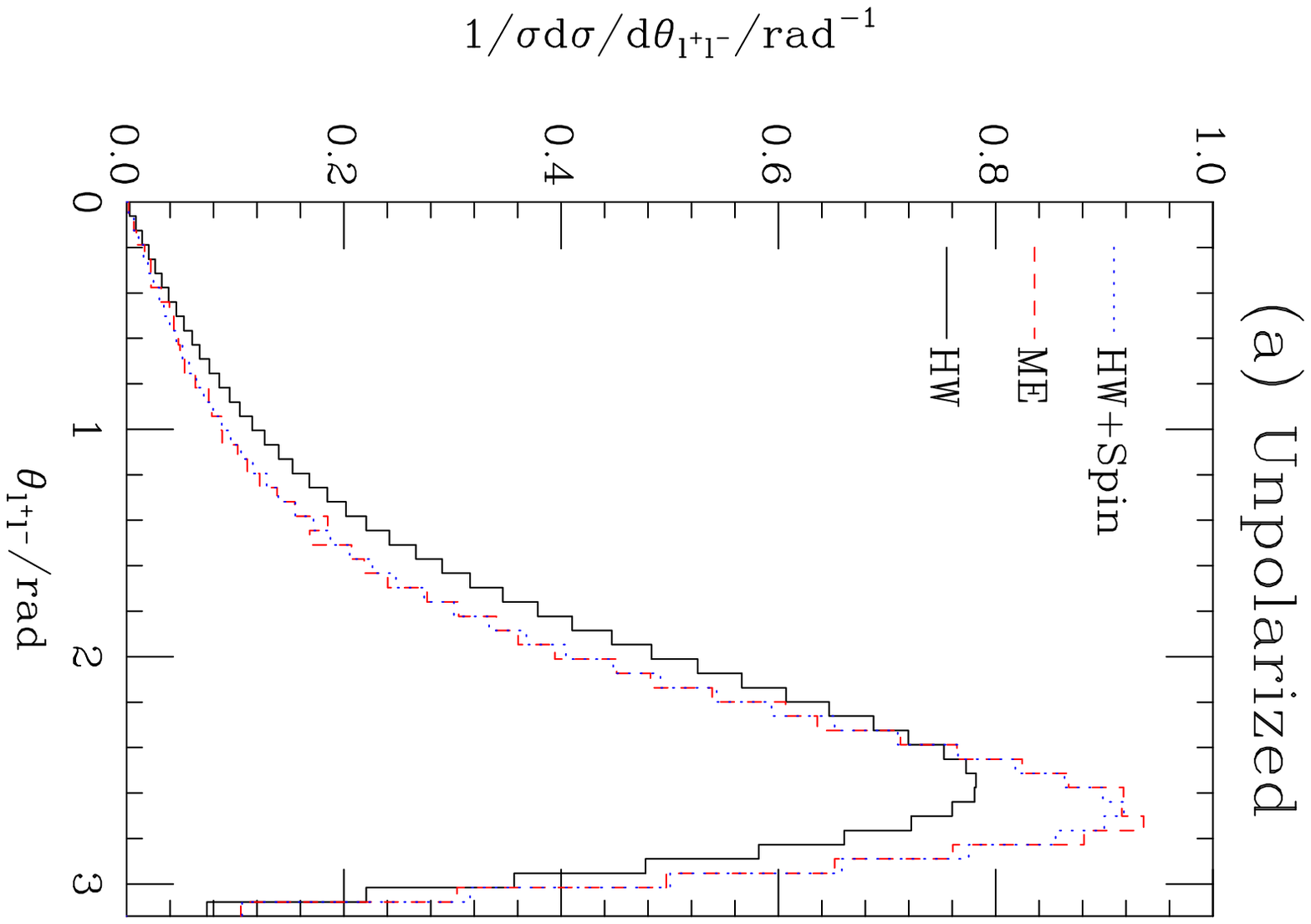}\hfill
\includegraphics[width=0.45\textwidth,angle=90]{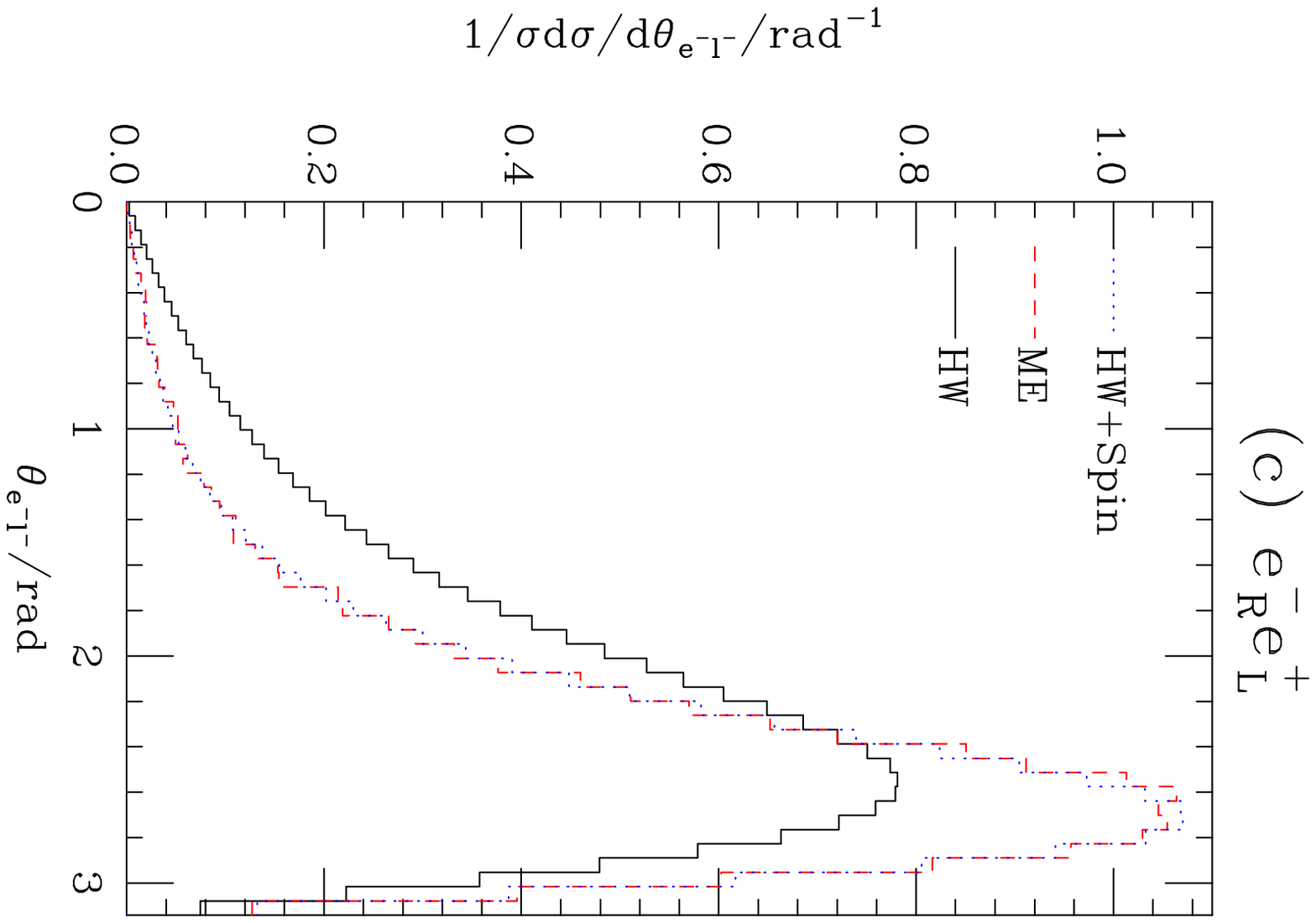}
\caption{Angle between the leptons produced in
 	\mbox{$\rm{e}^+\rm{e}^-\ra\cht^+_1\cht^-_1\ra\ell^+\nu_\ell\cht^0_1
	    \ell^-\bar{\nu}_\ell\cht^0_1$} in the laboratory frame for a centre-of-mass
         energy of
	500\gev\  with {\bf(a)} no polarization, {\bf(b)} negatively polarized electrons 
        and positively polarized positrons 
	and {\bf(c)} positively polarized electrons and negatively polarized positrons.
        The solid line shows the default result from HERWIG,
	the dashed line gives the full result from the 6-body
	matrix element and the dotted line the result of the spin correlation algorithm.}
\label{fig:chi+11e+e-angle}
\end{center}
\vspace{-1.5cm}
\end{figure}

  So far we have only considered processes in which the fermions have 
  small widths relative to their masses, the next-to-lightest neutralino
  in most SUSY models has a width of less than one hundred MeV and the top
  quark width is about one GeV. We can therefore neglect the
  width in top quark and neutralino production processes.
  In the limit that the width of the decaying fermion can be neglected
  the helicity amplitudes for the production processes, including the fermion
  decays, can be factorized into a production and decay helicity amplitude.
  The spin correlation algorithm is in good agreement with the full
  matrix element in these cases.

  However, if the decaying fermion is not on mass-shell the process cannot
  be factorized and the result of the spin correlation algorithm
  may not give such good agreement with the $2\to n$ body matrix elements.
  This should not be a serious problem because even in SUSY models the 
  electroweak gauginos tend to have widths of at most a few GeV, for the
  heavier gauginos. The gluino width tends to be larger, but the gluino is heavier,
  and therefore the effect is the same.
  The particles which usually have the largest widths in most SUSY models are
  the squarks and these effects can be taken into account because they are scalars.

  In order to study the effects of the widths of the decaying fermions
  it is easiest to study chargino production in ${\rm e}^+{\rm e}^-$ collisions
  because this allows us to consider the production of $\cht^\pm_1\cht^\mp_2$
  in which the lighter chargino has a very small width and the heavier chargino
  has a much greater width. Spin and polarization effects in chargino
  pair production in ${\rm e}^+{\rm e}^-$ collisions 
  have been previously considered in 
  \cite{Moortgat-Pick:1998sk,Moortgat-Pick:2000uz}.

  We used the following SUSY parameters in this study:
  \mbox{$M_1=78$\,GeV}, \linebreak\mbox{$M_2=158$\,GeV}, $\mu=330$\,GeV, $\tan\beta=3$,
  $M_{L_L}=171$\,GeV and $M_{e_R}=180$\,GeV, where
  the soft SUSY breaking masses for the gauginos and the $\mu$ parameter
  are given at the electroweak scale,
  $M_{L_L}$ is the soft SUSY breaking mass for the lepton doublet at the electroweak
  scale and $M_{e_R}$ is the soft SUSY breaking mass for the right-handed electron
  singlet at the electroweak scale.

  At this point the lightest neutralino mass is $M_{\cht^0_1}=71.6$\,GeV,
  the lightest chargino mass is $M_{\cht^+_1}=127.8$\,GeV, the
  heaviest chargino mass is $M_{\cht^+_2}=357.6$\,GeV,
  the sneutrino mass is $M_{\nut_L}=160.9$\,GeV,
  the left selectron mass is $M_{\elt_L}=176.1$\,GeV and
  the right selectron mass is $M_{\elt_R}=184.2$\,GeV.
  The width of the lightest chargino is $\Gamma_{\cht^+_1}=73$\,MeV 
  and the heaviest chargino is $\Gamma_{\cht^+_2}=2.9$\,GeV.

  This point was chosen to have a negligible lightest chargino width and
  a variety of leptonic 
  decay modes, \ie $\cht^+_2\ra\ell^+\nut_L\ra\ell^+\ell^-\cht^0_1$,
                   $\cht^+_2\ra\elt^+_L\nu\ra\ell^+\ell^-\cht^0_1$
  and $\cht^+_2\ra\cht^0_1W^+\ra\ell^+\ell^-\cht^0_1$, of the
  heaviest chargino.

  It is interesting to first consider $\cht^+_1\cht^-_1$ production, where
  there are no width effects. Fig.\,\ref{fig:chi+11eangle} shows the
  angle between the lepton produced in 
  ${\rm e}^+{\rm e}^-\ra\cht^+_1\cht^-_1\ra\cht^0_1\ell^+\nu\cht^0_1\ell^-\bar{\nu}$
  and the beam direction and Fig.\,\ref{fig:chi+11e+e-angle} shows the
  angle between the lepton and the antilepton. As before
  there is good agreement between the full result and the spin correlation 
  algorithm.

  In a Monte Carlo simulation however we need to include the widths of the
  unstable particles as these widths can potentially be measured.
  When we include the effect of the width in the spin correlation algorithm
  we have two choices. We can use the off-mass shell momentum in 
  Eqn.\,\ref{eqn:spinordef} for the spinor of the massive particle together
  with either the physical mass of the particle or the off-shell mass of the 
  particle.

  In the first case this gives
\begin{equation}
\sum_{\lam}u_\lam(p)\bar{u}_\lam(p) = p\sla+m +\frac{l\sla}{2p\cdot l}(m^2-p^2),
\end{equation}
  where $p$ is the off-shell four-momentum, $m$ is the on-shell mass and
  $l$ is the reference vector. This form reduces to the correct spin sum if the
  particle is on-mass shell. However for off-mass shell particles
  there is an additional component which depends on the choice of the
  reference vector and is related to how off-mass shell the particle is.

\begin{figure}
\begin{center}
\includegraphics[width=0.45\textwidth,angle=90]{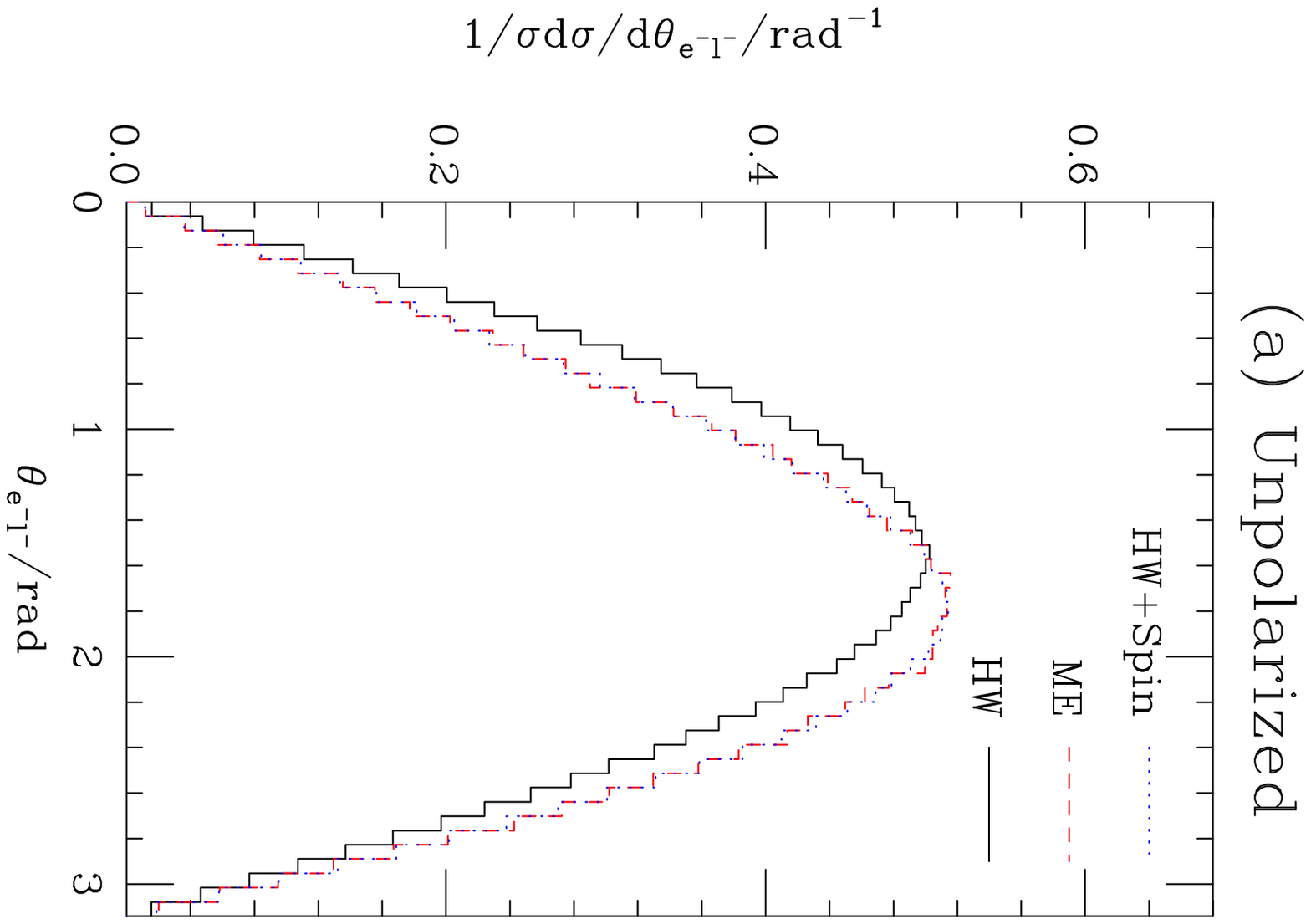}\hfill
\includegraphics[width=0.45\textwidth,angle=90]{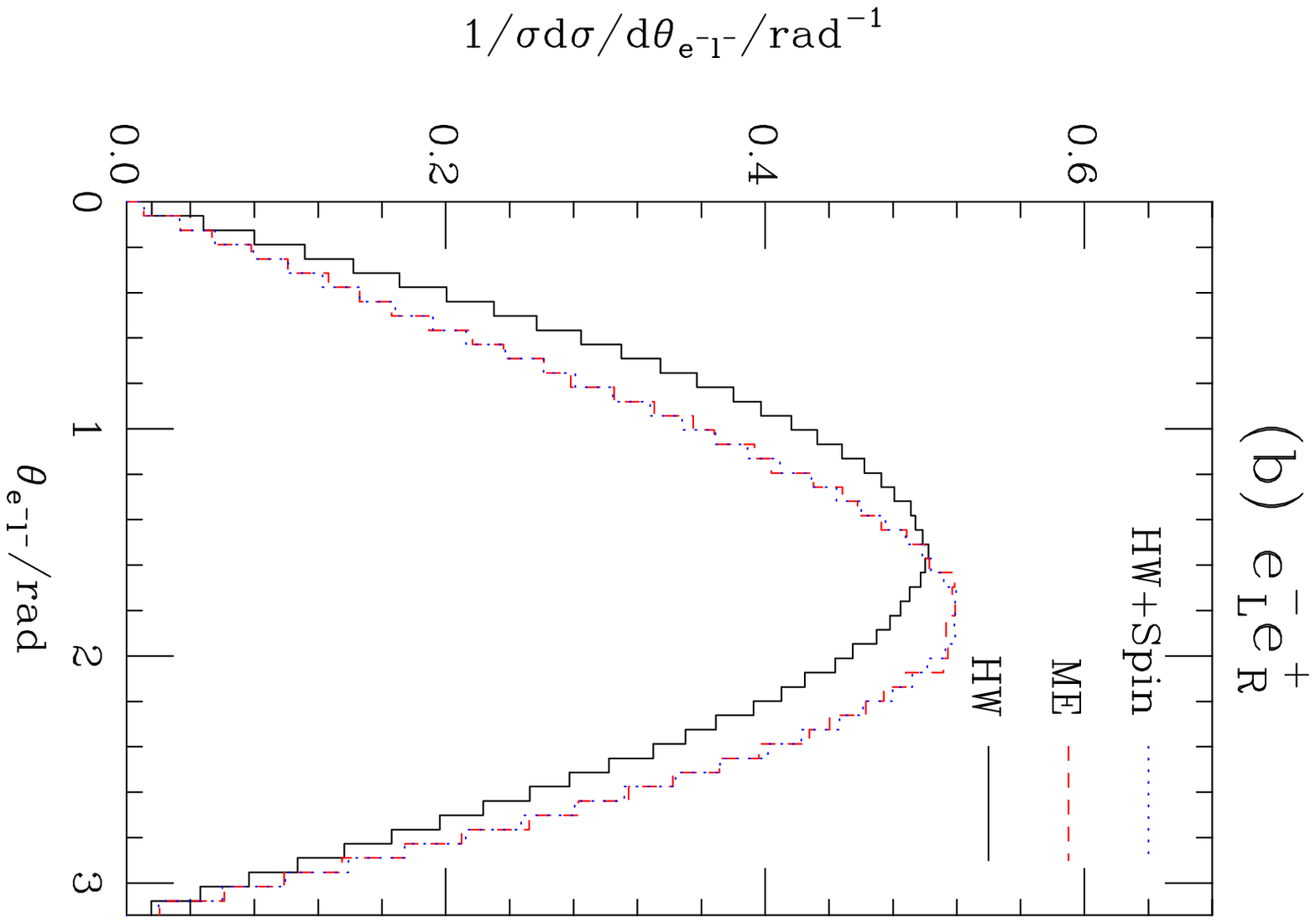}\hfill
\includegraphics[width=0.45\textwidth,angle=90]{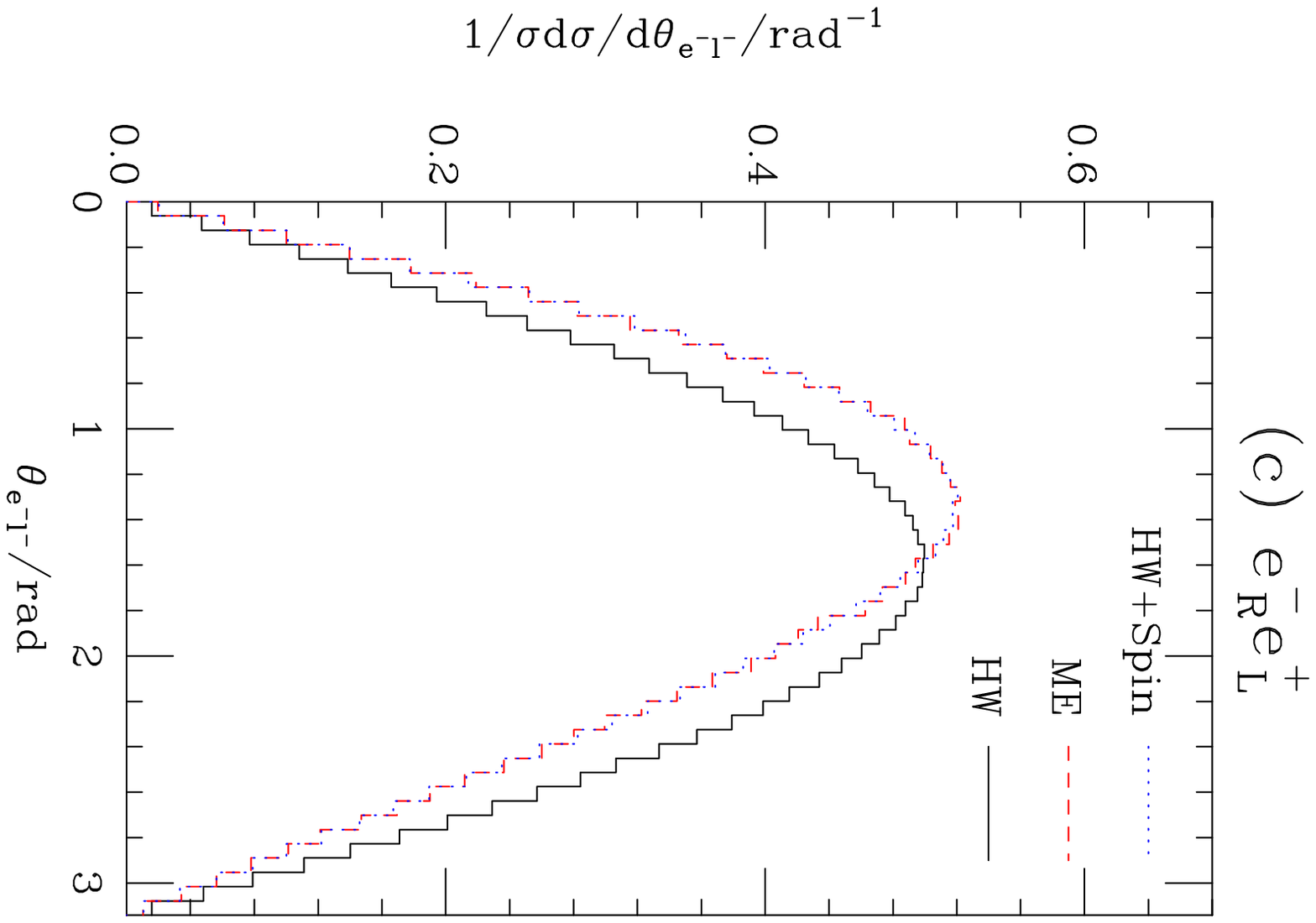}
\caption{Angle between the lepton produced in
 	\mbox{$\rm{e}^+\rm{e}^-\ra\cht^+_1\cht^-_2\ra\ell^+\nu_\ell\cht^0_1
	    \ell^-\bar{\nu}_\ell\cht^0_1$} and the incoming electron beam 
	in the laboratory frame for a centre-of-mass energy of
	500\gev\  with {\bf(a)} no polarization, {\bf(b)} negatively polarized electrons 
        and positively polarized positrons 
	and {\bf(c)} positively polarized electrons and negatively polarized positrons.
        The solid line shows the default result from HERWIG,
	the dashed line gives the full result from the 6-body
	matrix element and the dotted line the result of the spin correlation algorithm.}
\label{fig:widthA}
\end{center}
\end{figure}

\begin{figure}[h!!]
\begin{center}
\includegraphics[width=0.45\textwidth,angle=90]{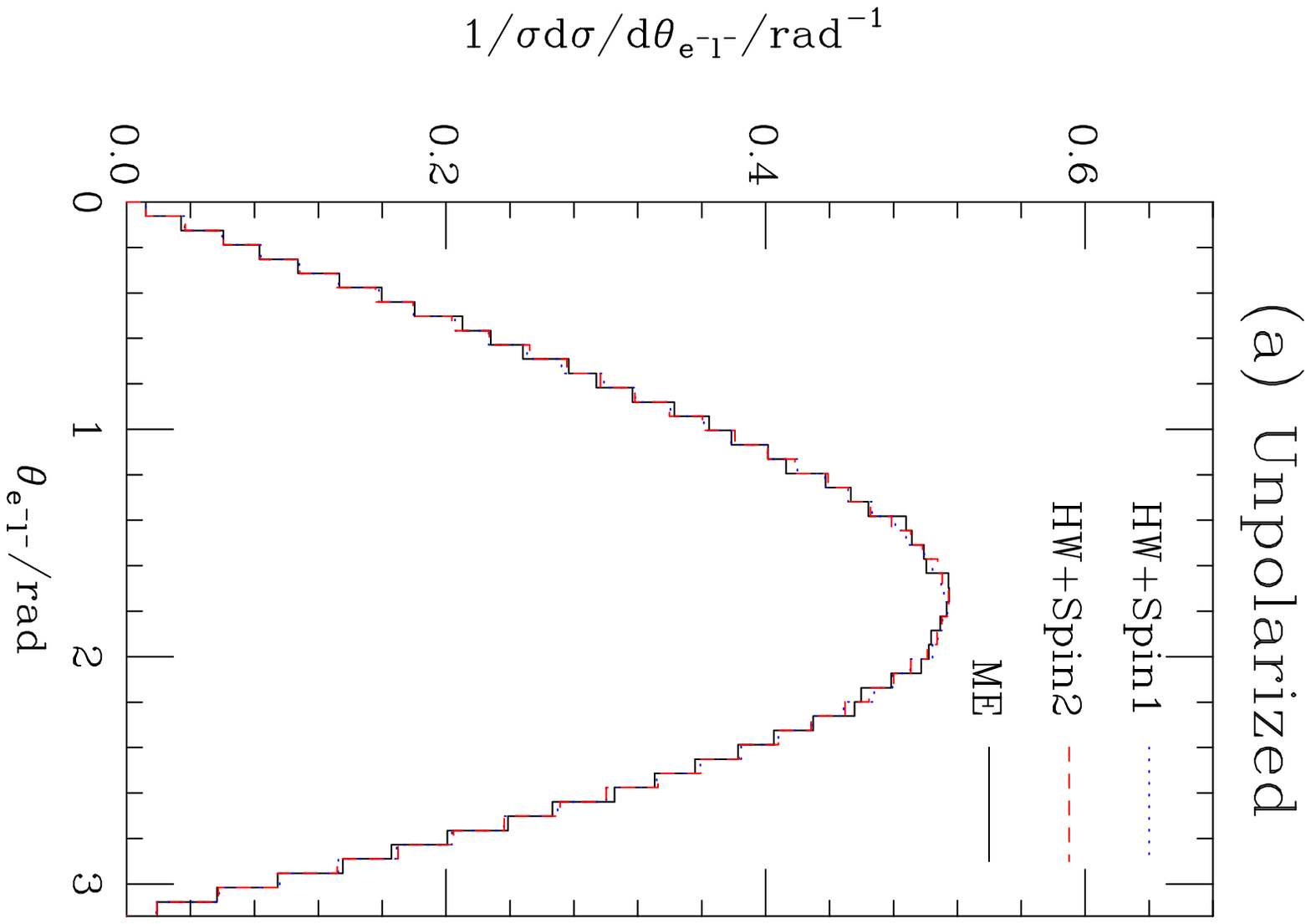}\hfill
\includegraphics[width=0.45\textwidth,angle=90]{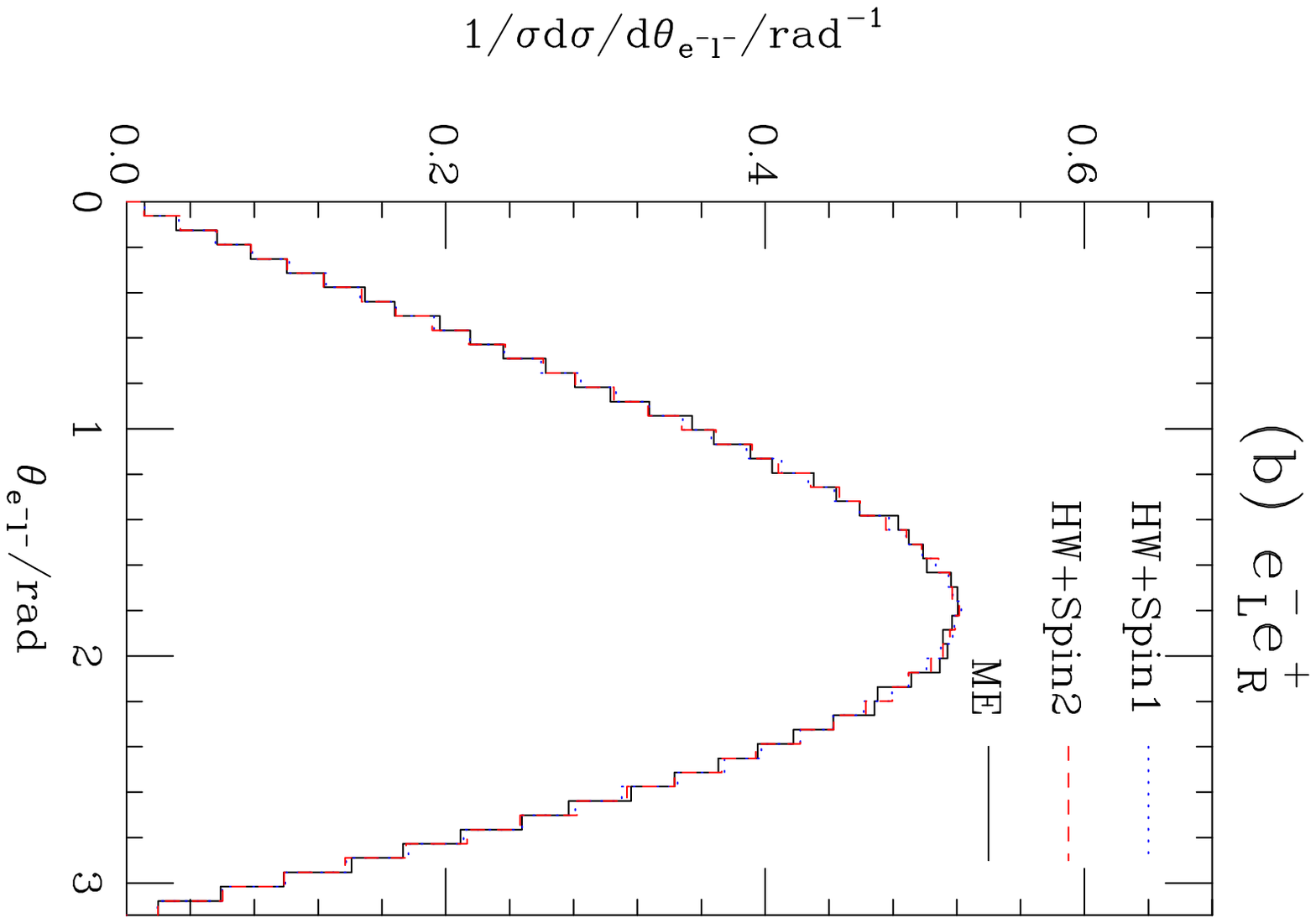}\hfill
\includegraphics[width=0.45\textwidth,angle=90]{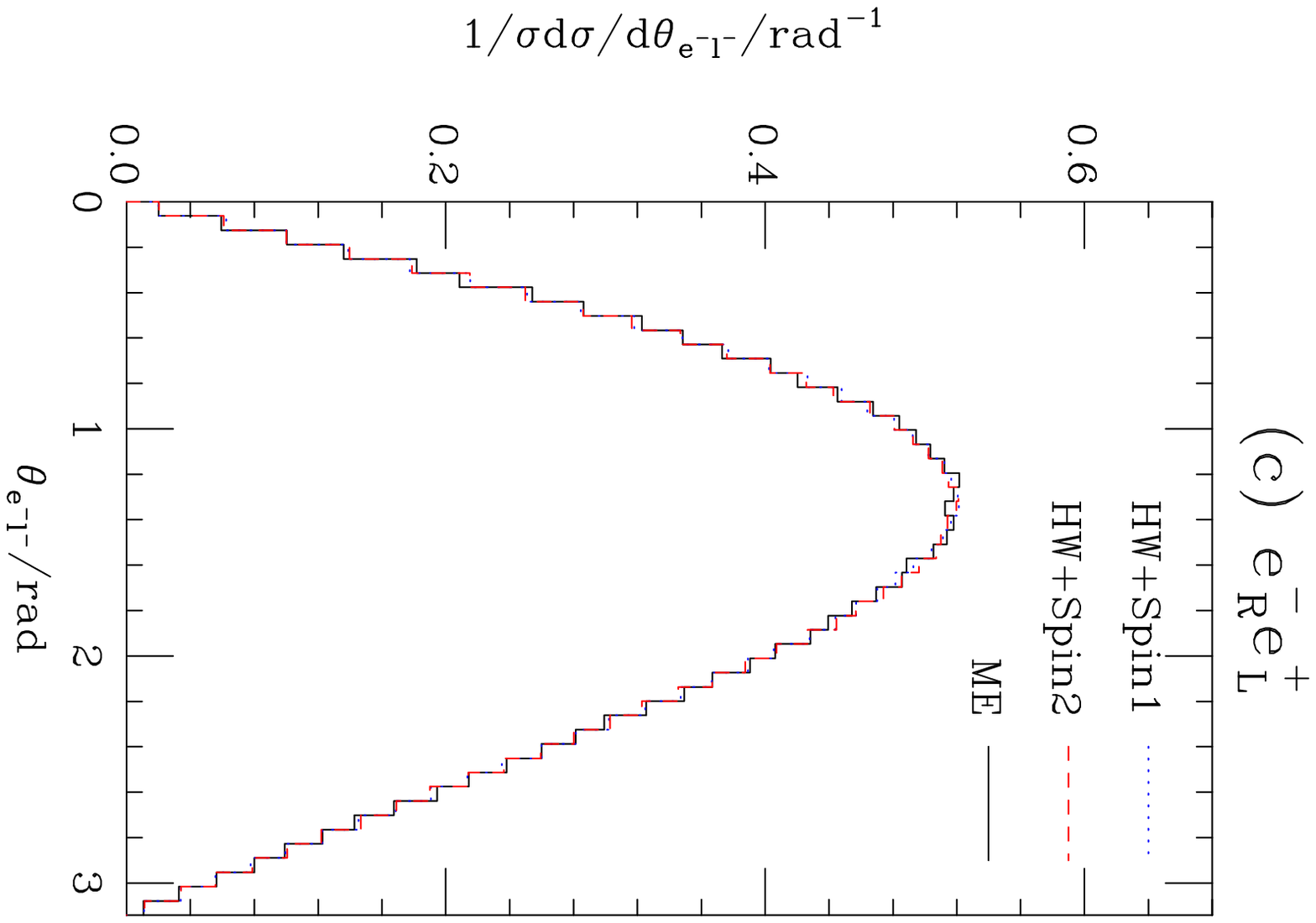}
\caption{Angle between the lepton produced in
 	\mbox{$\rm{e}^+\rm{e}^-\ra\cht^+_1\cht^-_2\ra\ell^+\nu_\ell\cht^0_1
	    \ell^-\bar{\nu}_\ell\cht^0_1$} and the incoming electron beam 
	in the laboratory frame for a centre-of-mass energy of
	500\gev\  with {\bf(a)} no polarization, {\bf(b)} negatively polarized electrons 
        and positively polarized positrons 
	and {\bf(c)} positively polarized electrons and negatively polarized positrons.
        The physical $\cht^\pm_2$ width was used.
        The solid line shows the full result, the dotted line shows the result of the
        spin correlation algorithm with the physical mass and the
        dashed line shows the result of the spin correlation algorithm with the
        off-shell mass of the chargino.}
\label{fig:widthB}
\end{center}
\begin{center}
\includegraphics[width=0.45\textwidth,angle=90]{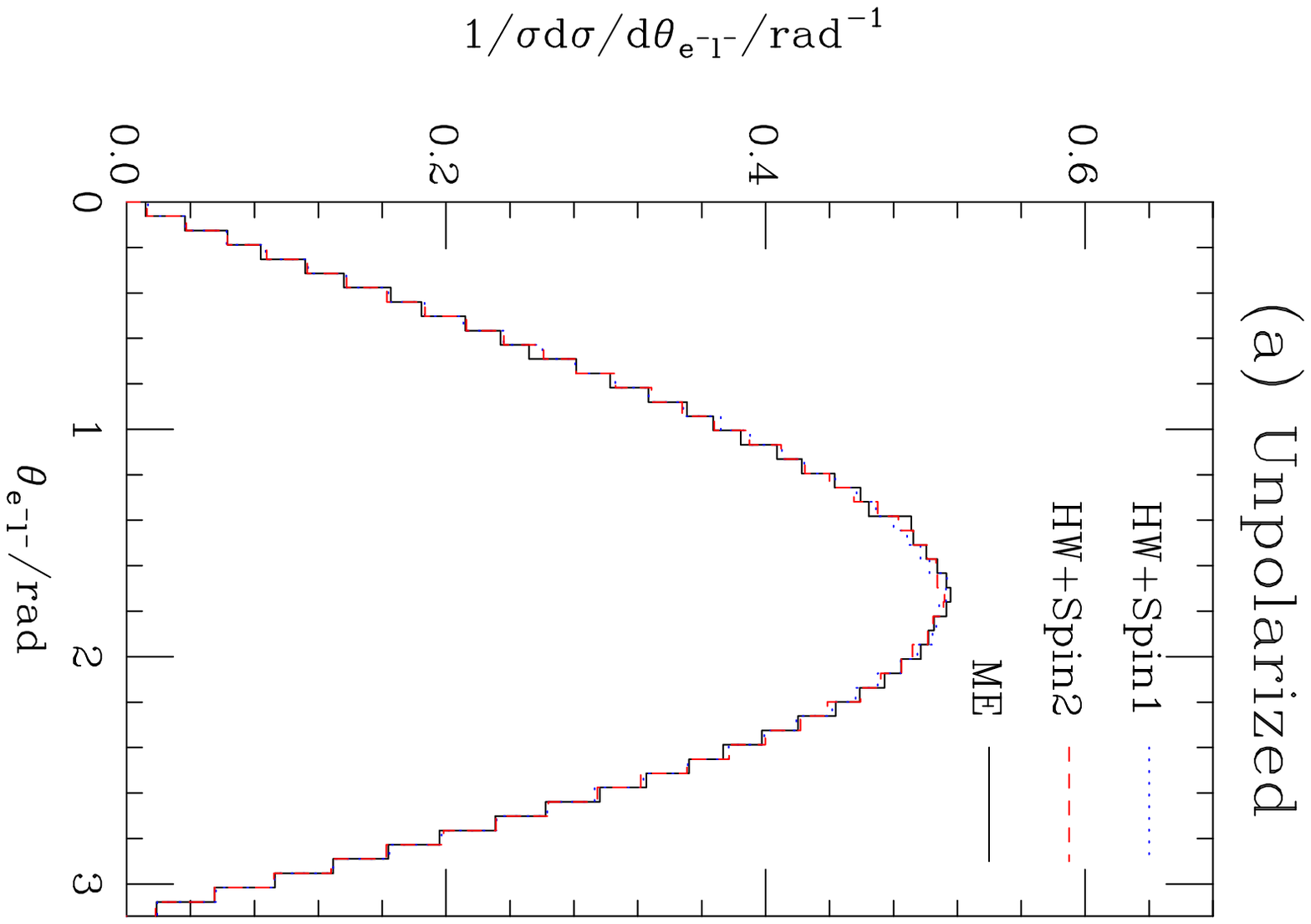}\hfill
\includegraphics[width=0.45\textwidth,angle=90]{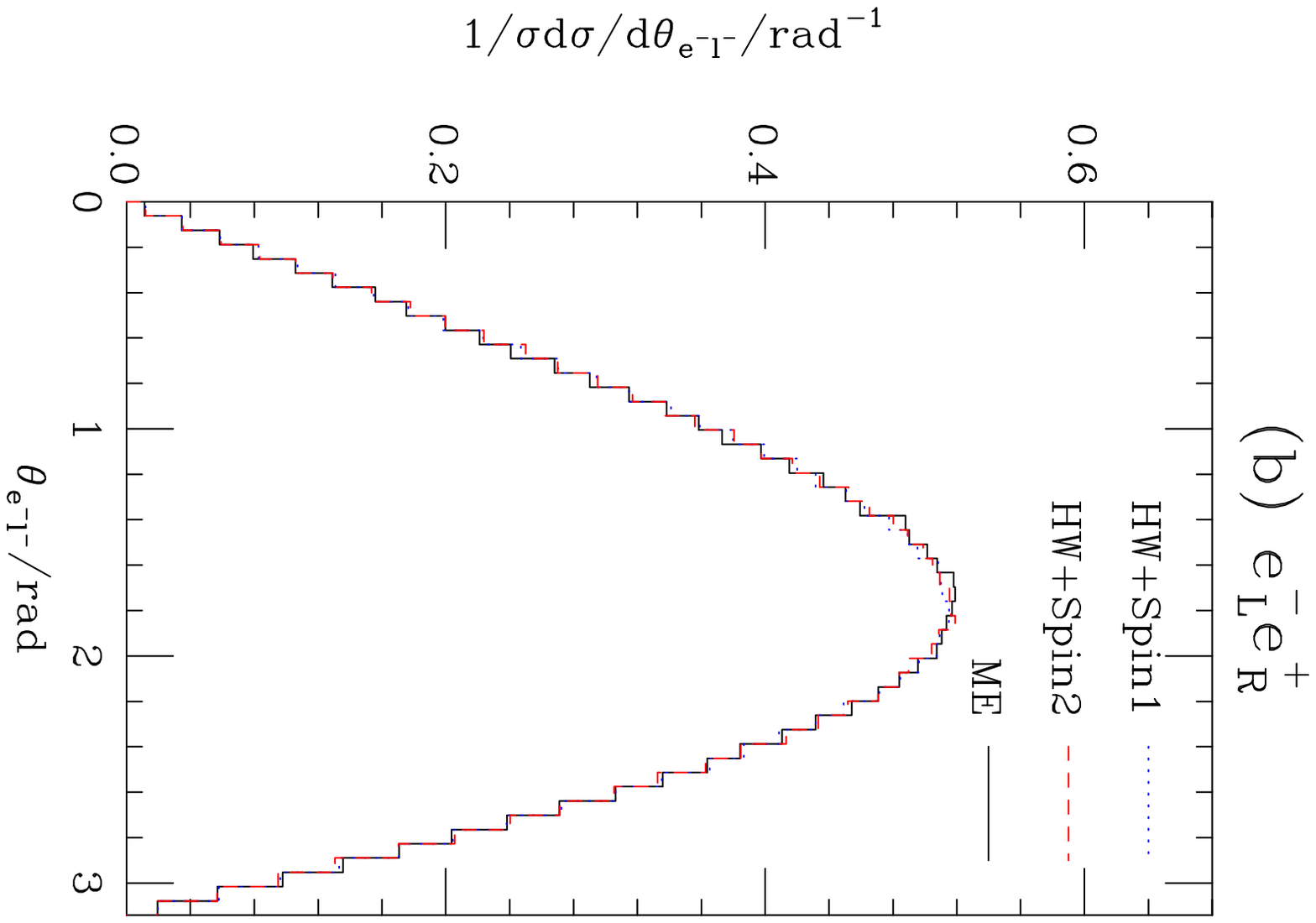}\hfill
\includegraphics[width=0.45\textwidth,angle=90]{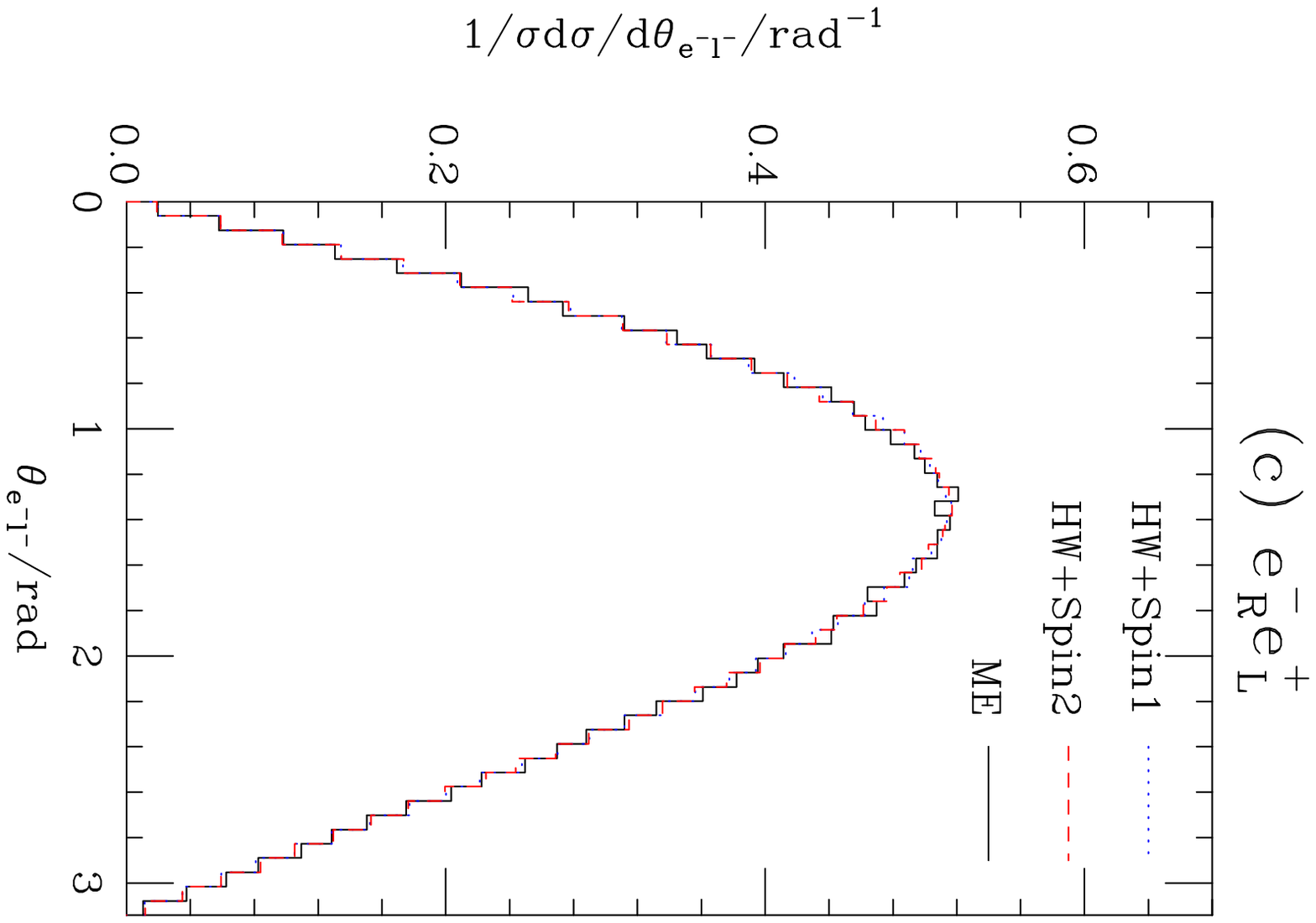}
\caption{Angle between the lepton produced in
 	\mbox{$\rm{e}^+\rm{e}^-\ra\cht^+_1\cht^-_2\ra\ell^+\nu_\ell\cht^0_1
	    \ell^-\bar{\nu}_\ell\cht^0_1$} and the incoming electron beam 
	in the laboratory frame for a centre-of-mass energy of
	500\gev\  with {\bf(a)} no polarization, {\bf(b)} negatively polarized electrons 
        and positively polarized positrons 
	and {\bf(c)} positively polarized electrons and negatively polarized positrons.
        Ten times the physical $\cht^\pm_2$ width was used.
        The lines are described in the caption on Fig.\,\ref{fig:widthB}.}
\label{fig:widthC}
\end{center}
\end{figure}

  If we choose to use the off-shell mass rather than the physical mass in 
  Eqn.\,\ref{eqn:spinordef} we obtain
\begin{equation}
\sum_{\lam}u_\lam(p)\bar{u}_\lam(p) = p\sla+m_{\rm off},
\end{equation}
  where $m_{\rm off}$ is the off-shell mass. Here the result again fails to reproduce
  the correct result by an amount which depends on the how far off-shell the particle
  is, however it does not depend on the arbitrary reference vector.
  We have studied the effects of both these choices.

  We considered $\cht^\pm_1\cht^\mp_2$ production followed by the decay
  of the lightest chargino $\cht^\pm_1\ra\ell^\pm\nu\cht^0_1$ via a three-body decay
  and heaviest chargino
  $\cht^\pm_1\ra {\rm W}^\pm\cht^0_1\ra\ell^\pm\nu\cht^0_1$.
  The angle of the lepton produced in the $\cht^\pm_2$ decay with respect to the
  beam direction is shown in Fig.\,\ref{fig:widthA} where we have neglected
  the widths of the decaying charginos.
  
  We then considered the effect of including the width of the decay chargino.
  Fig.\,\ref{fig:widthB} shows the result for the physical width of the
  heaviest chargino and Fig.\,\ref{fig:widthC} shows the effect of
  increasing the width by a factor of ten.

  These results show that there is still good agreement between the full result
  and the spin correlation algorithm even when particles with fairly large widths
  are included, despite the potential problems. The result does not significantly
  depend on the choice of the spinor for the unstable particle and therefore
  we will use the second choice in order to avoid any dependence on the reference
  vector used to define the particle's spin.

%
%   section on tau decays
%
\section{Tau Decays}

  So far we have only discussed spin correlations in the SM for the top quark 
  which is the only quark which decays before hadronization.
  However spin correlations can also be important for the decay of the tau.
  The tau will undergo a weak decay to a tau neutrino and
  either an electron/muon and its associated antineutrino or hadrons. Due to
  the relatively small mass of the tau only a small number of hadrons
  can be produced in its decay.
  
  The matrix elements for the leptonic decay of the tau and many of the
  hadronic decay modes are known, either theoretically or from
  experimental measurements of the distributions of the decay products. The
  helicity of the decaying tau can affect the distribution of the decay products.
  There is a package TAUOLA \cite{Jadach:1993hs} which includes the matrix elements
  for the leptonic and many of the hadronic modes and includes the helicity
  of the decaying tau.

  In order to implement the algorithm we have suggested in full for tau decays we would
  need to completely rewrite the TAUOLA package to use the spin density matrices
  for the decays and return the decay matrix after performing the decays.
  Given the number of decay modes and the sophisticated treatment of the
  decays in TAUOLA this would be a major project and is beyond the scope of this 
  study. 

  However as TAUOLA makes use of the helicity of the decay tau to perform the decays
  we can use the diagonal entries of the spin density matrices, which
  give the probability the tau has a given helicity, to select the helicity of the
  decaying tau.

%
%  Interface with the parton-shower
%
\section{Including Correlations with the Parton Shower}
\label{sect:parton}

  Given the similarities between the algorithm we are using in order to
  include spin correlations from heavy particle decays and the algorithm of 
  \cite{Collins:1988cp,Knowles:1988cu,Knowles:1988hu,Knowles:1988vs}
  for spin correlations in the QCD parton shower it should be possible
  to produce an algorithm which includes both effects, \ie correctly 
  includes  correlations from both heavy particle decays and in the QCD shower.

  In order to construct such an algorithm it is helpful to first
  review the algorithm of 
  \cite{Collins:1988cp,Knowles:1988cu,Knowles:1988hu,Knowles:1988vs} for
  correlations in the QCD parton shower. The full algorithm is given for
  both forward and backward evolution in \cite{Knowles:1988hu}.
  The algorithm proceeds in the following way:
\begin{enumerate}
\item The momenta of the particles in the hard collision process are generated
	according to Eqn.\,\ref{eqn:spin1}.
\item One of the outgoing partons is chosen at random and given a spin density matrix
      according to Eqn.\,\ref{eqn:spin2}.
\item The type of the branching, $i\to j+k$, together with
      the azimuthal angle $\phi$ and momentum fraction $z$, are generated according
      to 
\begin{equation}
\rho^i_{\lam_i\lam'_i}V^{jk}_{\lam_i\lam_j\lam_k}(z,\phi)
		      V^{jk*}_{\lam'_i\lam'_j\lam'_k}(z,\phi)
	D^j_{\lam_j\lam'_j}D^k_{\lam_k\lam'_k},	
\label{eqn:QCD1}
\end{equation} 
   	where $V^{jk}_{\lam_i\lam_j\lam_k}(z,\phi)$ is the splitting function
	for the branching given the helicities, $\lam_i,\lam_j,\lam_k$ of the 
	partons, $\rho^i_{\lam_i\lam'_i}$ is the spin density matrix of the
	parton entering the vertex and $D^j_{\lam_j\lam'_j}$ is the
	decay matrix for the partons leaving the vertex.
	As before in the first stage of the algorithm the decay matrix is taken
	to be $D^j_{\lam_j\lam'_j}=\delta_{\lam_j\lam'_j}$.

\item One of the outgoing partons is selected to be developed and a spin
	density matrix calculated for this parton using 
\begin{equation}
\rho^j_{\lam_j\lam'_j} = \frac1{N_{\rho\rm{QCD}}}
			\rho^i_{\lam_i\lam'_i}V^{jk}_{\lam_i\lam_j\lam_k}(z,\phi)
		      V^{jk*}_{\lam'_i\lam'_j\lam'_k}(z,\phi)
	D^k_{\lam_k\lam'_k},
\label{eqn:QCD2}
\end{equation}
        where the normalization
\begin{equation}
N_{\rho\rm{QCD}} = \rho^i_{\lam_i\lam'_i}V^{jk}_{\lam_i\lam_j\lam_k}(z,\phi)
		      V^{jk*}_{\lam'_i\lam_j\lam'_k}(z,\phi)D^k_{\lam_k\lam'_k},
\end{equation}
	is chosen so that the spin density matrix has unit trace.
 	This procedure is repeated until the parton to be developed
	reaches the cut-off scale. As before the decay matrix for this parton is
 	then set to $\delta_{\lam_i\lam'_i}$.
 
\item Another parton from the branching which produced the parton which 
	reached the cut-off scale is now developed with  a spin
	density matrix calculated for this parton using Eqn.\,\ref{eqn:QCD2}.
	When all the outgoing partons in a branching have been developed
        a decay matrix for that branching is calculated using
\begin{equation}
 D^i_{\lam_i\lam'_i} = \frac1{N_{D{\rm QCD}}}V^{jk}_{\lam_i\lam_j\lam_k}(z,\phi)
		      V^{jk*}_{\lam'_i\lam'_j\lam'_k}(z,\phi)
	D^j_{\lam_j\lam'_j}D^k_{\lam_k\lam'_k},
\label{eqn:QCD3}
\end{equation}
        where the normalization
\begin{equation}
N_{D{\rm QCD}} = V^{jk}_{\lam_i\lam_j\lam_k}(z,\phi)
		      V^{jk*}_{\lam_i\lam'_j\lam'_k}(z,\phi)
	D^j_{\lam_j\lam'_j}D^k_{\lam_k\lam'_k}
\end{equation}
    is chosen so that the trace of the decay matrix is one.
    The other outgoing partons from the previous branching are now developed
        with a spin density matrix calculated using Eqn.\,\ref{eqn:QCD2} and
		the calculated decay matrix
        rather than the identity for those partons which have been developed.
\item   This procedure is repeated until the hard process is reached. Another
	outgoing parton is selected and a spin density matrix calculated for
        it using Eqn.\,\ref{eqn:spin2} and the calculated decay matrices
        for those particles which have already been developed rather than the
        identity. Step three of the algorithm is then performed for this particle.
        This procedure is repeated until all the outgoing particles in the
	hard process have been developed.
\end{enumerate}

  This is the algorithm for the forward evolution of outgoing time-like partons.
  A similar algorithm \cite{Knowles:1988hu} can be used to develop the
  incoming space-like partons. This algorithm starts the development with
  a decay matrix calculated according to
\begin{equation} 
  D_{\kappa_j\kappa'_j}= \frac1{N_I}
 \rho^{i\neq j}_{\kappa_i\kappa'_i}
	\mathcal{M}_{\kappa_i\kappa_j;\lam_1\ldots\lam_n}
	\mathcal{M}^*_{\kappa'_i\kappa'_j;\lam'_1\ldots\lam'_n}
	\prod_{k=1,n}D^k_{\lam_k\lam'_k}.
\label{eqn:QCD4}
\end{equation}
      The normalization
\begin{equation}
N_I= \rho^{i\neq j}_{\kappa_i\kappa'_i}
	\mathcal{M}_{\kappa_i\kappa_j;\lam_1\ldots\lam_n}
	\mathcal{M}^*_{\kappa'_i\kappa_j;\lam'_1\ldots\lam'_n}
	\prod_{k=1,n}D^i_{\lam_k\lam'_k}
\end{equation}
  is again chosen such that the decay matrix has unit trace.
  We will not discuss the details of this backward evolution algorithm here,
  it is described in \cite{Knowles:1988hu}. It is sufficient to 
  note that when a space-like parton has been completely developed the algorithm
  returns a spin density matrix for the developed parton.
  So the full algorithm for spin correlations in the QCD shower is that
  the outgoing time-like partons are developed according to the algorithm
  described above and then using the decay matrices returned by this algorithm
  the incoming space-like partons are developed according to the algorithm of
  \cite{Knowles:1988hu}.

  In order to treat the spin correlations the best possible approach would be as
  follows:
\begin{enumerate}
\item The momenta of the particles in the hard collision are generated according
      to Eqn.\,\ref{eqn:spin1}.
\item One of the outgoing particles is selected at random and  a spin density 
      matrix calculated for it according to Eqn.\,\ref{eqn:spin2}.

\item If the particle is colourless and unstable step four is performed,
      if the particle is coloured step five is performed, otherwise
      the decay matrix for this particle is set to $\delta_{\lam_i\lam'_i}$,
      another particle selected and this step repeated.

\item The decay mode of the particle 
      is selected according to the branching ratios and the momenta of
      the particles produced in the $n-$body decay generated according to 
      Eqn.\,\ref{eqn:spin3}. 
\begin{enumerate}
  \item A particle from this decay is selected and a spin density matrix
        calculated for it using Eqn.\,\ref{eqn:spin3b}.
  \item If this particle is unstable and colourless we perform step four.
  \item If the particle is coloured step five is performed.
  \item If the particle is stable and colourless its decay matrix is set
        to $\delta_{\lam_i\lam'_i}$ and another particle from the decay
        which
	produced it is selected to be developed. Once all the particles
        in a given decay have been developed a decay matrix is calculated for
        the decay using Eqn.\,\ref{eqn:spin4} and another particle from
        the decay or branching which produced this particle selected to be
        developed. A spin density matrix is calculated for
        the development of this particle with the calculated decay matrix
        rather than the identity for those particles which have been
        developed.
	Step three is then performed for this particle.
\end{enumerate}
\item If the particle is coloured and has not reached the cut-off scale
      its branching, $i\ra jk$, is generated according
      to Eqn.\,\ref{eqn:QCD1}.
      If the parton has reached the cut-off scale then:
\begin{enumerate}
\item If the particle is stable its decay matrix is set to 
      $\delta_{\lam_i\lam'_i}$ and another particle from the branching or decay
	which produced this parton selected to be developed.
	If all the particles in a branching have been developed a decay matrix
        for the parton which branches is calculated using Eqn.\,\ref{eqn:QCD3}
        and another particle from the decay or branching which produced it selected
	to be developed. Similarly if all the particles in a decay have been developed
	a decay matrix is calculated for the decay using Eqn.\,\ref{eqn:spin4} and
	another particle from the decay or branching which produced it selected
	to be developed.  A spin density matrix is calculated for
        the development of this particle with the calculated decay matrix
        rather than the identity for those particles which have been
        developed. Step three is then performed for this particle.
\item If the particle is unstable step four is performed for the particle.
\end{enumerate}
    If a branching occurs the following procedure is performed:  
\begin{enumerate}
\item One of the outgoing
      particles is selected to be developed and a spin density matrix calculated
      using Eqn.\,\ref{eqn:QCD2}.
\item Step five is then performed for this particle.
\end{enumerate}

\item Eventually this algorithm will return a decay matrix for one of the particles
	produced in the hard process. If there are still outgoing particles which 
	have not been developed then another outgoing particle is selected
        to be developed and step two performed for this particle.

\item	If all the outgoing particles have been developed then one of the
	incoming particles is selected and a decay matrix calculated using
	Eqn.\,\ref{eqn:QCD4}. The algorithm described in \cite{Knowles:1988hu}
	is used to develop the incoming particle. This is repeated until
	all the incoming particles have been developed.
\end{enumerate}

  As with both the QCD and heavy particle decay algorithms we have already described
  the decay matrix
  for any particle which has not been developed is taken to be $\delta_{\lam_i\lam'_i}$.

  While this algorithm is the most efficient way in which to include the spin
  correlations it presents problems in generating the parton shower. This is
  because the parton shower algorithm needs to produce the QCD radiation from
  all the particles produced in the same hard collision or decay at the same time.
  There are several reasons for this:
\begin{enumerate}
\item The scale used for the strong coupling constant in the parton shower algorithm
      depends on the energy of the collision, or
      mass of the decaying particle.
\item In the angular-ordered parton shower used in HERWIG 
      \cite{Marchesini:1984bm,Marchesini:1988cf,Marchesini:1990yk} the maximum
      angle for the emission of QCD radiation from a parton is determined by
	the direction of its colour partner, which is defined by the colour
	 flow in the hard process or decay.
\item The parton-shower algorithm does not conserve energy and momentum. Therefore
      after the radiation from all the partons in a process has been generated
      the momenta of the resulting jets must be adjusted in order to 
      ensure energy and momentum conservation.
\end{enumerate}
    
  We need to consider an algorithm in which this is possible. While
  this algorithm will be less efficient in the way it implements the spin correlations
  it should prove much easier to interface with the QCD parton shower.
  The algorithm will proceed in the following way:
\begin{enumerate}
\item First the algorithm for spin correlations in the both the 
      initial- and final-state
      QCD parton shower is run as before.
\item Then if there are any unstable particles produced one of them is selected
      at random and a spin density matrix calculated in the following way:
\begin{enumerate}
\item The decay matrices for the other undecayed particles are set to
      $\delta_{\lam_i\lam'_i}$.
\item We then use the matrix elements of all the vertices to produce the decay matrices
      of the particles in the hard process, apart from the particle which 
      produced the one we have selected, and the spin density matrices for the
      incoming partons. This should be straightforward
      as these decay matrices will have already been calculated by the QCD
      spin correlation algorithm.
\item Using these decay matrices and Eqn.\,\ref{eqn:spin2} we construct the
      spin density matrix for the parton which initiated the shower which produced
      the selected particle.
\item We then move down the branch that produced the particle we have selected 
      using Eqn.\,\ref{eqn:QCD2} and the decay matrices which have already been
      calculated for the side-branches.
\item Eventually we will reach the particle we have selected with a spin density
      matrix calculated according to Eqn.\,\ref{eqn:QCD2}.
\end{enumerate}
\item The decay mode of this particle is selected according to the branching
      ratios and the momenta of the decay products generated according to 
      Eqn.\,\ref{eqn:spin3}.
\item If any coloured particles are produced the QCD parton-shower algorithm is
      run. If the decay produces unstable particles we perform 
      steps two and three for this decay,
      using Eqn.\,\ref{eqn:spin3b} rather than Eqn.\,\ref{eqn:spin2}.
\item Eventually we will perform a decay in which only stable particles are
      produced. When this happens we set the decay matrices for the
      stable particles to $\delta_{\lam_i\lam'_i}$
      and use the decay matrices calculated by the QCD algorithm and
      Eqn.\,\ref{eqn:spin4} to calculate a decay matrix for this decay.
\item We then select another unstable particle produced in the previous decay,
      or the parton shower initiated by that decay, and repeat the procedure
      with the decay matrix calculated from Eqn.\,\ref{eqn:spin4} for
      the particles for which the decays have been performed. If there are no
      further unstable particles to decay we use Eqn.\,\ref{eqn:spin4} to
      calculate the decay matrix and move up the chain.
\item Eventually we will reach a particle which was produced in the parton shower
      after the hard collision. We then select another unstable particle
      and repeat the process using the calculated decay matrix rather than the identity
      for those particles which have been decayed.
\item This procedure is repeated until all the unstable particles produced in the 
      parton shower initiated by the hard process have been decayed.
\end{enumerate}

  This procedure is more complicated as additional spin density and decay matrices
  have to 
  be stored and recalculated but it should prove easier to implement the parton shower
  with this procedure.

  The implementation of either of the algorithms we have described in this
  section would involve a major rewrite of the HERWIG, or any other, event generator.
  As there are currently projects under way to rewrite the current
  generation of FORTRAN event generators in C++\cite{Bertini:2000uh,ribon} it
  would be pointless to undertake this rewrite. Hopefully, one of the algorithms
  we have described here can be implemented in the next generation of C++ event
  generators.
%
%  Conclusions
%
\section{Conclusions}
\label{sect:conclude} 

  We have shown that it is possible to construct an algorithm for spin correlations
  in a Monte Carlo event generator that both allows us to generate the production
  and decay of the heavy particles as separate steps and has a complexity which 
  only grows linearly with the number of final-state particles.
  An additional advantage of this algorithm is that we only have to 
  calculate the helicity amplitudes for $2\to2$ collision processes and,
  at least for the processes we have studied, at most three body decays rather
  than the calculation of $n-$body matrix elements.

  This algorithm gives results that are in good agreement with the full calculations
  of the $n-$body matrix elements for all the processes we have studied.
  In the appendix we have given the helicity amplitudes
  necessary to implement this algorithm in a Monte Carlo event generator.
  These results have been incorporated into the HERWIG event generator and will
  be available in the next version.

  Finally, we have proposed an algorithm to include both the spin correlations
  in heavy particle decays which we have considered here as well the correlations
  in and between jets which are already included in the HERWIG event
  generator.
  This algorithm, which would require a major rewrite of the current
  generation of FORTRAN event generators, can hopefully be included in the next
  generation 
  of C++ event generators which are currently being written~\cite{Bertini:2000uh,ribon}.

\acknowledgments

  I would like to thank Mike Seymour and
  my colleagues in the Cambridge Supersymmetry working group,
  particularly Bryan Webber and Frank Krauss, for many helpful discussions.
  Kosuke Odagiri was very helpful in comparing the results of my helicity
  amplitude calculations with his original spin-averaged MSSM matrix elements
  which are implemented in HERWIG. This work was supported by PPARC.

\appendix
\section{Spinor Conventions}
\label{app:spin}
  In general we will use the conventions of \cite{Kleiss:1985yh,vanEijk:1990zp}
  for the spinors in the helicity amplitudes for the matrix elements.
  However it is more convenient for us to define
\begin{eqnarray}
  s_\lam(p_1,p_2) = \bar{u}_\lam(p_1)u_{-\lam}(p_2),
\end{eqnarray}
  rather than the $s$ and $t$ functions of \cite{Kleiss:1985yh}.

  We will use the following notation for massive fermions,
\begin{subequations}\label{eqn:spinordef}
\begin{eqnarray}
u(p,\lam) &=&\phantom{-}\frac1{\sqrt{2p\cdot\ell}}\left(p\sla+m\right)u_{-\lam}(l),\\
v(p,\lam) &=&- \frac1{\sqrt{2p\cdot\ell}}\left(p\sla-m\right)u_{-\lam}(l),
\end{eqnarray}
\end{subequations}
  where $p$ is the four-momentum of the fermion, $m$ is its mass, $\lam$ is the spin
  of the fermion and $\ell^\mu$ is a light-like four-vector.
  This choice for the massive fermions both reproduces the standard spin sums for 
  Dirac particles
\begin{subequations}
\begin{eqnarray}
\sum_\lam u_\lam(p)\bar{u}_\lam(p) &=& \left(p\sla+m\right),\\ 
\sum_\lam v_\lam(p)\bar{v}_\lam(p) &=& \left(p\sla-m\right),
\end{eqnarray}
\end{subequations}
  and the additional spin sums required in SUSY models which have Majorana fermions
  \cite{Haber:1985rc}
\begin{subequations}
\begin{eqnarray}
\sum_\lam u_\lam(p)v^T_\lam(p) &=&  \left(p\sla+m\right)C^T,\\
\sum_\lam \bar{u}^T_\lam(p)\bar{v}_\lam(p)&=&C^{-1}\left(p\sla-m\right),\\
\sum_\lam \bar{v}^T_\lam(p)\bar{u}_\lam(p) &=&C^{-1}\left(p\sla+m\right),\\
\sum_\lam v_\lam(p)u^T_\lam(p) &=&  \left(p\sla-m\right)C^T.
\end{eqnarray}
\end{subequations}
  Given this choice it is useful to define
\begin{equation}
p^\mu = \tilde{p}^\mu+\ell^\mu,
\end{equation}
  where $\tilde{p}^2=0$ and $\ell^2=0$. In this notation the spin vector for the
  particle is given by
\begin{equation}
s^\mu= \frac1m(\tilde{p}^\mu-l^\mu).
\end{equation}
  The choice of the spin vector is completely arbitrary in our case provided that we
  are consistent, \ie we use the same choice for the production and decay of
  the particle.

  In the expressions for the amplitudes it will often be convenient to express the
  results in terms of the function
\begin{equation}
F(\lam_1,p_1,k,m,\lam_2,p_2) = \bar{u}_{\lam_1}(p_1)\left(k\!\sla\,+m\right)u_{\lam_2}(p_2),
\end{equation}
  where $p_{1,2}$ are massless four vectors, $\lam_{1,2}$ are helicities and
  $k$ is an arbitrary four-vector.
  This function can be evaluated in terms of spinors as follows \cite{vanEijk:1990zp}
\begin{subequations}
\begin{eqnarray}
F(\lam,p_1,k,m,-\lam,p_2) &=&ms_\lam(p_1,p_2)\\
 F(\lam,p_1,k,m,\phantom{-}\lam,p_2) &=&s_\lam(p_1,\tilde{k})s_{-\lam}(\tilde{k},p_2),
\end{eqnarray}
\end{subequations}
   where
\begin{equation}
\tilde{k}^\mu = k^\mu- \frac{k^2}{2k\cdot p_2}p^\mu_2.
\end{equation}

  We will use the same notation as \cite{Kleiss:1985yh} for the polarization
  vectors of the massless gauge bosons
\begin{equation}
\epsilon_\lam^\mu(p) = \frac1{2\sqrt{p\cdot l}}\bar{u}_{\lam}(p)\gamma^\mu u_\lam(l),
\end{equation}
  where $p$ is the four-momentum of the gauge boson and $l$ is an arbitrary light-like four-vector
  which is not collinear to $p$. The choice of this light-like vector corresponds to
  the making a choice of gauge and therefore we can either chose $l^\mu$ in order to 
  simplify the calculation or vary it in order to test the gauge invariance of the 
  results.

  The choice of the representation for the polarizations of the massive gauge bosons is discussed
  in \cite{Kleiss:1985yh}. However, as this representation basically replaces the
  gauge boson with two massless fermions, into which the boson decays, we
  will not include any on-shell massive gauge bosons but replace them with their
  decay products.

\section{Couplings}
\label{app:couplings}
  In this section we will give the couplings for both the SM and MSSM in a format
  suitable for use in the helicity amplitudes for both the production and decay
  of the particles.

  In order to specify the matrix elements we need the couplings for the vertices involved.
  In general we have taken the form of a vector-fermion-fermion vertex to be
\begin{equation}
ia^\lam\gamma^\mu P_\lam.
\end{equation}
  The couplings of the gauge bosons to the Standard Model fermions are given in
  Table\,\ref{tab:SMbosonfermions}
  where $e$ is the magnitude of the electron's charge, $g$ the 
  weak coupling and $\theta_W$ the weak mixing angle.

\begin{table}
\begin{center}
\renewcommand{\arraystretch}{1.2}
\begin{tabular}{|c|c|c|}
\hline
$a_{{\rm B}{\rm f}{\rm \bar{f }}}$  & $a_{{\rm B}{\rm f}{\rm \bar{f}}}^+$&
		                      $a_{{\rm B}{\rm f}{\rm \bar{f}}}^-$\\
\hline
$a_{{\rm W}{\rm f}{\rm \bar{f'}}}$ & 0 & $-\frac{g}{\sqrt2}V_{{\rm f}{\rm f}'}$ \\
\hline
$a_{{\rm Z}{\rm f}{\rm \bar{f }}}$  & $\frac{ge_{\rm f}\ssw}{\cw}$ 
				    & $-\frac{g}{\cw}(t^3_{\rm f}-e_{\rm f}\ssw)$ \\
\hline
\end{tabular}
\end{center}
\caption{Couplings of the SM fermions to the gauge bosons, where
	 $t^3_{\rm f}$ is the weak isospin of the fermion and $e_{\rm f}$
	 is its electric
	 charge in units of the magnitude of the electron's charge.
	 The isospin partner of the fermion f is denoted by ${\rm f'}$.
	$V_{{\rm f}{\rm f}'}$ is the relevant element of the 
          Cabibbo-Kobayashi-Maskawa~(CKM) matrix.}
\label{tab:SMbosonfermions}
\end{table}

  In the MSSM there are a lot more couplings which must be specified and this
  is complicated by the mixing of the particles involved.
  The physical electroweak gauginos are mixings of the interaction eigenstates.
  There are mixing matrices: $N$ for the neutralinos; and $U$ and $V$ for the charginos.
  In general our conventions for the mixing of 
  the gauginos follow those of \cite{Haber:1985rc,Gunion:1986yn,Gunion:1992tq}.
  In performing the diagonalization to go from the interaction to mass
  eigenstates there is a choice: we can either take the mixing matrices to be
  complex, in which case the masses are positive; or the mixing matrices to be
  real and let the masses be positive or negative, in which case we
  must redefine the field $\cht\ra\gamma_5\cht$ if the mass is negative 
  \cite{Haber:1985rc,Gunion:1986yn,Gunion:1992tq}. As the calculations
  will be implemented numerically it is more convenient to deal with real
  mixing matrices and we must therefore keep track of the sign of the gaugino masses
  and change the sign of certain couplings. We will denote the sign of the
  mass of the $i$th neutralino by $\epsilon_i$ and the $i$th chargino by $\eta_i$. 
  The couplings of the gauge bosons to the gauginos of the MSSM are given in 
  Table\,\ref{tab:bosongauginos}.

\begin{table}
\begin{center}
\renewcommand{\arraystretch}{1.2}
\begin{tabular}{|c|c|c|}
\hline
$a_{{\rm B}\cht\cht}$ & $a^+_{{\rm B}\cht\cht}$ &$a^-_{{\rm B}\cht\cht}$\\
\hline
$a_{{\rm Z}\cht^+_i\cht^-_j}$ 
	& {\footnotesize$-\frac{g}{\cos\theta_W}\left(U^*_{i1}U_{j1}
	+\frac12U^*_{i2}U_{j2}-\delta_{ij}\sin^2\theta_W \right)$}	
        & {\footnotesize$-\frac{g\eta_i\eta_j}{\cos\theta_W}\left(V_{i1}V^*_{j1}
        +\frac12V_{i2}V^*_{j2}-\delta_{ij}\sin^2\theta_W \right)$}\\
\hline
$a_{{\rm Z}\cht^0_i\cht^0_j}$
        & {\footnotesize$\frac{g}{2\cos\theta_W}\left(N^*_{i3}N_{j3}-N^*_{i4}N_{j4}\right)$}
        & {\footnotesize$-\epsilon_i\epsilon_ja^{+*}_{Z\cht^0_i\cht^0_j}$}\\
\hline
$a_{{\rm W}\cht^0_i\cht^-_j}$
        & {\footnotesize$g\left( \frac1{\sqrt{2}}N^*_{i3}U_{j2}+N^*_{i2}U_{j1}\right)$}
        & {\footnotesize$-g\epsilon_i\eta_j\left(
	\frac1{\sqrt{2}}N_{i4}V^*_{j2}-N_{i2}V^*_{j1}\right)$}\\
\hline
\end{tabular}
\caption{Couplings of the gauge bosons to the electroweak gauginos. The
	 mixing matrices for the neutralinos, $N$, and charginos, $U$ and $V$,
	 are defined in the notation of
	 \cite{Haber:1985rc,Gunion:1986yn,Gunion:1992tq}.}
\label{tab:bosongauginos}
\end{center}
\end{table}
\begin{table}
\begin{center}
\renewcommand{\arraystretch}{1.2}
\begin{tabular}{|c|c|c|c|}
\hline
$a_{{\rm B}\ftl\ftl^*}$ & &$a_{{\rm B}\ftl\ftl^*}$ &\\
\hline
$a_{{\rm Z}{\rm \qkt}_{i\al}{\rm \qkt}^*_{i\be}}$ & $-\frac{g}{\cw}
	\left(t^3_qQ^i_{1\al}Q^{i*}_{1\be}-e_q\ssw\delta_{\al\be}\right)$&
$a_{{\rm W}{\rm \dnt}_{i\al}{\rm \upt}^*_{i\be}}$ &
	 $-\frac{g}{\sqrt2}Q^{2i-1}_{1\al}Q^{2i*}_{1\be}$\\
\hline
$a_{{\rm Z}\elt_{i\al}\elt^*_{i\be}}$ & $-\frac{g}{\cw}
	\left(t^3_\ell L^i_{1\al}L^{i*}_{1\be}-e_\ell\ssw\delta_{\al\be}\right)$&
$a_{{\rm W}\elt_{i\al}\nut_{iL}}$ & $-\frac{g}{\sqrt2}L^{2i-1}_{1\al}$\\
\hline
$a_{\gamma\ftl\ftl^*}$ & $-iee_{\rm f}$&&\\
\hline
\end{tabular}
\end{center}
\caption{Couplings of the sfermions to the electroweak gauge bosons. These
	 couplings are taken from \cite{Haber:1985rc}. The flavour of the
         sfermion is denoted by $i$ and $\al,\be$ gives the mass eigenstate.}
\label{tab:bosonsfermion}
\end{table}
  In general there can be mixing between all the sfermions with the same
  electric and colour charges. However, we will only consider mixing
  between the left and right scalar partners of the same fermion.  
  We use the same convention as the HERWIG event generator for this
  left/right mixing, this is discussed in
  \cite{Dreiner:1999qz,Richardson:2000nt,SUSYimplement}.
  In this notation the mixing matrix for the squarks is $Q^i_{\al\be}$,
  where $i=1-6$ for the d, u, s, c, b and t squarks, $\al$ is the left/right
  eigenstate and $\be$ is the mass eigenstate. Similarly the
  mixing matrix for the sleptons is $L^i_{\al\be}$, where 
  $i=1-6$ for the $\rm{e}$, $\nu_{\rm{e}}$, $\mu$, $\nu_\mu$, $\tau,\nu_\tau$ 
  sleptons, $\al$ is the left/right
  eigenstate and $\be$ is the mass eigenstate.

  The form of the boson-scalar-scalar vertex is taken to be
\begin{equation}
  ia(p+p')^\mu,
\end{equation}
  where the momentum $p$ is taken to be flowing into the vertex and $p'$
  is taken to be flowing out of the vertex. 
  The couplings of the sfermions to the electroweak gauge bosons are given in
  Table\,\ref{tab:bosonsfermion} and of the Higgs bosons to 
  the gauge bosons in Table\,\ref{tab:higgshiggsboson}.

\begin{table}
\begin{center}
\renewcommand{\arraystretch}{1.2}
\begin{tabular}{|c|c|c|c|}
\hline
$a_{{\rm B}{\rm H}{\rm H}}$ & &$a_{{\rm B}{\rm H}{\rm H}}$ &\\
\hline
$a_{{\rm W}^\pm{\rm H}^{\pm}{\rm h}^0}$ & $\mp \frac{g}2\cos(\beta-\al)$&
$a_{{\rm W}^\pm{\rm H}^{\pm}{\rm A}^0}$ & $-i\frac{g}2$\\
\hline
$a_{{\rm W}^\pm{\rm H}^{\pm}{\rm H}^0}$ & $\pm \frac{g}2\sin(\beta-\al)$&
$a_{{\rm Z}{\rm A}^0{\rm h}^0}$ & $-i\frac{g\cos(\be-\al)}{2\cw}$\\
\hline
$a_{{\rm Z}{\rm A}^0{\rm H}^0}$ & $i\frac{g\sin(\be-\al)}{2\cw}$&& \\
\hline
\end{tabular}
\caption{Couplings of the Higgs bosons to the electroweak gauge bosons. These
	 couplings are taken from \cite{Gunion:1989we,Gunion:1992hs}.
	The angle $\al$ is the mixing between interaction and mass eigenstates
	of the two neutral scalar Higgs bosons.}
\label{tab:higgshiggsboson}
\end{center}
\end{table}

  The form of the scalar-fermion-fermion vertex is
\begin{equation}
ia^\lam P_\lam.
\end{equation}
  The couplings of the neutralinos to the sfermions are given in 
  Table\,\ref{tab:neutralinosfermion}, of the charginos to the sfermions
  in Table\,\ref{tab:charginosfermion} and of
  the gluino to the squarks in Table\,\ref{tab:gluinosquark}.
  In order to define the couplings of the sfermions to the neutralinos
  it is useful to define the following functions
\begin{subequations}
\begin{eqnarray}
    S^+_{{\rm f}\cht^0_i} &=&         - N'_{i1}ee_{\rm f}+
				N'_{i2}a^+_{{\rm Z}{\rm f}{\rm \bar{f}}},\\
    S^-_{{\rm f}\cht^0_i} &=&\phantom{-}N^{\prime*}_{i1}ee_{\rm f}
			-N^{\prime*}_{i2}a^-_{{\rm Z}{\rm f}{\rm \bar{f}}},
\end{eqnarray}
\end{subequations}
  for the couplings of the gaugino part of the neutralinos to the fermions.
  The functions
\begin{eqnarray}
    H_{{\rm f}\cht^0_i} &=&\left\{\begin{array}{cl}
				\frac{gm_{\rm f}N_{i4}}{2\mw\sbe}&
			 \rm{for\  up-type\  quarks\  and\  neutrinos,}\\
&\\
				\frac{gm_{\rm f}N_{i3}}{2\mw\cbe}&
			 \rm{for\  down-type\  quarks\  and\  charged\  leptons,}\\
		           \end{array}\right.
\end{eqnarray}
  give the couplings of the fermions to the Higgsino part of the neutralinos.

\begin{table}
\begin{center}
\renewcommand{\arraystretch}{1.2}
\begin{tabular}{|c|c|c|}
\hline
$ a_{\cht^0_l\ftl {\rm f}}$ & $a^+_{\cht^0_l\ftl {\rm f}}$ &
			      $a^-_{\cht^0_l\ftl {\rm f}}$\\
\hline
$ a_{\cht^0_l{\rm \qkt}_{i\al} {\rm q}_i}$ & 
	$-\sqrt{2}\epsilon_l\left(Q^i_{1\al}H_{{\rm q}_i\cht^0_l} 
         +Q^i_{2\al}S^+_{{\rm q}_i\cht^0_l}\right)$&
	$-\sqrt{2}\left(Q^i_{2\al}H^*_{{\rm q}_i\cht^0_l}
         +Q^i_{1\al}S^-_{{\rm q}_i\cht^0_l}\right)$\\
\hline
$ a_{\cht^0_l\elt_{i\al} \ell_i}$ & 
	$-\sqrt{2}\epsilon_l\left(L^i_{1\al}H_{\ell_i\cht^0_l} 
	 +L^i_{2\al}S^+_{\ell_i\cht^0_l}\right)$&
	$-\sqrt{2}\left(L^i_{2\al}H^*_{\ell_i\cht^0_l}
	+L^i_{1\al}S^-_{\ell_i\cht^0_l}\right)$\\
\hline
\end{tabular}
\caption{Couplings of the neutralinos to the sfermions and fermions. The flavour of the
         sfermion is denoted by $i$ and $\al$ gives the mass eigenstate.}
\label{tab:neutralinosfermion}
\end{center}
\end{table}

\begin{table}
\begin{center}
\renewcommand{\arraystretch}{1.2}
\begin{tabular}{|c|c|c|}
\hline
$a_{\cht^+_l{\rm \tilde{f}}{\rm f}}$ &$a^+_{\cht^+_i{\rm \tilde{f}}{\rm f}}$ &
				      $a^-_{\cht^+_i{\rm \tilde{f}}{\rm f}}$\\
\hline
$a_{\cht^+_l{\rm \tilde{u}}_{i\al}{\rm d}_i}$
        &$\frac{g}{\sqrt{2}M_W\cos\beta}m_{{\rm d}_i}U_{l2}Q^{2i}_{1\al}$
	&$-g\eta_l\left(V^*_{l1}Q^{2i}_{1\al}
		  -\frac1{\sqrt{2}M_W\sbe}m_{{\rm u}_i}V^*_{l2}Q^{2i}_{2\al}\right)$\\
\hline
$a_{\cht^+_l{\rm \tilde{d}}_{i\al}{\rm u}_i}$
        &$\frac{g\eta_l}{\sqrt{2}M_W\sin\beta} m_{{\rm u}_i}V_{l2}Q^{2i-1}_{1\al}$
        &$-g\left(U^*_{l1}Q^{2i-1}_{1\al}
                 -\frac1{\sqrt{2}M_W\cos\beta}m_{{\rm d}_i}U^*_{l2}Q^{2i-1}_{2\al}\right)$\\
\hline
$a_{\cht^+_l\tilde{\nu}_{iL}\ell_i}$
        & $\frac{g}{\sqrt{2}M_W\cos\beta}m_{\ell_i}U_{l2}$
        & $-g\eta_lV^*_{l1}$\\
\hline
$a_{\cht^+_l\tilde{\ell}_{i\al}\nu_i}$
	& $0$
        & $-g\left(U^*_{l1}L^{2i-1}_{1\al}
         -\frac1{\sqrt{2}M_W\cos\beta}m_{\ell_i}U^*_{l2}L^{2i-1}_{2\al}\right)$\\
\hline
\end{tabular}
\caption{Couplings of the sfermions to the charginos. The generation of the sfermion is
	given by $i$ and $\al$ is its mass eigenstate.}
\label{tab:charginosfermion}
\end{center}
%\end{table}
%\begin{table}
\begin{center}
\renewcommand{\arraystretch}{1.2}
\begin{tabular}{|c|c|c|}
\hline
$a_{{\rm \glt}{\rm\tilde{q}}{\rm q}}$    
 &\ \ \ \ \ \ \ $a^+_{{\rm\glt}{\rm\tilde{q}}{\rm q}}$\ \ \ \ \ \ \  
 &\ \ \ \ \ \ \ $a^-_{{\rm\glt}{\rm\tilde{q}}{\rm q}}$\ \ \ \ \ \ \ \\
\hline
$a_{{\rm \glt}{\rm\tilde{u}}_i{\rm u}_i}$ &$g_s\sqrt{2}Q^{2i}_{2\al}$  
			                  &$-g_s\sqrt{2}Q^{2i}_{1\al}$   \\
\hline
$a_{{\rm \glt}{\rm\tilde{d}}_i{\rm d}_i}$ &$g_s\sqrt{2}Q^{2i-1}_{2\al}$
					  &$-g_s\sqrt{2}Q^{2i-1}_{1\al}$ \\
\hline
\end{tabular}
\end{center}
\caption{Couplings of the squarks to the gluino. The flavour of the squark is given
	by $i$ and $\al$ is the mass eigenstate of the squark.}
\label{tab:gluinosquark}
%\end{table}
%\begin{table}
\begin{center}
\begin{tabular}{|c|c|c|}
\hline
$a_{{\rm H}\cht\cht}$ &  $a^+_{{\rm H}\cht\cht}$&  $a^-_{{\rm H}\cht\cht}$\\
\hline
$a_{{\rm h}_0\cht^0_i\cht^0_j}$
             & $g\left(Q''_{ij}\sin\alpha+S''_{ij}\cos\alpha\right)$
             & $g\left(Q^{\prime\prime*}_{ji}\sin\alpha+S^{\prime\prime*}_{ji}\cos\alpha\right)$\\
\hline
$a_{{\rm H}_0\cht^0_i\cht^0_j}$ 
	     & $g\left(S''_{ij}\sin\alpha-Q''_{ij}\cos\alpha\right)$
	     & $g\left(S^{\prime\prime*}_{ji}\sin\alpha-Q^{\prime\prime*}_{ji}\cos\alpha\right)$\\
\hline
$a_{{\rm A}_0\cht^0_i\cht^0_j}$ 
             & $-ig\left(Q''_{ij}\sin\beta-S''_{ij}\cos\beta\right)$
             & $ig\left(Q^{\prime\prime*}_{ji}\sin\beta-S^{\prime\prime*}_{ji}\cos\beta\right)$\\
\hline
$a_{{\rm h}_0\cht^+_i\cht^-_j}$
             & $ g\left(Q_{ij}\sin\alpha-S_{ij}\cos\alpha\right)$
	     & $ g\left(Q^*_{ji}\sin\alpha-S^*_{ji}\cos\alpha\right)$\\
\hline
$a_{{\rm H}_0\cht^+_i\cht^-_j}$
             & $-g\left(Q_{ij}\cos\alpha+S_{ij}\sin\alpha\right)$
	     & $-g\left(Q^*_{ji}\cos\alpha+S^*_{ji}\sin\alpha\right)$\\
\hline
$a_{{\rm A}_0\cht^+_i\cht^-_j}$
             & $-ig\left(Q_{ij}\sin\beta+S_{ij}\cos\beta\right)$
	     & $ig\left(Q^*_{ji}\sin\beta+S^*_{ji}\cos\beta\right)$\\
\hline
$a_{{\rm H}^-\cht^0_i\cht^-_j}$
	     &  $-gQ^{\prime R}_{ij}\sin\beta$
             &  $-gQ^{\prime L}_{ij}\cos\beta$\\
\hline
\end{tabular}
\end{center}
\caption{Couplings of the MSSM Higgs bosons to the gauginos.}
\label{tab:higgsgaugino}
%\end{table}
%\begin{table}
\begin{center}
\renewcommand{\arraystretch}{1.2}
\begin{tabular}{|c|c|c|c|c|c|}
\hline
$a_{{\rm H}{\rm f}{\rm \bar{f}}}$ & $a^+_{{\rm H}{\rm f}{\rm \bar{f}}}$&
				    $a^-_{{\rm H}{\rm f}{\rm \bar{f}}}$&
$a_{{\rm H}{\rm f}{\rm \bar{f}}}$ & $a^+_{{\rm H}{\rm f}{\rm \bar{f}}}$&
				    $a^-_{{\rm H}{\rm f}{\rm \bar{f}}}$\\
\hline
$a_{{\rm h}^0{\rm u}_i{\rm \bar{u}}_i}$ & $-\frac{gm_{{\rm u}_i}\cos\al}{2\mw\sbe}$&
			$a^+_{{\rm h}^0{\rm u}_i{\rm \bar{u}}_i}$&
$a_{{\rm H}^0{\rm u}_i{\rm \bar{u}}_i}$ & $-\frac{gm_{{\rm u}_i}\sin\al}{2\mw\sbe}$&
			$a^+_{{\rm H}^0{\rm u}_i{\rm \bar{u}}_i}$\\
\hline
$a_{{\rm h}^0{\rm d}_i{\rm \bar{d}}_i}$ & $ \frac{gm_{{\rm d}_i}\sin\al}{2\mw\cbe}$& 
			$a^+_{{\rm h}^0{\rm d}_i{\rm \bar{d}}_i}$&
$a_{{\rm H}^0{\rm d}_i{\rm \bar{d}}_i}$ & $-\frac{gm_{{\rm d}_i}\cos\al}{2\mw\cbe}$& 
			$a^+_{{\rm H}^0{\rm d}_i{\rm \bar{d}}_i}$\\
\hline
$a_{{\rm h}^0\ell_i\bar{\ell}_i}$ & $ \frac{gm_{\ell_i}\sin\al}{2\mw\cbe}$ & 
				$a^+_{{\rm h}^0\ell_i\bar{\ell}_i}$&
$a_{{\rm H}^0\ell_i\bar{\ell}_i}$ &$-\frac{gm_{\ell_i}\cos\al}{2\mw\cbe}$ & 
				$a^+_{{\rm H}^0\ell_i\bar{\ell}_i}$\\
\hline
$a_{{\rm A}^0{\rm u}_i{\rm \bar{u}}_i}$ & $i\frac{gm_{{\rm u}_i}\cot\be}{2\mw}$          &
			$-a^+_{{\rm A}^0{\rm u}_i{\rm \bar{u}}_i}$&
$a_{{\rm H}^+{\rm u}_i{\rm \bar{d}}_i}$ & $\frac{gm_{{\rm d}_i}\tan\be}{\sqrt2\mw}$ &
			$\frac{gm_{{\rm u}_i}\cot\be}{\sqrt2\mw}$ \\
\hline
$a_{{\rm A}^0{\rm d}_i{\rm \bar{d}}_i}$ & $i\frac{gm_{{\rm d}_i}\tan\be}{2\mw}$& 
			$-a^+_{{\rm A}^0{\rm d}_i{\rm \bar{d}}_i}$&
$a_{{\rm H}^+\nu_i\bar{\ell}_i}$ & $\frac{gm_{\ell_i}\tan\be}{\sqrt2\mw}$ &
			$0$ \\
\hline
$a_{{\rm A}^0\ell_i\bar{\ell}_i}$ & $i\frac{gm_{\ell_i}\tan\be}{2\mw}$ & 
				$-a^+_{{\rm A}^0\ell_i\bar{\ell}_i}$&
 &  & 	\\
\hline
\end{tabular}
\caption{Couplings of the MSSM Higgs bosons to the Standard Model fermions.}
\label{tab:higgsfermions}
\vspace{-1.4cm}
\end{center}
\end{table}

    The couplings of the Higgs boson of the MSSM to the electroweak gauginos
  are given in Table~\ref{tab:higgsgaugino} where
\begin{subequations}
\begin{eqnarray}
      Q_{ij} &=& \frac{\eta_i}{\sqrt{2}}V_{i1}U_{j2},\\
      S_{ij} &=& \frac{\eta_i}{\sqrt{2}}V_{i2}U_{j1},\\
Q^{\prime L}_{ij}&=& \eta_j\left[N_{i4}V_{j1}+
	\frac1{\sqrt{2}}V_{j2}\left(N_{i2}+N_{i1}\tan\theta_W\right)\right],\\
Q^{\prime R}_{ij}&=&\epsilon_i\left[N_{i3}U_{j1}
	-\frac1{\sqrt{2}}U_{j2}\left(N_{i2}+N_{i1}\tan\theta_W\right)\right], \\
     Q''_{ij} &=& \frac{\epsilon_i}2\left[
		N_{i3}\left(N_{j2}-N_{j1}\tan\theta_W\right)
               +N_{j3}\left(N_{i2}-N_{i1}\tan\theta_W\right)
	\right],\\
     S''_{ij} &=&\frac{\epsilon_i}2\left[
		N_{i4}\left(N_{j2}-N_{j1}\tan\theta_W\right)
               +N_{j4}\left(N_{i2}-N_{i1}\tan\theta_W\right)
	\right],
\end{eqnarray}
  taken from \cite{Gunion:1989we}.
\end{subequations}

  The couplings of the Higgs bosons to the Standard Model fermions are given in
Table\,\ref{tab:higgsfermions}.

  We also need the couplings of the MSSM Higgs bosons to gauge boson
 pairs. The coupling has the form $iag^{\mu\nu}$ with the couplings $g\mw\sin(\be-\al)$
 for the lightest Higgs boson to a pair of either W or Z bosons. The coupling of the
 heavier scalar Higgs boson is $g\mw\cos(\be-\al)$.

\section{Production Matrix Elements}

  There are a large number of both SM and MSSM processes implemented in
  most general purpose Monte Carlo event generators. In the case of the Standard Model
  most of these processes do not involve the production of the top quark or
  tau lepton which are
  the only fermions for which the spin correlations are relevant.
  There are a number of other processes
  involving three or four particles in the final state, for example production of
  a gauge or Higgs boson in association with a $\rm{t}\rm{\bar{t}}$ pair which have much 
  smaller cross sections.

  A large number of the MSSM production processes involve the production of a pair
  of scalar particles, and therefore we do not need the matrix elements for these
  processes.\footnote{The matrix elements used for these processes in HERWIG
		are given in \cite{Odagiri:1998ep,SUSYimplement}.}
  This leaves a reasonably small number of $2\to2$ processes for 
  which we must calculate the matrix elements.

  In this section we will first present the helicity amplitude expressions for
  all the Feynman diagrams which occur in the $2\to2$ processes we are considering.
  These amplitudes can then be combined to give all the matrix elements we will
  need. In all the matrix elements $p_i$ and $m_i$ are  the four-momentum
  and mass of the $i$th particle, respectively. The vector $l_i$ is used to
  define the direction of the particle's spin and $\lam_i$ is the helicity
  of the $i$th particle.

\subsection{Feynman Diagrams}

%
%    f fbar --> gauge boson --> f fbar
%
\subsubsection{Diagram 1}
\begin{figure}[htp]
\begin{center}
\begin{picture}(120,80)
\ArrowLine(40,40)(0,0)
\ArrowLine(0,80)(40,40)
\Photon(40,40)(80,40){5}{5}
\ArrowLine(80,40)(120,80)
\ArrowLine(120,0)(80,40)
\Text(-2,80)[r]{1}
\Text(-2, 0)[r]{2}
\Text(123,80)[l]{3}
\Text(123, 0)[l]{4}
\Text(40,47)[b]{$a^\lam$}
\Text(80,47)[b]{$b^\lam$}
\end{picture}
\end{center}
\caption{Feynman diagram for \mbox{$\rm{f }(p_1)\rm{\bar{f }}(p_2)\ra
				    \rm{f'}(p_3)\rm{\bar{f'}}(p_4)$}
	 via $s$-channel gauge boson exchange, where ${\rm f}$ and
	 ${\rm f'}$ can be any fermion.}
\label{fig:proddia1}
\end{figure}

  The Feynman diagram for $\rm{f }\rm{\bar{f }}\ra\rm{f'}\rm{\bar{f'}}$ via $s$-channel 
  gauge boson exchange is shown in Fig.\,\ref{fig:proddia1}.
  The form of the first and second vertices are taken to be $ia^\lam\gamma^\mu P_\lam$
  and $ib^\lam \gamma^\mu P_\lam$. The amplitude is given by
\begin{eqnarray}
\mathcal{M}&=& -\frac{\delta_{\lam_1\lam_2}a^{\lam_1}}
		{\sqrt{p_3\cdot l_3p_4\cdot l_4}}
		\frac1{\hat{s}-M_B^2+i\Gamma_BM_B} \\
&& \left[\phantom{+}b^{-\lam_1}F(-\lam_3,l_3,p_3,m_3,\phantom{-}\lam_1,p_1)
	            F(\phantom{-}\lam_1,p_2,p_4,-m_4,-\lam_4,l_4)\right. \nonumber\\
&& \left.  \         + b^{ \lam_1\phantom{-}}F(-\lam_3,l_3,p_3,m_3,-\lam_1,p_2)
		    F(-\lam_1          ,p_1,p_4,-m_4,-\lam_4,l_4)\right],\nonumber
\end{eqnarray}
  where $\hat{s}=(p_1+p_2)^2$, $M_B$ is the mass of the exchanged boson and
  $\Gamma_B$ is the width of the exchanged boson.
  The masses of the incoming particles have been neglected.
  If we wish to consider both the outgoing particles to be fermions rather
  than a fermion and an antifermion, in for example gaugino pair production,
  the sign of $\lam_4$ should be changed.

%
%  f fbar --> f fbar via t-channel scalar exchange
%
\subsubsection{Diagram 2}
\begin{figure}[htp]
\begin{center}
\begin{picture}(120,70)
\ArrowLine(0,70)(60,70)
\ArrowLine(60,10)(0,10)
\DashArrowLine(60,70)(60,10){5}
\ArrowLine(60,70)(120,70)
\ArrowLine(120,10)(60,10)
\Text(-2,70)[r]{1}
\Text(-2,10)[r]{2}
\Text(123,70)[l]{3}
\Text(123,10)[l]{4}
\Text(60,72)[b]{$a^\lam$}
\Text(60,8)[t]{$b^\lam$}
\end{picture}
\end{center}
\caption{Feynman diagram for \mbox{$\rm{f }(p_1)\rm{\bar{f }}(p_2)\ra
				    \rm{f'}(p_3)\rm{\bar{f'}}(p_4)$}
	 via $t$-channel scalar exchange, where ${\rm f}$ and
	 ${\rm f'}$ can be any fermion.}
\label{fig:proddia2}
\end{figure}

  The Feynman diagram for $\rm{f }\rm{\bar{f }}\ra\rm{f'}\rm{\bar{f'}}$ via $t$-channel
  scalar exchange is shown in Fig.\,\ref{fig:proddia2}.
  We will take the form of the upper vertex to be $ia^\lam P_\lam$ and the
  lower vertex to be $ib^\lam P_\lam$.
  The matrix element is given by
\begin{eqnarray}
\mathcal{M} &=& -\frac{a^{\lam_1}b^{-\lam_2}}
		{2\sqrt{p_3\cdot l_3p_4\cdot l_4}}
		\frac1{\hat{t}-M^2_\Phi} \\
&& F(-\lam_3,l_3,p_3,m_3,\lam_1,p_1)F(\lam_2,p_2,p_4,-m_4,-\lam_4,l_4)\nonumber,
\end{eqnarray}
   where $\hat{t}=(p_1-p_3)^2$ and $M_\Phi$ is the mass of the exchanged scalar.
   The incoming particles have been taken to be  massless. As before if we
   wish to regard particle four as a fermion rather than an antifermion the 
   sign of its helicity should be changed.

%
%  f fbar --> f fbar via u-channel scalar exchange
%
\subsubsection{Diagram 3}
\begin{figure}[htp]
\begin{center}
\begin{picture}(120,70)
\ArrowLine(0,70)(60,70)
\ArrowLine(60,10)(0,10)
\DashArrowLine(60,70)(60,10){5}
\ArrowLine(120,70)(60,70)
\ArrowLine(60,10)(120,10)
\Text(-2,70)[r]{1}
\Text(-2,10)[r]{2}
\Text(123,10)[l]{3}
\Text(123,70)[l]{4}
\Text(60,72)[b]{$a^\lam$}
\Text(60,8)[t]{$a^\lam$}
\end{picture}
\end{center}
\caption{Feynman diagram for \mbox{$\rm{f }(p_1)\rm{\bar{f }}(p_2)\ra
				    \rm{f'}(p_3)\rm{\bar{f'}}(p_4)$}
	 via $u$-channel scalar exchange, where ${\rm f}$ and
	 ${\rm f'}$ can be any fermion.}
\label{fig:proddia3}
\end{figure}

  The Feynman diagram for $\rm{f }\rm{\bar{f }}\ra\rm{f'}\rm{\bar{f'}}$ via $u$-channel
  scalar exchange is shown in Fig.\,\ref{fig:proddia3}.
  As for the previous diagram we will take the form of the upper vertex to be
  $ia^\lam P_\lam$ and the lower vertex to be $ib^\lam P_\lam$.
  The amplitude is given by
\begin{eqnarray}
\mathcal{M} &=& \frac{a^{\lam_1}b^{-\lam_2}}
		{2\sqrt{p_3\cdot l_3p_4\cdot l_4}}
		\frac1{\hat{u}-M^2_\Phi} \\
&& F(\lam_4,l_4,p_4,m_4,\lam_1,p_1)F(p_2,\lam_2,p_3,-m_3,\lam_3,l_3)\nonumber,
\end{eqnarray}
   where $\hat{u}=(p_1-p_4)^2$, $M_\Phi$ is the mass of the exchanged sfermion
   and we have assumed the incoming particles are massless. The sign of $\lam_4$
   should be changed if we wish to consider both outgoing particles as fermions.
%
%  g g --> f fbar via t-channel fermion exchange
%
\subsubsection{Diagram 4}
\begin{figure}[htp]
\begin{center}
\begin{picture}(120,70)
\Gluon(0,70)(60,70){4}{5}
\Gluon(60,10)(0,10){4}{5}
\ArrowLine(60,10)(60,70)
\ArrowLine(60,70)(120,70)
\ArrowLine(120,10)(60,10)
\Text(-2,70)[r]{1}
\Text(-2,10)[r]{2}
\Text(123,10)[l]{4}
\Text(123,70)[l]{3}
\end{picture}
\end{center}
\caption{Feynman diagram for \mbox{$\rm{g}(p_1)\rm{g }(p_2)\ra
				    \rm{f}(p_3)\rm{\bar{f}}(p_4)$}
	 via $t$-channel fermion exchange.}
\label{fig:proddia4}
\end{figure}

  The Feynman diagram for $\rm{g}\rm{g }\ra\rm{f}\rm{\bar{f}}$ via $t$-channel
  fermion exchange is shown in Fig.\,\ref{fig:proddia4}.
  In this case the form of the gluon-quark-quark vertex is taken to be 
  $-ig_s\gamma^\mu$, where $g_s$ is the strong coupling. The colour matrices
  are not included as it is easier to handle the colour sums/averages for
  the diagrams separately. This would also allow us to use the
  same matrix elements for fermion-antifermion production in photon-photon
  collisions by replacing the strong coupling with the electric charge.
  The matrix element is given by
\begin{eqnarray}
\mathcal{M} &=& \frac{g_s^2}{2\sqrt{p_1\cdot l_1p_2\cdot l_2p_3\cdot l_3p_4\cdot l_4}}
	        \frac1{\hat{t}-m^2_3}\\
&& \left[\phantom{+}F(-\lam_3,l_3,p_3,m_3,\phantom{-}\lam_1,l_1)\right.\nonumber\\
&& \ \ \ \ \ \ \ \ \ \ \ \ \ \ \ \ \ 
	   \left\{\phantom{+}F(\phantom{-}\lam_1,p_1,p_2-p_4,m_3,\phantom{-}\lam_2,l_2)
	         F(\phantom{-}\lam_2,p_2,p_4,-m_4,-\lam_4,l_4)
\right.\nonumber\\
&&\ \ \ \ \ \ \ \ \ \ \ \ \ \ \ \ \ \ \left.
	\,	+F(\phantom{-}\lam_1,p_1,p_2-p_4,m_3,-\lam_2,p_2)
	         F(-\lam_2,l_2,p_4,-m_4,-\lam_4,l_4)\right\}
\nonumber\\
&&\ +F(-\lam_3,l_3,p_3,m_3,-\lam_1,p_1)\nonumber\\
&& \ \ \ \ \ \ \ \ \ \ \ \ \ \ \ \ \ 
	   \left\{\phantom{+}F(-\lam_1,l_1,p_2-p_4,m_3,\phantom{-}\lam_2,l_2)
	         F(\phantom{-}\lam_2,p_2,p_4,-m_4,-\lam_4,l_4)
\right.\nonumber\\
&& \ \ \ \ \ \ \ \ \ \ \ \ \ \ \ \ \ \, \left.\left.
		\,+F(-\lam_1,l_1,p_2-p_4,m_3,-\lam_2,p_2)
	         F(-\lam_2,l_2,p_4,-m_4,-\lam_4,l_4)\right\}\right].\nonumber
\end{eqnarray}
  As before if we wish to regard the outgoing antifermion as a fermion,
  in for example gluino pair production, then we must change the sign of $\lam_4$.

%
%  g g --> f fbar via t-channel fermion exchange
%
\subsubsection{Diagram 5}
\begin{figure}[htp]
\begin{center}
\begin{picture}(120,70)
\Gluon(0,70)(60,10){4}{6}
\Gluon(0,10)(60,70){-4}{6}
\ArrowLine(60,10)(60,70)
\ArrowLine(60,70)(120,70)
\ArrowLine(120,10)(60,10)
\Text(-2,70)[r]{1}
\Text(-2,10)[r]{2}
\Text(123,10)[l]{4}
\Text(123,70)[l]{3}
\end{picture}
\end{center}
\caption{Feynman diagram for \mbox{$\rm{g}(p_1)\rm{g }(p_2)\ra
				    \rm{f}(p_3)\rm{\bar{f}}(p_4)$}
	 via $u$-channel fermion exchange.}
\label{fig:proddia5}
\end{figure}

  The Feynman diagram for $\rm{g}\rm{g }\ra\rm{f}\rm{\bar{f}}$ via $u$-channel
  fermion exchange is shown in Fig.\,\ref{fig:proddia5}. As with the previous
  diagram we have not included the colour matrices in the expression for the
  helicity amplitude.
  In this case the form of the gluon-quark-quark vertex is taken to be 
  $-ig_s\gamma^\mu$.
  The matrix element is given by
\begin{eqnarray}
\mathcal{M} &=&-\frac{g_s^2}{2\sqrt{p_1\cdot l_1p_2\cdot l_2p_3\cdot l_3p_4\cdot l_4}}
	        \frac1{\hat{u}-m^2_3}\\
&& \left[\phantom{+}F(-\lam_3,l_3,p_3,m_3,\phantom{-}\lam_2,l_2)\right.\nonumber\\
&& \ \ \ \ \ \ \ \ \ \ \ \ \ \ \ \ \ 
	   \left\{\phantom{+}F(\phantom{-}\lam_2,p_2,p_1-p_4,m_3,\phantom{-}\lam_1,l_1)
	         F(\phantom{-}\lam_1,p_1,p_4,-m_4,-\lam_4,l_4)
\right.\nonumber\\
&&\ \ \ \ \ \ \ \ \ \ \ \ \ \ \ \ \ \,\left.
	\,	+F(\phantom{-}\lam_2,p_2,p_1-p_4,m_3,-\lam_1,p_1)
	         F(-\lam_1,l_1,p_4,-m_4,-\lam_4,l_4)\right\}
\nonumber\\
&&+F(-\lam_3,l_3,p_3,m_3,-\lam_2,p_2)\nonumber\\
&&\ \ \ \ \ \ \ \ \ \ \ \ \ \ \ \ \ \,
	   \left\{\phantom{+}F(-\lam_2,l_2,p_1-p_4,m_3,\phantom{-}\lam_1,l_1)
	         F(\phantom{-}\lam_1,p_1,p_4,-m_4,-\lam_4,l_4)
\right.\nonumber\\
&&\ \ \ \ \ \ \ \ \ \ \ \ \ \ \ \ \ \,\left.\left.
	\,	+F(-\lam_2,l_2,p_1-p_4,m_3,-\lam_1,p_1)
	         F(-\lam_1,l_1,p_4,-m_4,-\lam_4,l_4)\right\}\right].\nonumber
\end{eqnarray}
  As before if we wish to regard the outgoing antifermion as a fermion,
  in for example gluino pair production, then we must change the sign of $\lam_4$.
  The sign of this matrix element has been changed in order to simplify the
  combination of the matrix elements with the colour factors.

%
%  g g --> f fbar via s-channel gluon exchange
%
\subsubsection{Diagram 6}
\begin{figure}[htp]
\begin{center}
\begin{picture}(120,80)
\Gluon(40,40)(0,0){4}{5}
\Gluon(0,80)(40,40){4}{5}
\Gluon(40,40)(80,40){5}{5}
\ArrowLine(80,40)(120,80)
\ArrowLine(120,0)(80,40)
\Text(-2,80)[r]{1}
\Text(-2, 0)[r]{2}
\Text(123,80)[l]{3}
\Text(123, 0)[l]{4}
\end{picture}
\end{center}
\caption{Feynman diagram for \mbox{$\rm{g}(p_1)\rm{g }(p_2)\ra
				    \rm{f}(p_3)\rm{\bar{f}}(p_4)$}
	 via $s$-channel gluon exchange.}
\label{fig:proddia6}
\end{figure}

  The Feynman diagram for $\rm{g}\rm{g }\ra\rm{f}\rm{\bar{f}}$ via $s$-channel
  gluon exchange is shown in Fig.\,\ref{fig:proddia6}.
  In this case the form gluon-quark-quark vertex is taken to be 
  $-ig_s\gamma^\mu$ and the triple gluon vertex to be
  $ig_s\left[(p_1-p_2)^{\gamma}g^{\al\be}+(p_1+2p_2)^\al g^{\be\gamma}
             -(2p_1+p_2)^\be g^{\al\gamma}\right]$, where
  $p_1,\al$, $p_2,\be$ and $p_1+p_2,\gamma$ are the momenta and Lorentz indices of the
  three gluons.
  The colour matrices are not included in the expression for the amplitude in order
  to allow us to use the same matrix element for both quark and gluino production.
  The matrix element is given by\footnote{The polarization $\lam$ should be summed over.}
\begin{eqnarray}
\mathcal{M}&=& \frac{g_s^2}{4\sqrt{p_1\cdot l_1p_2\cdot l_2p_3\cdot l_3p_4\cdot l_4}}
\frac1{\hat{s}}
\label{eqn:prod6}\\
&&\left[ 
\left( \phantom{+}\delta_{\lam_1-\lam_2}s_{\lam_1}(p_1,l_2)s_{-\lam_1}(p_2,l_1)
	    +\delta_{\lam_1 \lam_2}s_{\lam_1}(p_1,p_2)s_{-\lam_1}(l_2,l_1)\right)\right.\nonumber\\
&&\ \ \ \ \ \ \ \ \ \ \ \ \ \ \ 
\left(\phantom{-}F(-\lam_3,l_3,p_3,m_3,\lam,p_1)F(\lam,p_1,p_4,-m_4,-\lam_4,l_4)\right.
       \nonumber\\
&&\ \ \ \ \ \ \ \ \ \ \ \ \ \ \ \,\left.
    \, -F(-\lam_3,l_3,p_3,m_3,\lam,p_2)F(\lam,p_2,p_4,-m_4,-\lam_4,l_4)\right)\nonumber\\
&& +2s_{\lam_1}(p_1,p_2)s_{-\lam_1}(p_2,l_1)\nonumber\\
&&\ \ \ \ \ \ \ \ \ \ \ \ \ \ \ \left(\phantom{+}
	F(-\lam_3,l_3,p_3,m_3,\phantom{-}\lam_2,l_2)
	F(\phantom{-}\lam_2,p_2,p_4,-m_4,-\lam_4,l_4)\right.\nonumber\\
&&\ \ \ \ \ \ \ \ \ \ \ \ \ \ \ \,\left.
       \,+F(-\lam_3,l_3,p_3,m_3,-\lam_2,p_2)F(-\lam_2,l_2,p_4,-m_4,-\lam_4,l_4)\right)\nonumber\\
&&-2s_{\lam_2}(p_2,p_1)s_{-\lam_2}(p_1,l_2)\nonumber\\
&&\ \ \ \ \ \ \ \ \ \ \ \ \ \ \ \left(\phantom{+}
    F(-\lam_3,l_3,p_3,m_3,\pmn\lam_1,l_1)F(\pmn\lam_1,p_1,p_4,-m_4,-\lam_4,l_4)\right.\nonumber\\
&& \ \ \ \ \ \ \ \ \ \ \ \ \ \ \ \,  \left.\left. 
   \,+F(-\lam_3,l_3,p_3,m_3,-\lam_1,p_1)F(-\lam_1,l_1,p_4,-m_4,-\lam_4,l_4)\right)\right].\nonumber	
\end{eqnarray}
  As before if we wish to regard the outgoing antifermion as a fermion,
  in for example gluino pair production, then we must change the sign of $\lam_4$.

  The expressions we have given for ${\rm g}{\rm g}\ra{\rm f}{\rm \bar{f}}$ via
  $t$- and $u$-channel fermion exchange and $s$-channel gluon exchange are given
  in a general gauge. While we used these results to check the gauge invariance of
  the various production processes, in HERWIG we made the gauge choice $l_1=p_2$
  and $l_2=p_1$ which simplifies the expression given in Eqn.\,\ref{eqn:prod6}
  for the $s$-channel gluon exchange which involves the triple gluon vertex 
  \cite{Kleiss:1985yh}.

%
%  q g --> f tilde{q} via s-channel gluon exchange
%
\subsubsection{Diagram 7}
\begin{figure}[htp]
\begin{center}
\begin{picture}(120,70)
\ArrowLine(0,70)(60,70)
\Gluon(0,10)(60,10){4}{6}
\DashArrowLine(60,70)(60,10){5}
\ArrowLine(60,70)(120,70)
\DashArrowLine(60,10)(120,10){5}
\Text(-2,70)[r]{1}
\Text(-2,10)[r]{2}
\Text(123,10)[l]{4}
\Text(123,70)[l]{3}
\end{picture}
\end{center}
\caption{Feynman diagram for \mbox{$\rm{f}(p_1)\rm{g }(p_2)\ra
				    \rm{\cht}(p_3)\rm{\tilde{f}}(p_4)$}
	 via $t$-channel sfermion exchange.}
\label{fig:proddia7}
\end{figure}

  The Feynman diagram for $\rm{f}\rm{g }\ra
				    \rm{\cht}\rm{\tilde{f}}$ via
  $t$-channel sfermion exchange, where
  $\cht$ can be any of the gauginos, is given in Fig.\,\ref{fig:proddia7}.
  The form of the fermion-gaugino-scalar coupling is \mbox{$ia^\lam P_\lam$} and the form
  of the scalar-scalar-gauge boson vertex is \mbox{$-ig_s(p+p')^\mu$}.
  As before the colour matrices are not included in the amplitude.
  The helicity amplitude is given by 
\begin{equation}
\mathcal{M}= \frac{g_s}{\sqrt{2p_2\cdot l_2 p_3\cdot l_3}}
	     \frac{ a^{\lam_1}}{\hat{t}-m_4^2}
             F(-\lam_3,l_3,p_3,m_3,\lam_1,p_1)F(\lam_2,p_2,p_4,0,\lam_2,l_2),
\end{equation}
  where the incoming fermion is assumed to be massless.
%
%  g g --> f fbar via s-channel gluon exchange
%
\subsubsection{Diagram 8}
\begin{figure}[htp]
\begin{center}
\begin{picture}(120,80)
\Gluon(40,40)(0,0){4}{5}
\ArrowLine(0,80)(40,40)
\ArrowLine(40,40)(80,40)
\ArrowLine(80,40)(120,80)
\DashArrowLine(80,40)(120,0){5}
\Text(-2,80)[r]{1}
\Text(-2, 0)[r]{2}
\Text(123,80)[l]{3}
\Text(123, 0)[l]{4}
\end{picture}
\end{center}
\caption{Feynman diagram for \mbox{$\rm{f}(p_1)\rm{g }(p_2)\ra
				    \rm{\cht}(p_3)\rm{\tilde{f}}(p_4)$}
	 via $s$-channel quark exchange.}
\label{fig:proddia8}
\end{figure}

  The Feynman diagram for $\rm{f}\rm{g }\ra
				    \rm{\cht}\rm{\tilde{f}}$ via
  $s$-channel quark exchange is shown in Fig.\,\ref{fig:proddia8}.
  The form of the fermion-fermion-gauge boson vertex is $-ig_s\gamma^\mu$
  and the form of the fermion-fermion-scalar vertex is $ia^\lam P_\lam$.
  Again $\cht$ can be any of the gauginos and the colour matrices are not 
  included in the amplitude.
  The helicity amplitude is given by
\begin{eqnarray}
\mathcal{M}&=& \frac{g_s}{\sqrt{2p_2\cdot l_2 p_3\cdot l_3}}\frac1{\hat{s}} \\
&&\left[ \phantom{+}\delta_{\lam_1\lam_2}a^{\lam_2}s_{\lam_2} (p_1,p_2)
		F(-\lam_3,l_3,p_3,m_3,\lam_2,p_1)
	  s_{-\lam_2}(l_2,p_1)\right.\nonumber\\
&&\,\,+\delta_{\lam_1-\lam_2}a^{-\lam_2}s_{\lam_2}(p_2,p_1)
\left(\phantom{+}F(-\lam_3,l_3,p_3,m_3,-\lam_2,p_1)s_{-\lam_2}(p_1,l_2)\right.
\nonumber\\
&&\left.\left.\,\,\phantom{+\delta_{\lam_1-\lam_2}a^{-\lam_2}s_{\lam_2}(p_2,p_1)}	
	       +F(-\lam_3,l_3,p_3,m_3,-\lam_2,p_2)s_{-\lam_2}(p_2,l_2)\right)
\right],
\nonumber
\end{eqnarray}
  where the incoming fermion is assumed to be massless.

%
%  q g --> f tilde{q} via s-channel gluon exchange
%
\subsubsection{Diagram 9}
\begin{figure}[htp]
\begin{center}
\begin{picture}(120,70)
\ArrowLine(0,70)(60,70)
\Gluon(0,10)(60,10){4}{6}
\ArrowLine(60,70)(60,10)
\DashArrowLine(60,70)(120,70){5}
\ArrowLine(60,10)(120,10)
\Text(-2,70)[r]{1}
\Text(-2,10)[r]{2}
\Text(123,10)[l]{3}
\Text(123,70)[l]{4}
\end{picture}
\end{center}
\caption{Feynman diagram for \mbox{$\rm{f}(p_1)\rm{g }(p_2)\ra
				    \rm{\cht}(p_3)\rm{\tilde{f}}(p_4)$}
	 via $u$-channel gaugino exchange.}
\label{fig:proddia9}
\end{figure}

  The Feynman diagram for $\rm{f}\rm{g }\ra
				    \rm{\cht}\rm{\tilde{f}}$ via
  $u$-channel gaugino exchange is shown in Fig.\,\ref{fig:proddia9}, due to
  the couplings this diagram only occurs when the outgoing gaugino is the gluino.
  The form of the fermion-fermion-scalar vertex is taken to be $ia^\lam P_\lam$
  and the form of the gaugino-gaugino-gauge boson vertex is taken to be
  $ig_S\gamma^\mu$. As before the colour matrices are not included in the amplitude.
  The helicity amplitude is given by
\begin{eqnarray}
\mathcal{M}&=&-\frac{g_s}{\sqrt{2p_2\cdot l_2 p_3 \cdot l_3}}
	       \frac{a^{\lam_1}}{\hat{u}-m_3^2}\\
&&\left[\phantom{+}F(-\lam_3,l_3,p_3,m_3,\phantom{-}\lam_2,l_2)
		   F(\phantom{-}\lam_2,p_2,p_1-p_4,m_3,\lam_1,p_1)\right.\nonumber\\
&&\left.\,+F(-\lam_3,l_3,p_3,m_3,-\lam_2,p_2)F(-\lam_2,l_2,p_1-p_4,m_3,\lam_1,p_1)\right],\nonumber
\end{eqnarray}
  where the incoming fermion is assumed to be massless.

  As with the diagrams for $\rm{g}\rm{g}\ra\rm{f}\rm{\bar{f}}$ the amplitudes
  were implemented with a general choice of the gauge vector for the
  incoming gluon and checked for gauge invariance.
  However it is again convenient to make the
  gauge choice $l_2=p_1$ in order to simplify the Feynman diagrams.

%
%  qbar g --> f tilde{q}* via t-channel sfermion exchange
%
\subsubsection{Diagram 10}

\begin{figure}[htp]
\begin{center}
\begin{picture}(120,70)
\ArrowLine(60,70)(0,70)
\Gluon(0,10)(60,10){4}{6}
\DashArrowLine(60,10)(60,70){5}
\ArrowLine(60,70)(120,70)
\DashArrowLine(120,10)(60,10){5}
\Text(-2,70)[r]{1}
\Text(-2,10)[r]{2}
\Text(123,10)[l]{4}
\Text(123,70)[l]{3}
\end{picture}
\end{center}
\caption{Feynman diagram for \mbox{$\rm{\bar{f}}(p_1)\rm{g }(p_2)\ra
				    \rm{\cht}(p_3)\rm{\tilde{f}}^*(p_4)$}
	 via $t$-channel sfermion exchange.}
\label{fig:proddia10}
\end{figure}
  The Feynman diagram for $\rm{\bar{f}}\rm{g }\ra
				    \rm{\cht}\rm{\tilde{f}}^*$ via
  $t$-channel sfermion exchange is shown in Fig.\,\ref{fig:proddia10}.  The
  outgoing gaugino can be any of the gauginos.
  The form of the fermion-fermion-scalar vertex is taken to be $ia^\lam P_\lam$ and the
  scalar-scalar-gauge boson vertex is taken to be $-ig_s(p+p')^\mu$, where
  we have not included the colour matrices in the amplitude.
  The helicity amplitude is given by
\begin{equation}
\mathcal{M} = \frac{g_s}{\sqrt{2p_2\cdot l_2 p_3 \cdot l_3}}
	      \frac{a^{-\lam_1}}{\hat{t}-m^2_4}
	      F(\lam_1,p_1,p_3,-m_3,\lam_3,l_3)F(\lam_2,p_2,p_4,0,\lam_2,p_2),
\end{equation}
  where the incoming antifermion is assume to be massless.

%
%  qbar g --> f tilde{q}* via s-channel gluon exchange
%
\subsubsection{Diagram 11}
\begin{figure}[htp]
\begin{center}
\begin{picture}(120,80)
\Gluon(40,40)(0,0){4}{5}
\ArrowLine(40,40)(0,80)
\ArrowLine(80,40)(40,40)
\ArrowLine(80,40)(120,80)
\DashArrowLine(120,0)(80,40){5}
\Text(-2,80)[r]{1}
\Text(-2, 0)[r]{2}
\Text(123,80)[l]{3}
\Text(123, 0)[l]{4}
\end{picture}
\end{center}
\caption{Feynman diagram for \mbox{$\rm{\bar{f}}(p_1)\rm{g }(p_2)\ra
				    \cht(p_3)\rm{\tilde{f}}^*(p_4)$}
	 via $s$-channel quark exchange.}
\label{fig:proddia11}
\end{figure}

  The Feynman diagram for $\rm{\bar{f}}\rm{g }\ra
				    \rm{\cht}\rm{\tilde{f}}^*$ via
  $s$-channel fermion exchange is shown in Fig.\,\ref{fig:proddia11}. The
  outgoing gaugino can be any of the gauginos. The form of the fermion-fermion-gauge
  boson vertex is taken to be $-ig_s\gamma^\mu$ and the form of the 
  fermion-fermion-scalar vertex is taken to be $ia^\lam P_\lam$. As before
  the colour matrices are not included in the amplitude. 
  The helicity amplitude is given by
\begin{eqnarray}
\mathcal{M}&=&\frac{g_s}{\sqrt{2p_2\cdot l_2p_3\cdot l_3}} \frac1{\hat{s}}\\
&&\left[\phantom{+}\delta_{\lam_1\lam_2}a^{-\lam_2}s_{\lam_1}(p_1,p_2)
\left(\phantom{+}s_{-\lam_2}(l_2,p_1)F(\lam_2,p_1,p_3,-m_3,\lam_3,l_3)
\right.\right.\nonumber\\
&&\phantom{+\delta_{\lam_1\lam_2}a^{-\lam_2}s_{\lam_1}(p_1,p_2)}\,\,\,\,\,\left.
       +s_{-\lam_2}(l_2,p_2)F(\lam_2,p_2,p_3,-m_3,\lam_3,l_3)\right)\nonumber\\
&&\left.\,\,+\delta_{\lam_1-\lam_2}a^{\lam_2}s_{\lam_1}(p_1,l_2)s_{\lam_2}(p_2,p_1)
	F(-\lam_2,p_1,p_3,-m_3,\lam_3,l_3)\right],\nonumber
\end{eqnarray}
  where the incoming antifermion is assumed to be massless.

%
%  qbar g --> f tilde{q}* via u-channel gluino exchange
%
\subsubsection{Diagram 12}
\begin{figure}[htp]
\begin{center}\begin{picture}(120,70)
\ArrowLine(60,70)(0,70)
\Gluon(0,10)(60,10){4}{6}
\ArrowLine(60,10)(60,70)
\DashArrowLine(120,70)(60,70){5}
\ArrowLine(60,10)(120,10)
\Text(-2,70)[r]{1}
\Text(-2,10)[r]{2}
\Text(123,10)[l]{3}
\Text(123,70)[l]{4}
\end{picture}
\end{center}
\caption{Feynman diagram for \mbox{$\rm{\bar{f}}(p_1)\rm{g }(p_2)\ra
				    \cht(p_3)\rm{\tilde{f}}^*(p_4)$}
	 via $u$-channel gaugino exchange.}
\label{fig:proddia12}
\end{figure}

  The Feynman diagram for $\rm{\bar{f}}\rm{g }\ra
				    \rm{\cht}\rm{\tilde{f}}^*$ via
  $u$-channel gaugino exchange is shown in Fig.\,\ref{fig:proddia12},
  due to the couplings in this case the outgoing gaugino must be the gluino.
  The form of the fermion-fermion-scalar vertex is taken to be $ia^\lam P_\lam$
  and the form of the gaugino-gaugino-gauge boson vertex is taken to be
  $ig_s\gamma^\mu$. The colour matrices are not included in the expression for the
  amplitude.
  The helicity amplitude is given by
\begin{eqnarray}
\mathcal{M}&=&-\frac{g_s}{\sqrt{2p_2\cdot l_2p_3\cdot l_3}}
	      \frac{a^{-\lam_1}}{\hat{u}-m^2_3}\\
&&\left[\phantom{+}F(\lam_1,p_1,p_1-p_4,-m_3,\phantom{-}\lam_2,l_2) F(\phantom{-}\lam_2,p_2,p_3,-m_3,\lam_3,l_3)\right.\nonumber\\
&&\left.+F(\lam_1,p_1,p_1-p_4,-m_3,-\lam_2,p_2)F(-\lam_2,l_2,p_3,-m_3,\lam_3,l_3)\right],\nonumber
\end{eqnarray}
  where the incoming antifermion is assumed to be massless.

  As with the diagrams for $\rm{g}\rm{g}\ra\rm{f}\rm{\bar{f}}$
  and $\rm{f}\rm{g }\ra\rm{\cht}\rm{\tilde{f}}$
  the amplitudes
  were implemented with a general choice of the gauge vector for the
  incoming gluon and checked for gauge invariance.
  However it is again convenient to make the
  gauge choice $l_2=p_1$ in order to simplify the Feynman diagrams.

%
%  First single top diagram
%
\subsubsection{Diagram 13}
\begin{figure}[htp]
\begin{center}
\begin{picture}(120,70)
\ArrowLine(0,70)(60,70)
\ArrowLine(0,10)(60,10)
\Photon(60,70)(60,10){5}{5}
\ArrowLine(60,70)(120,70)
\ArrowLine(60,10)(120,10)
\Text(-2,70)[r]{1}
\Text(-2,10)[r]{2}
\Text(123,70)[l]{3}
\Text(123,10)[l]{4}
\Text(60,72)[b]{$a^\lam$}
\Text(60,8)[t]{$b^\lam$}
\end{picture}
\end{center}
\caption{Feynman diagram for \mbox{$\rm{f}(p_1)\rm{f'}(p_2)\ra
				    \rm{f}(p_3)\rm{f'}(p_4)$}
	 via $t$-channel gauge boson exchange.}
\label{fig:proddia13}
\end{figure}

  The Feynman diagram for \mbox{$\rm{f}\rm{f'}\ra\rm{f}\rm{f'}$}
  via $t$-channel gauge boson exchange is shown in Fig.\,\ref{fig:proddia13}.
  This diagram occurs in single top quark production via $t$-channel W exchange.
  The helicity amplitude is
\begin{eqnarray}
\mathcal{M}&=& \frac{\sqrt{2}a^{\lam_1}b^{\lam_2}\delta_{\lam_2\lam_4}}{\hat{t}-\mw^2}
	\frac1{\sqrt{p_3\cdot l_3}}
\left[\phantom{+}\delta_{\lam_1-\lam_2}
      F(-\lam_3,l_3,p_3,m_3,\pmn\lam_2,p_2)s_{\lam_2}(p_4,p_1)\right.
\\
&&\left.\ \ \ \ \ \ \ \ \ \ \ \ \ \ \ \ \ \ \ \ \ \ \ \ \ \ \ \ \ \ \,
       +\delta_{\lam_1\lam_2}F(-\lam_3,l_3,p_3,m_3,-\lam_2,p_4)s_{-\lam_2}(p_2,p_1)	
\right],\nonumber
\end{eqnarray}
  where in addition to neglecting the masses of the incoming particles the
  mass of particle four, which will be a light quark in single top production,
  has also been neglected.
%
%  Second single top diagram
%
\subsubsection{Diagram 14}
\begin{figure}[htp]
\begin{center}
\begin{picture}(120,70)
\ArrowLine(0,70)(60,70)
\ArrowLine(60,10)(0,10)
\Photon(60,70)(60,10){5}{5}
\ArrowLine(60,70)(120,70)
\ArrowLine(120,10)(60,10)
\Text(-2,70)[r]{1}
\Text(-2,10)[r]{2}
\Text(123,70)[l]{3}
\Text(123,10)[l]{4}
\Text(60,72)[b]{$a^\lam$}
\Text(60,8)[t]{$b^\lam$}
\end{picture}
\end{center}
\caption{Feynman diagram for \mbox{$\rm{f}(p_1)\rm{\bar{f}'}(p_2)\ra
				    \rm{f}(p_3)\rm{\bar{f}'}(p_4)$}
	 via $t$-channel gauge boson exchange.}
\label{fig:proddia14}
\end{figure}

  The Feynman diagram for \mbox{$\rm{f}\rm{\bar{f}'}\ra\rm{f}\rm{\bar{f}'}$}
  via $t$-channel gauge boson exchange is shown in Fig.\,\ref{fig:proddia14}.
  This diagram occurs in single top quark production via $t$-channel W exchange.
  The helicity amplitude is
\begin{eqnarray}
\mathcal{M}&=& \frac{\sqrt{2}a^{\lam_1}b^{\lam_2}\delta_{\lam_2\lam_4}}{\hat{t}-\mw^2}
	\frac1{\sqrt{p_3\cdot l_3}}
\left[\phantom{+}
	\delta_{\lam_1-\lam_2}
F(-\lam_3,l_3,p_3,m_3,\pmn\lam_2,p_4)s_{\lam_2}(p_2,p_1)\right.\\
&&\left.\ \ \ \ \ \ \ \ \ \ \ \ \ \ \ \ \ \ \ \ \ \ \ \ \ \ \ \ \ \ 
       +\delta_{\lam_1\lam_2}F(-\lam_3,l_3,p_3,m_3,-\lam_2,p_2)s_{-\lam_2}(p_4,p_1)	
\right],\nonumber
\end{eqnarray}
  where in addition to neglecting the masses of the incoming particles the
  mass of particle four, which will be a light quark in single top production,
  has also been neglected.

%
%  Third single top diagram
%
\subsubsection{Diagram 15}
\begin{figure}[htp]
\begin{center}
\begin{picture}(120,70)
\ArrowLine(60,70)(0,70)
\ArrowLine(0,10)(60,10)
\Photon(60,70)(60,10){5}{5}
\ArrowLine(120,70)(60,70)
\ArrowLine(60,10)(120,10)
\Text(-2,70)[r]{1}
\Text(-2,10)[r]{2}
\Text(123,70)[l]{3}
\Text(123,10)[l]{4}
\Text(60,72)[b]{$a^\lam$}
\Text(60,8)[t]{$b^\lam$}
\end{picture}
\end{center}
\caption{Feynman diagram for \mbox{${\rm\bar{f}}(p_1)\rm{f'}(p_2)\ra
				    \rm{\bar{f}}(p_3)\rm{f'}(p_4)$}
	 via $t$-channel gauge boson exchange.}
\label{fig:proddia15}
\end{figure}

  The Feynman diagram for \mbox{$\rm{\bar{f}}\rm{f'}\ra\rm{\bar{f}}\rm{f'}$}
  via $t$-channel gauge boson exchange is shown in Fig.\,\ref{fig:proddia15}.
  This process occurs in single top quark production via $t$-channel W exchange.
  The helicity amplitude is
\begin{eqnarray}
\mathcal{M}&=& \frac{\sqrt{2}a^{\lam_1}b^{\lam_2}\delta_{\lam_2\lam_4}}{\hat{t}-\mw^2}
	\frac1{\sqrt{p_3\cdot l_3}}
\left[
	\phantom{+}\delta_{\lam_1-\lam_2}s_{\lam_1}(p_1,p_2)
F(\pmn\lam_2,p_4,p_3,-m_3,-\lam_3,l_3)\right.\ \ \ \ \ \\
&&\left.\ \ \ \ \ \ \ \ \ \ \ \ \ \ \ \ \ \ \ \ \ \ \ \ \ \ \ \ \ \ 
       +\delta_{\lam_1 \lam_2}s_{\lam_1}(p_1,p_4)F(-\lam_2,p_2,p_3,-m_3,-\lam_3,l_3)
\right],\nonumber
\end{eqnarray}
  where in addition to neglecting the masses of the incoming particles the
  mass of particle four, which will be a light quark in single top production,
  has also been neglected.

%
%  Fourth single top diagram
%
\subsubsection{Diagram 16}
\begin{figure}[htp]
\begin{center}
\begin{picture}(120,70)
\ArrowLine(60,70)(0,70)
\ArrowLine(60,10)(0,10)
\Photon(60,70)(60,10){5}{5}
\ArrowLine(120,70)(60,70)
\ArrowLine(120,10)(60,10)
\Text(-2,70)[r]{1}
\Text(-2,10)[r]{2}
\Text(123,70)[l]{3}
\Text(123,10)[l]{4}
\Text(60,72)[b]{$a^\lam$}
\Text(60,8)[t]{$b^\lam$}
\end{picture}
\end{center}
\caption{Feynman diagram for \mbox{${\rm\bar{f}}(p_1)\rm{\bar{f}'}(p_2)\ra
				    \rm{\bar{f}}(p_3)\rm{\bar{f}'}(p_4)$}
	 via $t$-channel gauge boson exchange.}
\label{fig:proddia16}
\end{figure}

  The Feynman diagram for \mbox{$\rm{\bar{f}}\rm{\bar{f}'}\ra\rm{\bar{f}}\rm{\bar{f}'}$}
  via $t$-channel gauge boson exchange is shown in Fig.\,\ref{fig:proddia16}.
  This diagram occurs in single top quark production via $t$-channel W exchange.
  The helicity amplitude is
\begin{eqnarray}
\mathcal{M}&=& \frac{\sqrt{2}a^{\lam_1}b^{\lam_2}\delta_{\lam_2\lam_4}}{\hat{t}-\mw^2}
	\frac1{\sqrt{p_3\cdot l_3}}
\left[
	\phantom{+}\delta_{\lam_1-\lam_2}s_{\lam_1}(p_1,p_4)
F( \pmn\lam_2,p_2,p_3,-m_3,-\lam_3,l_3)\right.\ \ \ \ \ \\
&&\left.\ \ \ \ \ \ \ \ \ \ \ \ \ \ \ \ \ \ \ \ \ \ \ \ \ \ \ \ \ \ 
       +\delta_{\lam_1 \lam_2}s_{\lam_1}(p_1,p_2)F(-\lam_2,p_4,p_3,-m_3,-\lam_3,l_3)
\right],\nonumber
\end{eqnarray}
  where in addition to neglecting the masses of the incoming particles the
  mass of particle four, which will be a light quark in single top production,
  has also been neglected.
%
%  now the matrix elements
%
\subsection{Matrix Elements}

  We can now combine the amplitudes given in the previous section to give the
  matrix elements for the various processes.

  The matrix elements for the $2\to2$ Standard Model top quark and tau lepton
  production processes
  are given in Table\,\ref{tab:SMcross}. It is easiest to extract the colour
  matrices from the diagrams and perform the traces separately.
  This leads to matrix elements for the different colour flows in the process and
  colour factors of the squares of the helicity amplitudes for the individual colour
  flows and the interferences between them.
  The colour flows for $\rm{g}\rm{g}\ra\rm{t}\rm{\bar{t}}$ are discussed in
  Section~\ref{sect:egtophadron}.

\begin{table}
\renewcommand{\arraystretch}{1.2}
\begin{center}
\begin{tabular}{|c|c|c|c|c||c|c|c|c|c|}
\hline
    Process    & $N_{CF}$& \multicolumn{3}{c||}{Colour Factors}& ID & CF & VP &  $a^\lam$      &  $b^\lam$   \\
\cline{3-5}
        &    &$C_{11}$ &$C_{22}$ & $C_{12}$   &&&&  &\\
\hline
% process and colour flow
$\rm{f}\rm{\bar{f}}\ra\rm{n}\rm{\bar{n}}$ & 1 & 1 & - &  - &
% diagrams
         1 & 1 & $\gamma$& $-ee_{\rm f}$ & $-ee_{{\rm n}}$\\
&&&&&	 1 & 1 &    Z    &$a^\lam_{{\rm Z}{\rm f}{\rm\bar{f}}}$ & 
			  $a^\lam_{{\rm Z}{\rm n}{\rm\bar{n}}}$\\
\hline
%process and colour flow
${\rm q}{\rm \bar{q}'}\ra{\rm f}\rm{\bar{f}'}$ & 1 & 1 & - & - &
%diagrams
         1 & 1 & ${\rm W}^\pm$ & $a^\lam_{{\rm W} {\rm q}{\rm \bar{q}'}}$ &
	                         $a^\lam_{{\rm W}{\rm f}\rm{\bar{f}'}}$ \\
\hline
%process and colour flow
${\rm q}\rm{\bar{q}}\ra\rm{t}\rm{\bar{t}}$ & 1 & $\frac{N^2_c-1}{4N^2_c}$ & - &  - &
% diagrams
	 1 & 1 & g&$1$ & $1$\\
\hline
%process and colour flow
$\rm{g}\rm{g}\ra\rm{t}\rm{\bar{t}}$ & 2 & $\frac1{4N_c}$ & $\frac1{4N_c}$ & $\frac1{4N_c(N_c^2-1)}$ &
%diagrams
         4 & 1 &t& - & - \\\cline{6-10}
&&&&&    5 & 2 &t& - & - \\\cline{6-10}
&&&&&    6 & 1 &g& - & - \\\cline{6-10}
&&&&&    6 & 2 &g& - & - \\
\hline
%process and colour flow
${\rm b}{\rm u}\ra{\rm t}{\rm d}$ & 1 & 1 & - & - &
        13 & 1 & W & $a^\lam_{{\rm W}{\rm t}{\rm\bar{b}}}$
		   & $a^\lam_{{\rm W}{\rm d}{\rm\bar{u}}}$\\
\hline
${\rm b}{\rm\bar{d}}\ra{\rm t}{\rm{\bar{u}}}$ & 1 & 1 & - & - &
        14 & 1 & W & $a^\lam_{{\rm W}{\rm t}{\rm\bar{b}}}$
		   & $a^\lam_{{\rm W}{\rm d}{\rm\bar{u}}}$\\
\hline
${\rm\bar{b}}{\rm d}\ra{\rm\bar{t}}{\rm u}$ & 1 & 1 & - & - &
        15 & 1 & W & $a^\lam_{{\rm W}{\rm t}{\rm\bar{b}}}$
		   & $a^\lam_{{\rm W}{\rm d}{\rm\bar{u}}}$\\
\hline
${\rm\bar{b}}{\rm\bar{u}}\ra{\rm\bar{t}}{\rm\bar{d}}$& 1 & 1 & - & - &
        16 & 1 & W & $a^\lam_{{\rm W}{\rm t}{\rm\bar{b}}}$
		   & $a^\lam_{{\rm W}{\rm d}{\rm\bar{u}}}$\\
\hline
\end{tabular}
\caption{Matrix elements for the Standard Model production processes.
         The number of different colour flows is given by $N_{CF}$,
         ID gives the type of diagram, CF gives the colour flow for a 
	 given diagram
	 and VP gives the virtual particle exchanged
         in the $s$-, $t$- or $u$-channels depending on the diagram.
         The isospin partner of a fermion is denoted with a prime and ${\rm n}$
	 can be any fermion.
	 The colour factor given for ${\rm f}{\rm\bar{f}}\ra{\rm n}{\rm \bar{n}}$
	 is for tau production in lepton collisions or electroweak production
	 of top quark pairs in ${\rm q}{\rm\bar{q}}$ annihilation. The corresponding
	 colour factor is $1/{N_c}$ for tau production in
	 ${\rm q}{\rm\bar{q}}$ annihilation. The colour factor
	 given for ${\rm f}{\rm \bar{f}'}$ production is for the
	 production of ${\rm t}{\rm\bar{b}}$ or ${\rm b}{\rm\bar{t}}$,
	 the corresponding colour factor for $\tau^-\bar{\nu}_\tau$ or
	 $\tau^+\nu_\tau$ production is $1/{N_c}$.
	 In the single top quark production processes the up and down
	 quarks can be any of the quarks with the same weak isospin.}
\label{tab:SMcross}
\end{center}
\end{table}
  The matrix elements for the $2\to2$ MSSM electroweak 
  gaugino pair, electroweak gaugino and gluino, and gluino pair
  production processes
  are given in Tables\,\ref{tab:MSSMcrossgaugino},~\ref{tab:MSSMcrossgauginogluino}~and~\ref{tab:MSSMcrossgluino}, respectively.
  The discussion in Section~\ref{sect:egtophadron} 
  on the colour flows also applies to
  $\rm{q}\rm{\bar{q}}\ra\rm{\glt}\rm{\glt}$ which has the same colour structure
  as $\rm{g}\rm{g}\ra\rm{t}\rm{\bar{t}}$, after crossing.

  The situation is slightly different for $\rm{g}\rm{g}\ra\rm{\glt}\rm{\glt}$.
  As with $\rm{q}\rm{\bar{q}}\ra\rm{\glt}\rm{\glt}$ the $s$-channel diagram which
  again involves the triple gluon vertex contributes to both colour flows.
  In this case the colour matrices which occur in the $s$-channel diagram
  can be rewritten using the Jacobi identity
\begin{equation}
f^{abe}f^{ecd} = f^{bce}f^{aed}-f^{bed}f^{ace},
\end{equation}
  where $a$ is the colour of the first incoming gluon, $b$ is the colour of the
  second incoming gluon, $c$ is the colour of the first outgoing gluino and
  $d$ is the colour of the second outgoing gluino. 
  This allows us to write the
  colour matrices for the $s$-channel diagram
  in terms of those for the $t$- and $u$-channel diagrams which 
  means we can write the $s$-channel diagram as a piece which contributes to the
  $t$-channel colour flow and one which contributes to the $u$-channel colour flow.

%
%
%   Electroweak gaugino pair production cross sections
%
{\renewcommand{\arraystretch}{1.2}
\TABULAR[h!!]{|c|c|c|c|c||c|c|c|c|c|}{
\hline
        & \ \ \ \  $N_{CF}$\ \ \ \ & \multicolumn{3}{c||}{Colour Factors}& ID &CF &VP &       &   \\
Process &          & \multicolumn{3}{c||}{}              &    &   &   & $a^\lam$ & $b^\lam$ \\
\cline{3-5}
        & &$C_{11}$ &$C_{22}$ & $C_{12}$      & & &  &  &\\
\hline
% process and colour flow
${\rm f}{\rm\bar{f}}\ra\cht^0_i\cht^0_j$ & 1 & $1$&-&-&
% diagrams
       1 & 1 & Z            & $a^\lam_{{\rm Z}{\rm f}{\rm\bar{f}}}$ &
			      $a^\lam_{{\rm Z}\cht^0_i\cht^0_j}$\\
\cline{6-10}
&&&&&
       2 & 1 & $\ftl_{\al}$ & $a^\lam_{\cht^0_i\ftl_\al {\rm f}}$ &
                              $a^{*-\lam}_{\cht^0_j\ftl_\al {\rm f}}$\\
\cline{6-10}
&&&&&  3 & 1 & $\ftl_{\al}$ & $\epsilon_ja^\lam_{\cht^0_j\ftl_\al {\rm f}}$ &
			      $\epsilon_ia^{*-\lam}_{\cht^0_i\ftl_\al {\rm f }}$\\
\hline
%process and colour flow
${\rm f}{\rm\bar{f}}\ra\cht^+_i\cht^-_j$ & 1 & $1$&-&-&
% diagrams
      1 & 1 & $\gamma$& $-ee_{\rm f}$              & $-e\delta_{ij}$\\
\cline{6-10}
&&&&& 1 & 1 & Z       & $a^\lam_{{\rm Z}{\rm f}{\rm \bar{f}}}$&
		        $a^\lam_{{\rm Z}\cht^+_i\cht^-_j}$\\
\cline{2-10}
&\multicolumn{4}{l||}{only u-type quarks}& 2 & 1 & $\ftl'_\al$ & 
		        $a^{\lam}_{\cht^+_i{\rm\tilde{f}}'_\al{\rm f}}$&
		        $a^{*-\lam}_{\cht^+_j{\rm\tilde{f}}'_\al{\rm f}}$\\
\cline{2-10}
&\multicolumn{4}{l||}{only d-type quarks/leptons}& 3 & 1 & $\ftl'_\al$ &
			 $a^{  \lam}_{\cht^+_j{\rm\tilde{f}}'_\al{\rm f}}$&
			 $a^{*-\lam}_{\cht^+_i{\rm\tilde{f}}'_\al{\rm f}}$\\
\hline
% process and colour flow
${\rm u}{\rm\bar{d}}\ra\cht^+_i\cht^0_j$ &1&$\frac1{N_c}$&-&-&
%diagrams
       1 & 1 & W      & $a^\lam_{{\rm W}{\rm u}{\rm\bar{d}}}$ &
		        $a^{*\lam}_{{\rm W}\cht^0_j\cht^-_i}$ \\
\cline{6-10}
&&&&&  2 & 1 & ${\rm\dnt}_{\al}$  & $a^\lam_{\cht^+_i{\rm\dnt}_\al{\rm u}}$ &
			       $a^{*-\lam}_{\cht^0_j{\rm\dnt}_\al{\rm d}}$\\
\cline{6-10}
&&&&&  3 & 1 & ${\rm\upt}_{\al}$ & $\epsilon_ja^{\lam}_{\cht^0_j{\rm\upt}_\al{\rm u}}$ &
			      $a^{*-\lam}_{\cht^+_i{\rm \upt}_\al{\rm d}}$ \\
\hline
% process and colour flow
${\rm d}{\rm\bar{u}}\ra\cht^0_j\cht^-_i$ &1&$\frac1{N_c}$&-&-&
%diagrams
       1 & 1 & W      & $a^\lam_{{\rm W}{\rm d}{\rm\bar{u}}}$ &
			 $a^{\lam}_{{\rm W}\cht^0_j\cht^-_i}$ \\
\cline{6-10}
&&&&&  3 & 1 & ${\rm\upt}_{\al}$  & $a^\lam_{\cht^+_i{\rm\upt}_\al{\rm d}}$ &
			       $a^{*-\lam}_{\cht^0_j{\rm\upt}_\al{\rm u}}$\\
\cline{6-10}
&&&&&  2 & 1 & ${\rm\dnt}_{\al}$ & $\epsilon_ja^{\lam}_{\cht^0_j{\rm\dnt}_\al{\rm d}}$ &
			      $a^{*-\lam}_{\cht^+_i{\rm\dnt}_\al{\rm u}}$ \\
\hline}
{Matrix elements for the Minimal Supersymmetric Standard Model 
 electroweak gaugino pair production
	 processes.
         The number of different colour flows is given by $N_{CF}$,
         ID gives the type of diagram, CF gives the colour flow for a 
	 given diagram
	 and VP gives the virtual particle exchanged
         in the $s$-, $t$- or $u$-channels depending on the diagram. 
         The colour factors given for $\cht^0\cht^0$ and $\cht^+\cht^-$ production
         are for incoming leptons, the colour factor is $1/N_c$ for incoming quarks.
         The mass eigenstate of the sfermion is given by $\al$ and where
	 the sfermions appear in $u$- and $t$-channel propagators the mass
	 eigenstates should be summed over.
         The isospin partner of a fermion, or sfermion, is denoted by a prime. 
         An identical particle symmetry factor of one half must also
         be included for the production of $\cht^0_i\cht^0_i$.
\label{tab:MSSMcrossgaugino}}}
%
%   electroweak gaugino and gluino production
%
%
%
\begin{table}[h!!]
\begin{center}
\renewcommand{\arraystretch}{1.2}
\begin{tabular}{|c|c|c|c|c||c|c|c|c|c|}
\hline
        & $N_{CF}$ & \multicolumn{3}{c||}{Colour Factors}& ID &CF &VP &       &   \\
Process &          & \multicolumn{3}{c||}{}              &    &   &   & $a^\lam$ & $b^\lam$ \\
\cline{3-5}
        & &$C_{11}$ &$C_{22}$ & $C_{12}$      & & &  &  &\\
\hline
% process and colour flow
${\rm q}{\rm\bar{q}}\ra\cht^0_i{\rm\glt}$ &1& $\frac{(N^2_c-1)}{2N^2_c}$&-&-&
% diagrams
       2 & 1 & ${\rm\qkt}_{\al}$ & $a^\lam_{\cht^0_i{\rm\qkt}_\al {\rm q}}$ & 
	                      $a^{*-\lam}_{{\rm\glt}{\rm\qkt}_\al{\rm q}}$\\
\cline{6-10}
 &&&&& 3 & 1 & ${\rm\qkt}_{\al}$ & $a^\lam_{{\rm\glt}{\rm\qkt}_\al{\rm q}}$
			     & $\epsilon_ia^{*-\lam}_{\cht^0_i{\rm\qkt}_\al {\rm q} }$\\
\hline
% process and colour flow
${\rm u}{\rm\bar{d}}\ra\cht^+_i{\rm\glt}$ &1&$\frac{(N^2_c-1)}{2N^2_c}$&-&-&
% diagrams
         2 & 1 & ${\rm\dnt}_{\al}$ & $a^\lam_{\cht^+_i{\rm\dnt}_\al {\rm u}}$ &
				 $a^{*-\lam}_{{\rm\glt}{\rm\dnt}_\al{\rm d}}$\\
\cline{6-10}
&&&&&    3 & 1 & ${\rm\upt}_{\al}$ & $a^\lam_{{\rm\glt}{\rm\upt}_\al{\rm u}}$ &
				 $a^{*-\lam}_{\cht^+_i {\rm\upt}_\al{\rm d}}$\\
\hline
${\rm d}{\rm\bar{u}}\ra\cht^-_i{\rm\glt}$ &1&$\frac{(N^2_c-1)}{2N^2_c}$&-&-&
% diagrams
         2 & 1 & ${\rm\upt}_{\al}$ & $a^\lam_{\cht^+_i{\rm\upt}_\al{\rm d}}$ &
				 $a^{*-\lam}_{{\rm\glt}{\rm \upt}_\al{\rm u}}$\\
\cline{6-10}
&&&&&    3 & 1 & ${\rm\dnt}_{\al}$ & $a^\lam_{{\rm\glt}{\rm \dnt}_\al{\rm d}}$ &
				 $a^{*-\lam}_{\cht^+_i {\rm\dnt}_\al{\rm u}}$\\
\hline
\end{tabular}
\end{center}
\caption{Matrix elements for the Minimal Supersymmetric Standard Model 
	 electroweak gaugino production in association with a gluino.
         The number of different colour flows is given by $N_{CF}$,
         ID gives the type of diagram, CF gives the colour flow for a 
	 given diagram
	 and VP gives the virtual particle exchanged
         in the $s$-, $t$- or $u$-channels depending on the diagram. 
         The mass eigenstate of the squark is given by $\al$ and when
	 the squarks appear in $u$- and $t$-channel propagators the mass
	 eigenstates should be summed over.
         The isospin partner of a fermion, or sfermion, is denoted by a prime.}
\label{tab:MSSMcrossgauginogluino}
%\end{table}
%
%
%
%\begin{table}[h!!]
\begin{center}
\renewcommand{\arraystretch}{1.2}
\begin{tabular}{|c|c|c|c|c||c|c|c|c|c|}
\hline
        & $N_{CF}$ & \multicolumn{3}{c||}{Colour Factors}& ID &CF &VP &       &   \\
Process &          & \multicolumn{3}{c||}{}              &    &   &   & $a^\lam$ & $b^\lam$ \\
\cline{3-5}
        & &$C_{11}$ &$C_{22}$ & $C_{12}$      & & &  &  &\\
\hline
% process and colour flow
${\rm q}{\rm\bar{q}}\ra{\rm\glt}{\rm\glt}$ &2&
	 $\frac{(N^2_c-1)^2}{8N^3_c}$& $\frac{(N^2_c-1)^2}{8N^3_c}$&
	 $-\frac{(N^2_c-1)}{8N^3_c}$ &
% diagrams
         1 & 1 & g & $1$ & $1$ \\
\cline{6-10}
&&&&&    1 & 2 & g & $-1$ & $1$ \\
\cline{6-10}
&&&&&    2 & 1 & ${\rm\qkt}_\al$ & $a^\lam_{{\rm\glt} {\rm\qkt}_\al{\rm q}}$ &
				 $a^{*-\lam}_{{\rm\glt} {\rm\qkt}_\al {\rm q}}$ \\
\cline{6-10}
&&&&&    3 & 1 & ${\rm\qkt}_\al$ & $a^\lam_{{\rm\glt} {\rm\qkt}_\al {\rm q}}$ &
				 $a^{*-\lam}_{{\rm\glt} {\rm\qkt}_\al{\rm q}}$ \\
\hline
% process and colour flow
$\rm{g}\rm{g}\ra\glt\glt$ &2& $\frac{N^2_c}{2(N^2_c-1)}$&
			$\frac{N^2_c}{2(N^2_c-1)}$&$-\frac{N^2_c}{4(N^2_c-1)}$&
%diagrams
      4 & 1 & ${\rm\glt}$ & - & -\\
\cline{6-10}
&&&&& 5 & 2 & ${\rm\glt}$ & - & -\\
\cline{6-10}
&&&&& 6 & 1 &   g    & - & -\\
\cline{6-10}
&&&&& 6 & 2 &   g    & - & -\\
\cline{6-10}
\hline
\end{tabular}
\end{center}
\caption{Matrix elements for the Minimal Supersymmetric
	 Standard Model gluino pair production
	 processes.
         The number of different colour flows is given by $N_{CF}$,
         ID gives the type of diagram, CF gives the colour flow for a 
	 given diagram
	 and VP gives the virtual particle exchanged
         in the $s$-, $t$- or $u$-channels depending on the diagram. 
         The mass eigenstate of the squark is given by $\al$ and where
	 the squarks appear in $u$- and $t$-channel propagators the mass
	 eigenstates should be summed over.
         The isospin partner of a quark, or squark, is denoted by a prime. 
         The identical particle symmetry factor for ${\rm\glt}{\rm\glt}$ production 
	 has been included in the colour factor.}
\label{tab:MSSMcrossgluino}
\vspace{-2cm}
\end{table}

  The matrix elements for the sfermion-gaugino production processes are given
  in Table\,\ref{tab:MSSMcrossscalar}.
  The colour flows of the gluino-squark production processes are the
  same, after crossing, as those in $\rm{g}\rm{g}\ra\rm{t}\rm{\bar{t}}$.

{\renewcommand{\arraystretch}{1.2}
\TABULAR{|c|c|c|c|c||c|c|c|c|}{
\hline
        & $N_{CF}$ & \multicolumn{3}{c||}{Colour Factors}& ID &CF &VP &         \\
Process &          & \multicolumn{3}{c||}{}              &    &   &   & $a^\lam$ \\
\cline{3-5}
        & &$C_{11}$ &$C_{22}$ & $C_{12}$      & &  &  &\\
\hline
% process and colour flow
${\rm q}{\rm g}\ra\cht^0_i{\rm\qkt}_{\al}$ &1&$\frac1{2N_c}$&-&-&
%diagrams
      7 & 1 & ${\rm\qkt}_\al$ & $a^{\lam}_{\cht^0_i{\rm\qkt}_\al{\rm q}}$ \\
\cline{6-9}
&&&&& 8 & 1 & q          & $a^{\lam}_{\cht^0_i{\rm\qkt}_\al {\rm q}}$  \\
\hline
% process and colour flow
${\rm u}{\rm g}\ra\cht^{+}_i{\rm\dnt}_{\al}$ &1&$\frac1{2N_c}$&-&-&
%diagrams
      7 & 1 & ${\rm\dnt}_\al$ & $a^\lam_{\cht^+_i {\rm\dnt}_\al{\rm u}}$ \\
\cline{6-9}
&&&&& 8 & 1 & q          & $a^\lam_{\cht^+_i {\rm\dnt}_\al {\rm u}}$ \\
\hline
% process and colour flow
${\rm d}{\rm g}\ra\cht^{-}_i{\rm \upt}_{\al}$ &1&$\frac1{2N_c}$&-&-&
%diagrams
      7 & 1 & ${\rm\upt}_\al$ & $a^\lam_{\cht^+_i {\rm\upt}_\al {\rm d}}$ \\
\cline{6-9}
&&&&& 8 & 1 & q          & $a^\lam_{\cht^+_i {\rm\upt}_\al {\rm d}}$ \\
\hline
% process and colour flow
${\rm q}{\rm g}\ra{\rm \glt}{\rm\qkt}_{\al}$ &2&$\frac{(N^2_c-1)}{4N^2_c}$&
		$\frac{(N^2_c-1)}{4N^2_c}$&$-\frac1{4N^2_c}$&
%diagrams
      7 & 1 & ${\rm\qkt}_{\al}$ & $a^\lam_{{\rm\glt}{\rm\qkt}_\al{\rm q}}$  \\
\cline{6-9}
&&&&& 8 & 2 & q            & $a^\lam_{{\rm\glt} {\rm\qkt}_\al {\rm q}}$  \\
\cline{6-9}
&&&&& 9 & 1 & ${\rm\glt}$       & $a^\lam_{{\rm\glt} {\rm\qkt}_\al {\rm q}}$  \\
\cline{6-9}
&&&&& 9 & 2 & ${\rm\glt}$       & $-a^\lam_{{\rm\glt} {\rm\qkt}_\al {\rm q}}$  \\
\hline
% process
${\rm\bar{q}}{\rm g}\ra\cht^0_i{\rm\qkt}^*_{\al}$ &1&$\frac1{2N_c}$&-&-&
%diagrams
      10 & 1 & ${\rm\qkt}^*_\al$ & $a^{*-\lam}_{\cht^0_i {\rm\qkt}_\al {\rm q}}$  \\
\cline{6-9}
&&&&& 11 & 1 & ${\rm \bar{q}}$    & $a^{*-\lam}_{\cht^0_i {\rm\qkt}_\al {\rm q}}$  \\
\hline
% process
${\rm\bar{u}}{\rm g}\ra\cht^{-}_i {\tilde{\rm{d}}}^*_{\al}$ &1&$\frac1{2N_c}$&-&-&
% diagrams
      10 & 1 & ${\tilde{\rm{d}}}^*_{\al}$ & 
	$a^{*-\lam}_{\cht^+_i{\rm\dnt}_\al {\rm u}}$ \\
\cline{6-9}
&&&&& 11 & 1 & $\rm{\bar{u}}$         
     & $a^{*-\lam}_{\cht^+_i{\rm\dnt}_\al{\rm u}}$ \\
\hline
% process
${\rm\bar{d}}{\rm g}\ra\cht^{+}_i {\tilde{\rm{u}}}^*_{\al}$ &1&$\frac1{2N_c}$&-&-&
% diagrams
      10 & 1 & $\tilde{\rm{u}}^*_{\al}$ & $a^{*-\lam}_{\cht^+_i{\rm\upt}_\al{\rm d}}$ \\
\cline{6-9}
&&&&& 11 & 1 & $\rm{\bar{d}}$         & $a^{*-\lam}_{\cht^+_i{\rm\upt}_\al {\rm d}}$ \\
\hline
% process and colour flows
${\rm\bar{q}}{\rm g}\ra{\rm\glt}{\rm\qkt}^*_{\al}$ 
	&2&$\frac{(N^2_c-1)}{4N^2_c}$&$\frac{(N^2_c-1)}{4N^2_c}$&$-\frac1{4N^2_c}$&
      10 & 1 & ${\rm\qkt}^*_{\al}$ & $a^{*-\lam}_{{\rm\glt} {\rm\qkt}_\al {\rm q}}$  \\
\cline{6-9}
&&&&& 11 & 2 & ${\rm\bar{q}}$      & $a^{*-\lam}_{{\rm\glt} {\rm\qkt}_\al {\rm q}}$  \\
\cline{6-9}
&&&&& 12 & 1 & ${\rm\glt}$         & $a^{*-\lam}_{{\rm\glt} {\rm\qkt}_\al {\rm q}}$ \\
\cline{6-9}
&&&&& 12 & 2 & ${\rm\glt}$         & $-a^{*-\lam}_{{\rm\glt} {\rm\qkt}_\al {\rm q}}$ \\
\hline}
{Matrix elements for the Minimal Supersymmetric Standard Model gaugino squark production
	 processes.
         The number of different colour flows is given by $N_{CF}$,
         ID gives the type of diagram, CF gives the colour flow for a 
	 given diagram
	 and VP gives the virtual particle exchanged
         in the $s$-, $t$- or $u$-channels depending on the diagram.
	As before the mass eigenstate of the squark is given by $\al$.
\label{tab:MSSMcrossscalar}}}

%
%  the decays
%
\section{Decay Matrix Elements}

  In the Standard Model, for the processes we are considering, there
  are only two decay modes, \ie the
  decays of top and antitop quarks via a W boson.
  In the MSSM however, there are a large number of decay modes which must be calculated.
  As before we will only calculate those decay modes for which there are fermions
  and the use of the spin correlation algorithm becomes important.

  Despite the large number of decay modes
  there are only a small number of Feynman diagrams. We have calculated these
  with arbitrary couplings which can then be specified for particular decay modes.
  We will first give the helicity amplitudes for these processes and
  then the diagrams and couplings involved for particular decay modes.

\subsection{Two Body Decay Feynman Diagrams}

  There are only five two body decay processes for which we need the helicity amplitudes
  and most of these processes are relatively simple as they involve scalars.
  The four-momentum, mass, helicity and
  reference vector used to define the $i$th particle's
  spin are $p_i$, $m_i$, $\lam_i$ and $l_i$, respectively. 

\subsubsection{Diagram 1}

\begin{figure}[htp]
\begin{center}
\begin{picture}(80,60)
\SetOffset(0,-10)
\ArrowLine(0,30)(40,30)
\ArrowLine(40,30)(80,60)
\DashArrowLine(80,0)(40,30){5}
\Text(-2,30)[r]{0}
\Text(82,60)[l]{1}
\Text(82, 0)[l]{2}
\end{picture}
\end{center}
\caption{Two body decay of a fermion to a fermion and a scalar.}
\label{fig:2bodydia1}
\end{figure}

  The decay of a fermion to a fermion and a scalar is shown in Fig.\,\ref{fig:2bodydia1}.
  This process is common in SUSY models involving the decays of the gauginos
  to either a fermion and an antisfermion or to a gaugino and a Higgs boson.
  As before we will take the form of the vertex to be $ia^\lam P_\lam$.
  The helicity amplitude is given by
\begin{eqnarray}
\mathcal{M} &=& \frac1{2\sqrt{p_0\cdot l_0 p_1\cdot l_1}}\\
&& \left[a^{-\lam_1}s_{-\lam_1}(l_1,\tilde{p}_1)
 	F(\lam_1,\tilde{p}_1,p_0,m_0,-\lam_0,l_0)
         +a^{\lam_1}m_1F(-\lam_1,l_1,p_0,m_0,-\lam_0,l_0)\right].\nonumber 
\end{eqnarray}

\subsubsection{Diagram 2}
\begin{figure}[htp]
\begin{center}
\begin{picture}(80,60)
\ArrowLine(40,30)(0,30)
\ArrowLine(80,60)(40,30)
\DashArrowLine(40,30)(80,0){5}
\Text(-2,30)[r]{0}
\Text(82,60)[l]{1}
\Text(82, 0)[l]{2}
\end{picture}
\end{center}
\caption{Two body decay of an antifermion to an antifermion and a scalar.}
\label{fig:2bodydia2}
\end{figure}

  The decay of an antifermion to an antifermion and a scalar is
  shown in Fig.\,\ref{fig:2bodydia2}. In the MSSM this process occurs in the
  decay of a gaugino to an antifermion and a sfermion when we regard the
  incoming gaugino as an antifermion. The helicity amplitude is given by
\begin{eqnarray}
\mathcal{M}&=& \frac1{2\sqrt{p_0\cdot l_0 p_1\cdot l_1}}
\left[\phantom{-}a^{-\lam_0}s_{-\lam_0}(l_0,\tilde{p}_0)
		F(\lam_0,\tilde{p}_0,p_1,-m_1,-\lam_1,l_1)\right.\\
&&\ \ \ \ \ \ \ \ \ \ \ \ \ \ \ \ \ \ \  \,\ \left.
	 -a^{\lam_0}m_0F(-\lam_0,l_0,p_1,-m_1,-\lam_1,l_1)
\right].\nonumber
\end{eqnarray}
As we will usually consider the incoming gaugino to be a particle rather
than an antiparticle the sign of $\lam_0$ will be opposite.

\subsubsection{Diagram 3}
\begin{figure}[htp]
\begin{center}
\begin{picture}(80,60)
\ArrowLine(0,30)(40,30)
\ArrowLine(40,30)(80,60)
\Photon(80,0)(40,30){5}{4.5}
\Text(-2,30)[r]{0}
\Text(82,60)[l]{1}
\Text(82, 0)[l]{2}
\end{picture}
\end{center}
\caption{Two body decay of a fermion to a fermion and a scalar.}
\label{fig:2bodydia3}
\end{figure}

  The decay of a fermion to a fermion and a massless gauge boson, 
  Fig.\,\ref{fig:2bodydia3},
  cannot occur in either the Standard or Minimal Supersymmetric Standard Models
  at tree level. However, the processes $\cht^0_i\ra\cht^0_j\gamma$
  and $\glt\ra\cht^0_i{\rm g}$ via loop diagrams can be important in the MSSM.

  The matrix elements for these processes must have the following
  form \cite{Haber:1989px}
\begin{equation}
\mathcal{M} = \frac{ig_{\rm eff}}{M_0}\bar{u}(p_1)
		\left(P_R-\epsilon_0\epsilon_1P_L\right)
		\sigma^{\mu\nu}p_{2\mu}\epsilon^*_\nu u(p_0),
\end{equation}
  by gauge invariance, 
  where $\epsilon_0$ is the sign of the mass of the decaying fermion,
  $\epsilon_1$ is the sign of the mass of the fermion produced in the decay,
  and 
  \mbox{$\sigma^{\mu\nu}=\frac{i}{2}\left(\gamma^\mu\gamma^\nu
		-\gamma^\nu\gamma^\mu\right)$}.
  The effective coupling $g_{\rm eff}$ is given in \cite{Haber:1989px} for the decays
  of the neutralino to a neutralino and a photon, and in \cite{Baer:1990sc}
  for the decay of a gluino to a neutralino and a gluon.

  We can write the helicity amplitude for this decay in terms of the effective coupling
giving
\begin{eqnarray}
\mathcal{M}&=& \frac{g_{\rm eff}}{2M_0\sqrt{p_0\cdot l_0 p_1\cdot l_1 p_2\cdot l_2}}\\
&&\left[\delta_{\lam_2-}F(-\lam_1,l_1,p_1,m_1,+,p_2)s_+(p_2,l_2)
			F(-,p_2,p_0,m_0,-\lam_0,l_0)\right.\nonumber\\
&&\left.-\delta_{\lam_2+}\epsilon_0\epsilon_1
	F(-\lam_1,l_1,p_1,m_1,-,p_2)s_-(p_2,k_2)F(+,p_2,p_0,m_0,-\lam_0,l_0)\right].\nonumber
\end{eqnarray}

\subsubsection{Diagram 4}
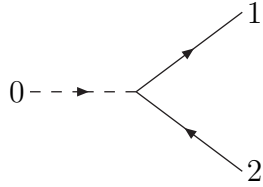
\begin{figure}[htp]
\begin{center}
\begin{picture}(80,60)
\DashArrowLine(0,30)(40,30){5}
\ArrowLine(40,30)(80,60)
\ArrowLine(80,0)(40,30)
\Text(-2,30)[r]{0}
\Text(82,60)[l]{1}
\Text(82, 0)[l]{2}
\end{picture}
\end{center}
\caption{Two body decay of a scalar to a fermion and an antifermion.}
\label{fig:2bodydia4}
\end{figure}

  The decay of a scalar to a fermion and an antifermion occurs in the MSSM
  in either the decays of the Higgs bosons or the sfermions.
  The matrix element is given by
\begin{eqnarray}
\mathcal{M}&=& -\frac1{2\sqrt{p_1\cdot l_1 p_2\cdot l_2}}
\left[\phantom{+}a^{-\lam_1}s_{-\lam_1}(l_1,\tilde{p}_1)
		F(\lam_1,\tilde{p}_1,p_2,-m_2,-\lam_2,l_2)\right.\nonumber\\
&&\left.\,\ \ \ \ \ \ \ \ \ \ \ \ \ \ \ \ \ \ \  \  \ \ 
	 +a^{\lam_1}m_1F(-\lam_1,l_1,p_2,-m_2,-\lam_2,l_2)\right].\nonumber
\end{eqnarray}
  For the decay of the Higgs bosons to two gauginos we will wish to treat both outgoing
  gauginos as particles in which case the sign of $\lam_4$ must be changed.

\subsubsection{Diagram 5}
\begin{figure}[htp]
\begin{center}
\begin{picture}(120,60)
\SetOffset(0,20)
\ArrowLine(0,20)(40,20)
\ArrowLine(40,20)(80,40)
\Photon(40,20)(80,0){-5}{5}
\GOval(120,-20)(15,5)(-26.57){0.7}
\ArrowLine(80,0)(126.71, -6.58)
\ArrowLine(113.29,-33.42)(80,0)
\Text(0,20)[r]{$\cht^\pm_j$}
\Text(80,40)[l]{$\cht^0_i$}
\Text(130,-20)[l]{$\pi^\pm$}
\end{picture}
\end{center}
\caption{Feynman diagram for the decay $\cht^\pm_j\ra\cht^0_i\pi^\pm$.}
\label{fig:2bodydia5}
\end{figure}

  In anomaly mediated SUSY breaking (AMSB) models \cite{Randall:1998uk,Giudice:1998xp}
  the mass splitting between the lightest neutralino and chargino
  can be very small. The dominant decay mode of the lightest chargino
  is \mbox{$\cht^\pm_1\ra\cht^0_1\pi^\pm$}. The Monte Carlo event generators
  usually contain the three body decay modes \mbox{$\cht^+\ra\cht^0{\rm u}{\rm\bar{d}}$}
  and \mbox{$\cht^-\ra\cht^0{\rm d}{\rm\bar{u}}$} using the constituent
  quark masses. In AMSB models this decay mode, with constituent quark masses,
  is not kinematically accessible which means that the decay of the
  chargino to a pion cannot be described by 
  the three body decay mode followed by the hadronization of the quarks.

  The decay $\cht^\pm_1\ra\cht^0_1\pi^\pm$ must therefore be included as a
  $1\to2$ process using chiral perturbation theory.
  The matrix element for this process is
\begin{equation}
\mathcal{M}= \frac{gf_\pi}{2\mw^2}\bar{u}(p_1)p\sla_2a^\lam P_\lam u(p_0), 
\end{equation}
  where $f_\pi$
	 is the pion decay constant.\footnote{We have defined the pion decay constant in an isospin basis
			 and therefore the pion decay constant in 
			 \cite{Groom:2000in} should be divided by $\sqrt2$.} The helicity amplitude for this
   process can be written as
\begin{eqnarray}
\mathcal{M}&=& \frac{gf_\pi}{4\mw^2\sqrt{p_0\cdot l_0 p_1\cdot l_1}}\\
&&\left[m_0a^{-\lam_0}\left\{s_{-\lam_1}(l_1,\tilde{p}_1)F(\lam_1,\tilde{p}_1,p_2,0,-\lam_0,l_0)+m_1F(-\lam_1,l_1,p_2,0,-\lam_0,l_0)\right\}\right.\nonumber\\
&&\left.+a^{\lam_0}s_{\lam_0}(\tilde{p}_0,l_0)\left\{
s_{-\lam_1}(l_1,\tilde{p}_1)F(\lam_1,\tilde{p}_1,p_2,0,\lam_0,\tilde{p}_0)
+m_1F(-\lam_1,l_1,p_2,0,\lam_0,\tilde{p}_0)
\right\}\right].\nonumber
\end{eqnarray}

\subsection{Three Body Decay Feynman Diagrams}

  There are six Feynman diagrams we will need for the SM and MSSM three body decays
  we consider. Four of these diagrams are for the decays of fermions via either
  virtual gauge boson or scalar exchange, one diagram for the decay of an
  antifermion via gauge boson exchange and one diagram for scalar decay via gauge
  boson exchange. As we are considering all the gauginos to be particles this means
  we do not need the diagrams for antifermion decay via virtual scalar exchange.
  In all these diagrams $p_i$ is the four-momentum of the $i$th particle,
  $l_i$ is the reference vector used to define the direction of the $i$th particle's
  spin, $m_i$ is the the mass of the $i$th particle,
  $\lam_i$ is the helicity of the $i$th particle and $m_{ij}^2=(p_i+p_j)^2$.

\subsubsection{Diagram 1}

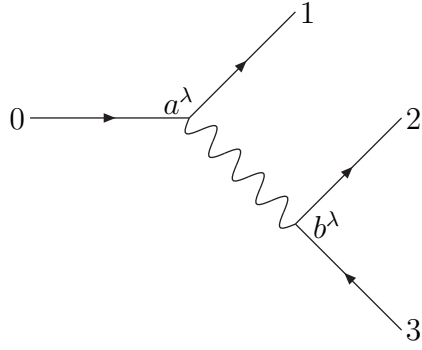
\begin{figure}[htp]
\begin{center}
\begin{picture}(120,90)
\ArrowLine(-20,60)(40,60)
\ArrowLine(40,60)(80,100)
\Photon(40,60)(80,20){-5}{4.5}
\ArrowLine(80,20)(120,60)
\ArrowLine(120,-20)(80,20)
\Text(-22,60)[r]{$0$}
\Text(82,100)[l]{$1$}
\Text(122,60)[l]{$2$}
\Text(122,-20)[l]{$3$}
\Text(42,62)[br]{$a^\lam$}
\Text(87,20)[l]{$b^\lam$}
\end{picture}
\end{center}
\caption{Feynman diagram for fermion decay via gauge boson exchange.}
\label{fig:3bodydia1}
\end{figure}

  The Feynman diagram for fermion decay via gauge boson exchange is shown in
  Fig.\,\ref{fig:3bodydia1}.
  This process occurs for both the top quark, via W exchange, and the MSSM
  electroweak gauginos, via either W or Z exchange.
  The form of the first  and second vertices are
  $ia^\lam\gamma^\mu P_\lam$ and $ib^\lam\gamma^\mu P_\lam$, respectively.
  It is useful to define a number of functions in order to simplify the expression
  for the amplitude
\begin{subequations}
\begin{eqnarray}
b^{12}_{++}&=& \delta_{\lam_1 \lam_2}b^{ \lam_2}s_{-\lam_1}(l_1,\tilde{p}_1)
				s_{\lam_1}(\tilde{p}_2,l_2),\\
b^{12}_{+-}&=& \delta_{\lam_1-\lam_2}b^{ \lam_2}m_1
				s_{-\lam_1}(\tilde{p}_2,l_2),\\
b^{12}_{-+}&=&-\delta_{\lam_1-\lam_2}b^{-\lam_2}m_2
				s_{-\lam_1}(l_1,\tilde{p}_1),\\
b^{12}_{--}&=&-\delta_{\lam_1 \lam_2}b^{-\lam_2}m_1m_2,
\end{eqnarray}
\end{subequations}
   and
\begin{subequations}
\begin{eqnarray}
A^{01} &=&
\phantom{+}\delta_{\lam_0\lam_1}\left\{
	\phantom{+}a^{\lam_0}
		\left[m^2_0s_{-\lam_1}(l_1,\tilde{p}_1)s_{\lam_1}(\tilde{p}_1,l_0)
		-m^2_1s_{-\lam_1}(l_1,\tilde{p}_0)s_{\lam_1}(\tilde{p}_0,l_0)\right]	
\right.\\
&&\phantom{+} \left.\phantom{\delta_{\lam_0\lam_1}}\,\,
	+a^{-\lam_0}m_0m_1\left[
	 s_{-\lam_1}(l_1,\tilde{p}_0)s_{\lam_1}(\tilde{p}_0,l_0)
        -s_{-\lam_1}(l_1,\tilde{p}_1)s_{\lam_1}(\tilde{p}_1,l_0)\right]\right\}
\nonumber\\
&&	+\delta_{\lam_0-\lam_1}\left\{
	\phantom{+}a^{\lam_0}m_1\left[
	-s_{-\lam_1}(l_1,\tilde{p}_1)s_{\lam_1}(\tilde{p}_1,\tilde{p}_0)
	s_{-\lam_1}(\tilde{p}_0,l_0)+m^2_0s_{-\lam_1}(l_1,l_0)\right]
\right.\ \ \ \ \  \ \ \ \ \nonumber\\
&&\left.\phantom{+\delta_{\lam_0-\lam_1}}\,\,
	+a^{-\lam_0}m_0\left[
	s_{-\lam_1}(l_1,\tilde{p}_1)s_{\lam_1}(\tilde{p}_1,\tilde{p}_0)
	s_{-\lam_1}(\tilde{p}_0,l_0)-m^2_1s_{-\lam_1}(l_1,l_0)
	\right]\right\},\nonumber\\
B^{12} &=&-\frac1{\mw^2}\left[\rule{0mm}{5mm}\right.
\delta_{\lam_1\lam_2}\left\{
	b^{\lam_2}\left[
	m^2_1s_{-\lam_1}(l_1,\tilde{p}_2)s_{\lam_1}(\tilde{p}_2,l_2)
       +m^2_2s_{-\lam_1}(l_1,\tilde{p}_1)s_{\lam_1}(\tilde{p}_1,l_2)
			\right]
\right.\\
&&\left.\ \ \ \ \ \ \ \ \ \ \ \phantom{\delta_{\lam_1\lam_2}}\,\,   
	-b^{-\lam_2}m_1m_2\left[
	s_{-\lam_1}(l_1,\tilde{p}_1)s_{\lam_1}(\tilde{p}_1,l_2)+
	s_{-\lam_1}(l_1,\tilde{p}_2)s_{\lam_1}(\tilde{p}_2,l_2)\right]\right\}
\nonumber\\
&&\ \ \ \ \ \ \ \ \ \ \ +\delta_{\lam_1-\lam_2}\left\{
\phantom{+}b^{\lam_2}m_1\left[
	s_{-\lam_1}(l_1,\tilde{p}_1)s_{\lam_1}(\tilde{p}_1,\tilde{p}_2)
	s_{-\lam_1}(\tilde{p}_2,l_2)+m^2_2s_{-\lam_1}(l_1,l_2)\right]	
\right.\nonumber\\
&&\ \ \ \ \ \ \  \ \ \left.\left.\phantom{\delta_{\lam_1-\lam_2}}\,\,
	-b^{-\lam_2}m_2\left[
	s_{-\lam_1}(l_1,\tilde{p}_1)s_{\lam_1}(\tilde{p}_1,\tilde{p}_2)
	s_{-\lam_1}(\tilde{p}_2,l_2)
	+m^2_1s_{-\lam_1}(l_1,l_2)\right]\right\}\rule{0mm}{5mm}\right],\nonumber
\end{eqnarray}
\end{subequations}
  where $M_B$ and $\Gamma_B$ are the mass and width the of the boson, respectively. 

  The amplitude is given by
\begin{eqnarray}
\mathcal{M} &=&-\frac1{2\sqrt{p_0\cdot l_0 p_1\cdot l_1 p_2\cdot l_2 p_3\cdot l_3}}
		\frac1{m^2_{23}-M^2_B+i\Gamma_BM_B}\left[\rule{0mm}{7mm}\right. \\
&& \phantom{+}\,
  b^{23}_{++}\left\{\phantom{+}
	a^{-\lam_2}F(-\lam_1,l_1,p_1,m_1,\pmn\lam_2,\tilde{p}_3)
		   F(\pmn\lam_2,\tilde{p}_2,p_0,m_0,-\lam_0,l_0)
\right.\nonumber\\ 	
&& \left.\phantom{+b^{23}_{++}}\,\,\,\,
        +a^{\lam_2\pmn}F(-\lam_1,l_1,p_1,m_1,-\lam_2,\tilde{p}_2)
		   F(-\lam_2,\tilde{p}_3,p_0,m_0,-\lam_0,l_0)	
\right\}\nonumber\\
&& \,\,\,+b^{23}_{+-}\left\{\phantom{+}
	a^{\lam_2\pmn}F(-\lam_1,l_1,p_1,m_1,-\lam_2,\tilde{p}_3)
		   F(-\lam_2,l_2,p_0,m_0,-\lam_0,l_0)
	\right.\nonumber\\
&& \left.\phantom{+b^{23}_{+-}}\,\,\,\,
	+a^{-\lam_2}F(-\lam_1,l_1,p_1,m_1,\pmn\lam_2,l_2)
		    F(\pmn\lam_2,\tilde{p}_3,p_0,m_0,-\lam_0,l_0)
\right\}\nonumber\\	
&& \,\,\,+b^{23}_{-+}\left\{\phantom{+}
	a^{-\lam_2}F(-\lam_1,l_1,p_1,m_1,\pmn\lam_2,l_3)
		   F(\pmn\lam_2,\tilde{p}_2,p_0,m_0,-\lam_0,l_0)	
\right.\nonumber\\
&& \left.\phantom{+b^{23}_{+-}}\,\,\,\,
	+a^{\lam_2\pmn}F(-\lam_1,l_1,p_1,m_1,-\lam_2,\tilde{p}_2)
		   F(-\lam_2,l_3,p_0,m_0,-\lam_0,l_0)
\right\}\nonumber\\
&& \,\,\,+b^{23}_{--}\left\{\phantom{+}
	a^{\lam_2\pmn}F(-\lam_1,l_1,p_1,m_1,-\lam_2,l_3)
		   F(-\lam_2,l_2,p_0,m_0,-\lam_0,l_0)
\right.\nonumber\\
&&\left.\phantom{+b^{23}_{+-}}\,\,\,\,
	+a^{-\lam_2}F(-\lam_1,l_1,p_1,m_1,\pmn\lam_2,l_2)
		    F(\pmn\lam_2,l_3,p_0,m_0,-\lam_0,l_0)
\right\}\nonumber\\	
&&\left.+\frac1{2}A^{01}B^{23}\right].\nonumber
\end{eqnarray}

\subsubsection{Diagram 2}

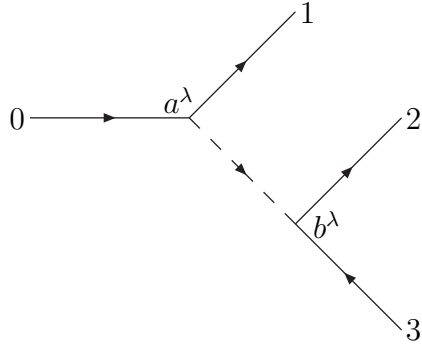
\begin{figure}[htp]
\begin{center}
\begin{picture}(120,90)
\ArrowLine(-20,60)(40,60)
\ArrowLine(40,60)(80,100)
\DashArrowLine(40,60)(80,20){5}
\ArrowLine(80,20)(120,60)
\ArrowLine(120,-20)(80,20)
\Text(-22,60)[r]{$0$}
\Text(82,100)[l]{$1$}
\Text(122,60)[l]{$2$}
\Text(122,-20)[l]{$3$}
\Text(42,62)[br]{$a^\lam$}
\Text(87,20)[l]{$b^\lam$}
\end{picture}
\end{center}
\caption{Feynman diagram for fermion decay via Higgs boson exchange.}
\label{fig:3bodydia2}
\end{figure}

  The Feynman diagram for fermion decay via Higgs boson exchange is given in
  Fig.\,\ref{fig:3bodydia2}. This process occurs in the MSSM for electroweak gaugino
  decay into a different electroweak gaugino and SM fermions.
  The form of the first vertex is $ia^\lam P_\lam$ and the
  second vertex $ib^\lam P_\lam$.
  It is easiest to express the matrix element for this diagram in terms of a function
  of each of the two vertices
\begin{subequations}
\begin{eqnarray}
V_1^{\lam_0\lam_1} &=&  a^{\lam_0}F(-\lam_1,l_1,p_1,m_1,\lam_0,\tilde{p}_0)
			s_{\lam_0}(\tilde{p}_0,l_0)\\
&&	              +a^{-\lam_0}m_0F(-\lam_1,l_1,p_1,m_1,-\lam_0,l_0),\nonumber\\
V_2^{\lam_2\lam_3} &=&  b^{\lam_3}F(-\lam_2,l_2,p_2,m_2,\lam_3,\tilde{p}_3)
			s_{\lam_3}(\tilde{p}_3,l_3)\\
&&		      -b^{-\lam_3}m_3F(-\lam_2,l_2,p_2,m_2,-\lam_3,l_3).\nonumber
\end{eqnarray}
\end{subequations}
  The matrix element is given by
\begin{equation}
\mathcal{M} =-\frac1{4\sqrt{p_0\cdot l_0 p_1\cdot l_1 p_2\cdot l_2 p_3\cdot l_3}}
	\frac1{m_{23}^2-M^2_\Phi+i\Gamma_\Phi M_\Phi}
	V_1^{\lam_0\lam_1}V_2^{\lam_2\lam_3},
\end{equation}
  where $M_\Phi$ and $\Gamma_\Phi$ are the mass and width of the exchanged scalar,
  respectively.

\subsubsection{Diagram 3}

\begin{figure}[htp]
\begin{center}
\begin{picture}(120,90)
\ArrowLine(-20,60)(40,60)
\ArrowLine(40,60)(80,100)
\DashArrowLine(80,20)(40,60){5}
\ArrowLine(80,20)(120,60)
\ArrowLine(120,-20)(80,20)
\Text(-22,60)[r]{$0$}
\Text(82,100)[l]{$2$}
\Text(122,60)[l]{$1$}
\Text(122,-20)[l]{$3$}
\Text(42,62)[br]{$a^\lam$}
\Text(87,20)[l]{$b^\lam$}
\end{picture}
\end{center}
\caption{Feynman diagram for fermion decay via antisfermion exchange.}
\label{fig:3bodydia3}
\end{figure}
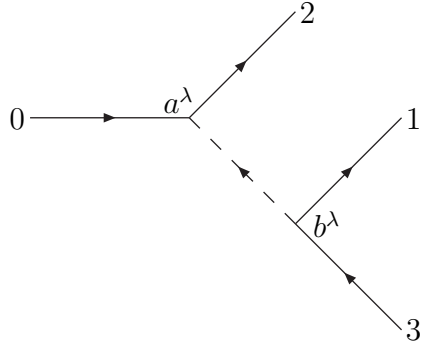

  The Feynman diagram for fermion decay via antisfermion exchange is shown
  in Fig.\,\ref{fig:3bodydia3}. This process occurs in the MSSM for the decay of
  a gaugino to a different gaugino and SM fermions.
  As before we will take the first vertex to be $ia^\lam P_\lam$ and the second vertex to be
  $ib^\lam P_\lam$. As this diagram involves scalar exchange it can be written
  in terms of a function for each of the vertices
\begin{subequations}
\begin{eqnarray}
V_1^{\lam_0\lam_2}&=& a^{\lam_0}F(-\lam_2,l_2,p_2,m_2,\lam_0,\tilde{p}_0)
	s_{\lam_0}(\tilde{p}_0,l_0)\\
&&			+a^{-\lam_0}m_0F(-\lam_2,l_2,p_2,m_2,-\lam_0,l_0),\nonumber\\
V_2^{\lam_1\lam_3}&=& b^{\lam_3}F(-\lam_1,l_1,p_1,m_1,\lam_3,\tilde{p}_3)
	s_{\lam_3}(\tilde{p}_3,l_3)\\
&&			-b^{-\lam_3}m_3F(-\lam_1,l_1,p_1,m_1,-\lam_3,l_3).\nonumber
\end{eqnarray}
\end{subequations}
  The matrix element is given by
\begin{equation}
\mathcal{M} =-\frac1{4\sqrt{p_0\cdot l_0 p_1\cdot l_1 p_2\cdot l_2 p_3\cdot l_3}}
	\frac1{m_{13}^2-M^2_\Phi+i\Gamma_\Phi M_\Phi} V_1^{\lam_0\lam_2}
						      V_2^{\lam_1\lam_3},
\end{equation}
  where $M_\Phi$ and $\Gamma_\Phi$ are the mass and width of the exchanged scalar,
  respectively.
\subsubsection{Diagram 4}

\begin{figure}[htp]
\begin{center}
\begin{picture}(120,90)
\ArrowLine(-20,60)(40,60)
\ArrowLine(80,100)(40,60)
\DashArrowLine(40,60)(80,20){5}
\ArrowLine(80,20)(120,60)
\ArrowLine(80,20)(120,-20)
\Text(-22,60)[r]{$0$}
\Text(82,100)[l]{$3$}
\Text(122,60)[l]{$2$}
\Text(122,-20)[l]{$1$}
\Text(42,62)[br]{$a^\lam$}
\Text(87,20)[l]{$b^\lam$}
\end{picture}
\end{center}
\caption{Feynman diagram for fermion decay via sfermion exchange.}
\label{fig:3bodydia4}
\end{figure}
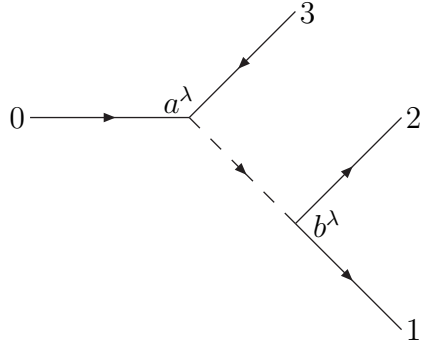

  The Feynman diagram for fermion decay via sfermion exchange is shown
  in Fig.\,\ref{fig:3bodydia4}. This process occurs in the decay of
  a gaugino to SM fermions and a different gaugino.
  As before we will take the first vertex to be $ia^\lam P_\lam$ and the second vertex to be
  $ib^\lam P_\lam$. As this diagram involves scalar exchange it can be written
  in terms of a function for each of the vertices
\begin{subequations}
\begin{eqnarray}
V_1^{\lam_0\lam_3}&=& a^{\lam_3}F(\lam_0,l_0,p_0,-m_0,\lam_3,\tilde{p}_3)
	s_{\lam_3}(\tilde{p}_3,l_3)\\
&&			-a^{-\lam_3}m_3F(\lam_0,l_0,p_0,-m_0,-\lam_3,l_3),\nonumber\\
V_2^{\lam_1\lam_2}&=& b^{-\lam_1}F(-\lam_2,l_2,p_2,m_2,-\lam_1,\tilde{p}_1)
	s_{-\lam_1}(\tilde{p}_1,l_1)\\
&&			-b^{\lam_1}m_1F(-\lam_2,l_2,p_2,m_2,\lam_1,l_1).\nonumber
\end{eqnarray}
\end{subequations}
  The matrix element is given by
\begin{equation}
\mathcal{M} =\frac1{4\sqrt{p_0\cdot l_0 p_1\cdot l_1 p_2\cdot l_2 p_3\cdot l_3}}
	\frac1{m_{12}^2-M^2_\Phi+i\Gamma_\Phi M_\Phi} V_1^{\lam_0\lam_3}
						V_2^{\lam_1\lam_2},
\end{equation}
  where $M_\Phi$ and $\Gamma_\Phi$ are the mass and width of the exchanged scalar,
  respectively.

\subsubsection{Diagram 5}

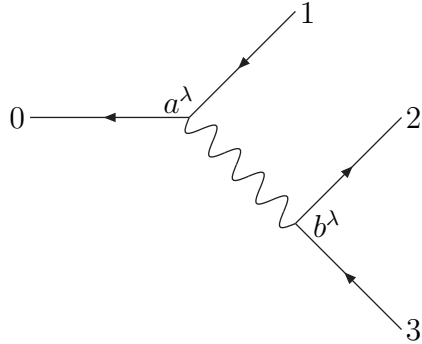
\begin{figure}[htp]
\begin{center}
\begin{picture}(120,90)
\ArrowLine(40,60)(-20,60)
\ArrowLine(80,100)(40,60)
\Photon(40,60)(80,20){-5}{4.5}
\ArrowLine(80,20)(120,60)
\ArrowLine(120,-20)(80,20)
\Text(-22,60)[r]{$0$}
\Text(82,100)[l]{$1$}
\Text(122,60)[l]{$2$}
\Text(122,-20)[l]{$3$}
\Text(42,62)[br]{$a^\lam$}
\Text(87,20)[l]{$b^\lam$}
\end{picture}
\end{center}
\caption{Feynman diagram for antifermion decay via gauge boson exchange.}
\label{fig:3bodydia5}
\end{figure}

  The Feynman diagram for antifermion decay via gauge boson exchange is
  shown in Fig.\,\ref{fig:3bodydia5}. This diagram only occurs in the decay of 
  a top antiquark via W exchange because we are treating all decaying gauginos as
  particles.
  The amplitude for this diagram can be expressed using many of the same functions
as for diagram one except for 
\begin{eqnarray}
E^{01}&=& \phantom{+}\delta_{\lam_0\lam_1}\left\{
	\phantom{+}a^{\lam_1}\left[
	m^2_0s_{-\lam_0}(l_0,\tilde{p}_1)s_{\lam_0}(\tilde{p}_1,l_1)
       -m^2_1s_{-\lam_0}(l_0,\tilde{p}_0)s_{\lam_0}(\tilde{p}_0,l_1)\right]
\right.\\
&&\phantom{+\delta_{\lam_0\lam_1}}\left.\,\,
+a^{-\lam_1}m_0m_1\left[
	 s_{-\lam_0}(l_0,\tilde{p}_0)s_{\lam_0}(\tilde{p}_0,l_1)
        -s_{-\lam_0}(l_0,\tilde{p}_1)s_{\lam_0}(\tilde{p}_1,l_1)
\right]\right\}\nonumber\\
&&+\delta_{\lam_0-\lam_1}\left\{
\phantom{+}a^{\lam_1}m_0\left[
	m^2_1s_{-\lam_0}(l_0,l_1)-s_{-\lam_0}(l_0,\tilde{p}_0)
	s_{\lam_0}(\tilde{p}_0,\tilde{p}_1)s_{-\lam_0}(\tilde{p}_1,l_1)\right]
\right.\nonumber\\
&&\phantom{+\delta_{\lam_0\lam_1}}\left.\,\,
\ -a^{-\lam_1}m_1\left[
	m^2_0s_{-\lam_0}(l_0,l_1)-s_{-\lam_0}(l_0,\tilde{p}_0)
	s_{\lam_0}(\tilde{p}_0,\tilde{p}_1)s_{-\lam_0}(\tilde{p}_1,l_1)	
\right]
\right\}.\nonumber
\end{eqnarray}
  The amplitude is given by
\begin{eqnarray}
\mathcal{M}&=&-\frac1{2\sqrt{p_0\cdot l_0 p_1\cdot l_1 p_2\cdot l_2 p_3\cdot l_3}}
		\frac1{m^2_{23}-M^2_B+i\Gamma_BM_B} \left[\rule{0mm}{7mm}\right.\\
&&\phantom{+}b^{23}_{++}\left[
	\phantom{+}a^{-\lam_2}F(-\lam_0,l_0,p_0,-m_0,\pmn\lam_2,\tilde{p}_3)
			      F(\pmn\lam_2,\tilde{p}_2,p_1,-m_1,-\lam_1,l_1)
\right.\nonumber\\
&&\left.\phantom{+b^{23}_{++}}
	+a^{\lam_2\pmn}F(-\lam_0,l_0,p_0,-m_0,-\lam_2,\tilde{p}_2)
		   F(-\lam_2,\tilde{p}_3,p_1,-m_1,-\lam_1,l_1)
\right]\nonumber\\
&&+b^{23}_{+-}\left[
	\phantom{+}a^{\lam_2\pmn}F(-\lam_0,l_0,p_0,-m_0,-\lam_2,\tilde{p}_3)
			     F(-\lam_2,l_2,p_1,-m_1,-\lam_1,l_1)
\right.\nonumber\\
&&\phantom{+b^{23}_{+-}}\left.\,
	+a^{-\lam_2}F(-\lam_0,l_0,p_0,-m_0,\pmn\lam_2,l_2)
		    F(\pmn\lam_2,\tilde{p}_3,p_1,-m_1,-\lam_1,l_1)
\right]\nonumber\\
&&+b^{23}_{-+}\left[
\phantom{+}a^{-\lam_2}F(-\lam_0,l_0,p_0,-m_0,\pmn\lam_2,l_3)
		      F(\pmn\lam_2,\tilde{p}_2,p_1,-m_1,-\lam_1,l_1)
\right.\nonumber\\
&&\phantom{+b^{23}_{-+}}\,\left.
	+a^{\lam_2\pmn}F(-\lam_0,l_0,p_0,-m_0,-\lam_2,\tilde{p}_2)
		   F(-\lam_2,l_3,p_1,-m_1,-\lam_1,l_1)
\right]\nonumber\\
&&+b^{23}_{--}\left[
\phantom{+}a^{ \lam_2\pmn}F(-\lam_0,l_0,p_0,-m_0,-\lam_2,l_3)
		    	F(-\lam_2,l_2,p_1,-m_1,-\lam_1,l_1)
\right.\nonumber\\
&&\phantom{+b^{23}_{--}}\left.\,
	+a^{-\lam_2}F(-\lam_0,l_0,p_0,-m_0,\pmn\lam_2,l_2)
		    F(\pmn\lam_2,l_3,p_1,-m_1,-\lam_1,l_1)
\right]\nonumber\\
&&\left.+\frac12E^{01}B^{23}\right],\nonumber
\end{eqnarray}
  where $M_B$ and $\Gamma_B$ are the mass and width of the gauge boson,
  respectively.

\subsubsection{Diagram 6}

\begin{figure}[htp]
\begin{center}
\begin{picture}(120,90)
\DashArrowLine(-20,60)(40,60){5}
\DashArrowLine(40,60)(80,100){5}
\Photon(40,60)(80,20){-5}{4.5}
\ArrowLine(80,20)(120,60)
\ArrowLine(120,-20)(80,20)
\Text(-22,60)[r]{$0$}
\Text(82,100)[l]{$1$}
\Text(122,60)[l]{$2$}
\Text(122,-20)[l]{$3$}
\Text(42,62)[br]{$a$}
\Text(87,20)[l]{$b^\lam$}
\end{picture}
\end{center}
\caption{Feynman diagram for scalar decay via gauge boson exchange.}
\label{fig:3bodydia6}
\end{figure}
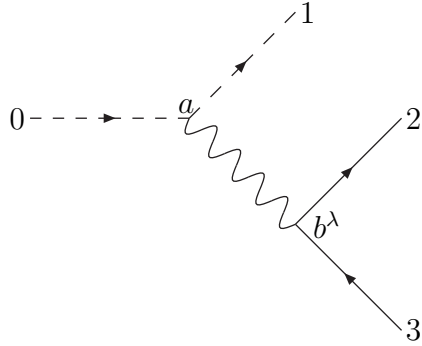

  The Feynman diagram for scalar decay via gauge boson exchange is shown in 
  Fig.\,\ref{fig:3bodydia6}. This process occurs in the decay of both
  the Higgs bosons of the MSSM, to another Higgs boson and SM fermions,
  and of the sfermions, to another sfermion and SM fermions.
  Again the amplitude for this diagram can be written using many of the same functions
as for diagram one. The form of the first vertex is $ia(p_0+p_1)^\mu$ and the
second vertex $ib^\lam\gamma^\mu P_\lam$.
The amplitude is
\begin{eqnarray}
\mathcal{M} &=&-\frac{a}{2\sqrt{p_2\cdot l_2 p_3\cdot l_3}}
		\frac1{m^2_{23}-M^2_B+i\Gamma_BM_B}\\
&&\left[\phantom{+}b^{23}_{++}F(\lam_2,\tilde{p}_2,p_0+p_1,0,\lam_2,\tilde{p}_3)
        +b^{23}_{+-}F(-\lam_2,l_2,p_0+p_1,0,-\lam_2,\tilde{p}_3)\right.\nonumber\\
&&   \,\,\,  +b^{23}_{-+}F(\lam_2,\tilde{p}_2,p_0+p_1,0,\lam_2,l_3)
        +b^{23}_{--}F(-\lam_2,l_2,p_0+p_1,0,-\lam_2,l_3)\nonumber\\
&&\left.\,\,\,+(p_0+p_1)\cdot(p_2+p_3)B^{23}\right].\nonumber
\end{eqnarray}

\subsection{Four Body Decay Feynman Diagrams}

  There is only one four body decay process which we consider. This is the decay
  of the MSSM Higgs bosons to gauge boson pairs, either W or Z, followed by 
  the decay of the gauge bosons to fermions. As before $p_i$ is the four-momentum of
  the $i$th particle, $l_i$ is the reference vector for the $i$th particle,
  $m_i$ is the mass of the $i$th particle, $\lam_i$ is the
  helicity of the $i$th particle  and $m^2_{ij}=(p_i+p_j)^2$.

\begin{figure}[htp]
\begin{center}
\begin{picture}(140,140)
\DashArrowLine(0,70)(40,70){5}
\Photon(40,70)(80,110){5}{4.5}
\Photon(40,70)(80,30){-5}{4.5}
\ArrowLine(80,110)(120,140)
\ArrowLine(120,80)(80,110)
\ArrowLine(80,30)(120,60)
\ArrowLine(120,0)(80,30)
\Text(-2,70)[r]{$0$}
\Text(122,140)[l]{$1$}
\Text(122,80)[l]{$2$}
\Text(122,60)[l]{$3$}
\Text(122,0)[l]{$4$}
\end{picture}
\end{center}
\caption{Four body Higgs boson decay
 to a gauge pair followed by the decay of the gauge bosons.}
\label{fig:4bodydia1}
\end{figure}
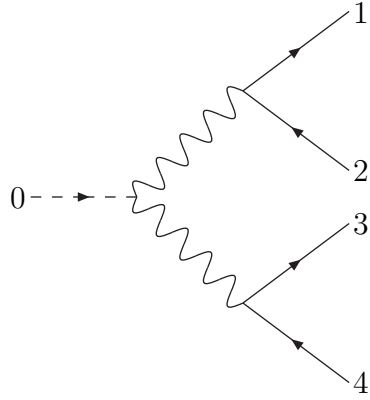

  The Feynman diagram for this process is shown in Fig.\,\ref{fig:4bodydia1}.
  We will take the form of the Higgs boson-gauge boson pair vertex to be $iag^{\mu\nu}$,
  the form of the vertex between the first gauge boson and fermion-antifermion to 
  be $ib^\lam\gamma^\mu P_\lam$ and the form of the vertex between the second
  gauge boson and the fermion-antifermion to be $ic^\lam\gamma^\mu P_\lam$.
  It is useful to define the functions $c^{12}_{\pm\pm}$ and $C^{12}$ which
  are the  functions $b^{12}_{\pm\pm}$ and $B^{12}$ with the
  coupling $b^\lam$ replaced by $c^\lam$.

  The helicity amplitude is given by
\begin{eqnarray}
\lefteqn{\mathcal{M}=}&\\
&& \frac{a}{2\sqrt{p_1\cdot l_1p_2\cdot l_2p_3\cdot l_3p_4\cdot l_4 }}
   \frac1{m^2_{12}-M^2_{B_1}+i\Gamma_{B_1}M_{B_1}}
   \frac1{m^2_{34}-M^2_{B_2}+i\Gamma_{B_2}M_{B_2}}
\nonumber\\
&&\left\{
   \delta_{\lam_1\lam_3}\left[
	\phantom{+}	b^{12}_{++}\left(
 \phantom{+}s_{\lam_1}(\tilde{p}_1,\tilde{p}_3)\left[
	c^{34}_{++}s_{-\lam_1}(\tilde{p}_4,\tilde{p}_2)
       +c^{34}_{-+}s_{-\lam_1}(l_4,\tilde{p}_2)
\right]\right.\right.\right.\nonumber\\
&&\phantom{\delta_{\lam_1\lam_3+}b^{12}_{++}\ \  \ \ \ }\left.\,
+s_{-\lam_1}(l_3,\tilde{p_2})\left[
        c^{34}_{+-}s_{\lam_1}(\tilde{p}_1,\tilde{p}_4)
       +c^{34}_{--}s_{\lam_1}(\tilde{p}_1,l_4)
\right]\right)\nonumber\\
&&\ \ \   \phantom{\delta_{\lam_1\lam_3}} +b^{12}_{+-}\left(
    \phantom{+}s_{\lam_1}(\tilde{p}_3,\tilde{p}_2)\left[
	c^{34}_{++}s_{-\lam_1}(l_1,\tilde{p}_4)
       +c^{34}_{-+}s_{-\lam_1}(l_1,l_4)
\right]\right.\nonumber\\
&&\ \ \ \phantom{\delta_{\lam_1\lam_3}+b^{12}_{++}\ }   \,
\left.
             +s_{-\lam_1}(l_1,l_3)\left[
	c^{34}_{+-}s_{\lam_1}(\tilde{p}_4,\tilde{p}_2)
       +c^{34}_{--}s_{\lam_1}(l_4,\tilde{p}_2)
\right]\right)\nonumber\\
&&\ \ \ \phantom{\delta_{\lam_1\lam_3}}+b^{12}_{-+}\left(
       \phantom{+}s_{\lam_1}(\tilde{p}_1,\tilde{p}_3)\left[
        c^{34}_{++}s_{-\lam_1}(\tilde{p}_4,l_2)
       +c^{34}_{-+}s_{-\lam_1}(l_4,l_2)
\right]
\right.\nonumber\\
&&\ \ \ \left.\phantom{\delta_{\lam_1\lam_3}+b^{12}_{++}\ } 
                +s_{-\lam_1}(l_3,l_2)\left[
        c^{34}_{+-}s_{\lam_1}(\tilde{p}_1,\tilde{p}_4)
       +c^{34}_{--}s_{\lam_1}(\tilde{p}_1,l_4)
\right]   \right)\nonumber\\
&&\ \ \ \phantom{\delta_{\lam_1\lam_3}}+b^{12}_{--}\left(
      \phantom{+}s_{\lam_1}(\tilde{p}_3,l_2)\left[
          c^{34}_{++}s_{-\lam_1}(l_1,\tilde{p}_4)
         +c^{34}_{-+}s_{-\lam_1}(l_1,l_4)
\right]\right.\nonumber\\
&&\ \ \ \phantom{\delta_{\lam_1\lam_3}+b^{12}_{++}\  } \,\left.\left.
                +s_{-\lam_1}(l_1,l_3)\left[
          c^{34}_{+-}s_{\lam_1}(\tilde{p}_4,l_2)
         +c^{34}_{--}s_{\lam_1}(l_4,l_2)
\right]\right)\right]\nonumber
\end{eqnarray}
\begin{eqnarray}
&&
   +\delta_{\lam_1-\lam_3}\left[
	\phantom{+}	b^{12}_{++}\left(
 \phantom{+}s_{-\lam_1}(\tilde{p}_3,\tilde{p}_2)\left[
	c^{34}_{++}s_{\lam_1}(\tilde{p}_1,\tilde{p}_4)
       +c^{34}_{-+}s_{\lam_1}(\tilde{p}_1,l_4)
\right]\right.\right.     \nonumber\\
&&\ \ \ \  \ \,  \left.\phantom{\delta_{\lam_1\lam_3} +b^{12}_{+-}}
    +s_{\lam_1}(\tilde{p}_1,l_3)\left[
	c^{34}_{+-}s_{-\lam_1}(\tilde{p}_4,\tilde{p}_2)
       +c^{34}_{--}s_{-\lam_1}(l_4,\tilde{p}_2)
\right]\right)\nonumber\\
&&\ \ \   \ \ \,   \phantom{\delta_{\lam_1\lam_3}} +b^{12}_{+-}\left(
    \phantom{+}s_{-\lam_1}(l_1,\tilde{p}_3)\left[
	c^{34}_{++}s_{\lam_1}(\tilde{p}_4,\tilde{p}_2)
       +c^{34}_{-+}s_{\lam_1}(l_4,\tilde{p}_2)
\right]\right.\nonumber\\
&&\ \ \ \ \ \,   \phantom{\delta_{\lam_1\lam_3}+b^{12}_{++} \ }   \,
\left.
             +s_{\lam_1}(l_3,\tilde{p}_2)\left[
	c^{34}_{+-}s_{-\lam_1}(l_1,\tilde{p}_4)
       +c^{34}_{--}s_{-\lam_1}(l_1,l_4)
\right]
\right)\nonumber\\
&&\ \ \   \ \ \,  \phantom{\delta_{\lam_1\lam_3}}+b^{12}_{-+}\left(
       \phantom{+}s_{-\lam_1}(\tilde{p}_3,l_2)\left[
        c^{34}_{++}s_{\lam_1}(\tilde{p}_1,\tilde{p}_4)
       +c^{34}_{-+}s_{\lam_1}(\tilde{p}_1,l_4)
\right]
\right.\nonumber\\
&&\ \ \ \ \ \,   \left.\phantom{\delta_{\lam_1\lam_3}+b^{12}_{++} \ } \,
                +s_{\lam_1}(\tilde{p}_1,l_3)\left[
        c^{34}_{+-}s_{-\lam_1}(\tilde{p}_4,l_2)
       +c^{34}_{--}s_{-\lam_1}(l_4,l_2)
\right]   \right)\nonumber\\
&&\ \ \ \ \ \,   \phantom{\delta_{\lam_1\lam_3}}+b^{12}_{--}\left(
      \phantom{+}s_{-\lam_1}(l_1,\tilde{p}_3)\left[
          c^{34}_{++}s_{\lam_1}(\tilde{p}_4,l_2)
         +c^{34}_{-+}s_{\lam_1}(l_4,l_2)
\right]\right.\nonumber\\
&&\ \ \ \ \ \,   \phantom{\delta_{\lam_1\lam_3}+b^{12}_{++} \ }\, \left.\left.
                +s_{\lam_1}(l_3,l_2)\left[
          c^{34}_{+-}s_{-\lam_1}(l_1,\tilde{p}_4)
         +c^{34}_{--}s_{-\lam_1}(l_1,l_4)
\right]\right)\right]\nonumber\\
&&+\frac12B^{12}\left[
\phantom{+} c^{34}_{++}F(\lam_3,\tilde{p}_3,p_1+p_2,0,\lam_3,\tilde{p}_4)
+c^{34}_{+-}F(-\lam_3,l_3,p_1+p_2,0,-\lam_3,\tilde{p}_4)
\right.\nonumber\\
&&\phantom{+B^{12}}\left.\ \ \ 
+c^{34}_{-+}F(\lam_3,\tilde{p}_3,p_1+p_2,0,\lam_3,l_4)
+c^{34}_{--}F(-\lam_3,l_3,p_1+p_2,0,-\lam_3,l_4)
\right]\nonumber\\
&&+\frac12C^{34}\left[
\phantom{+}b^{12}_{++}F( \lam_1,\tilde{p}_1,p_3+p_4,0, \lam_1,\tilde{p}_2)
+b^{12}_{+-}F(-\lam_1,        l_1,p_3+p_4,0,-\lam_1,\tilde{p}_2)
\right]\nonumber\\
&&\phantom{+C^{34}}\left.\ \ \ 
+b^{12}_{-+}F( \lam_1,\tilde{p}_1,p_3+p_4,0, \lam_1,l_2)
+b^{12}_{--}F(-\lam_1,        l_1,p_3+p_4,0,-\lam_1,l_2)
\right]\nonumber\\
&&+\left.\frac12(p_1+p_2)\cdot(p_3+p_4)B^{12}C^{34}
\right\},\nonumber
\end{eqnarray}
  where $M_{B_i}$ and $\Gamma_{B_i}$ are the mass and width of the $i$th gauge
  boson.

\subsection{Matrix Elements}

{\renewcommand{\arraystretch}{1.2}
\TABULAR[t]{|c|c|c|c|}
{\hline
Decay & Colour Factor & Diagram Type &$a^\lam$ \\
\hline
${\rm \glt} \ra {\rm q}{\rm \qkt}^*_{\al}$     &
	 $\frac12$ & 1 &$a^{*-\lam}_{{\rm \glt} {\rm\qkt}_\al {\rm q}}$\\
\hline
${\rm \glt} \ra {\rm \bar{q}} {\rm \qkt}_{\al}$ &
	 $\frac12$ & 2 &$a^{\lam}_{{\rm \glt} {\rm\qkt}_\al {\rm q}}$ \\
\hline
${\rm \glt} \ra \cht^0_i {\rm g}$        & $\frac14$ & 3 &- \\
\hline}
{Two body decays of the gluino. As we are treating the gluino as a particle
 	 the sign of the gluino's helicity should be changed in the expression
	 for the second decay amplitude. In the decays of the gluino
	 to squarks $\al$ gives the mass eigenstate of the squark.
\label{tab:decaygluino2}}}

  We can now use the amplitudes for the two and three body decays in order to express
  the matrix elements for decay processes we are interested in.
  As with the hard processes this is simply a matter of listing the Feynman diagrams
  which contribute to a given decay and the various colour factors.
  This is somewhat easier here because the SM and MSSM decays we are considering
  have only one colour flow so we do not have the additional complication
  of dealing with different colour flows as we did in the hard production processes.

  The two body decays are particularly simple to specify because there is only one
  Feynman diagram for each decay mode. The two body decay modes of the gluino,
  neutralinos and charginos
  are given in 
  Tables\,\ref{tab:decaygluino2},~\ref{tab:decayneutralino2}~and~\ref{tab:decaychargino2},
  respectively.
  The two body decays of the Higgs bosons are given in Table\,\ref{tab:decayhiggs2} and of
  the sfermions in Table.\,\ref{tab:decaysfermion2}.

\begin{table}
\renewcommand{\arraystretch}{1.2}
\begin{center}
\begin{tabular}{|c|c|c|c|}
\hline
Decay & Colour Factor & Diagram Type &$a^\lam$ \\
\hline
$\cht^0_j\ra\cht^0_i {\rm h}^0$ & 1 & 1 &$a^\lam_{{\rm h}^0\cht^0_i\cht^0_j}$ \\
\hline
$\cht^0_j\ra\cht^0_i {\rm H}^0$ & 1 & 1 &$a^\lam_{{\rm H}^0\cht^0_i\cht^0_j}$ \\
\hline
$\cht^0_j\ra\cht^0_i {\rm A}^0$ & 1 & 1 &$a^\lam_{{\rm A}^0\cht^0_i\cht^0_j}$ \\
\hline
$\cht^0_i\ra\cht^+_j {\rm H}^-$ & 1 & 1 &$a^{*-\lam}_{{\rm H}^-\cht^0_i\cht^-_j}$ \\
\hline
$\cht^0_i\ra\cht^-_j {\rm H}^+$ & 1 & 1 &$a^\lam_{{\rm H}^-\cht^0_i\cht^-_j}$ \\
\hline
$\cht^0_i\ra {\rm f}     \ftl^*_\al$ & 1 & 1 &$a^{*-\lam}_{\cht^0_i \ftl_\al {\rm f}}$ \\
\hline
$\cht^0_i\ra {\rm \bar{f}} \ftl_\al$ & 1 & 2 &$a^{\lam}_{\cht^0_i \ftl_\al {\rm f}}$ \\
\hline
$\cht^0_j\ra \cht^0_i \gamma$ & 1 & 3 & - \\
\hline
\end{tabular}
\end{center}
\caption{Two body decays of the neutralinos. As we are treating the neutralino
         as a particle the sign of the neutralino's helicity should be changed in the
	 expression for the second type of decay amplitude.
	 In the decay modes of the neutralino to sfermions $\al$ gives
	 the mass eigenstate of the sfermion.
         The colour factor given for sfermion production is for the 
	 production of sleptons, the corresponding colour factor is $N_c$ for the
	 production of squarks.\label{tab:decayneutralino2}}
\end{table}
\begin{table}
\begin{center}
\renewcommand{\arraystretch}{1.2}
\begin{tabular}{|c|c|c|c|}
\hline
Decay & Colour Factor & Diagram Type &$a^\lam$ \\
\hline
$\cht^+_j\ra\cht^+_i{\rm h}^0$ & 1 & 1 & $a^\lam_{{\rm h}^0\cht^+_i\cht^-_j}$ \\
\hline
$\cht^+_j\ra\cht^+_i{\rm H}^0$ & 1 & 1 & $a^\lam_{{\rm H}^0\cht^+_i\cht^-_j}$ \\
\hline
$\cht^+_j\ra\cht^+_i{\rm A}^0$ & 1 & 1 & $a^\lam_{{\rm A}^0\cht^+_i\cht^-_j}$ \\
\hline
$\cht^+_j\ra\cht^0_i{\rm H}^+$ & 1 & 1 & $a^\lam_{{\rm H}^-\cht^0_i\cht^-_j}$ \\
\hline
$\cht^+_i\ra {\rm f} {\rm \tilde{f'}}^*_{\al}$ & 1 & 1 &
	 $a^{*-\lam}_{\cht^+_i \ftl'_\al {\rm  f}}$ \\
\hline
$\cht^+_i\ra {\rm \bar{f}} {\rm \tilde{f'}}_{\al}$ & 1 & 2 & 
	$a^{\lam}_{\cht^+_i \ftl'_\al {\rm f}}$ \\
\hline
$\cht^+_j\ra \cht^0_i \pi^+$ & $N_c$ & 5 & $a^\lam_{{\rm W}\cht^0_i\cht^-_j}$ \\ 
\hline
$\cht^-_j\ra\cht^-_i{\rm h}^0$ & 1 & 1 & $a^{*-\lam}_{{\rm h}^0\cht^+_i\cht^-_j}$ \\
\hline
$\cht^-_j\ra\cht^-_i{\rm H}^0$ & 1 & 1 & $a^{*-\lam}_{{\rm H}^0\cht^+_i\cht^-_j}$ \\
\hline
$\cht^-_j\ra\cht^-_i{\rm A}^0$ & 1 & 1 & $a^{*-\lam}_{{\rm A}^0\cht^+_i\cht^-_j}$ \\
\hline
$\cht^-_j\ra\cht^0_i{\rm H}^+$ & 1 & 1 & $a^{*-\lam}_{{\rm H}^-\cht^0_i\cht^-_j}$ \\
\hline
$\cht^-_i\ra {\rm f} {\rm \tilde{f'}}^*_{\al}$ & 1 & 1 &
	 $a^{*-\lam}_{\cht^+_i \ftl'_\al {\rm f}}$ \\
\hline
$\cht^-_i\ra {\rm \bar{f}} {\rm \tilde{f'}}_{\al}$ & 1 & 2 &
	 $a^{\lam}_{\cht^+_i \ftl'_\al {\rm f}}$ \\
\hline
$\cht^-_j\ra \cht^0_i \pi^-$ & $N_c$ & 5 & $-a^\lam_{{\rm W}\cht^0_i\cht^-_j}$ \\ 
\hline
\end{tabular}
\end{center}
\caption{Two body decays of the charginos. As we are treating the chargino
         as a particle the sign of the chargino's helicity should be changed in the
	 expression for the second decay amplitude.
        As before the isospin partner of a fermion or sfermion is denoted with a
	 prime and $\al$ gives the mass eigenstate of the sfermion.
         The colour factor given for sfermion production is for the 
	 production of sleptons, the corresponding colour factor is $N_c$ for the
	 production of squarks.     
\label{tab:decaychargino2}}
\end{table}

{\renewcommand{\arraystretch}{1.2}
\TABULAR{|c|c|c|c|}
{\hline
Decay & Colour Factor & Diagram Type &$a^\lam$ \\
\hline
${\rm h}^0\ra\cht^0_i\cht^0_j$ & 1 & 4 & $a^\lam_{{\rm h}^0\cht^0_i\cht^0_j}$ \\
\hline
${\rm H}^0\ra\cht^0_i\cht^0_j$ & 1 & 4 & $a^\lam_{{\rm H}^0\cht^0_i\cht^0_j}$ \\
\hline
${\rm A}^0\ra\cht^0_i\cht^0_j$ & 1 & 4 & $a^\lam_{{\rm A}^0\cht^0_i\cht^0_j}$ \\
\hline
${\rm h}^0\ra\cht^+_i\cht^-_j$ & 1 & 4 & $a^\lam_{{\rm h}^0\cht^+_i\cht^-_j}$ \\
\hline
${\rm H}^0\ra\cht^+_i\cht^-_j$ & 1 & 4 & $a^\lam_{{\rm H}^0\cht^+_i\cht^-_j}$ \\
\hline
${\rm A}^0\ra\cht^+_i\cht^-_j$ & 1 & 4 & $a^\lam_{{\rm A}^0\cht^+_i\cht^-_j}$ \\
\hline
${\rm h}^0\ra {\rm f} {\rm \bar{f}}$       & 1 & 4 &
	 $a^\lam_{{\rm h}^0 {\rm f} {\rm \bar{f}}}$ \\
\hline
${\rm H}^0\ra {\rm f} {\rm \bar{f}}$       & 1 & 4 &
	 $a^\lam_{{\rm H}^0 {\rm f} {\rm \bar{f}}}$ \\
\hline
${\rm A}^0\ra {\rm f} {\rm \bar{f}}$       & 1 & 4 & 
	$a^\lam_{{\rm A}^0 {\rm f} {\rm \bar{f}}}$ \\
\hline
${\rm H}^+\ra \cht^+_j \cht^0_i$ & 1 & 4 & $a^{*-\lam}_{{\rm H}^-\cht^0_i\cht^-_j }$ \\
\hline
${\rm H}^+ \ra {\rm f} {\rm \bar{f'}}$       & 1 & 4 &
	 $a^{*-\lam}_{{\rm H}^- {\rm f} {\rm \bar{f'}}}$ \\
\hline
${\rm H}^-\ra \cht^-_j \cht^0_i$ & 1 & 4 & $a^\lam_{{\rm H}^-\cht^0_i\cht^-_j }$ \\
\hline
${\rm H}^- \ra {\rm f} {\rm \bar{f'}}$       & 1 & 4 & 
	$a^{\lam}_{{\rm H}^- {\rm f }{\rm \bar{f'}}}$ \\
\hline}
{Two body decays of the MSSM Higgs bosons. As we are treating the electroweak gauginos
 as particles the sign of the second outgoing particle's helicity should
 be changed for the
 Higgs boson to two gaugino decay modes.
  The colour factor given for fermion production is
 for the production of leptons, the corresponding colour factor is $N_c$ for quark
 production.\label{tab:decayhiggs2}}}

{\renewcommand{\arraystretch}{1.2}
\TABULAR{|c|c|c|c|}
{\hline
Decay & Colour Factor & Diagram Type &$a^\lam$ \\
\hline
${\rm\qkt}_\al \ra {\rm q} {\rm\glt}$ & $\frac{(N^2_c-1)}{2N_c}$ & 4 & $a^{*-\lam}_{{\rm \glt} {\rm\qkt}_\al {\rm q}}$ \\
\hline
${\rm\qkt}^*_\al \ra {\rm\glt} {\rm\bar{q}}$ & $\frac{(N^2_c-1)}{2N_c}$ & 4 & $a^{\lam}_{{\rm\glt} {\rm\qkt}_\al {\rm q}}$ \\
\hline
$\ftl_\al\ra {\rm f}\cht^0_i$ & 1 & 4 & $a^{*-\lam}_{\cht^0_i \ftl_\al {\rm f}}$ \\
\hline
$\ftl^*_\al\ra \cht^0_i{\rm\bar{f}}$ & 1 & 4 & $a^{\lam}_{\cht^0_i \ftl_\al {\rm f}}$ \\
\hline
$\ftl_\al\ra {\rm f'}\cht^{\pm}_i$ & 1 & 4 &$a^{*-\lam}_{\cht^+_i \ftl_\al {\rm f}'}$ \\
\hline
$\ftl^*_\al\ra \cht^{\pm}_i\bar{f}'$ & 1 & 4 &$a^{\lam}_{\cht^+_i \ftl_\al {\rm f}'}$ \\
\hline}{Two body decays of the sfermions. Again as we are treating
        the gauginos as particles the sign of the  second particle's
        helicity should be changed in the sfermion to fermion gaugino processes. The
	mass eigenstate of the decaying sfermion is given by $\al$ and ${\rm f}'$ is
	the isospin partner of ${\rm f}$.
	\label{tab:decaysfermion2}}}

  The three body decays are more complicated as there is often more than one diagram 
  contributing to a given decay mode.
  The three body SM decay modes of the top quarks and antiquarks are given in 
  Table\,\ref{tab:decaytop3}.
  The three body decays of the gluinos to quarks and electroweak gauginos are given in
  Table\,\ref{tab:decaygluino3}.
  The three body decays of the neutralinos to fermions and gauginos are given
  in Table\,\ref{tab:decayneutralino3}.
  The chargino three body decay modes to gauginos are given in 
  Table\,\ref{tab:decaychargino3}.

  The decays of the MSSM Higgs bosons via gauge boson exchange are given in
  Table\,\ref{tab:decayhiggs3}.
  The decays of the sfermions via gauge boson exchange are given in 
  Table\,\ref{tab:decaysfermion3}. It should be noted that for processes in which 
  the fermions produced in these sfermion decays have the same flavour as the
  decaying sfermion there can be additional diagrams involving gaugino
  exchange which we have neglected. 
\begin{table}
\begin{center}
\begin{tabular}{|c|c||c|c|c|c|}
\hline
Decay & Colour  Factor & Diagram Type & Virtual Particle & $a^\lam$ & $b^\lam$\\
\hline
${\rm t}\ra {\rm b} {\rm f}{\rm\bar{f'}}$ & 1 & 1 & W 
	&$a^\lam_{{\rm W}{\rm t}{\rm \bar{b}}}$ &
	 $a^\lam_{{\rm W}{\rm f}{\rm \bar{f'}}}$\\
\hline
${\rm \bar{t}} \ra {\rm\bar{b}}  {\rm f}{\rm\bar{f'}}$ & 1 & 5 & W&
	$a^\lam_{{\rm W}{\rm t}{\rm \bar{b}}}$ & 
	$a^\lam_{{\rm W}{\rm f}{\rm \bar{f'}}}$\\
\hline
\end{tabular}
\end{center}
\caption
{Three body top quark decay modes. The colour factors given are for the production of
leptons in the decays, the colour factor is $N_c$ for the production of quarks. 
 The isospin partner of ${\rm f}$ is ${\rm f'}$.
\label{tab:decaytop3}}
\end{table}

\begin{table}
\begin{center}
\begin{tabular}{|c|c||c|c|c|c|}
\hline
Decay & Colour  Factor & Diagram Type & Virtual Particle &$a^\lam$ & $b^\lam$\\
\hline
% decay
${\rm\glt}\ra\cht^0_l{\rm q}{\rm\bar{q}}$ & $\frac12$ &
% diagrams
      3 & ${\rm\qkt}^*_{\al}$ & $a^{*-\lam}_{{\rm\glt}{\rm\qkt}_\al {\rm q}}$ &
			         $a^\lam_{\cht^0_l{\rm\qkt}_\al {\rm q}}$\\
\cline{3-6}
&&    4 & ${\rm\qkt}_{\al}$   & $a^\lam_{{\rm\glt}{\rm\qkt}_\al {\rm q}}$ 
			       & $\epsilon_la^{*-\lam}_{\cht^0_l{\rm \qkt}_\al {\rm q}}$\\
\hline
% decay
${\rm\glt}\ra\cht^+_l{\rm q}{\rm\bar{q}'}$ & $\frac12$ &
% diagrams
      3 & ${\rm\qkt}^*_{\al}$  & $a^{*-\lam}_{{\rm\glt}{\rm\qkt}_\al {\rm q}}$ 
			       & $a^\lam_{\cht^+_l{\rm \qkt}_\al {\rm q}'}$
\\
\cline{3-6}
&&    4 & ${\rm{\tilde{q}}}^{\prime*}_{\al}$ & 
		$a^\lam_{{\rm\glt}{\rm\qkt}'_\al {\rm q}}$    & 
		$a^{*-\lam}_{\cht^+_l{\rm\qkt}' {\rm q}}$\\
\hline
% decay
${\rm\glt}\ra\cht^-_l{\rm q}{\rm\bar{q}'}$ & $\frac12$ &
% diagrams
      3 & ${\rm\qkt}^*_{\al}$  & $a^{*-\lam}_{{\rm\glt}{\rm\qkt}_\al {\rm q}}$ 
			       & $a^\lam_{\cht^+_l{\rm\qkt}_\al {\rm q}'}$\\
\cline{3-6}
&&    4 & ${\rm{\tilde{q}}}^{\prime*}_{\al}$ &
	 $a^\lam_{{\rm\glt}{\rm\qkt}'_\al {\rm q}}$    &
	 $a^{*-\lam}_{\cht^+_l{\rm\qkt}'_\al {\rm q}}$\\
\hline
\end{tabular}
\end{center}
\caption{Three body gluino decay modes. As with the hard production cross sections
 the mass eigenstates, $\al$, of the intermediate squarks should be summed over.
The isospin partner of q is ${\rm q'}$.
\label{tab:decaygluino3}}
\end{table}
\begin{table}
\begin{center}
\renewcommand{\arraystretch}{1.2}
\begin{tabular}{|c|c||c|c|c|c|}
\hline
Decay & Colour  Factor & Diagram Type & Virtual Particle &$a^\lam$ & $b^\lam$\\
\hline
% decay
$\cht^0_j\ra\cht^0_i{\rm f}{\rm\bar{f}}$ & 1 leptons & 
% diagrams
                 1 &   Z          & $a^\lam_{{\rm Z}\cht^0_i\cht^0_j}$ 
				  & $a^\lam_{{\rm Z} {\rm f} {\rm \bar{f}}}$\\
\cline{3-6}
& $N_c$ quarks & 2 & $\rm{h}^0$   & $a^\lam_{{\rm h}^0\cht^0_i\cht^0_j}$ & 
				    $a^\lam_{{\rm h}^0{\rm f}{\rm\bar{f}}}$\\
\cline{3-6}
&&               2 & $\rm{H}^0$   & $a^\lam_{{\rm H}^0\cht^0_i\cht^0_j}$ &
				    $a^\lam_{{\rm H}^0{\rm f}{\rm\bar{f}}}$\\
\cline{3-6}
&&               2 & $\rm{A}^0$   & $a^\lam_{{\rm A}^0\cht^0_i\cht^0_j}$ &
				    $a^\lam_{{\rm A}^0{\rm f}{\rm\bar{f}}}$\\
\cline{3-6}
&&               3 & $\ftl^*_\al$ & $a^{*-\lam}_{\cht^0_j\ftl_\al {\rm f}}$ &
		                    $a^\lam_{\cht^0_i\ftl_\al {\rm f}}$\\
\cline{3-6}
&&               4 & $\ftl_\al$   & $\epsilon_ja^\lam_{\cht^0_j\ftl_\al {\rm f}}$     &
				    $\epsilon_ia^{*-\lam}_{\cht^0_i\ftl_\al {\rm f}}$\\
\hline
% decay
$\cht^0_i\ra\cht^+_j{\rm f}{\rm \bar{f'}}$ & 1 leptons &
% diagrams
                 1 &   W          & $a^{*\lam}_{{\rm W}\cht^0_i\cht^-_j}$ &
				    $a^\lam_{{\rm W}{\rm f}{\rm \bar{f}}'}$\\
\cline{3-6}
& $N_c$ quarks & 3 & $\ftl^*_\al$ & $a^{*-\lam}_{\cht^0_i\ftl_\al {\rm f}}$ &
				    $a^\lam_{\cht^+_j \ftl_\al {\rm f}'}$\\
\cline{3-6}			                                 
&              & 4 & $\ftl'_\al$   & $\epsilon_ia^\lam_{\cht^0_i\ftl'_\al {\rm f}'}$
			           & $a^{*-\lam}_{\cht^+_j \ftl'_\al {\rm f}}$\\
\hline
% decay
$\cht^0_i\ra\cht^-_j{\rm f}{\rm \bar{f'}}$ & 1 leptons &
% diagrams
                 1 &   W          & $-a^\lam_{{\rm W}\cht^0_i\cht^-_j}$ &
				    $a^\lam_{{\rm W}{\rm f} {\rm\bar{f}}'}$\\
\cline{3-6}
& $N_c$ quarks & 3 & $\ftl^*_\al$ & $a^{*-\lam}_{\cht^0_i\ftl_\al {\rm f}}$ &
				    $a^\lam_{\cht^+_j \ftl_\al {\rm f}'}$\\
\cline{3-6}			                                 
&              & 4 & $\ftl'_\al$   & $\epsilon_ia^\lam_{\cht^0_i\ftl'_\al {\rm f}'}$
			           & $a^{*-\lam}_{\cht^+_j \ftl'_\al {\rm f}}$\\
\hline
\end{tabular}
\end{center}
\caption{Three body neutralino decay modes. We have only included the Higgs boson 
 exchange diagrams
 for neutralino decays. These diagrams are only important for the production of
 bottom and tau due to the Higgs couplings, the production of the top quark is
 is usually kinematically forbidden. In most SUSY models the gluino is heavier
 than the electroweak gauginos and we have therefore not included the modes
 $\cht^0_l\ra{\rm\glt}{\rm q}{\rm\bar{q}}$. 
 As with the hard production processes the mass eigenstates of the intermediate sfermions
 must be summed over. The isospin partner of f is ${\rm f}'$.
\label{tab:decayneutralino3}}
\end{table}

\begin{table}[h!!]
\begin{center}
\begin{tabular}{|c|c||c|c|c|c|}
\hline
Decay & Colour  Factor & Diagram Type & Virtual Particle &$a^\lam$ & $b^\lam$\\
\hline
% decay
$\cht^+_j\ra\cht^0_i{\rm f}{\rm\bar{f}'}$ & 1 leptons 
%diagrams
               & 1 &   W                  & $a^\lam_{{\rm W}\cht^0_i\cht^-_j}$
				          & $a^\lam_{{\rm W}{\rm f}{\rm\bar{f}'}}$\\
\cline{3-6}
& $N_c$ quarks & 3 & $\ftl^{\prime*}_\al$ & $a^{*-\lam}_{\cht^+_j\ftl'_\al{\rm f}}$ 
					  & $a^\lam_{\cht^0_i \ftl'_\al {\rm f'}}$\\
\cline{3-6}
&              & 4 & $\ftl_\al$           & $a^\lam_{\cht^+_j \ftl_\al {\rm f}'}$
			& $\epsilon_ia^{*-\lam}_{\cht^0_i \ftl_\al {\rm f}}$ \\
\hline
% decay
$\cht^+_j\ra\cht^+_i {\rm f}{\rm\bar{f}}$ & 1 leptons
%diagrams
               & 1 &  Z                   & $a^\lam_{{\rm Z}\cht^+_i\cht^-_j}$  & 
			$a^\lam_{{\rm Z}{\rm f}{\rm \bar{f}}}$ \\
\cline{3-6}
& $N_c$ quarks & 3 & $\ftl^{\prime*}_\al$ & $a^{*-\lam}_{\cht^+_j\ftl'_\al{\rm f}}$ & 
				           $a^\lam_{\cht^+_i \ftl'_\al{\rm f}}$ \\
\cline{3-6}
&              & 4 & $\ftl'_\al$          & $a^\lam_{\cht^+_j \ftl'_\al {\rm f}}$   &
				 $a^{*-\lam}_{\cht^+_i \ftl'_\al {\rm f}}$\\
\hline
% decay
$\cht^-_j\ra\cht^0_i{\rm f}{\rm\bar{f}'}$ & 1 leptons 
%diagrams
               & 1 &   W                  & $-a^{-\lam}_{{\rm W}\cht^0_i\cht^-_j}$  & 
					    $a^\lam_{{\rm W}{\rm f}{\rm\bar{f}}'}$\\
\cline{3-6}
& $N_c$ quarks & 3 & $\ftl^{\prime*}_\al$ & $a^{*-\lam}_{\cht^+_j\ftl'_\al{\rm f}}$ &
				 $a^\lam_{\cht^0_i \ftl'_\al {\rm f}'}$\\
\cline{3-6}
&              & 4 & $\ftl_\al$           & $a^\lam_{\cht^+_j \ftl_\al {\rm f}'}$   & 
					$\epsilon_ia^{*-\lam}_{\cht^0_i \ftl_\al {\rm f}}$ \\
\hline
% decay
$\cht^-_j\ra\cht^-_i {\rm f}{\rm\bar{f}}$ & 1 leptons
%diagrams
               & 1 &  Z                   & $-a^{-\lam}_{{\rm Z}\cht^+_j\cht^-_i}$  & 
					  $a^\lam_{{\rm Z}{\rm f}{\rm\bar{f}}}$ \\
\cline{3-6}
& $N_c$ quarks & 3 & $\ftl^{\prime*}_\al$ & $a^{*-\lam}_{\cht^+_j\ftl'_\al{\rm f}}$ &
					 $a^\lam_{\cht^+_i \ftl'_\al{\rm f}}$ \\
\cline{3-6}
&              & 4 & $\ftl'_\al$          & $a^\lam_{\cht^+_j \ftl'_\al {\rm f}}$   & 
					$a^{*-\lam}_{\cht^+_i \ftl'_\al {\rm f}}$\\
\hline
\end{tabular}
\end{center}
\caption{Three body chargino decay modes. The positive chargino can decay to 
 $\cht^0_l{\rm u}{\rm\bar{d}}$ or $\cht^0_l\nu\ell^+$
 and the negative chargino to $\cht^0_l{\rm d}{\rm\bar{u}}$ or
 $\cht^0_l\ell^-\bar{\nu}$. The gluinos are usually heavier than the charginos in
 SUSY models and we have therefore not included the decay modes 
 $\cht^\pm_j\ra{\rm\glt}{\rm q}{\rm\bar{q}}'$.
 The mass eigenstate, $\al$, of the intermediate sfermions must be summed over.
 As before the isospin partner of a fermion or sfermion
 is denoted with a prime.
 \label{tab:decaychargino3}}
\begin{center}
\begin{tabular}{|c|c||c|c|c|c|}
\hline
Decay & Colour  Factor & Diagram Type & Virtual Particle &$a$ & $b^\lam$\\
\hline
$\rm{A}_0\ra\rm{h}^0{\rm f}{\rm\bar{f}}$  & 1 & 6 & Z &
 $-a_{{\rm A}^0{\rm h}^0{\rm Z}}$ & $a^\lam_{{\rm Z}{\rm f}{\rm\bar{f}}}$ \\
\hline 
$\rm{A}_0\ra\rm{H}^0{\rm f}{\rm\bar{f}}$  & 1 & 6 & Z &
 $-a_{{\rm A}^0{\rm H}^0{\rm Z}}$ & $a^\lam_{{\rm Z}{\rm f}{\rm \bar{f}}}$ \\
\hline 
$\rm{h}_0\ra\rm{A}^0{\rm f}{\rm\bar{f}}$  & 1 & 6 & Z &
 $a_{{\rm A}^0{\rm h}^0{\rm Z}}$ & $a^\lam_{{\rm Z}{\rm f}{\rm \bar{f}}}$ \\
\hline 
$\rm{H}_0\ra\rm{A}^0{\rm f}{\rm\bar{f}}$  & 1 & 6 & Z &
 $a_{{\rm A}^0{\rm H}^0{\rm Z}}$ & $a^\lam_{{\rm Z}{\rm f}{\rm\bar{f}}}$ \\
\hline 
$\rm{H}^+\ra\rm{h}^0{\rm f}{\rm\bar{f}'}$ & 1 & 6 & ${\rm W}^+$ &
 $-a_{{\rm H}^-{\rm h}^0{\rm W}^-}$ & $a^\lam_{{\rm W}{\rm f}{\rm\bar{f}'}}$ \\
\hline
$\rm{H}^+\ra\rm{H}^0{\rm f}{\rm\bar{f}'}$ & 1 & 6 & ${\rm W}^+$ &
 $-a_{{\rm H}^-{\rm H}^0{\rm W}^-}$ & $a^\lam_{{\rm W}{\rm f}{\rm \bar{f}'}}$ \\
\hline
$\rm{H}^+\ra\rm{A}^0{\rm f}{\rm\bar{f}'}$ & 1 & 6 & ${\rm W}^+$ & 
$-a_{{\rm H}^-{\rm A}^0{\rm W}^-}$ & $a^\lam_{{\rm W}{\rm f}{\rm \bar{f}'}}$ \\
\hline
$\rm{H}^-\ra\rm{h}^0{\rm f}{\rm\bar{f}'}$ & 1 & 6 & ${\rm W}^-$ &
$-a_{{\rm H}^+{\rm h}^0{\rm W}^+}$ & $a^\lam_{{\rm W}{\rm f}{\rm \bar{f}}'}$ \\
\hline
$\rm{H}^-\ra\rm{H}^0{\rm f}{\rm\bar{f}'}$ & 1 & 6 & ${\rm W}^-$ &
$-a_{{\rm H}^+{\rm H}^0{\rm W}^+}$ & $a^\lam_{{\rm W}{\rm f}{\rm \bar{f}}'}$ \\
\hline
$\rm{H}^-\ra\rm{A}^0{\rm f}{\rm\bar{f}'}$ & 1 & 6 & ${\rm W}^-$ &
$-a_{{\rm H}^+{\rm A}^0{\rm W}^+}$ & $a^\lam_{{\rm W}{\rm f}{\rm \bar{f}}'}$ \\
\hline
$\rm{h}^0\ra\rm{H}^\pm {\rm f}{\rm\bar{f}'}$ & 1 & 6 & ${\rm W}^\pm$ &
$a_{{\rm H}^\pm {\rm h}^0{\rm W}^\pm}$ & $a^\lam_{{\rm W}{\rm f}{\rm \bar{f}}'}$ \\
\hline
$\rm{H}^0\ra\rm{H}^\pm {\rm f}{\rm\bar{f}'}$ & 1 & 6 & ${\rm W}^\pm$ &
$a_{{\rm H}^\pm {\rm H}^0{\rm W}^\pm}$ & $a^\lam_{{\rm W}{\rm f}{\rm \bar{f}}'}$ \\
\hline
$\rm{A}^0\ra\rm{H}^\pm {\rm f}{\rm\bar{f}'}$ & 1 & 6 & ${\rm W}^\pm$ &
$a_{{\rm H}^\pm {\rm A}^0{\rm W}^\pm}$ & $a^\lam_{{\rm W}{\rm f}{\rm \bar{f}}'}$ \\
\hline
\end{tabular}
\end{center}
\caption
{Three body MSSM Higgs boson decay modes.
 The colour factors given are for the decay of the Higgs boson
 to a different Higgs boson and leptons. 
 For the decay of the Higgs boson to another Higgs boson 
 and quarks the colour factor is $N_c$. As before ${\rm f}'$ is the isospin partner of f.
\label{tab:decayhiggs3}}
\vspace{-2.7cm}
\end{table}

  We have only considered the four body decays of Higgs bosons, via either real or
  virtual gauge boson pairs. These decay modes are given in Table\,\ref{tab:decayhiggs4}.

\begin{table}
\begin{center}
\renewcommand{\arraystretch}{1.2}
\begin{tabular}{|c|c||c|c|c|c|}
\hline
Decay & Colour  Factor & Diagram Type & Virtual Particle &$a$ & $b^\lam$\\
\hline
$\ftl_\al\ra\ftl_\be {\rm n}{\rm \bar{n}}$ & 1 & 6 & Z &
 $a_{{\rm Z}\ftl_\be\ftl^*_\al}$ & $a^\lam_{{\rm Z}{\rm n}{\rm \bar{n}}}$\\
\hline
$\ftl^*_\al\ra\ftl^*_\be {\rm n}{\rm \bar{n}}$ & 1 & 6 & Z &
 $a_{{\rm Z}\ftl_\al\ftl^*_\be}$ & $a^\lam_{{\rm Z}{\rm n}{\rm \bar{n}}}$\\
\hline
$\ftl_\al\ra\ftl'_\be {\rm n}{\rm \bar{n}}'$ & 1 & 6 & W & 
$a_{{\rm W}\ftl'_\be\ftl^*_\al}$ & $a^\lam_{{\rm W}{\rm n}{\rm \bar{n}}'}$ \\
\hline
$\ftl^*_\al\ra\ftl^{\prime*}_\be {\rm n}{\rm \bar{n}}'$ & 1 & 6 & W &
 $a_{{\rm W}\ftl_\al\ftl^{\prime*}_\be}$ & $a^\lam_{{\rm W}{\rm n}{\rm \bar{n}}'}$ \\
\hline
\end{tabular}
\end{center}
\caption{Three body decays modes of the sfermions. The fermion ${\rm n}$ and antifermion
 $\rm{\bar{n}}$
 produced in the sfermion decay need not be the same as the flavour decaying sfermion.
 The isospin partner of ${\rm n}$ is denoted by ${\rm n}'$.
 In fact if the produced fermion is of the same flavour as the decaying sfermion
 there are
 additional Feynman diagrams which we have not included.
 As before the colour factors given are for the production of leptons, the colour factor
 is $N_c$ for the production of quarks.
\label{tab:decaysfermion3}}
\vspace{2mm}
\begin{center}
\begin{tabular}{|c|c|c|c|c|c|}
\hline
Decay & First    & Second   & $a$ & $b^\lam$ & $c^\lam$ \\
      & Virtual  & Virtual  &     &          &          \\
      & Particle &Particle  &     &          &          \\
\hline
$\rm{h}^0\ra\rm{W}^+\rm{W}^-\ra\rm{f}\rm{\bar{f}'}{\rm n}\rm{\bar{n}'}$ &
     $\rm{W}^+$ & $\rm{W}^-$ & $g\mw\sin(\be-\al)$
     & $a^\lam_{W\rm{f}\rm{\bar{f}'}}$ & $a^\lam_{W{\rm n}\rm{\bar{n}'}}$\\
$\rm{H}^0\ra\rm{W}^+\rm{W}^-\ra\rm{f}\rm{\bar{f}'}{\rm n}\rm{\bar{n}'}$ &
     $\rm{W}^+$ & $\rm{W}^-$ & $g\mw\cos(\be-\al)$
     & $a^\lam_{W\rm{f}\rm{\bar{f}'}}$ & $a^\lam_{W{\rm n}\rm{\bar{n}'}}$\\
$\rm{h}^0\ra\rm{Z}\rm{Z}\ra\rm{f}\rm{\bar{f}}{\rm n}\rm{\bar{n}}$ &
     $\rm{Z}$ & $\rm{Z}$ & $g\mw\sin(\be-\al)$
     & $a^\lam_{Z\rm{f}\rm{\bar{f}}}$ & $a^\lam_{Z{\rm n}\rm{\bar{n}}}$\\
$\rm{H}^0\ra\rm{Z}\rm{Z}\ra\rm{f}\rm{\bar{f}}{\rm n}\rm{\bar{n}}$ &
     $\rm{Z}$ & $\rm{Z}$ & $g\mw\cos(\be-\al)$
     & $a^\lam_{Z\rm{f}\rm{\bar{f}}}$ & $a^\lam_{Z{\rm n}\rm{\bar{n}}}$\\
\hline
\end{tabular}
\end{center}
\caption{Four body Higgs boson decay modes. The colour factor is 1 if all the produced
 fermions are leptons, $N_c$ if a lepton and neutrino and a quark/antiquark pair
 are produced and $N^2_c$ is all the produced particles are quarks. 
 The isospin partners of ${\rm n}$ and ${\rm f}$ are denoted by
 ${\rm n}'$ and ${\rm f}'$, respectively.
\label{tab:decayhiggs4}}
\end{table}
\providecommand{\href}[2]{#2}\begingroup\raggedright\endgroup

\end{document}